\documentclass[]{interact}

\usepackage{epstopdf}
\usepackage[caption=false]{subfig}
\usepackage{xcolor}
\usepackage{soul}
\usepackage{hyperref}

\usepackage[numbers,sort&compress]{natbib}
\bibpunct[, ]{[}{]}{,}{n}{,}{,}
\renewcommand\bibfont{\fontsize{10}{12}\selectfont}
\makeatletter
\def\NAT@def@citea{\def\@citea{\NAT@separator}}
\makeatother

\theoremstyle{plain}

\theoremstyle{definition}

\theoremstyle{remark}

\begin{document}

\title{THIRTY YEARS OF 2D MIT}

\author{
\name{Sergey V. Kravchenko\textsuperscript{a}\thanks{CONTACT S.~V. Kravchenko. Email: s.kravchenko@northeastern.edu},
Vladimir M. Pudalov\textsuperscript{b}, and Alexander A. Shashkin\textsuperscript{c}}
\affil{\textsuperscript{a}Physics Department, Northeastern University, Boston, MA 02115, USA\\
\textsuperscript{b}P. N. Lebedev Physical Institute, Moscow 119991, Russia\\
\textsuperscript{c}Institute of Solid State Physics, Chernogolovka, Moscow District 142432, Russia}
}

\maketitle

\begin{abstract}

The zero-magnetic-field metal-insulator transition in strongly correlated two-dimensional electron systems was experimentally discovered three decades ago, thereby triggering extensive experimental and theoretical investigations. Prior to this discovery, the occurrence of such a transition --- as well as the very existence of a metallic phase in two dimensions --- was regarded as forbidden within the framework of the one-parameter scaling theory of localization. In this work, we provide a comprehensive review of the transport and thermodynamic properties of two-dimensional electron and hole systems in which carrier-carrier interactions constitute the dominant energy scale. We present and critically analyze both experimental observations and theoretical developments accumulated over the past thirty years.

\end{abstract}

\begin{keywords}
Metal-insulator transition, two-dimensional electron systems, strong correlations, Wigner crystal
\end{keywords}

\tableofcontents

\section{Introduction}

The metal-insulator transition (MIT) in three-dimensional bulk materials was pioneered by Sir Nevill Mott, who established the foundations of our understanding of this phenomenon. It remains one of the most challenging problems in condensed matter physics,
especially in lower dimensional systems that have become accessible by advances in technology since Mott's time.

The discovery in the mid-1990s of the metallic-like transport in two-dimensional (2D) strongly interacting electron systems and of its critical behavior at the verge of the transition to the hopping-type transport sparked an explosion of interest among theorists and experimentalists. Furthermore, it revealed a critical deficiency in the existing harmonious structure of knowledge developed primarily by Mott \cite{mott1974metal}, Anderson \cite{anderson1958absence}, and Abrahams \textit{et al.} \cite{abrahams1979scaling}, as well as in the concepts of quantum transport in strongly correlated systems. Understanding the mechanisms of metal-like transport, the nature of the ground state, and the properties of strongly interacting electron systems is essential for modeling the observed phenomena.  The identified knowledge gap prompted a marked increase in both theoretical and experimental research, leading to a comprehensive revision of existing concepts in this field. This has led to a new understanding of the quantum physics of strongly interacting 2D electron systems. As a consequence of this influx of research, new theoretical ideas and experimental methods have been developed, revealing several unique features in the thermodynamics and kinetics of low-dimensional systems comprising charged fermions.

The purpose of this review is to introduce the reader to the world of vivid, exciting, and sometimes unforeseen effects that manifest themselves in strongly interacting systems, as well as to the foundations of the theoretical description of their properties. Indeed, this discovery for 2D systems forced us to re-examine the properties not only of transport but, more generally, of quantum kinetics and thermodynamics in strongly correlated electron systems, far beyond the frameworks of the Fermi liquid model.

We will focus on several interconnected phenomena in strongly correlated 2D electron systems. These include the temperature-dependent metallic conductivity, the transition between metal and insulator, a strong increase in spin susceptibility and effective mass near the transition, and collective transport in the insulating state.  We will also review several theoretical approaches proposed to explain the observed phenomena.

The review is organized as follows. In Section 2, we present the main theoretical concepts and experimental results on quantum transport prior to the discovery of the 2D MIT.  In Section 3, the major experimental results on metallic transport and the zero-field metal-insulator transition in strongly interacting 2D electron and hole systems are presented.  Section 4 is devoted to the suppression of the coherent backscattering on the metallic side near the metal-insulator transition.  In Section 5, we show that two independent approaches to determining the critical electron density for the MIT yield the same result.  In Sections 6 and 7, the effects of the parallel and perpendicular magnetic fields are considered.  In Section 8, we examine the significant increase and potential divergence of both spin susceptibility and effective electron mass, as indicated by transport and thermodynamic measurements. In Section 9, we present the argument that, on the insulating side of the metal-insulator transition, there exists an electron solid that closely resembles a pinned Wigner crystal in many aspects. Finally, Section 10 reviews modern theoretical approaches to the 2D MIT.

\section{Weak and strong localization: early theory and experiments}

In 2D electron systems, the electrons are confined to move in a plane in the presence of a random potential.  The end of the 70s and the beginning of the 80s witnessed a triumph of the one-parameter scaling theory of localization \cite{abrahams1979scaling, wegner1979disordered, gorkov1979particle} and the theory of electron-electron interactions in disordered conductors \cite{altshuler1979contribution, altshuler1980interaction} (for a review, see Ref.~\cite{lee1985disordered}).  Charge carriers are always localized in one dimension; electronic states can be either localized or extended in three dimensions, and the absence of extended states is predicted in two-dimensional systems in the limit of zero temperature.  In two dimensions, electrons may conduct well at room temperature, but a weak logarithmic increase in resistance is expected as the temperature is reduced in the metallic-conductivity regime at high electron densities.  This is because, when scattered by impurities back to their starting point, electron waves interfere constructively with their time-reversed counterparts. Although these effects are negligible at high temperatures due to inelastic scattering, quantum interference becomes significant as the temperature is reduced, leading to the so-called ``weak localization''.  The terms ``weak localization'' and ``strong localization'' are commonly used to denote states that exhibit logarithmic (in the 2D case) and exponential (in any dimension) temperature dependence of resistance.

\begin{figure}
\centering
\includegraphics[width=8 cm]{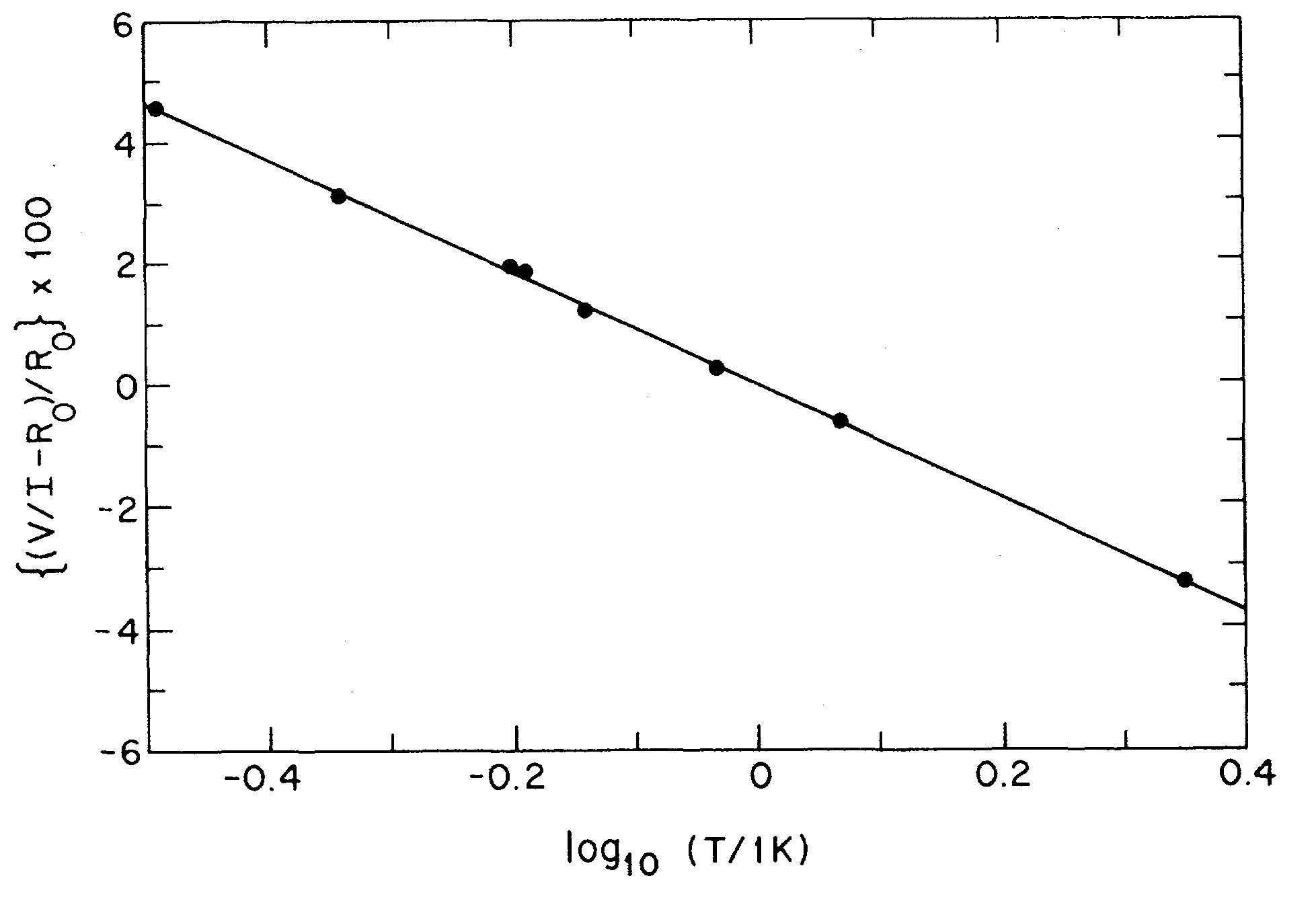}
\caption{($R-R_0)/R_0)$ in a thin metal film vs.\ $\log T$ ($R_0\approx1k\Omega$ is the wire resistance at $T$=1.0 K). Adapted from Ref.~\cite{dolan1979nonmetallic}.}\label{Dolan-Osheroff}
\end{figure}

\begin{figure}
\centering
\includegraphics[width=8 cm]{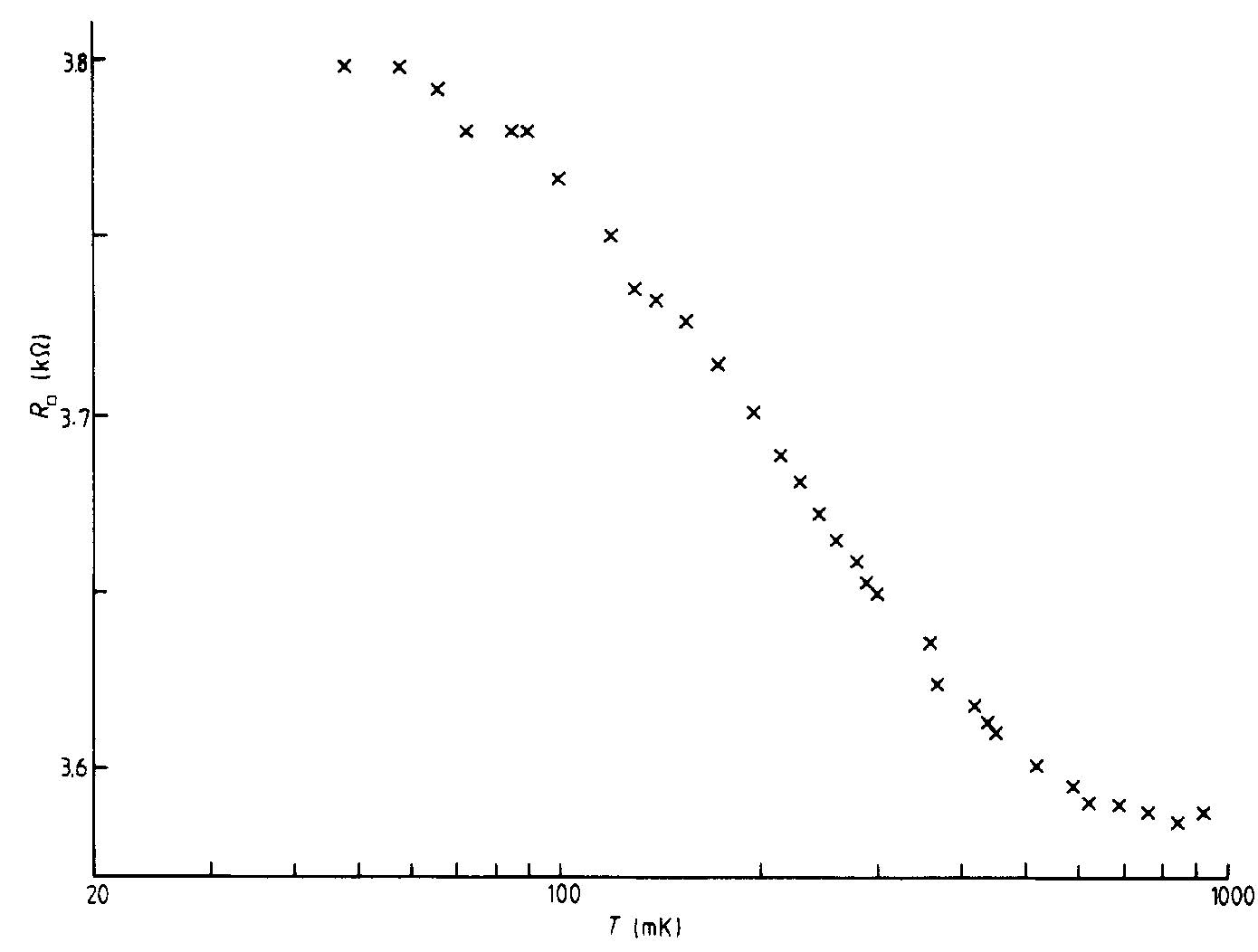}
\caption{The resistance is plotted against the logarithm of the temperature for a Si MOSFET device with a maximum mobility of 0.65 m$^2$/Vs.  Adapted from Ref.~\cite{uren1981it}.}\label{pepper logarithmic}
\end{figure}

Furthermore, weak electron-electron interactions introduce a separate set of corrections to conductivity $\sigma$ that can either enhance or counteract the effect depending on the value of the interaction constant $F_0^\sigma$ in the triplet channel, which depends on the interaction strength \cite{altshuler1980interaction,finkelstein1983influence,finkelstein1984weak,castellani1984interaction,lee1985disordered}:
\begin{equation}
\delta\sigma=-
\frac{e^2}{2\pi^2\hbar}\ln\left(\frac{\hbar}{k_BT\tau}\right)
\Big[1 +  3\left(1-\frac{\ln(1+F_0^\sigma)}{F_0^\sigma}
\right) \Big],
\label{introdiff}
\end{equation}
where $k_B$ is the Boltzmann constant, $T$ is the temperature, and $\tau$ is the mean free time.  Indeed, thin metallic films and 2D electron systems fabricated on semiconductor interfaces were found to display the predicted logarithmic increase in resistivity with decreasing temperature \cite{dolan1979nonmetallic, uren1980the, uren1981it}, providing support for the weak localization theory (Figs.~\ref{Dolan-Osheroff} and \ref{pepper logarithmic}).

\begin{figure}
\centering
\includegraphics[width=6 cm]{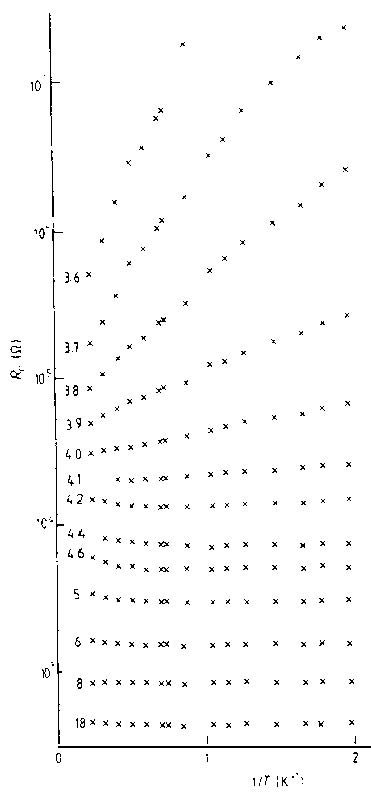}
\caption{The temperature dependence of the conductivity of a Si MOSFET device with a maximum mobility of 0.65 m$^2$/Vs is plotted against $1/T$. No magnetic field was applied. From Ref.~\cite{uren1981it}.}\label{pepper activated}
\end{figure}

In 2D electron systems with significant disorder, which inevitably arises at very low electron densities, the localization becomes strong. The resistance displays an activation temperature dependence (see, e.g., Ref.~\cite{uren1981it} and Fig.~\ref{pepper activated}) that at very low temperatures switches to a $R\propto \exp((T_0/T)^{1/2})$ temperature dependence \cite{efros1975coulomb} corresponding to a variable-range hopping regime.

Given the remarkable agreement between theory and experiment, the description of the electronic properties of 2D electron systems seemed well-defined at that time.  However, from time to time, indications appeared that the accepted view might not always be correct, as discussed in the next section.

\section{Zero-magnetic-field metal-insulator transition in strongly interacting 2D electron systems}
\subsection{Temperature dependence of the resistivity in Si MOSFETs}

Curiously, applying these ideas to the 2D electron liquid in silicon metal-oxide-semiconductor field-effect transistors (MOSFETs) --- one of the most ubiquitous 2D systems --- remained a challenge.  In the early experiments  \cite{bishop1980nonmetallic} with low-mobility $(\sim0.1$~m$^2$/Vs) structures in the regime of relatively high electron densities ($n_s>10^{12}$~cm$^{-2}$), a decrease of the conductivity with cooling was observed, apparently in qualitative agreement with the ideas of weak localization and electron-electron interactions \cite{abrahams1979scaling, wegner1979disordered, gorkov1979particle, altshuler1979contribution, altshuler1980interaction, lee1985disordered}. However, quantitatively describing this behavior was problematic \cite{burdis1988anomalous}.  The disagreement with the theory became qualitative with the advent of moderate-mobility (up to $\approx 2\,$m$^2$/Vs) structures: the low-$T$ conductivity $\sigma$ in such Si~MOSFETs increased as temperature decreased \cite{cham1980temperature, smith1986temperature, vyrodov1988temperature}.  This seemingly metallic behavior of the conductivity is especially pronounced at low electron densities $n_s \sim (1-2)\times 10^{11}$~cm$^{-2}$, where a five-fold decrease of $\rho$ with cooling was observed \cite{zavaritskaya1987metal}.

The strength of the electron-electron interactions is characterized by the ratio between the Coulomb and Fermi energies, which in a single-valley case is equal to
\begin{equation}r_s=1/(\pi n_s)^{1/2}a_B\end{equation}
(here $a_B$ is the Bohr radius in a semiconductor).  In the early experiments \cite{dolan1979nonmetallic,bishop1980nonmetallic,uren1980the,uren1981it}, the Coulomb interactions are relatively weak. As the electron density decreases, the ratio of Coulomb energy to Fermi energy increases. Consequently, interactions become the primary energy scale in the system.  No comprehensive analytical theory existed for the strongly-interacting limit ($r_s\gg1$) at the time when the 2D metal-insulator transition was experimentally observed. To our knowledge, only Finkel'stein \cite{finkelstein1983influence,finkelstein1984weak} and Castellani \textit{et al.} \cite{castellani1984interaction} have proposed that a two-dimensional system with weak disorder and sufficiently strong interactions will be in a conducting state as the temperature decreases.

However, this scaling process increases the effective interaction strength, leading to a divergent spin susceptibility. As a result, the perturbative approach breaks down as the temperature approaches zero. Therefore, the possibility of a stable 2D metallic ground state driven by strong electron-electron interactions was not given serious consideration, particularly because of the lack of experimental confirmation.

Advances in semiconductor technology have enabled the fabrication of high-quality two-dimensional samples with very low disorder, thereby allowing measurements at extremely low carrier densities. As a result, the strongly interacting regime ($r_s \gg 1$) became experimentally accessible.  The initial report on the substantial metallic temperature dependence of resistance and the two-dimensional metal-insulator transition in $n$- and $p$-Si MOSFETs (Zavaritskaya and Zavaritskaya \cite{zavaritskaya1987metal}) received little attention. However, numerous experimental results from various research groups suggested that a metallic state may be possible in two dimensions. Pudalov \textit{et al}. \cite{pudalov1993hall, pudalov1994termination, kravchenko1995global} and Shashkin \textit{et al}. \cite{shashkin1993floating,shashkin1994insulating,shashkin1994percolation} studied the phase diagram of the quantum Hall effect (QHE) in low-disordered Si MOSFETs and GaAs/AlGaAs heterostructures and have come to the conclusion that a metal-insulator transition exists in clean samples. In a strong magnetic field, each Landau level comprises a set of states with an energy spread due to disorder. The conventional view is that these states are all localized except for those at the center of each Landau level, which are extended. In the limit of a vanishing magnetic field, these extended states are expected to float up indefinitely in energy \cite{khmelnitskii1984quantum, laughlin1984levitation}, thus making the 2D system insulating. However, the authors of Refs.~\cite{pudalov1993hall, shashkin1993floating, shashkin1994insulating, pudalov1994termination, kravchenko1995global} have found, contrary to expectations, that the extended states do not float up indefinitely in the limit of zero magnetic field, but coalesce instead at finite energies, thus allowing for a metallic state at $B=0$.

\begin{figure}[t]
\centering
\includegraphics[width=8 cm]{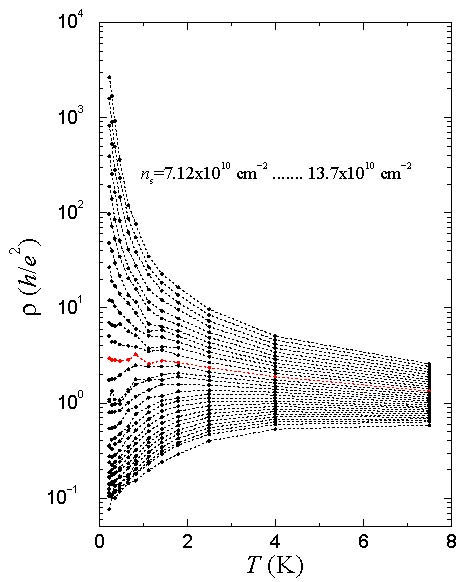}
\caption{Temperature dependences of the resistivity in a Si MOSFET with a peak electron mobility of $3\times10^4$~cm$^2$/Vs for different electron densities (designated by different symbols) at $B = 0$.  From Ref.~\cite{kravchenko1995scaling}.}\label{RT_silicon}
\end{figure}

Later, in 1994-1996, strong metallic temperature dependences with the sevenfold decrease in resistivity at temperatures $\lesssim 2.5$~K were observed in high-mobility Si MOSFET samples \cite{kravchenko1994possible,kravchenko1995scaling}, revealing dramatic differences in the behavior of strongly interacting 2D systems with $r_s > 10$ compared to weakly interacting systems. As the electron density increased, the systems transitioned from a regime in which resistance diverged as temperature decreased (indicating insulating behavior) to one in which resistance decreased with decreasing temperature (indicating strongly metallic behavior) (see Fig.~\ref{RT_silicon}).  The two regimes are separated by the line at a critical electron density, $n_c$ (dashed red line in Fig.~\ref{RT_silicon}), which has a weak temperature dependence and extrapolates to approximately $3h/e^2$ as the temperature tends to zero.
\footnote[1]{Care should be taken to prevent the electron system from overheating at millikelvin temperatures due to external electromagnetic interference. Common experimental techniques include dedicated grounding, proper shielding and screening of the measurement circuits, filtering out pickup and noise, and using optocouplers and battery-operated preamplifiers.}

\subsubsection{High- and low-mobility samples}

\begin{figure}
\centering
\includegraphics[width=7 cm]{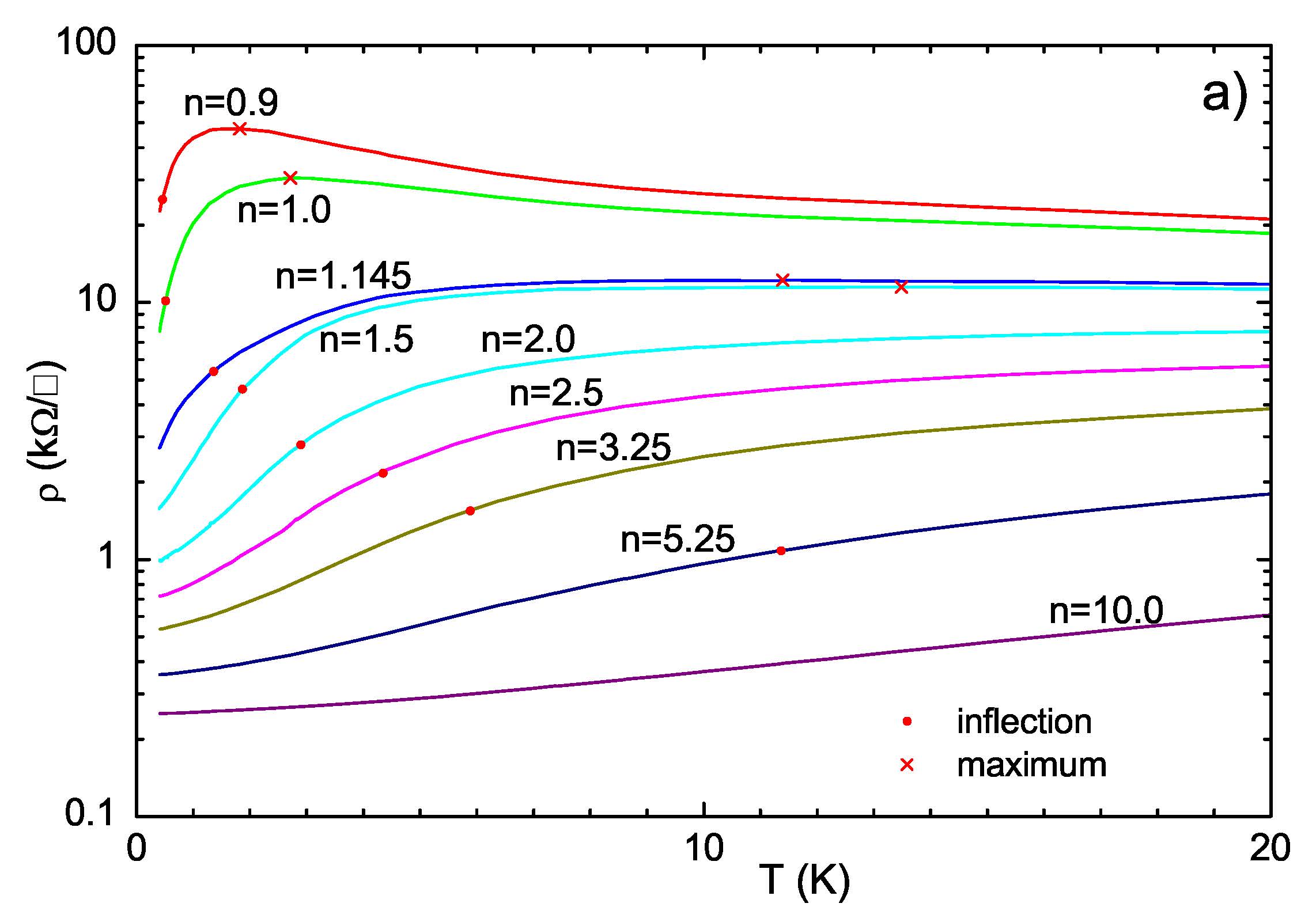}
\includegraphics[width=7 cm]{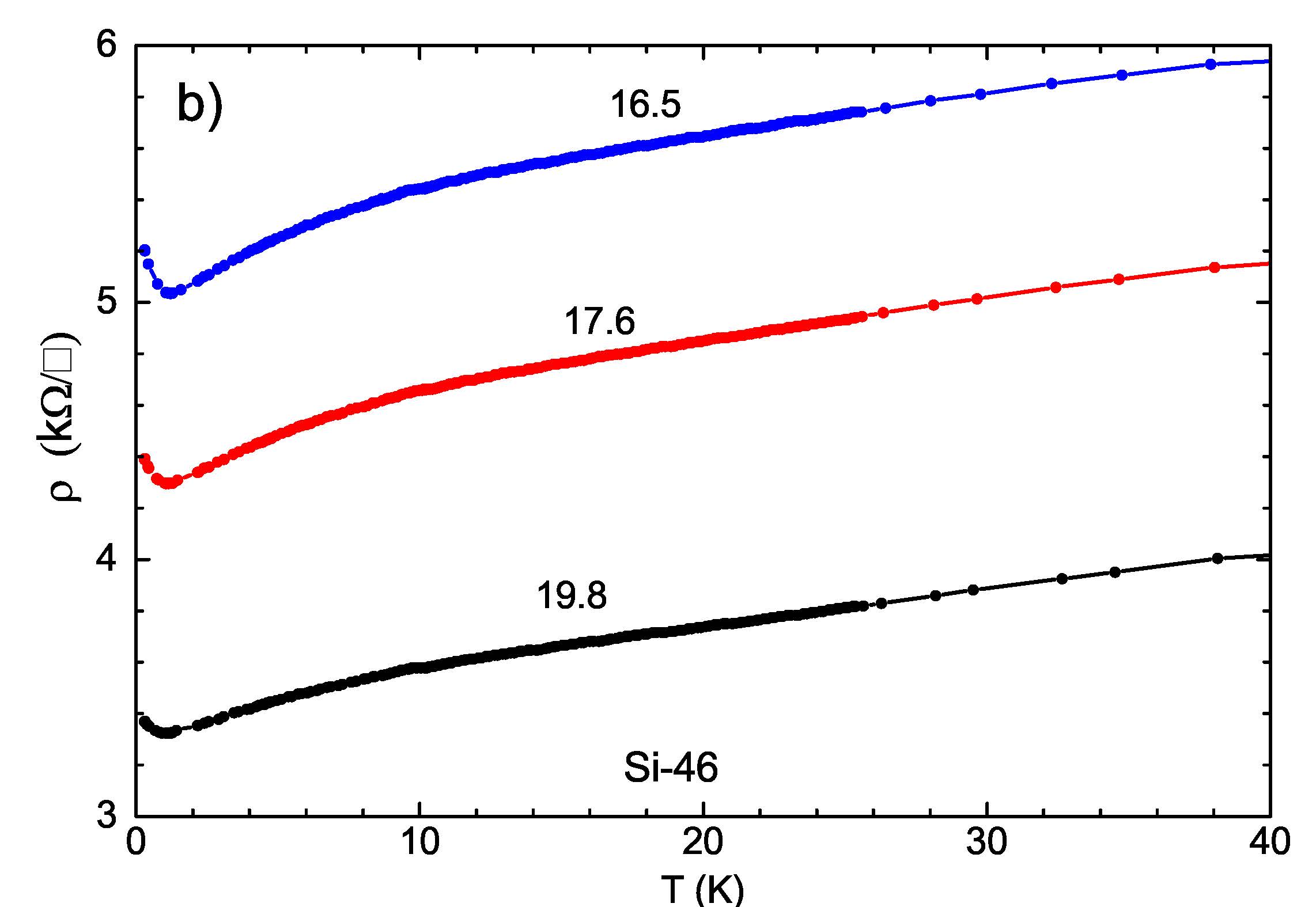}
\caption{Temperature dependence of resistivity in two Si~MOSFET samples at zero field: (a)~high-mobility ($32\times10^3$~cm$^2$/Vs) sample, and (b)~low-mobility ($1.5\times10^3$~cm$^2$/Vs) sample.  Carrier densities are shown in units of $10^{11}$~cm$^{-2}$. Crosses and dots on the left panel mark the $\rho(T)$ maxima and the inflection points, respectively. From Ref.~\cite{klimov2008interaction}.}
\label{high low}
\end{figure}

In zero magnetic field, the temperature dependences of the resistivity in high- and low-mobility samples differ drastically, as shown in Fig.~\ref{high low}. In the ``metallic'' range of densities, $n_s > n_c$, the resistivity of high-mobility ($\sim30\times10^3$~cm$^2$/Vs) samples varies sharply by a factor of about 7 \cite{kravchenko1994possible,kravchenko1995scaling}. In contrast, for low-mobility samples ($\sim(1-2)\times10^3$~cm$^2$/Vs), $\rho(T)$ varies by only $\approx 15\%$, and the variation occurs at much higher temperatures and densities \cite{pudalov1998metal}. These well-known features have been explored and explained in Refs.~\cite{gold1986temperature, pudalov1998metal, proskuryakov2002hole, kuntsevich2007intervalley, klimov2008interaction}.

The upturn at low temperatures in Fig.~\ref{high low}(b) is due to quantum corrections, which for low-mobility samples have an ``insulating'' sign at all densities (see Fig.~\ref{high low}(b) and Ref.~\cite{pudalov1998metal}). For high-mobility samples, the upturn is expected to set in upon lowering the temperature below $T= h/k_B\tau$ and $T = h/k_B\tau_v$ (here $\tau_v$ is the intervalley scattering time). This low-temperature regime is outside the range shown in Fig.~\ref{high low}(a).

\subsection{Scaling of the resistivity}

\begin{figure}
\centering
\includegraphics[width=8 cm]{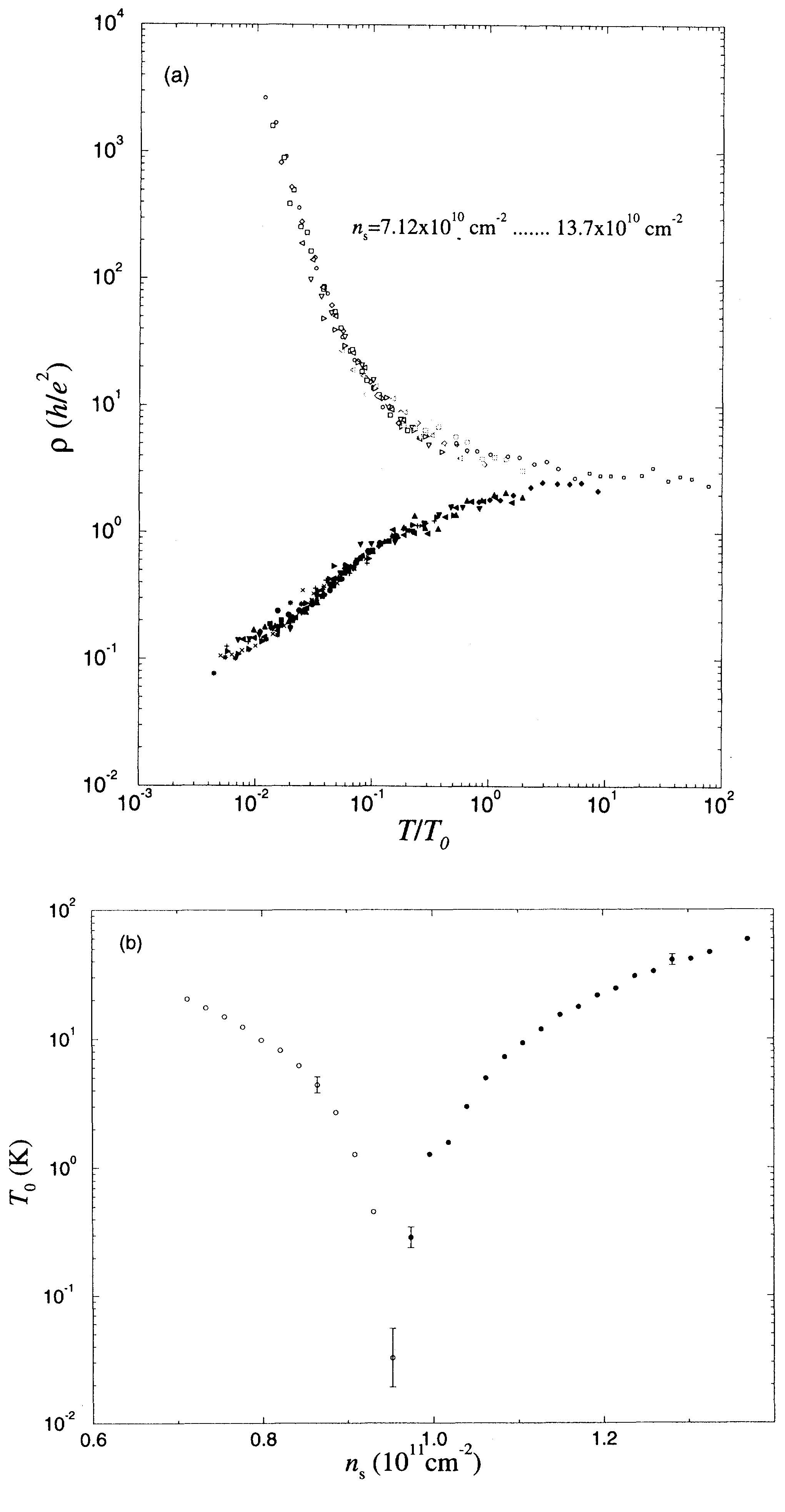}
\caption{(a) Resistivity \textit{vs.} $T/T_0$ for the same sample as in Fig.~\ref{RT_silicon}.  (b) Scaling parameter $T_0$ \textit{vs.} electron density. Open symbols correspond to the insulating side of the transition, and closed symbols correspond to the metallic one.
From Ref.~\cite{kravchenko1995scaling}.}\label{Scaling T}
\end{figure}

A notable feature of the resistivity $\rho(T)$ dependences across various electron densities is that the data can be made to overlap by scaling along the temperature axis. In other words, resistivity can be expressed as a function of $T/T_0$, where $T_0$ depends solely on the carrier density, $n_s$. This scaling is valid over a wide interval of electron densities around $n_c\approx0.95\times10^{11}$~cm$^{-2}$, specifically for $|\delta n_s|\lesssim 2.5 \times 10^{10}$ cm$^{-2}$ (here $\delta n_s=n_s-n_c$), and within the temperature interval of 0.2 to 3~K \cite{kravchenko1995scaling}.  The results of the scaling are illustrated in Fig.~\ref{Scaling T}(a), which shows $\rho$ plotted as a function of $T/T_0$. The data collapse into two distinct curves: the upper curve, represented by open symbols, corresponds to the insulating side of the transition, whereas the lower curve, shown with filled symbols, corresponds to the metallic side. Line thickness primarily reflects the within-dataset scatter, highlighting the high quality of the scaling results. The scaling parameter, $T_0$, is shown as a function of $n_s$ in Fig.~\ref{Scaling T}(b).  On both sides of the transition, the scaling parameter obeys a power law dependence $T_0\propto|\delta n_s|^\gamma$ with $\gamma=1.60\pm0.1$ for the insulating side of the transition and $\gamma=1.62\pm0.1$ for the metallic side, and tends to zero as $n_s\rightarrow n_c$. It is noteworthy that a similar scaling behavior was previously reported in Ref.~\cite{liu1991scaling} for the insulator-superconductor transition in thin films of bismuth.

\begin{figure}
\centering
\includegraphics[width=8 cm]{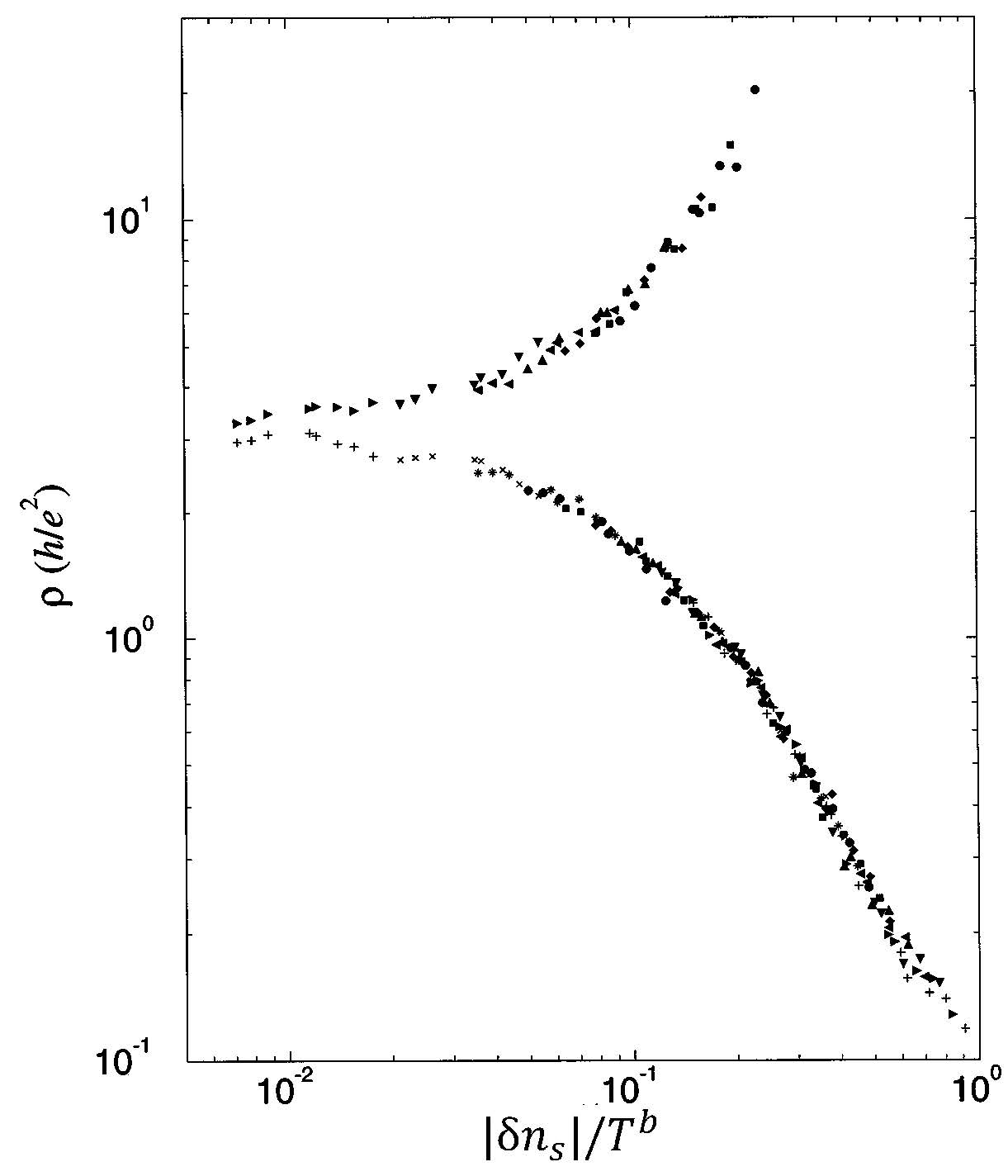}
\caption{Demonstrating scaling with temperature in the temperature interval 0.3 to 1 K, the resistivity is shown as a function of $|\delta n_s|/T^b$ for $b=1/z\kappa=0.83$.  Electron densities are in the range $7.81-10.78\times10^{10}$~cm$^{-2}$.  From Ref.~\cite{kravchenko1996electric}.}\label{RT scaling}
\end{figure}

An alternative way to scale the $R(T)$ curves is to plot them as a function of $|\delta n_s|/T^b$, as shown in Fig.~\ref{RT scaling}:
\begin{equation}
\rho(T,n_s)=f_1(|\delta n_s|/T^{b})
\end{equation}
Here $b=1/z\kappa$, $z$ is the dynamical exponent, and $\kappa$ is the correlation length exponent \cite{sondhi1997continuous}. This method allows one to find the product of the dynamical and the correlation exponents \cite{kravchenko1996electric}.  More on this will be discussed below.

The remarkable reflection symmetry of the two branches in the scaled $\rho(T)$ dependences shown in Figs.~\ref{Scaling T} and \ref{RT scaling} relative to $n_c$ indicates an approximately symmetrical empirical description of $\rho(T,n_s)$, as suggested by Pudalov \cite{pudalov1997unconventional}:
\begin{eqnarray}
	\rho(T) &=&\rho_0 +\rho_1\exp(-\Delta(n_s)/T),\nonumber \\
	\Delta (n_s)&=& \lambda(n_s-n_c),
	\label{eq:rho-exp}
\end{eqnarray}
where $\lambda$ is a phenomenological parameter. This empirical form is discussed in more detail in Section \ref{sec:Phase Separation}.

Note that there are both similarities and differences between the metal-insulator transition and the superconductor-insulator transition. For superconductor-insulator transitions, two scenarios are considered: the fermionic scenario, in which the order parameter amplitude vanishes, and the bosonic scenario, in which the amplitude of phase fluctuations reaches a critical value. This diversity is absent in the case of the metal-insulator transition, at least in the current state of knowledge. On the other hand, similarities include, for example, mirror symmetry between the two phases in the critical regime and a tendency toward phase separation that strengthens as the critical point is approached by varying the control parameter (disorder or carrier density).

\subsection{Nonlinear effects}

\begin{figure}
\centering
\includegraphics[width=8 cm]{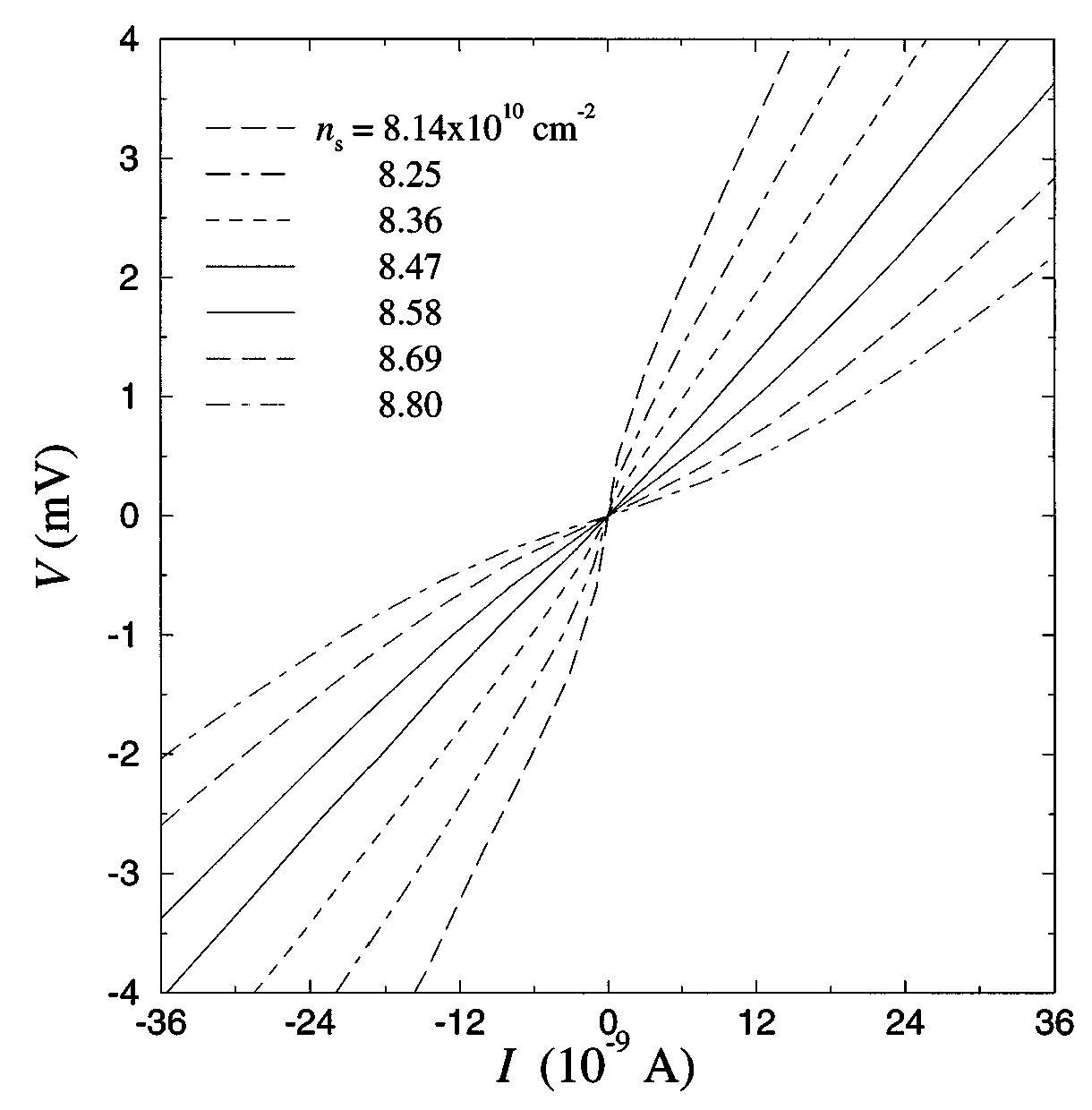}
\caption{Nonlinear current-voltage curves at several electron densities around the metal-insulator transition. $T=0.22$~K.  From Ref.~\cite{kravchenko1996electric}.}\label{IV}
\end{figure}

\begin{figure}
\centering
\includegraphics[width=8 cm]{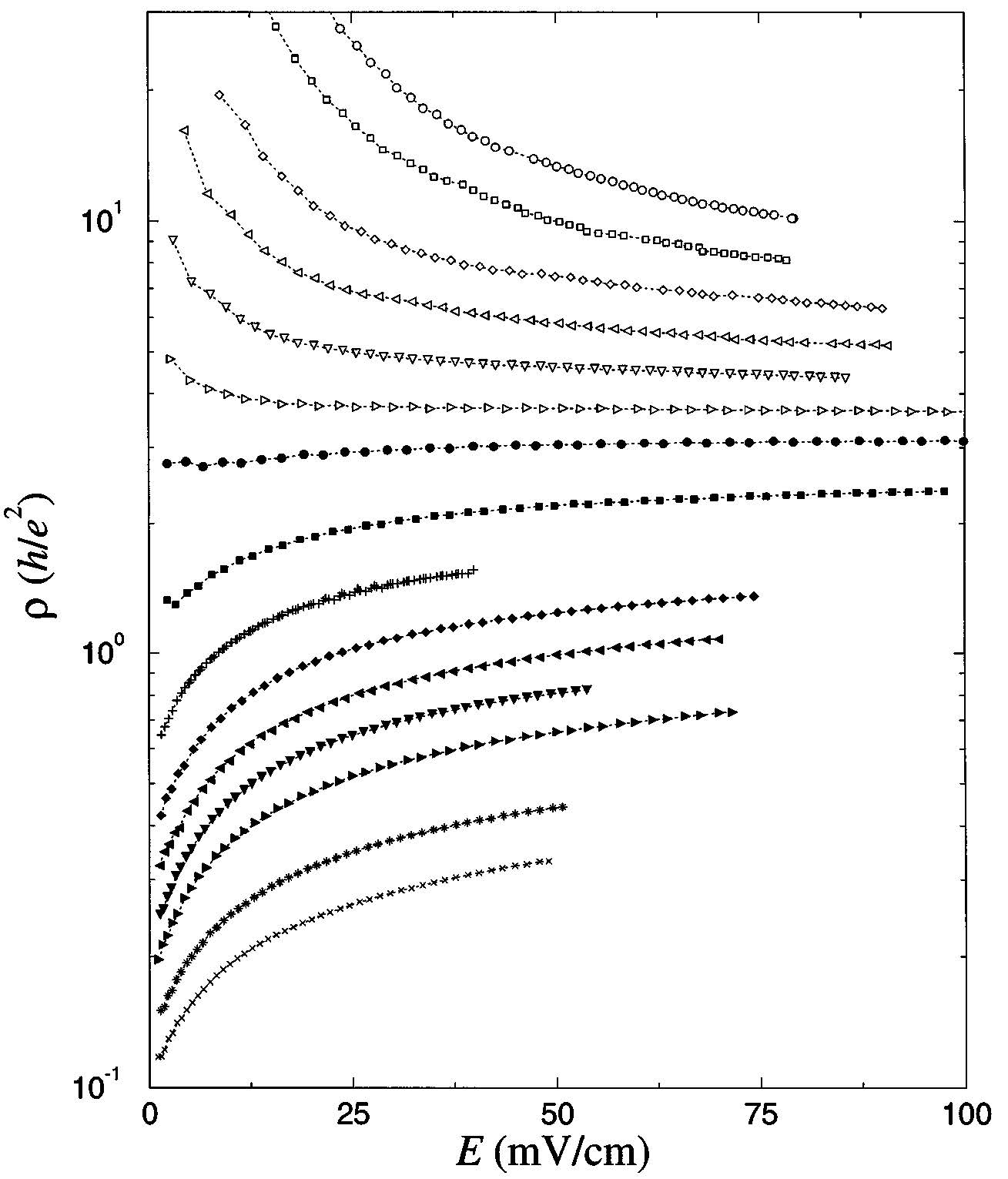}
\caption{Resistivity as a function of electric field at $n_s =$7.81, 7.92, 8.03, 8.14, 8.25, 8.36, 8.47, 8.70, 8.91, 9.79, 10.34, and $10.78\times10^{10}$~cm$^{-2}$ (top to bottom) at $T=0.22$~K. From Ref.~\cite{kravchenko1996electric}.}\label{RE}
\end{figure}

The data reported in Refs.~\cite{kravchenko1994possible,kravchenko1995scaling} were obtained in the linear regime, \textit{i.e.}, in the limit of the zero electric field, $E\rightarrow0$. Typical current-voltage ($I-V$) curves across the metal-insulator transition are shown in Fig.~\ref{IV}.  The middle, solid curve is close to linear and separates all $I-V$ curves into two groups exhibiting different types of nonlinear behavior. Note that there is a remarkable symmetry about the middle curve. This symmetry is reminiscent of that observed for the $I-V$ curves in the vicinity of the quantum Hall liquid to insulator phase transition \cite{shahar1995universal}.  Curves of resistivity \textit{vs.}\ the electric field for different electron densities are shown in Fig.~\ref{RE}. Again, there is clearly a critical electron density that separates two distinct density regions characterized by different types of resistivity behavior as a function of the electric field. At the critical point, the resistivity (the solid curve in Fig.~\ref{IV}, closed circles in Fig.~\ref{RE}) is virtually independent of the electric field and close to $3h/e^2$. The curves below this line are characterized by $dR/dE>0$, while all curves above are characterized by $dR/dE<0$.  The observed critical behavior of the resistivity as a function of $E$ is very similar to the critical behavior of the resistivity with temperature (Fig.~\ref{RT_silicon}) with the only difference being that the resistivity at the critical point is independent of the electric field while the metallic $R(T)$ curves in the close vicinity of the critical electron density are non-monotonic.

\begin{figure}
\centering
\includegraphics[width=8 cm]{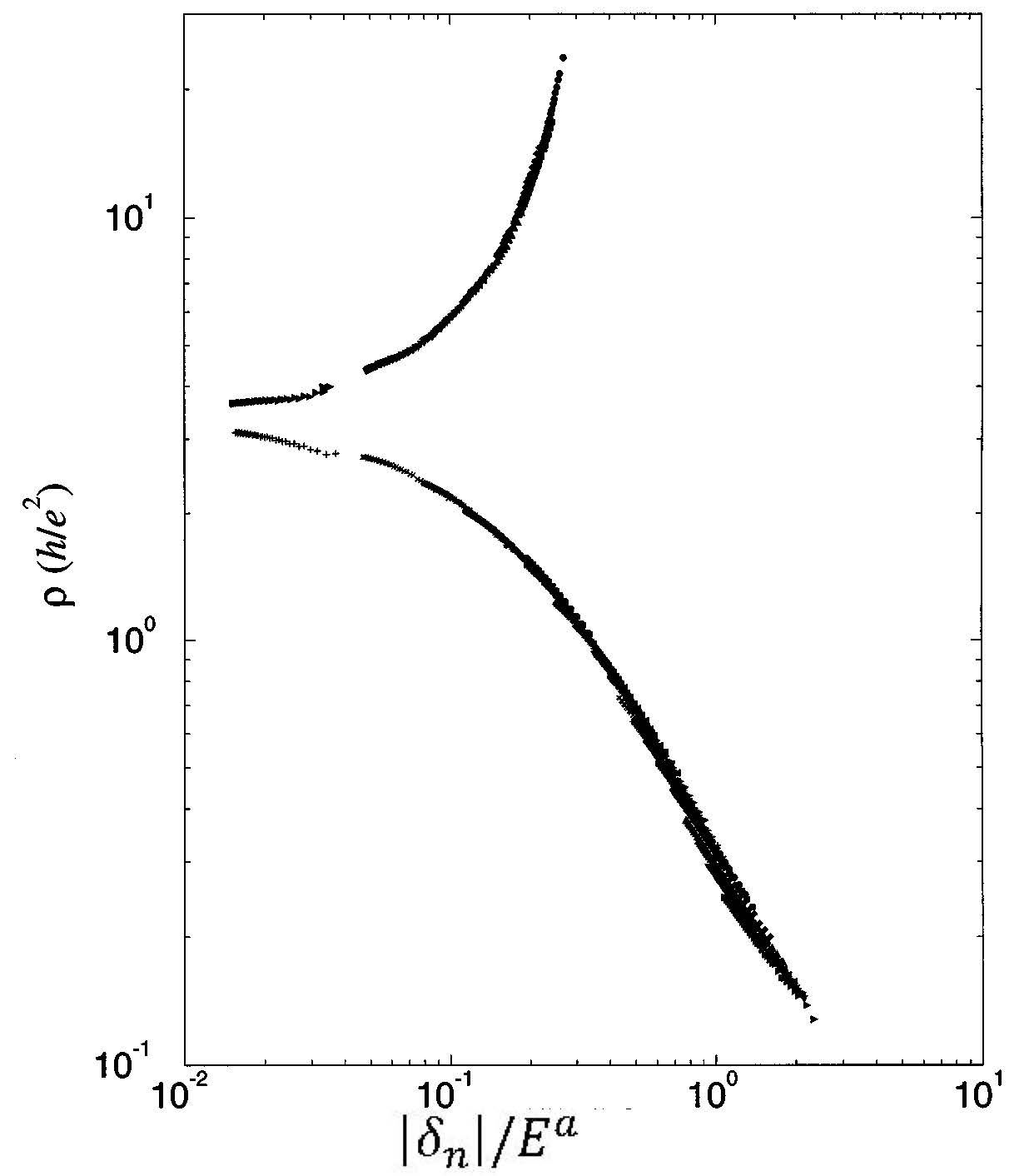}
\caption{Demonstrating scaling with electric field, the resistivity at 0.22 K is plotted as a function of $|\delta n_s|/E^a$ for $a=1/(z+1)\kappa=0.37$.  Electron densities are in the range $7.81-10.78$ in units of $10^{10}$~cm$^{-2}$. From Ref.~\cite{kravchenko1996electric}.}\label{Scaling E}
\end{figure}

A general scaling analysis of the Coulomb-interacting quantum critical point \cite{sondhi1997continuous} suggests that the resistivity should also scale with the electric field:
\begin{equation}
\rho(E,n_s)=f_2(|\delta n_s|/E^a)
\end{equation}
Indeed, plotting resistivity against the scaling variable, $|\delta n_s|/E^a$, collapses all curves onto two distinct branches \cite{kravchenko1996electric}, as shown in Fig.~\ref{Scaling E} (here $a=1/(z+1)\kappa$). At the transition, the resistivity is close to $3h/e^2$. The quality of the electric-field scaling is considerably higher than that of the temperature scaling. The best collapse was achieved for $a=0.37\pm0.01$.

Knowing both $z\kappa$ and $(z+1)\kappa$, it is possible to determine the exponents $z$ and $\kappa$ separately, as was done for the superconductor-insulator transition in thin disordered films \cite{yazdani1995superconducting}. The values obtained in Ref.~\cite{kravchenko1996electric} are $z=0.8\pm0.1$ and $\kappa=1.5\pm0.1$. Although the correlation length exponent has not been theoretically predicted for this system, the dynamical exponent in a strongly interacting 2D system was predicted to be $z=1$, which is close to what was reported in Ref.~\cite{kravchenko1996electric}.  In the majority of interacting 2D systems where it has been determined, the exponent $z$ has also been found to be close to 1 (Ref.~\cite{yazdani1995superconducting} for the case of a superconductor-insulator transition or Ref.~\cite{wei1994current} for the transition between two neighboring quantum Hall effect plateaus).

\subsection{Experiments on other strongly-interacting 2D systems}

\begin{figure}[h]
\centering
\includegraphics[width=8 cm]{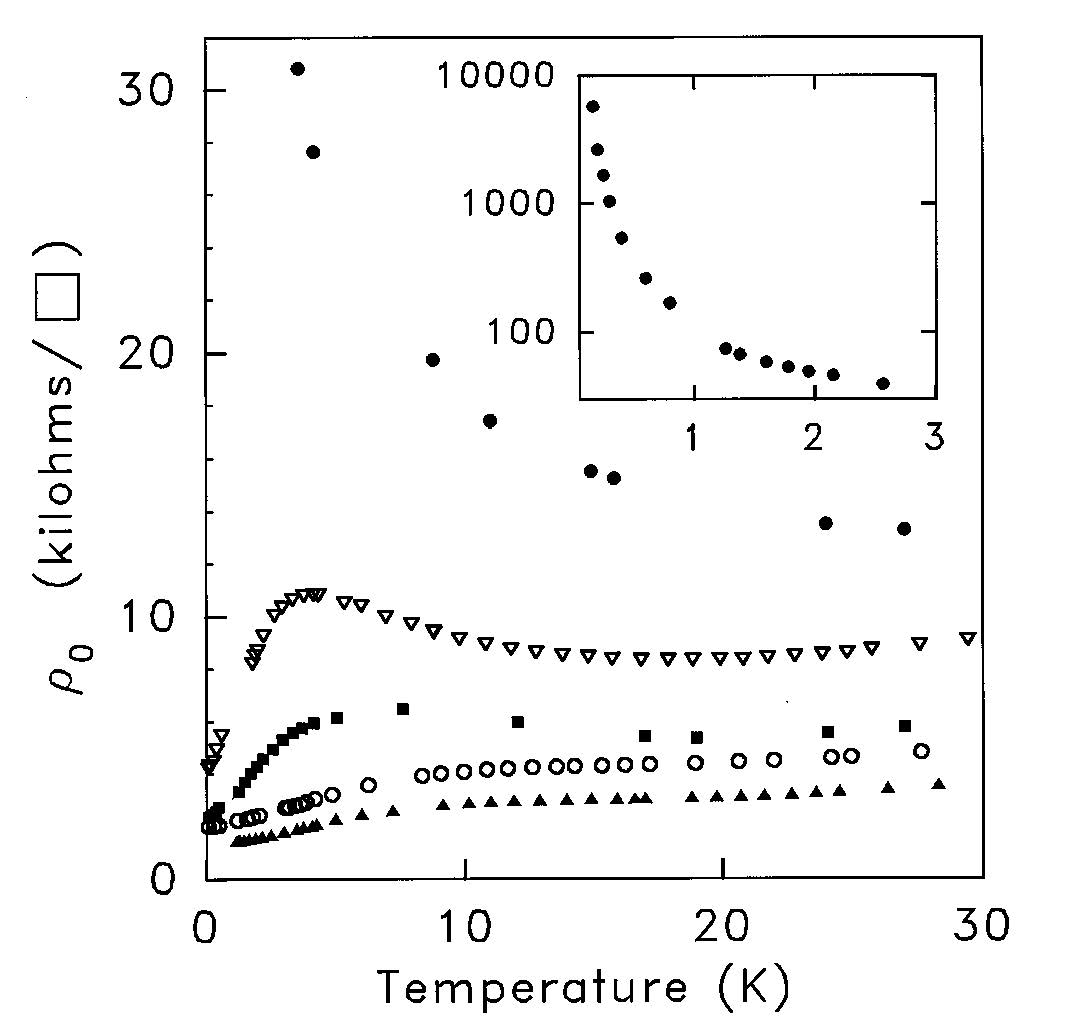}
\caption{Temperature dependence of the zero-field resistivity for $p$-SiGe samples.  Hole densities are (top to bottom): 0.48, 1.2, 1.5, 2.8, and 3.05 in units of $10^{15}$~m$^{-2}$. The inset shows low-temperature data for a sample with a hole density of $0.48\times10^{15}$~m$^{-2}$ on a logarithmic scale; same units as in the main panel. From Ref.~\cite{coleridge1997metal}.}\label{Coleridge}
\end{figure}

\begin{figure}
\centering
\includegraphics[width=8 cm]{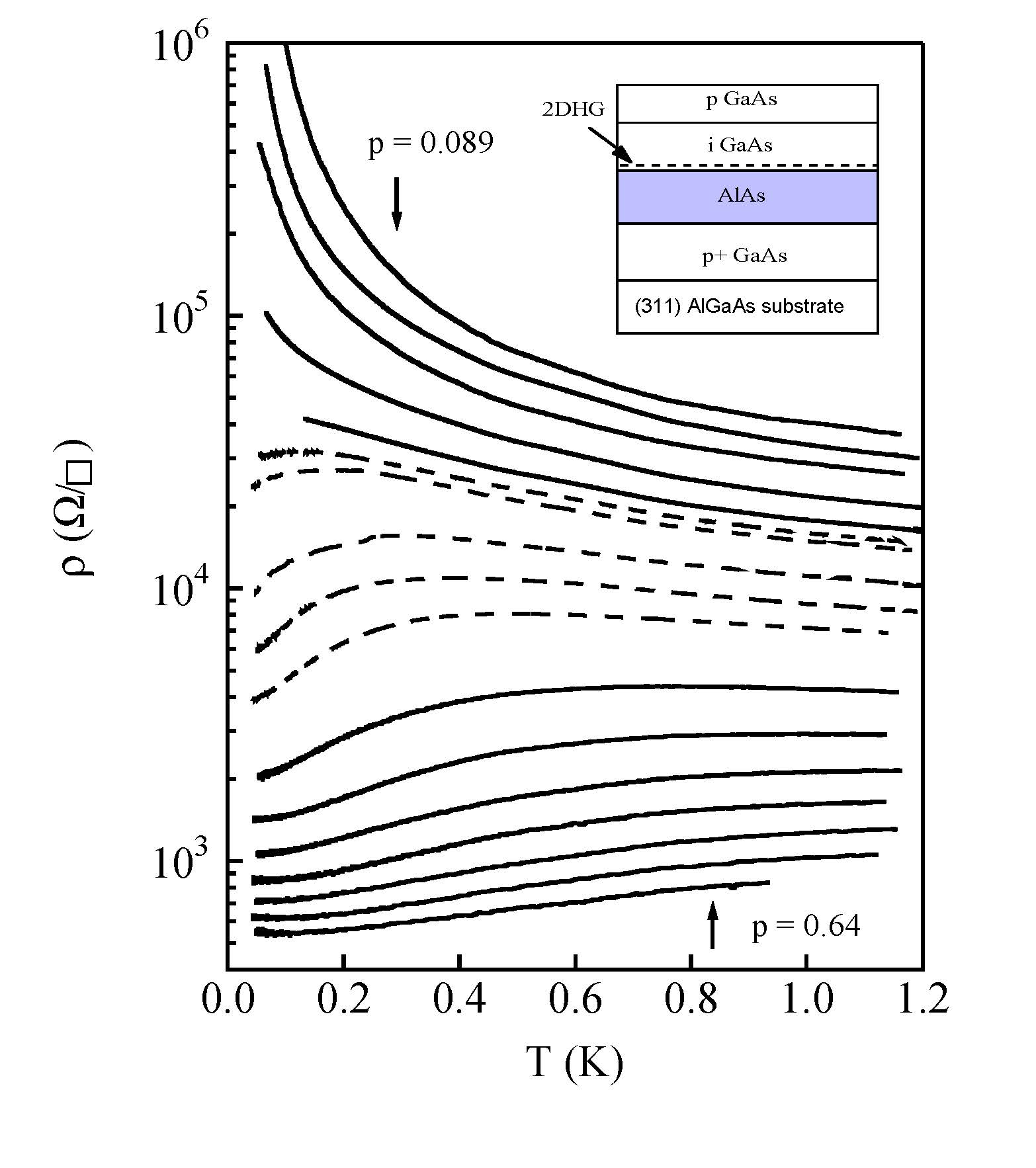}
\caption{$\rho$ as a function of $T$ data obtained on $p$-type GaAs 2D system at $B = 0$ at various fixed densities (top to bottom), $p$ = 0.089, 0.94, 0.99, 0.109, 0.119, 0.125, 0.13, 0.15, 0.17, 0.19, 0.25, 0.32, 0.38, 0.45, 0.51, 0.57, $0.64\times10^{11}$~cm$^{-2}$.  Note the three distinct regimes: an insulating regime at low densities, a mixed regime at intermediate densities indicated by dashed lines, and a metallic regime at high densities.  Inset: Schematic presentation of the $p$-type structure grown on semi-insulating (311) GaAs substrate, consisting of a thick $p$+ buffer, a 300~nm undoped AlAs barrier, a 150 nm undoped GaAs channel layer, and a top 50~nm GaAs layer which is $p$-doped.  The 2D hole system forms at the lower interface of the channel layer upon application of a negative bias to the $p$+ conducting layer.  From Ref.~\cite{hanein1998the}.}\label{Hanein}
\end{figure}

\begin{figure}
\centering
\includegraphics[width=8 cm]{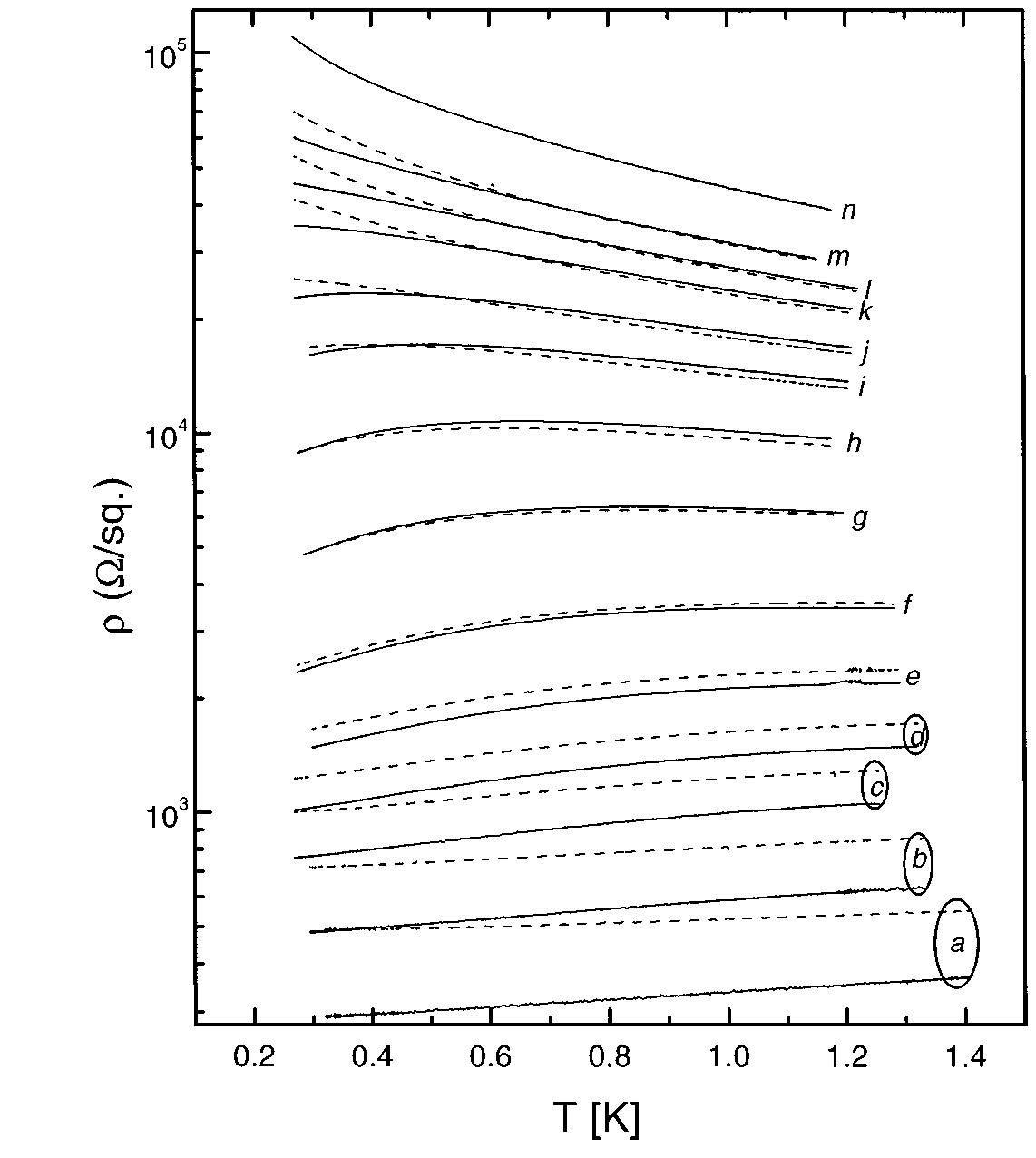}
\caption{Resistivity vs temperature data for AlAs 2DES. The high- and low-mobility direction data are shown by solid/dashed curves, respectively. The densities, in units of 10$^{11}$~cm$^{-2}$, are a: 2.73, b: 2.08, c: 1.42, d: 1.22, e: 1.02, f: 0.82, g: 0.74, h: 0.72, i: 0.70, j: 0.65, k: 0.64, l: 0.63, m: 0.60, n: 0.59. The low-mobility curve for case n is not shown because the Ohmic contacts failed at such a low density. From Ref.~\cite{papadakis1998apparent}.}\label{Papadakis}
\end{figure}

\begin{figure}
\centering
\includegraphics[width=8 cm]{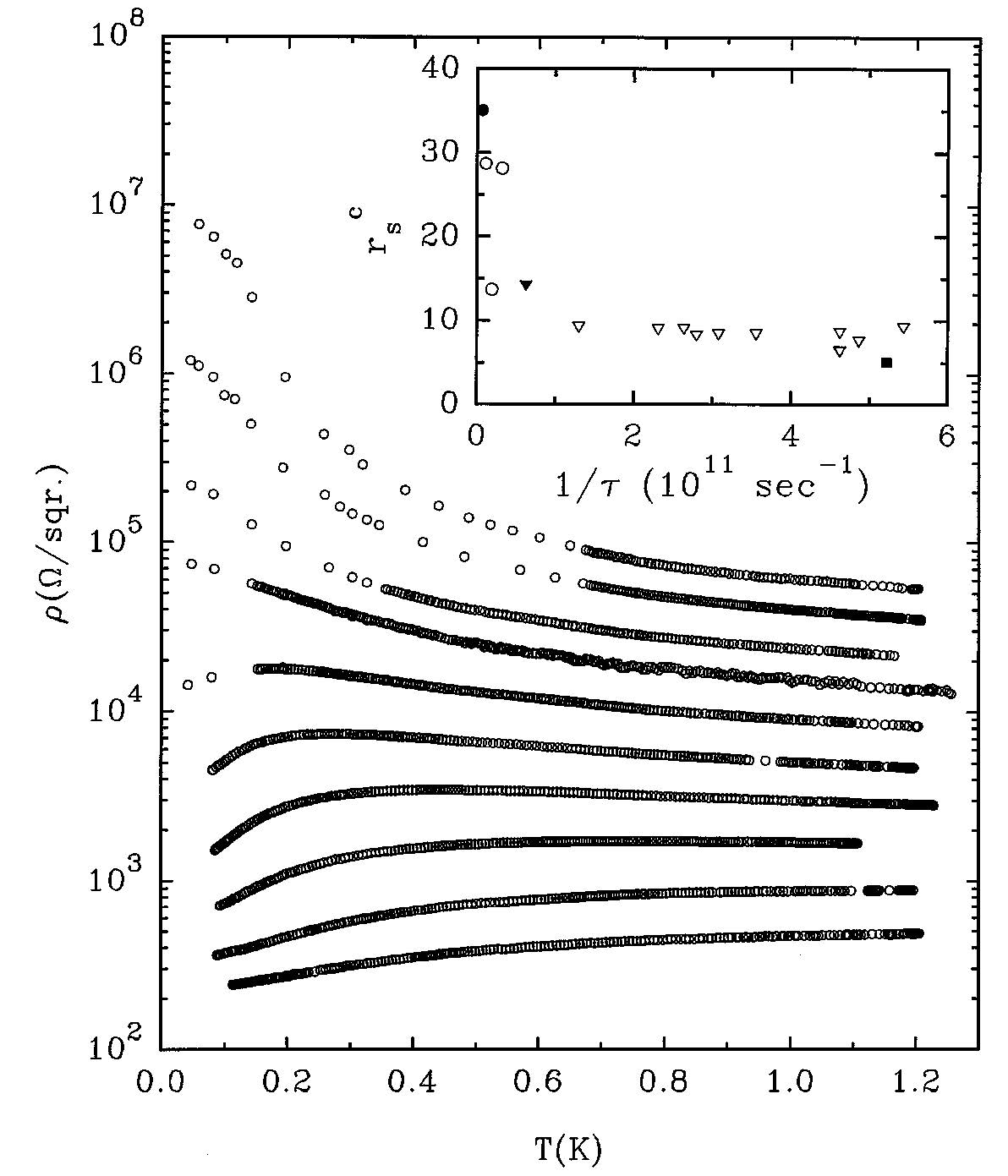}
\caption{$T$ dependence of the resistivity in $p$-type GaAs/AlGaAs heterostructure: from the top curve, $p=$ 0.48, 0.55, 0.64, 0.72, 0.90, 1.02, 1.27, 1.98, 2.72, and 3.72 in units of $10^{10}$ cm$^2$. The inset shows $r_s$ \textit{vs}.\ the inverse scattering time at the critical carrier density for the MIT in various 2D systems; this work (filled circle), 2D hole system in GaAs/AlGaAs (open circles), 2D electron system in AlAs/GaAlAs (filled triangle), 2D hole system in SiGe (filled square), and 2D electron system in Si MOSFET (open triangles). From Ref.~\cite{yoon1999wigner}.}\label{Yoon}
\end{figure}

\begin{figure}
\centering
\includegraphics[width=9 cm]{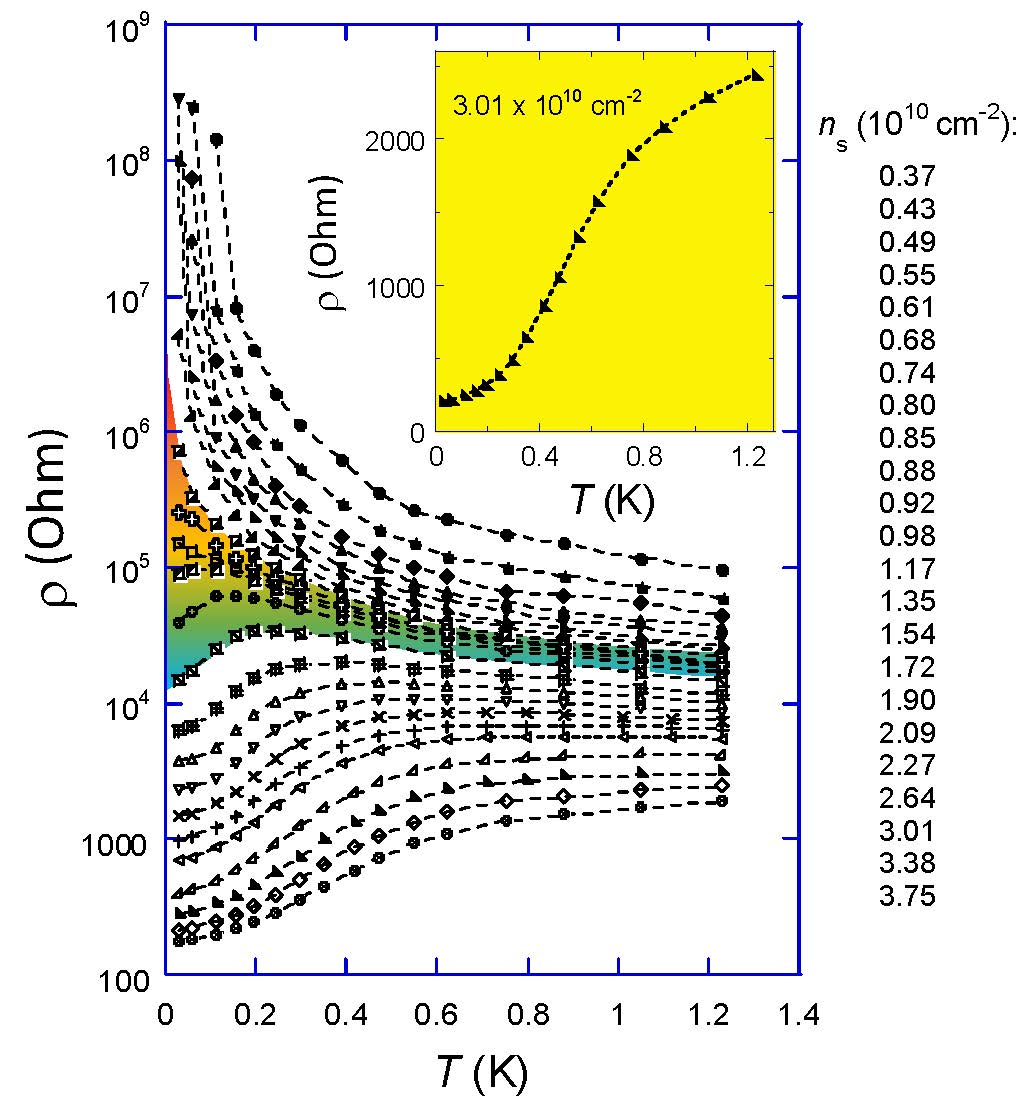}
\caption{Temperature dependences of the resistivity in a SiGe/Si/SiGe quantum well at different electron densities in zero magnetic field. The inset shows a close-up of $\rho(T)$, which exhibits a drop in resistivity by a factor of 12. From Ref.~\cite{melnikov2019quantum}.}\label{SiGe RT}
\end{figure}

Initially, these results were met with great skepticism and largely overlooked until 1997, when they were almost simultaneously experimentally confirmed in other strongly interacting 2D electron and hole systems: silicon MOSFETs from a different research group \cite{popovic1997metal}, $p$-type SiGe structures \cite{coleridge1997metal}, $p$-GaAs/AlGaAs heterostructures \cite{hanein1998the,yoon1999wigner,mills1999nonmonotonic}, $n-$type GaAs/AlGaAs heterostructures \cite{hanein1998observation}, AlAs systems \cite{papadakis1998apparent}, and, more recently, in unprecedentedly clean SiGe quantum wells \cite{melnikov2019quantum} with the drop of the resistance by a factor of 12 below $T\approx1.2$~K and ZnO heterostructures with a fivefold drop of the resistance below $\approx100$~mK \cite{falson2022competing}.  Some of the examples are shown in Figs.~\ref{Coleridge}, \ref{Hanein}, \ref{Papadakis}, \ref{Yoon}, \ref{SiGe RT}. The values of $r_s$ in these studies varied from $\approx4$ \cite{hanein1998observation} to $>40$ \cite{yoon1999wigner}.

Scaling analysis has been applied to a variety of 2D systems with different degrees of success. Several experiments have produced scaling exponents \( b = 1/z\kappa \) ranging from 1.25 to 1.6 for Si MOSFETs \cite{popovic1997metal} and \( p \)-type SiGe systems \cite{coleridge1997metal}.  In the 2D systems where both temperature and the electric field scaling were attempted, the results for the correlation length exponent ranged between 1.5 and 1.9 \cite{kravchenko1995scaling,heemskerk1998nonlinear}. The dynamical exponent varied between 0.8 and 1.2 in Si MOSFETs \cite{kravchenko1995scaling,heemskerk1998nonlinear} and was found equal to 1 in $p$-SiGe heterostructures \cite{senz1999metal}. However, attempts to scale the resistivity of \( p \)-type GaAs/AlGaAs heterostructures have either resulted in significantly larger exponents \cite{simmons1998metal} or have entirely failed \cite{yoon1999wigner}. Generally, it seems that scaling analysis yields larger exponents or breaks down in systems that exhibit relatively weak metallic temperature behavior, as observed in some \( p \)-GaAs/AlGaAs heterostructures and low-mobility Si MOSFETs. Conversely, the more pronounced the metallic temperature behavior, the better the scaling fits.

\begin{figure}
\centering
\includegraphics[width=8 cm]{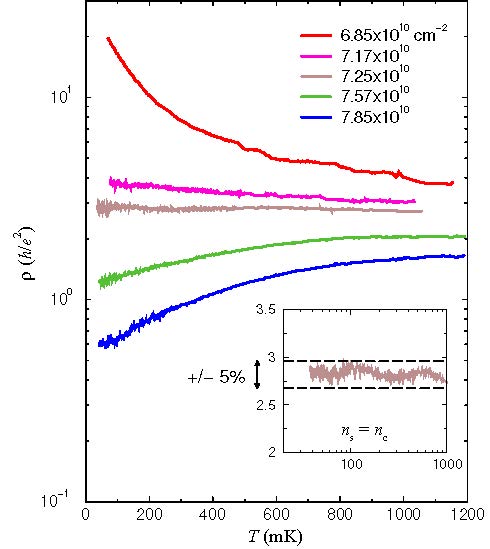}
\caption{Resistivity vs.\ temperature in a Si MOSFET sample at five electron densities as labeled. The inset shows that the middle curve ($n_s=7.25\times10^{10}$~cm$^{-2}$) changes by less than $\pm5$\% in the entire temperature range. From Ref.~\cite{kravchenko2000metallic}.}\label{Flat separatrix}
\end{figure}

In the most clean and uniform Si MOSFET samples, the resistivity at $n_s=7.25\times10^{10}$~cm$^{-2}$ has been found to be essentially independent of temperature from approximately 1~K down to about 35~mK \cite{kravchenko2000metallic}, the lowest temperature reached in that experiment.  It is tempting to identify the corresponding electron density with the critical one for the MIT.  Just below this ``separatrix'' (for ``just-metallic'' electron densities, $n_s\gtrsim n_c$), the resistivity decreases with decreasing temperature down to the lowest accessed temperatures with no indication of an upturn at $T\rightarrow0$ (see Fig.~\ref{Flat separatrix}).  Later, a symmetry-based analysis of the resistivity near the critical fixed point \cite{knyazev2008metal} suggested a more rigorous definition of the separatrix. This critical $\rho(T)$ curve must have no exponential temperature term, no inflection points, and no maxima.
According to such a definition, the critical electron density is $n_c\approx 0.88\pm{0.02} \times 10^{10}$~cm$^{-2}$ in Fig.~\ref{SiGe RT} and $n_c\approx 7.25\times 10^{10}$~cm$^{-2}$ in Fig.~\ref{Flat separatrix}.  We emphasize that the metallic temperature dependence overpowers quantum localization.  However, the transport behavior in the limit as \( T \) approaches 0 requires more thorough consideration, which will be presented in Section~\ref{suppression of WL}.

\begin{figure}
\centering
\includegraphics[width=8 cm]{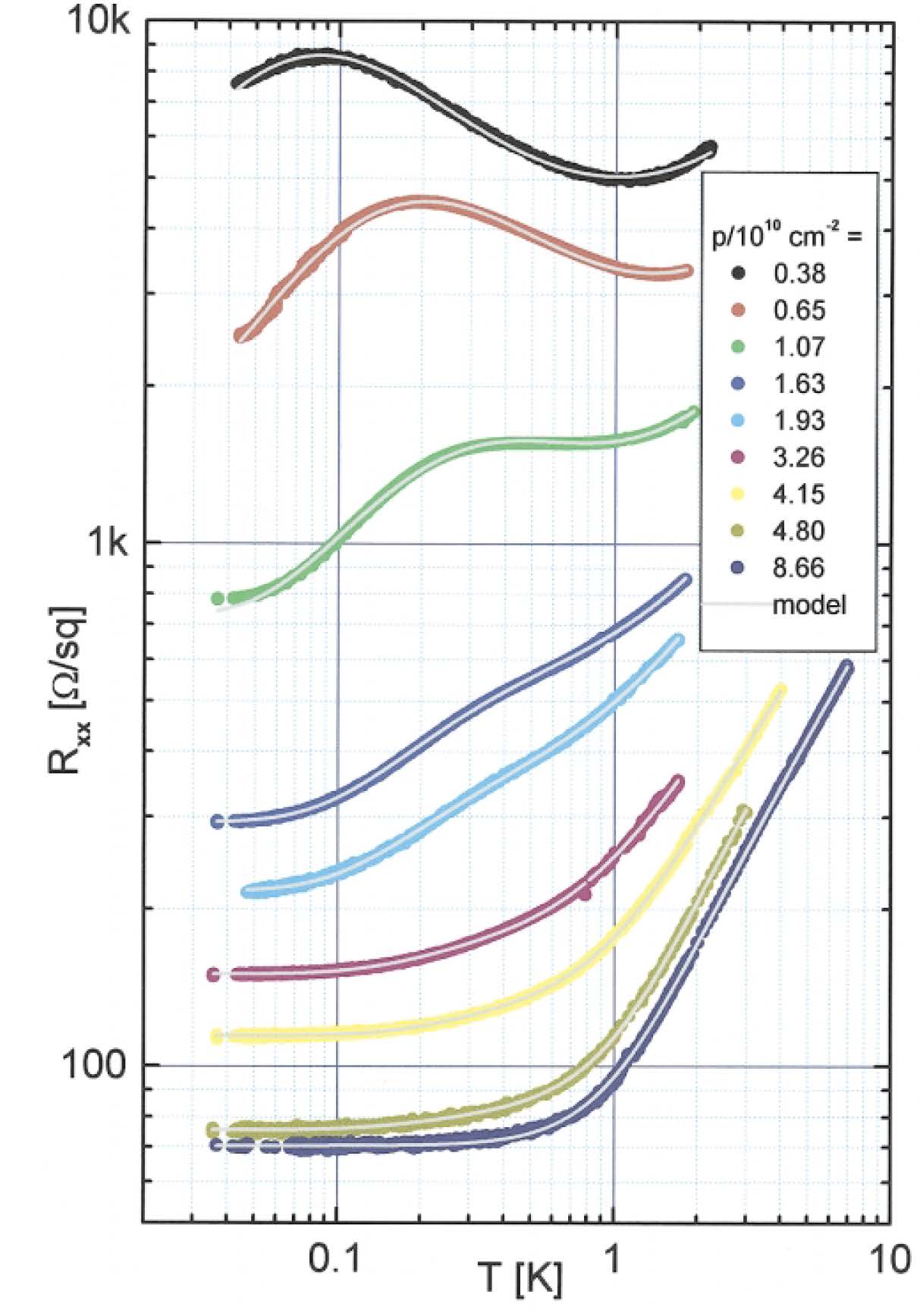}
\caption{Longitudinal resistance per square vs.\ temperature for ultra-high mobility p-GaAs/AlGaAs samples of various hole densities $p$. From Ref.~\cite{mills1999nonmonotonic}.}\label{Mills low T}
\end{figure}

In ultra-clean $p$-GaAs 2D systems, the resistivity shows no upturns at least down to 25~mK \cite{mills1999nonmonotonic}.  Fig.~\ref{Mills low T} shows $\rho(T)$ over the range 25~mK - 10~K for 9 different hole densities.  In the limit of low temperatures, the resistance becomes either constant (7 lower curves) or continues to decrease (two upper curves).

\section{Suppression of the coherent backscattering near the MIT}\label{suppression of WL}

\begin{figure}
\centering
\includegraphics[width=8 cm]{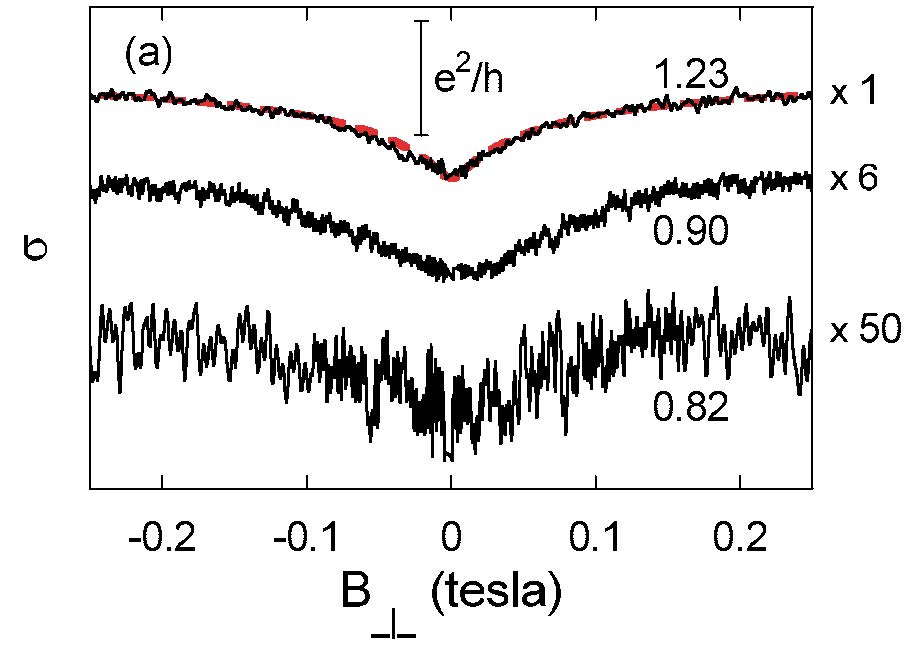}
\caption{Magnetoconductivity in a Si MOSFET sample at $T=42$~mK and at different electron densities indicated near the curves in units of $10^{11}$~cm$^{-2}$. The curves are vertically shifted for clarity, and the two lower curves are multiplied by 6 (the middle one) and 50 (the lower one), making them quantitatively similar to the upper (unmodified) curve. From Ref.~\cite{rahimi2003coherent}.}\label{no WL}
\end{figure}

It turns out that the weak localization corrections, which exist deep in the metallic phase, weaken as one approaches the transition and practically vanish at the critical density. In Ref.~\cite{rahimi2003coherent}, the weak-field negative magnetoresistance was studied in Si MOSFETs (see Fig.~\ref{no WL}), and its magnitude was found to decrease by a factor of 50 compared to its ``noninteracting'' value upon approaching $n_c$ from the metallic side of the transition. The same conclusion was reached for ultraclean $p$-GaAs/AlGaAs heterostructures with very low carrier densities (down to $7\times10^9$~cm$^{-2}$; $r_s\approx24$), suggesting that the effects of coherent backscattering are suppressed at such low densities and, as a consequence, in the presence of strong interactions \cite{mills2001suppression}. Relatively far from the transition, the corrections to the conductivity due to coherent backscattering are consistent with their non-interacting values, as was reported earlier \cite{brunthaler2001exclusion}. However, once the transition is approached ($\delta n_s/n_c\lesssim0.1$; $r_s\approx20$), the corrections become progressively weaker, and in the immediate vicinity of the critical density, they practically vanish.

An alternative theoretical explanation for the decrease in the weak-localization prefactor as the metal-insulator transition (MIT) is approached was proposed in Ref.~\cite{minkov2004magnetoresistance}. This explanation is based on the failure of the conventional Hikami-Larkin first-order expansion in \(1/G\) as the dimensionless conductance \(G\) decreases. Including second-order corrections has been shown to reduce the weak-localization prefactor.

\section{Equivalence of two approaches for determining the critical point in zero magnetic field}

To determine the limiting behavior at zero temperature experimentally, one can follow two principal approaches: extrapolating the temperature dependence of the conductivity or resistance to $T=0$ or analyzing temperature-independent characteristics (e.g., the correlation length). Since zero temperature is inaccessible, comparing different criteria provides an experimental test to assess whether a true MIT exists. As previously mentioned, the separatrix between metal and insulator in high-mobility silicon and $p$-GaAs samples remains either slightly tilted or flat down to the lowest accessible temperatures, with no signs of divergence. It is tempting to extrapolate the separatrix to $T=0$ and identify the corresponding electron density as the critical density for the metal-insulator transition. However, like any extrapolation to \( T=0 \), this approach should be justified. Therefore, an independent determination of the critical point is required. This is why in Ref.~\cite{shashkin2001metal}, the critical density obtained by an extrapolation criterion was compared with that obtained by a temperature-independent method.

\begin{figure}
\centering
\includegraphics[width=7 cm]{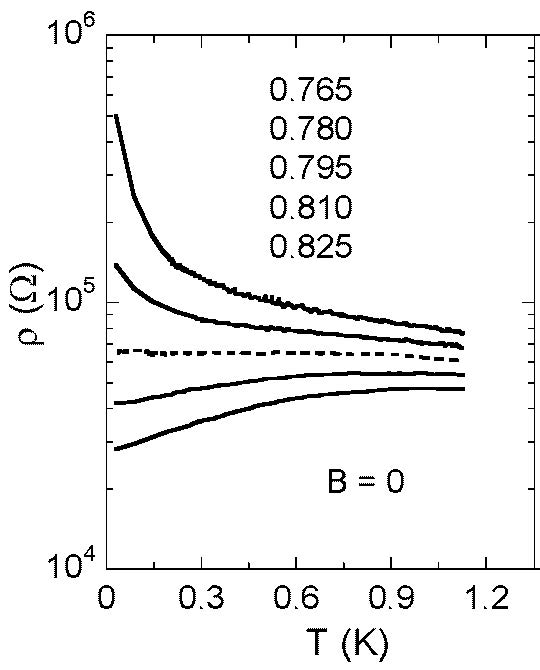}
\caption{Temperature dependence of the resistivity of a Si MOSFET at different electron densities near the MIT in zero magnetic field. The electron densities are indicated in units of $10^{11}$~cm$^{-2}$. From Ref.~\cite{shashkin2001metal}.}\label{Equivalence 1}
\end{figure}

The first criterion based on the extrapolation to zero temperature is a change in sign of the temperature derivative of the resistivity, $d\rho/dT$. A positive (negative) sign of the derivative at the lowest achievable temperatures is empirically associated with a metallic (insulating) phase. The second criterion is based on a vanishing activation energy, $E_a$, combined with a vanishing nonlinearity (threshold voltage, $V_c$) of the current-voltage characteristics when extrapolated from the insulating phase \cite{shashkin1994insulating}. The activation energy and the threshold voltage are connected via the localization length, which is temperature-independent. While the derivative method deals with the vicinity of the MIT in which the dependence $\rho(T)$ is flat, the $I-V$ method is related to the insulating phase with an exponential $\rho(T)$. These two methods have been applied simultaneously to the 2D MIT in Ref.~\cite{shashkin2001metal}.

The resistivity as a function of temperature is shown in Fig.~\ref{Equivalence 1} for several electron densities on both sides of the metal-insulator transition. The resistivity of the middle (dashed) curve shows virtually no temperature dependence over a wide range of temperatures and separates those with positive and negative $d\rho/dT$ nearly symmetrically (at temperatures above 0.2~K) as reported earlier \cite{simonian1997reflection,kravchenko2000metallic}. Assuming that the middle curve remains flat down to $T=0$, we obtain the MIT critical point at $0.795\times 10^{11}$~cm$^{-2}$ which corresponds to a resistivity $\rho\approx 3h/e^2$, as in other experiments on Si MOSFETs \cite{sarachik1999novel,kravchenko2000metallic}. We designate the so-determined electron density as the critical density, $n_{c1}$.

\begin{figure}
\centering
\includegraphics[width=7 cm]{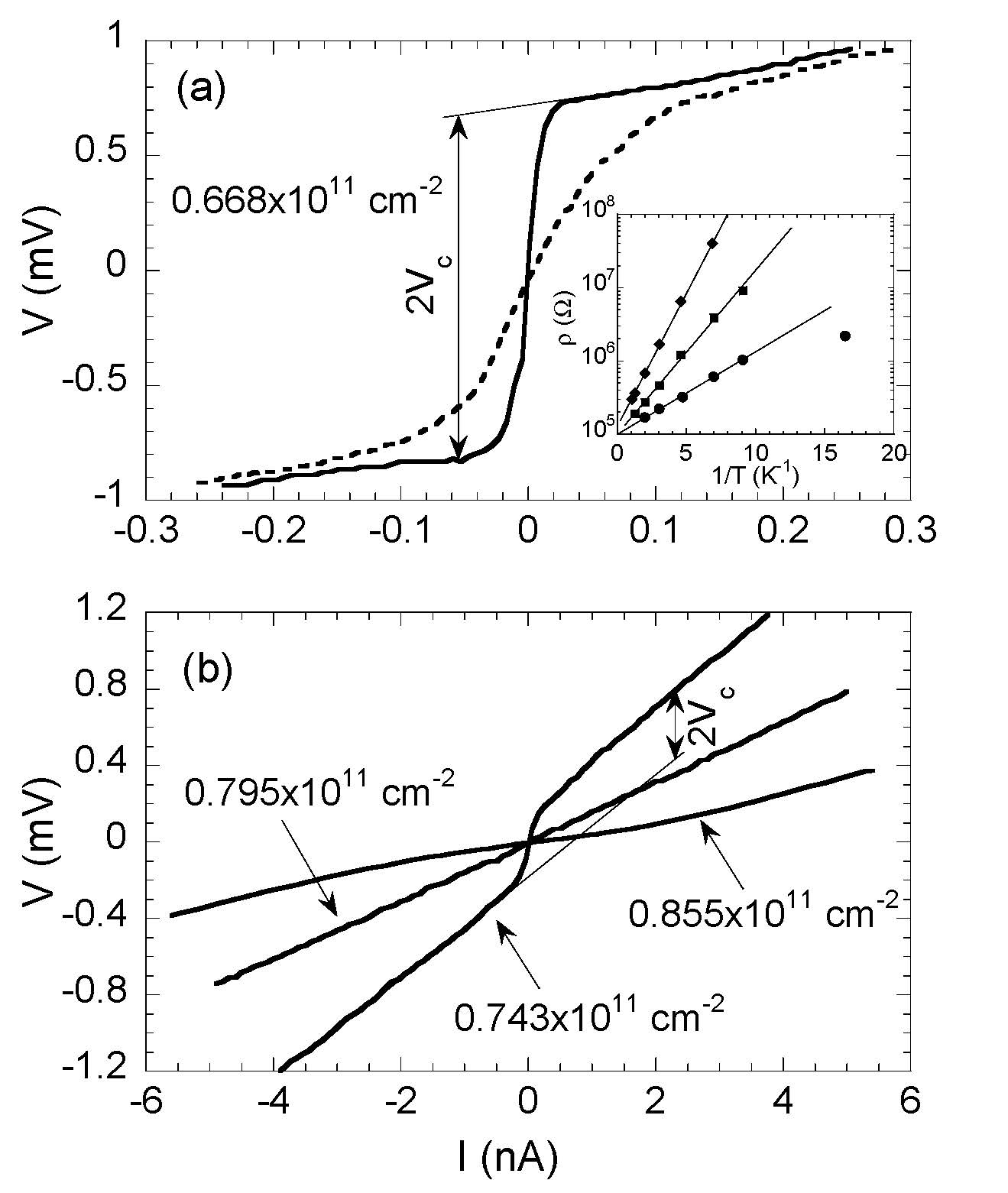}
\caption{Current-voltage characteristics at a temperature of $\approx 30$~mK in zero magnetic field. (The aspect ratio is equal to 2.4 for this sample.) In case (a), the $I-V$ curve obtained at a higher temperature (211~mK; dashed line) is also shown for comparison; note that the threshold voltage is practically temperature-independent. An Arrhenius plot of the resistivity in the insulating phase is displayed in the inset for $n_s=0.741\times10^{11}$~cm$^{-2}$.  Adapted from Ref.~\cite{shashkin2001metal}.}\label{Equivalence 2}
\end{figure}

\begin{figure}
\centering
\includegraphics[width=8 cm]{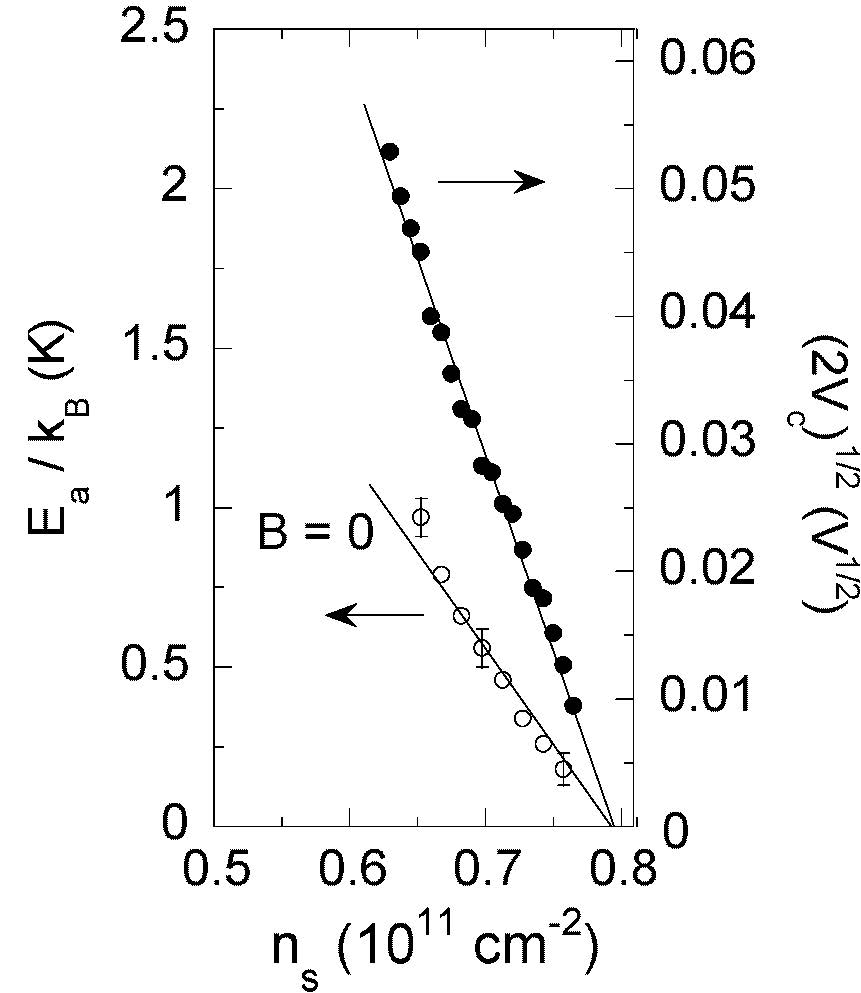}
\caption{Activation energy (open circles) and square root of the threshold voltage (closed circles) as a function of electron density in a zero magnetic field.  Adapted from Ref.~\cite{shashkin2001metal}.}\label{Equivalence 3}
\end{figure}

The MIT point has also been determined by studying the behavior of nonlinear $I-V$ characteristics on the insulating side of the transition. Deep in the insulating state ($n_s<n_{c1}$), a typical low-temperature $I-V$ curve is close to a step-like function: the voltage rises abruptly at low current and then saturates, as shown in Fig.~\ref{Equivalence 2}. The magnitude of the step is $2\, V_c$. At higher temperatures, the curve becomes less sharp, yet the threshold voltage, $V_c$, remains practically unchanged. Closer to the MIT, the $I-V$ curves still show a nonlinear step-like behavior provided that $n_s<n_{c1}$ (see the curve in Fig.~\ref{Equivalence 2}(b) corresponding to $n_s=0.743\times 10^{11}$~cm$^{-2}$). Exactly at $n_s=n_{c1}=0.795\times 10^{11}$~cm$^{-2}$, the $I-V$ curve is strictly linear; at $n_s>n_{c1}$, it becomes superlinear, which is characteristic of the metallic state \cite{kravchenko1996electric}. Figure~\ref{Equivalence 3} (closed circles) shows that the square root of the threshold voltage is a linear function of electron density. Extrapolation of the $V_c^{1/2}(n_s)$ dependence to zero threshold value yields the critical electron density $n_{c2}$. Note that $n_{c2}$ is equal to $n_{c1}$ with high accuracy.

Not too deep on the insulating side of the transition, and at not too low temperatures, the resistance has an activated form \cite{shashkin1994insulating, diorio1992reentrant}, as shown in the inset to Fig.~\ref{Equivalence 2}(a). (At lower temperatures, the activated $\rho(T)$ transforms into the Efros-Shklovskii hopping \cite{efros1975coulomb}. Note also that in the immediate vicinity of the transition, $\rho(T)$ dependences are not activated but rather reflection-symmetric about the transition.) Such a form of temperature dependence was interpreted in Ref.~\cite{adkins1976the} as a result of the thermal activation of carriers to the mobility edge, $E_c$; in this case, the activation energy is $E_a=E_c-E_F$. Since in the insulating regime, the $I-V$ curves are strongly nonlinear, the resistivity was determined from $dV/dI$ in the linear interval of $I-V$ curves, {\it i.e.}, at $I\rightarrow0$. Figure~\ref{Equivalence 3} shows $E_a$ as a function of the electron density (open circles). In previous detailed studies \cite{shashkin1994insulating}, this dependence was found to be linear near the mobility edge (deeper in the insulating state, the density of states decreases as compared to its value in the vicinity of the transition, which results in an increase in the slope of the $E_a(n_s)$ dependence). The present data can also be approximated by a linear function which yields, within the experimental uncertainty, the same critical electron density $n_{c2}$ as the square root of the threshold voltage. The reciprocal slope of $E_a(n_s)$, $D^*$, can be interpreted \cite{adkins1976the} as the thermodynamic density of states near the transition point.

The breakdown occurs when the localized electrons at the Fermi level gain enough energy to reach the mobility edge in an electric field, $V_c/d$, over a distance given by the localization length, $L$: $eV_cL/d=E_c-E_F$, where $d$ is the distance between the potential probes \cite{shashkin1994insulating}. The values $E_a$ and $V_c$ are related through the localization length, which is temperature-independent and diverges near the transition as $L(E_F)\propto (E_c-E_F)^{-s}$ with exponent $s$ close to unity \cite{shashkin1994insulating}. This corresponds to a linear dependence $V_c^{1/2}(n_s)$ \cite{pudalov1993zero, pudalov1994character, shashkin1994insulating} near the MIT, as seen in Fig.~\ref{Equivalence 3}.

We stress that, in a zero magnetic field, both methods --- extrapolating $\rho(T)$ to zero temperature and using the behavior of the temperature-independent localization length --- yield the same critical electron density. This adds confidence that the curve with zero derivative $d\rho/dT$ will remain ``flat'' (or at least will retain a finite resistivity value) down to zero temperature.  It is important to note that in Ref.~\cite{pudalov1994character}, the density dependence of the threshold electric field, given by \( E_t \propto (\delta n_s)^{3/2} \), was explained within the framework of the pinned Wigner solid based on the quantum tunneling of dislocations. For more details on the nonlinear threshold transport in the pinned Wigner solid, see section \ref{sec:WS}.

\section{The effect of a parallel magnetic field}

\begin{figure}
\centering
\includegraphics[width=8 cm]{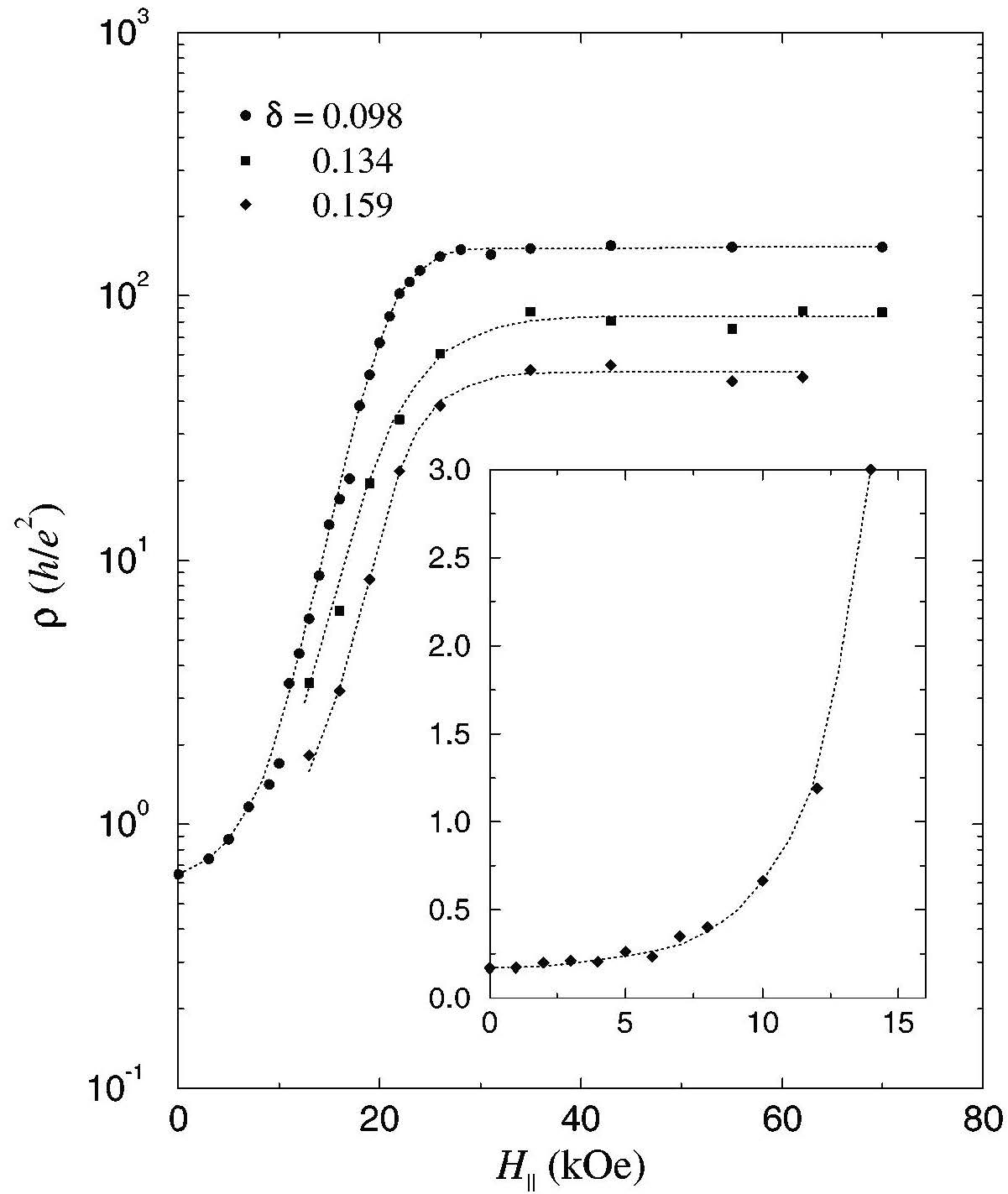}
\caption{Resistivity of a Si MOSFET sample on a logarithmic scale as a function of a magnetic field applied parallel to the plane at $T=250$~mK for three electron densities corresponding to (top to bottom) $\delta n_s=0.78, 1.07,\, \rm{and }\, 1.27 \times10^{10}$~cm$^{-2}$. The inset shows the resistivity on a linear scale in a low magnetic field for $\delta n_s=1.2\times10^{10}$~cm$^{-2}$ and $T = 130$~mK. From Ref.~\cite{simonian1997magnetic}.}\label{magnetoresistance Si}
\end{figure}

\begin{figure}
\centering
\includegraphics[width=8 cm]{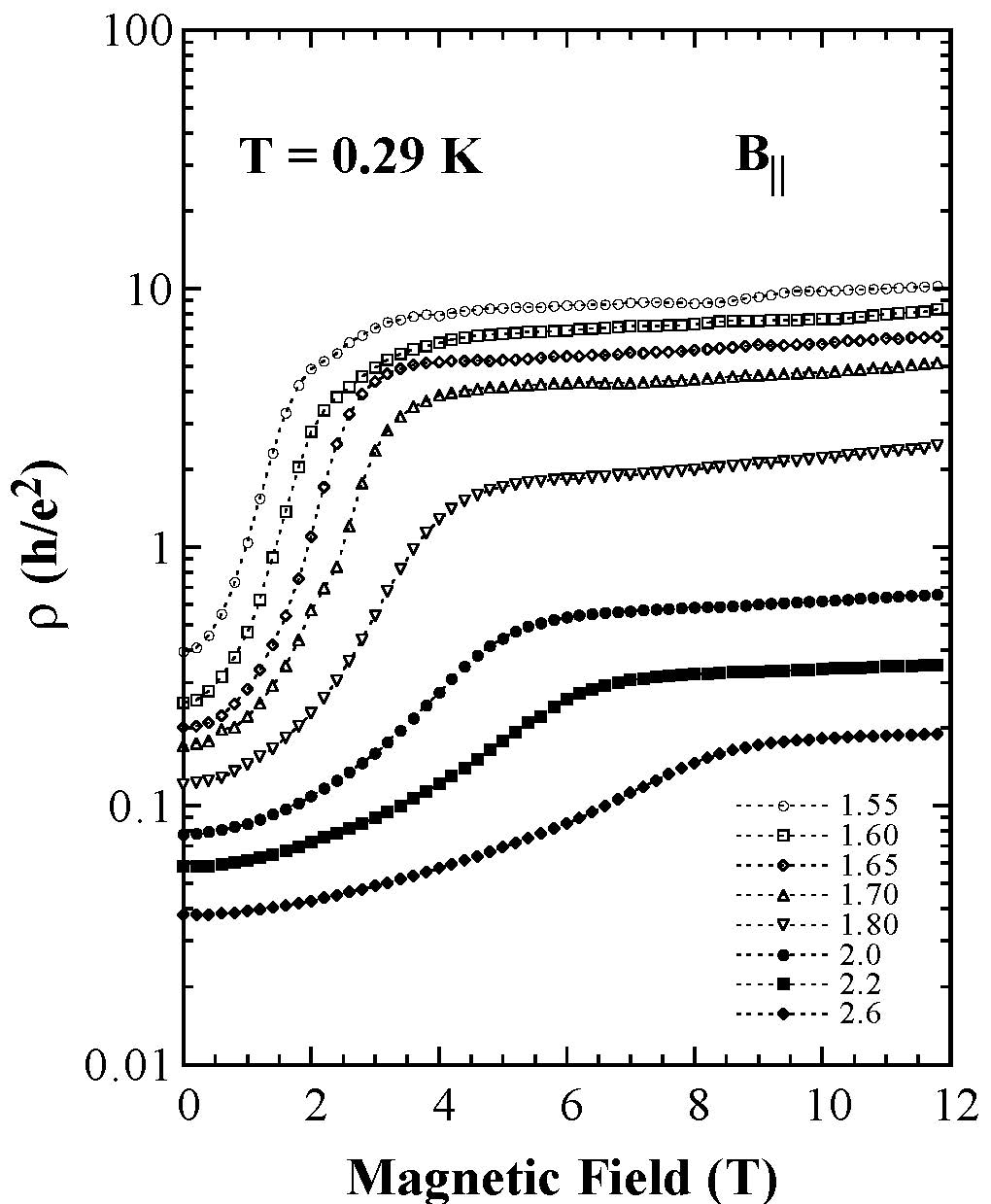}
\caption{Resistivity vs.\ parallel magnetic field, measured at $T=0.29$~K in a Si MOSFET sample.  Different symbols correspond to densities from 1.01 to 2.17 in units of $10^{11}$~cm$^{-2}$. From Ref.~\cite{pudalov1997instability}.}\label{magnetoresistance Pudalov}
\end{figure}

We now consider the effect of a field $B_\parallel$ applied parallel to the 2D plane. A parallel field couples only to electron spins and not to their orbital motion (provided that the fields are not so high that the magnetic length becomes comparable to the thickness of the 2D system). The first observation of a suppression of conductivity in low-density Si MOSFETs by $B_\parallel$ was reported in Ref.~\cite{dolgopolov1992properties}, in which the cut-off criterion (the resistivity either smaller or higher than 100 k$\Omega$) was used to separate the metallic and insulating states on the ($B,n_s$) plane. Magnetoresistance in a parallel magnetic field was studied in detail in Refs.~\cite{simonian1997magnetic, pudalov1997instability} in Si MOSFETs (see Figs.~\ref{magnetoresistance Si}, \ref{magnetoresistance Pudalov}).  With increasing $B_\parallel$, the resistance increases by a factor of a few deep in the metallic state and then saturates at a constant value at a density-dependent field $B_{sat}$ (see three bottom curves in Fig.~\ref{magnetoresistance Pudalov}), but when the application of $B_\parallel$ promoted the insulating state, the resistance increased by orders of magnitude before saturating (Figs.~\ref{magnetoresistance Si},\;\ref{magnetoresistance Pudalov}).  Later, in Refs.~\cite{okamoto1999spin, vitkalov2000small}, it was shown that $B_{sat}$ corresponds to the onset of a complete spin polarization: measurements in magnetic fields applied at small angles relative to the 2D plane indicated a factor of two increase in the frequency of Shubnikov-de~Haas (SdH) oscillations at $B\geq B_{sat}$.

\begin{figure}
\centering
\includegraphics[width=8 cm]{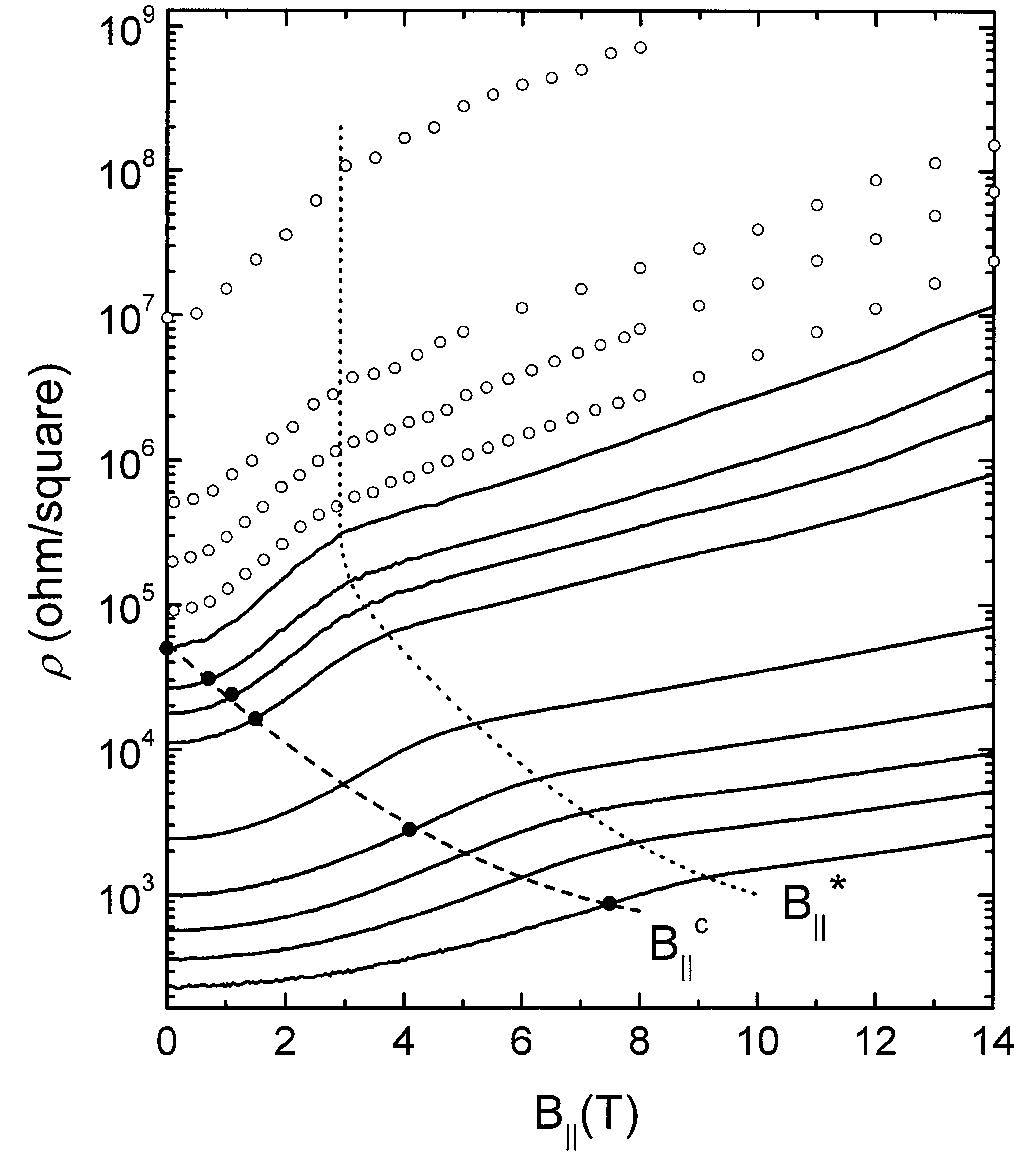}
\caption{The $B_\parallel$ dependence of $\rho$ in $p$-GaAs/AlGaAs heterostructure at 50~mK and hole densities $p=$~4.11, 3.23, 2.67, 2.12, 1.63, 1.10, 0.98, 0.89, 0.83, 0.79, 0.75, 0.67, and 0.57 in units of $10^{10}$~cm$^{-2}$ from the bottom. The solid lines are for $p$ above the critical density for the metal-insulator transition, $p_c$, and the open circles are for $p<p_c$.  The solid circles denote measured $B_\parallel^c$, the magnetic field below which the temperature dependences $\rho(T)$ are metallic and above which they are insulating.  The dashed line is a guide to the eye.  $B_\parallel^*$, the complete spin-polarization field, is marked by the dotted line. From Ref.~\cite{yoon2000parallel}.}\label{magnetoresistance GaAs}
\end{figure}

The effect of a parallel magnetic field is qualitatively similar in $p$-GaAs/AlGaAs heterostructures \cite{yoon2000parallel,papadakis2000anisotropic}.  An example is shown in Fig.~\ref{magnetoresistance GaAs}.  However, $\rho(B_\parallel)$ does not saturate to a constant value as in Si MOSFETs, but instead continues to increase with increasing field, albeit at a considerably slower rate.  As in Si MOSFETs, a distinct knee demarcates behavior in low and high fields.  Unlike in Si MOSFETs, magnetoresistance in $p$-GaAs/AlGaAs heterostructures is anisotropic, depending on the relative orientations of the measuring current, magnetic field, and crystal axes.  These effects were studied in detail in Ref.~\cite{papadakis2000anisotropic}.  In this system, there may be a contribution to the magnetoresistance anisotropy depending upon the angle between the current and the magnetic field due to the finite thickness of the 2D layer, as proposed in Ref.~\cite{dassarma2000parallel}.

\begin{figure}
\centering
\includegraphics[width=8 cm]{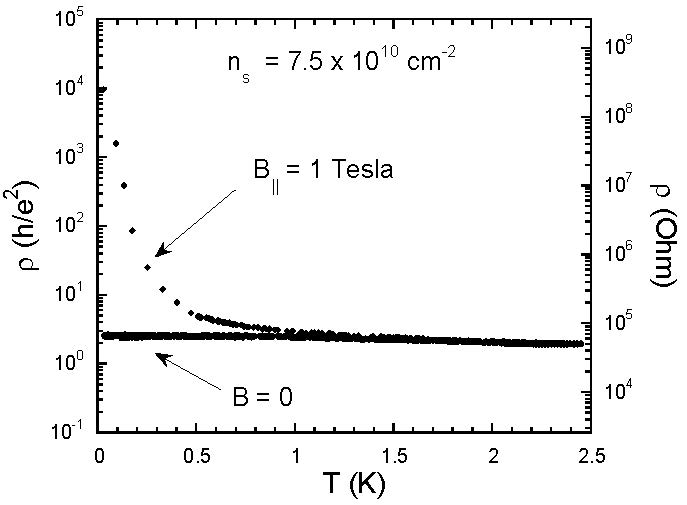}
\caption{For the electron density corresponding to the critical density $n_s=n_c=7.5\times10^{10}$~cm$^{-2}$ in zero field, the resistivity of a silicon MOSFET is shown as a function of temperature in zero field and in a parallel magnetic field of 1~tesla. From Ref.~\cite{abrahams2001metallic}.}\label{localization B}
\label{magnetorsistance Si}
\end{figure}

As shown in Fig.~\ref{localization B}, a parallel magnetic field suppresses the metallic behavior and eventually turns the zero-field metal into a high-field insulator for densities up to at least $1.4\;n_c$ \cite{abrahams2001metallic}.  The separatrix resistance, which is temperature-independent and close to $3h/e^2$ in zero field, acquires a strongly (by orders of magnitude) insulating temperature dependence at $T\lesssim1$~K in a parallel field of only 1~tesla. Thus, the localized behavior, which is absent at $n_s\geq n_c$ in $B=0$, is ``restored'' in a parallel magnetic field \cite{dolgopolov1992properties,simonian1997magnetic,pudalov1997instability,yoon2000parallel,papadakis2000anisotropic}.  In contrast, the same field has a negligible effect on the resistance at temperatures higher than 1~K where the resistance can be described by the Drude formula.

If the electron densities significantly exceed $n_c$, where $R<<h/e^2$ at $B=0$, the complete spin polarization causes the increase of the resistance by a factor of 4 \cite{dolgopolov2000magnetoresistance}, not leading to localization of electrons as long as $R(B_{sat})$ remains below $h/e^2$.  In general, the critical electron density for the Anderson MIT increases with the parallel magnetic field due to the effects of exchange and correlations \cite{dolgopolov2017spin}. This is in agreement with the conclusions of Refs.~\cite{punnoose2001dilute, punnoose2005metal, fleury2008many} where it was shown that the increase in the number of spin and valley degrees of freedom promotes the metallicity.

Finally, we would like to mention that the metallic temperature dependence of the resistance, although weaker than that in the $B=0$ case, still survives in a completely spin-polarized (``spinless''), two-valley electron system in ultra-clean SiGe/Si/SiGe heterostructures \cite{melnikov2020metallic}. This is in contrast to the above-described results obtained on more disordered two-valley silicon samples, in which a spin-polarizing magnetic field completely quenches metallic temperature behavior.  The fact that the spinless electrons behave differently in ultra-clean systems in SiGe/Si/SiGe quantum wells as compared to those in Si MOSFETs can be attributed to different intervalley scattering rates: while the intervalley scattering due to the irregularities of the interface in Si MOSFETs mixes the two valleys to effectively produce a single valley at low temperatures, the level of short-range disorder potential in SiGe/Si/SiGe samples is some two orders of magnitude lower than that in the least disordered Si MOSFETs, hence the intervalley scattering rate is small compared to that in Si MOSFETs, corresponding to the case of two distinct valleys. The critical electron density, $n_c(B_{sat})$, for the MIT in the spinless electron system in SiGe/Si/SiGe quantum wells is higher by a factor of about 1.2 compared to $n_c(0)$, which is consistent with theoretical calculations \cite{dolgopolov2017spin}.  The observed metallic temperature behavior in spinless SiGe/Si/SiGe systems is comparable to that in strongly interacting, spin-unpolarized single-valley 2D systems in the cleanest $p$-type GaAs/AlGaAs heterostructures, which indicates the same role of spins and distinct valleys with respect to the existence of the metallic state and the MIT.

\section{The effect of quantizing perpendicular magnetic fields}

\begin{figure}
\centering
\includegraphics[width=8 cm]{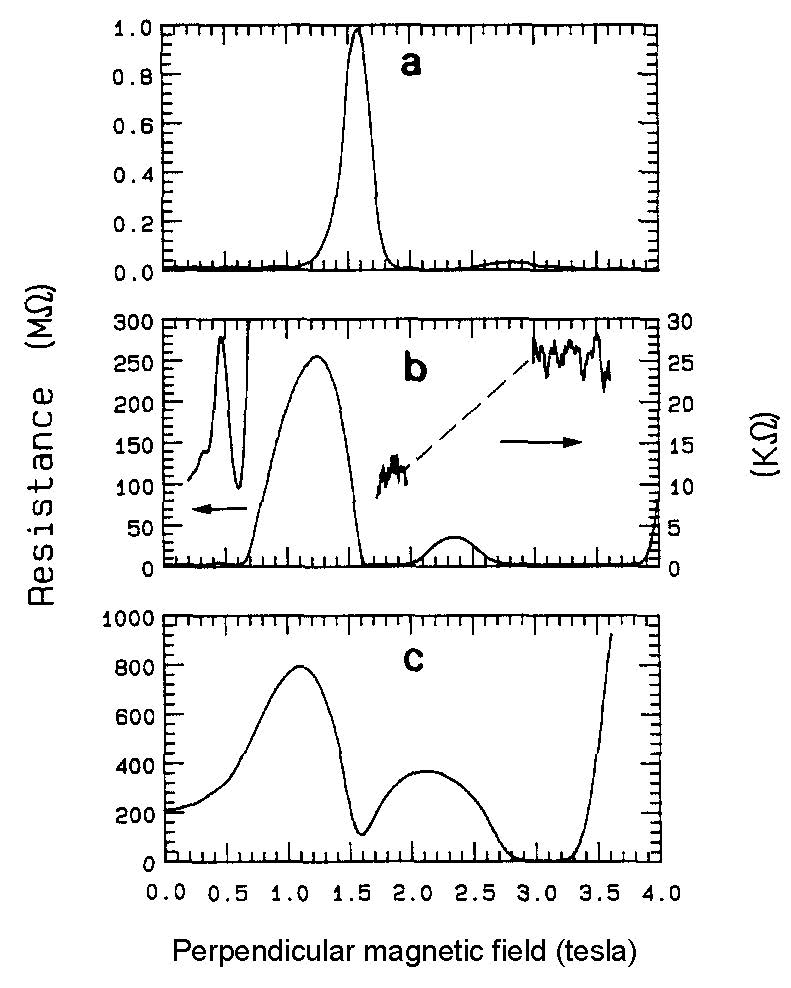}
\caption{Magnetoresistance of a highly resistive dilute 2D gas in a perpendicular magnetic field in a Si MOSFET. $T=18$~mK.  Electron densities are 9.4~(a), 8.2~(b), and 7.2~(c) in units of $10^{10}$~cm$^{-2}$. The insets in (b) show the development and growth of novel peaks around $\nu=7$ magnified by a factor of 30 (left) and the $R_{xy}$ (right).  From Ref.~\cite{diorio1990magnetic}.}
\label{Diorio 1990}
\end{figure}

\begin{figure}
\centering
\includegraphics[width=7 cm]{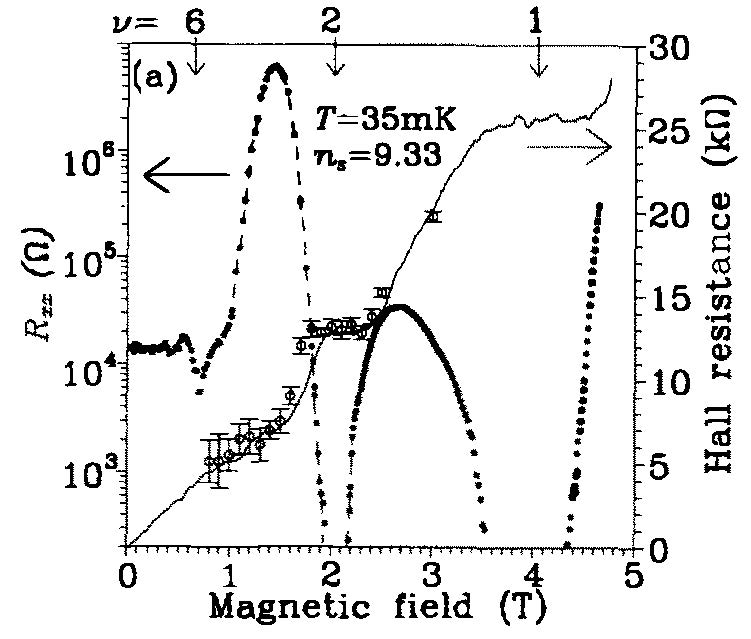}
\includegraphics[width=7 cm]{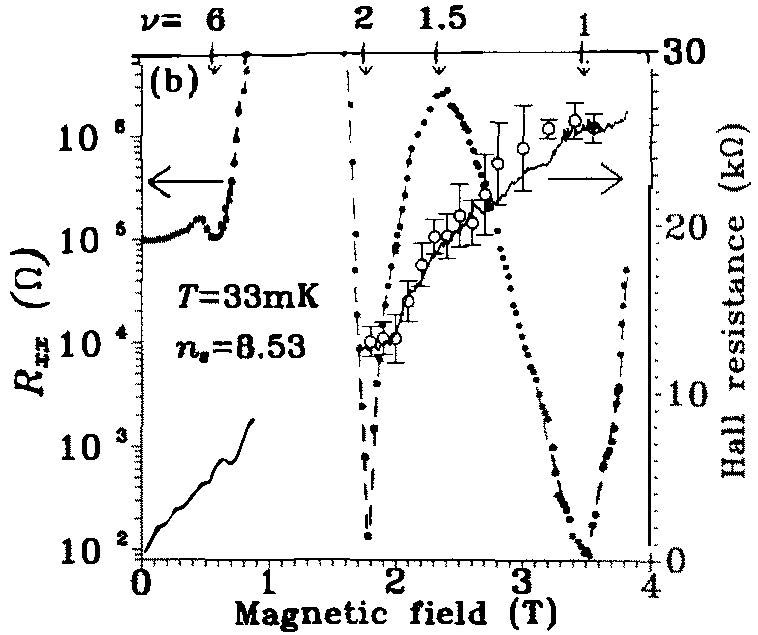}
\caption{Diagonal and Hall resistances as a function of a magnetic field in the reentrant insulating regime for electron densities $9.326\times10^{10}$~cm$^{-2}$~(a) and $8.53\times10^{10}$~cm$^{-2}$~(b).  Full circles represent $R_{xx}$; solid curves represent $R_{xy}$ measured at 10~nA. Open circles with error bars represent $R_{xy}$, measured at 0.25 nA. Arrows indicate the positions of the filling factors 6, 2, and 1. The dashed lines are a guide to the eye. From Ref.~\cite{pudalov1994termination}.}\label{Pudalov Surface Science}
\end{figure}

\begin{figure}
\centering
\includegraphics[width=8 cm]{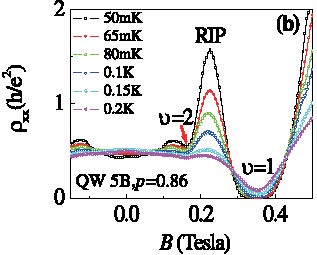}
\caption{Magnetoresistivity of a $p$-GaAs quantum well with $p=0.86\times10^{10}$~cm$^{-2}$ at different temperatures. The reentrant insulating phase (RIP) resides between the $\nu=1$ and 2 quantum Hall states. From Ref.~\cite{qiu2012connecting}.}
\label{Gao reentrant}
\end{figure}

In a strong perpendicular magnetic field, $B_\bot$, samples at electron densities below and just above $n_c$ exhibit rather complex magnetoresistance. As reported in Refs.~\cite{pudalov1994termination, diorio1990magnetic, diorio1992reentrant}, the diagonal resistance first increases strongly with $B_\bot$, sometimes by orders of magnitude (see Figs.~\ref{Diorio 1990} and \ref{Pudalov Surface Science}). At higher $B_\bot$, the magnetoresistance turns negative, and eventually the system exhibits giant resistance oscillations with deep minima at magnetic fields corresponding to Landau level filling factors $\nu=1$ and $2$ and a maximum in between (often referred to as the ``reentrant behavior''). However, the Hall resistance follows the classical dependence $R_{xy}(B)$ (the so-called Hall insulator \cite{viehweger1990the,kivelson1992global}) despite $R_{xx}$ reaching values exceeding $h/e^2$ by orders of magnitude between the integer filling factors 6 and 2 and between 2 and 1 (Fig.~\ref{Pudalov Surface Science}(a), (b)).  At $\nu=2$ and $\nu=1$, $R_{xy}$ takes on quantized integer quantum Hall effect values of $h/2e^2$ and $h/e^2$, respectively \cite{diorio1990magnetic, pudalov1999weak}. Similar features in $\rho_{xx}(B)$ dependences have been observed in $p$-GaAs quantum wells \cite{qiu2012connecting}, as shown in Fig.~\ref{Gao reentrant}.

\begin{figure}
\centering
\includegraphics[width=8 cm]{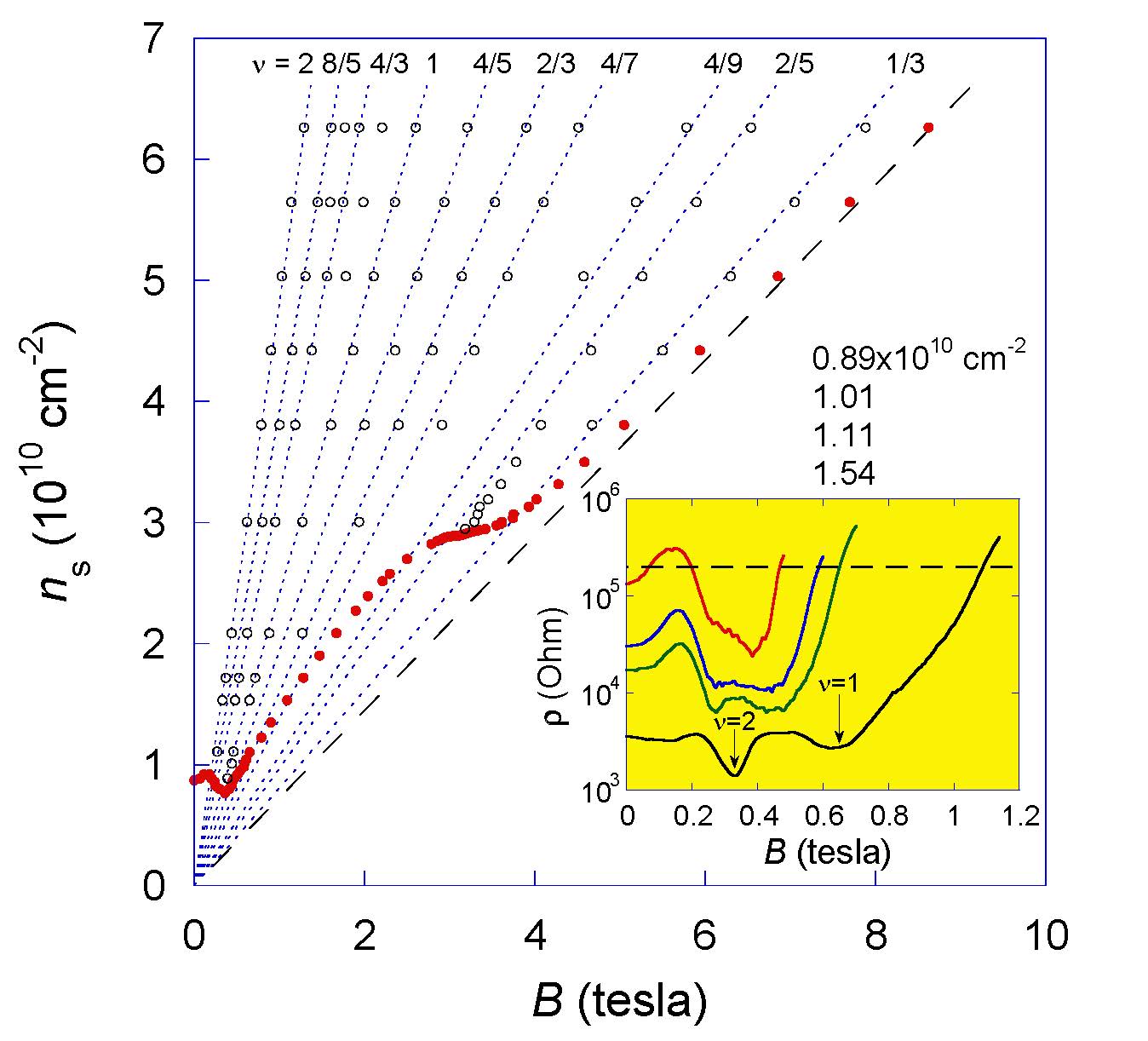}
\caption{Critical electron density for the MIT as a function of the magnetic field perpendicular to the 2D plane (solid red points) in a SiGe/Si/SiGe heterostructure. The dashed line indicates the slope of the $n_{c}(B)$ dependence in the high-field region. Open symbols and dotted lines correspond to the locations of the quantum Hall effect minima in this system. In the inset, $\rho(B)$ dependences used to define the phase boundary are plotted; the cut-off value of the resistivity $\rho=200$~k$\Omega$ is indicated by the long-dashed horizontal line. From Ref.~\cite{melnikov2019quantum}.}
\label{SiGe perpendicular field}
\end{figure}

The application of the perpendicular magnetic field affects the critical density of the MIT. Magnetic field dependences of the longitudinal resistivity in SiGe/Si/SiGe heterostructures are shown in the inset to Fig.~\ref{SiGe perpendicular field}. The resistivity minimum at the Landau level filling factor $\nu=n_{s}hc/eB=1$ survives down to electron densities near the MIT. This is similar to the re-entrant behavior that was discussed in the previous paragraph, observed earlier in Si MOSFETs and GaAs/AlGaAs heterostructures \cite{diorio1990magnetic,dolgopolov1992metal,jiang1993observation,kravchenko1995global,qiu2012connecting}. The resulting phase diagram is shown in Fig.~\ref{SiGe perpendicular field}. The critical electron density increases with $B_\bot$ at low magnetic fields and then, at the Landau filling factor $\nu=1$, decreases to a value below that for $B=0$. At higher magnetic fields (in the extreme quantum limit), it monotonically grows and exhibits a knee at $\nu=2/5$. Indeed, in this electron system, the longitudinal resistance minimum at $\nu=2/5$ is stronger than that at $\nu=1/3$ (see Ref.~\cite{dolgopolov2018fractional}), in contrast to the strongly-interacting 2D hole system in a GaAs/AlGaAs heterostructure \cite{santos1992observation}. The boundary then continues to grow with a slope corresponding to $\nu\approx0.3$. This is different from the slope of $\nu\approx0.5$ observed in Si MOSFETs in the extreme quantum limit \cite{dolgopolov1992metal} and interpreted as a consequence of the localization of electrons below half-filling of the lowest Landau level. For comparison, the slope of the high-field boundary in $p$-type GaAs/AlGaAs heterostructures is intermediate (\textit{i.e.}, $\nu\approx0.38$; see Ref.~\cite{qiu2012connecting}). Based on the results obtained in these strongly interacting carrier systems, the authors of Ref.~\cite{melnikov2019quantum} arrived at the conclusion that the critical density for the MIT in the extreme quantum limit is likely to be determined by the level of disorder.

\section{Measurements of the renormalized electron parameters: spin susceptibility and the effective mass}

As discussed in the previous section, the behaviors of spin-unpolarized and spin-polarized strongly interacting 2D systems differ drastically. This underscores the essential role of electron spins in two-dimensional metal-insulator transitions and encourages the investigation of magnetic properties.

\subsection{Transport measurements}

In Ref.~\cite{diorio1990magnetic, diorio1992reentrant,kravchenko2000shubnikov}, Shubnikov-de~Haas oscillations were studied in Si MOSFET samples in perpendicular magnetic fields in a wide range of electron densities.  The authors reported that near the metal-insulator transition, only ``spin'' minima in the resistance are observed at Landau filling factors $\nu=$~2, 6, 10, and 14, whereas the ``cyclotron'' minima at $\nu=$~4, 8, and 12 disappear.  This behavior was observed in a rather broad range of magnetic fields.  It was interpreted as a giant (by a factor of $\sim5$) enhancement of the product, $g^* m^*$, of the $g$-factor and the effective mass near the metal-insulator transition.
\footnote[2]{Throughout this paper, we prefer to use quasiparticle parameters such as the renormalized effective mass $m^*$, spin susceptibility $\chi$, \textit{etc}., even though these parameters differ substantially from the band values in a strongly correlated electron system. This is done within a standard approach without discussing the applicability of the Fermi liquid concept when it can successfully describe the observed phenomena.}

\begin{figure}
\centering
\includegraphics[width=8 cm]{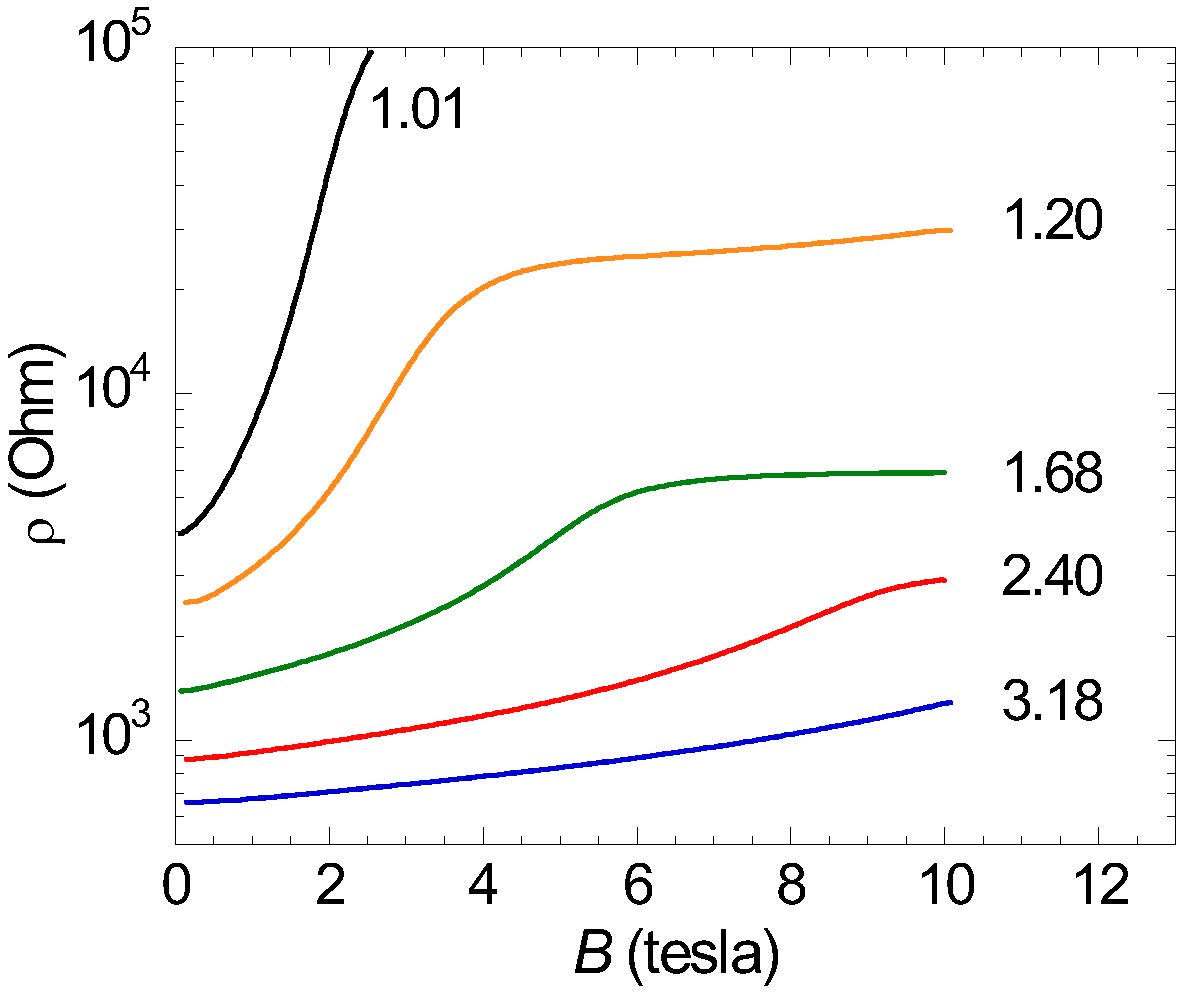}
\caption{Low-temperature magnetoresistance in parallel magnetic fields at different electron densities above the critical density for the $B=0$ metal-insulator transition. From Ref.~\cite{shashkin2001indication}.}
\label{BR}
\end{figure}

Indirect measurements of the spin susceptibility of 2D electron systems in Si MOSFETs were made in Ref.~\cite{shashkin2001indication}. The authors studied low-temperature magnetoresistance $\rho(B)$ in a parallel magnetic field (see Fig.~\ref{BR}) and found, in agreement with earlier measurements \cite{pudalov1997instability, simonian1997magnetic, okamoto1999spin, vitkalov2000small}, that the resistivity increases approximately quadratically with the field until it saturates at a constant value above a certain density-dependent magnetic field, $B_{sat}$, corresponding to the onset of a complete spin polarization. In the vicinity of the metal-insulator transition, magnetoresistance strongly depends on temperature, as reported, {\it e.g.}, in Ref.~\cite{simonian1998h}. However, as one moves away from the transition deep into the metallic state, this temperature dependence weakens and eventually disappears at very low temperatures. The data discussed in Ref.~\cite{shashkin2001indication} were obtained in this low-temperature limit, where the magnetoresistance becomes temperature-independent.

\begin{figure}
\centering
\includegraphics[width=8 cm]{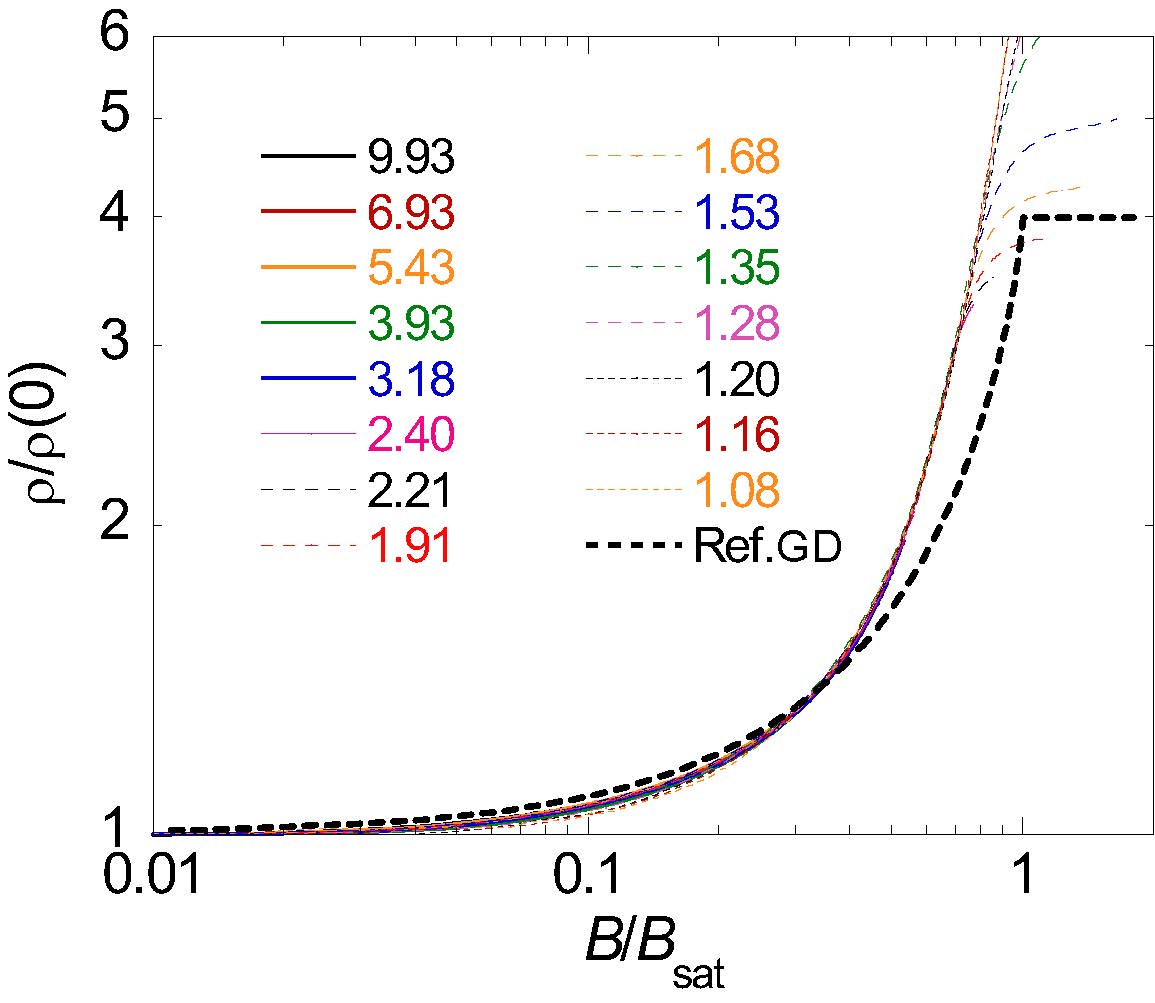}
\caption{Scaled curves of the normalized magnetoresistance at different $n_s$ vs.\ $B/B_{sat}$. The electron densities are indicated in units of $10^{11}$~cm$^{-2}$. Also shown by a dashed line is the normalized magnetoresistance calculated in
Ref.~\cite{dolgopolov2000magnetoresistance}. From Ref.~\cite{shashkin2001indication}.}
\label{Scaling B}
\end{figure}

In Fig.~\ref{Scaling B}, we illustrate how the magnetoresistance curves, shown in Fig.~\ref{BR}, collapse onto a single curve after being normalized by $\rho(0)$ and plotted as a function of $B/B_{sat}$. The scaling parameter, $B_{sat}$, has been normalized to the magnetic field at which the magnetoresistance saturates and, hence, the electrons reach full spin polarization \cite{okamoto1999spin,vitkalov2000small}. The observed scaling is remarkably good for $B/B_{sat}\le 0.7$ in the electron density range $1.08\times 10^{11}- 1\times10^{12}$~cm$^{-2}$, although with increasing $n_s$ the scaled experimental data occupy progressively shorter intervals on the resulting curve. At both $B/B_{sat}>0.7$ and outside the indicated range of electron densities, the scaled data start to noticeably deviate from the universal curve. In particular, the scaling breaks down as one approaches the metal-insulator transition ($n_s<1.3\,n_c$), which in this sample occurs at zero magnetic field at $n_c=8\times 10^{10}$~cm$^{-2}$. We note that the observed scaling dependence is described reasonably well by the theoretical dependence of $\rho/\rho(0)$ on the degree of spin polarization

\begin{equation}
\xi=g^* m^* \mu_BB/\pi\hbar^2n_s=B/B_{sat}
\end{equation}
(where both the $g$-factor $g^*$ and the effective mass $m^*$ may be enhanced due to interactions) as predicted in Ref.~\cite{dolgopolov2000magnetoresistance} based on the spin polarization-dependent screening of a random potential deep within the metallic regime.

\begin{figure}
\centering
\includegraphics[width=8 cm]{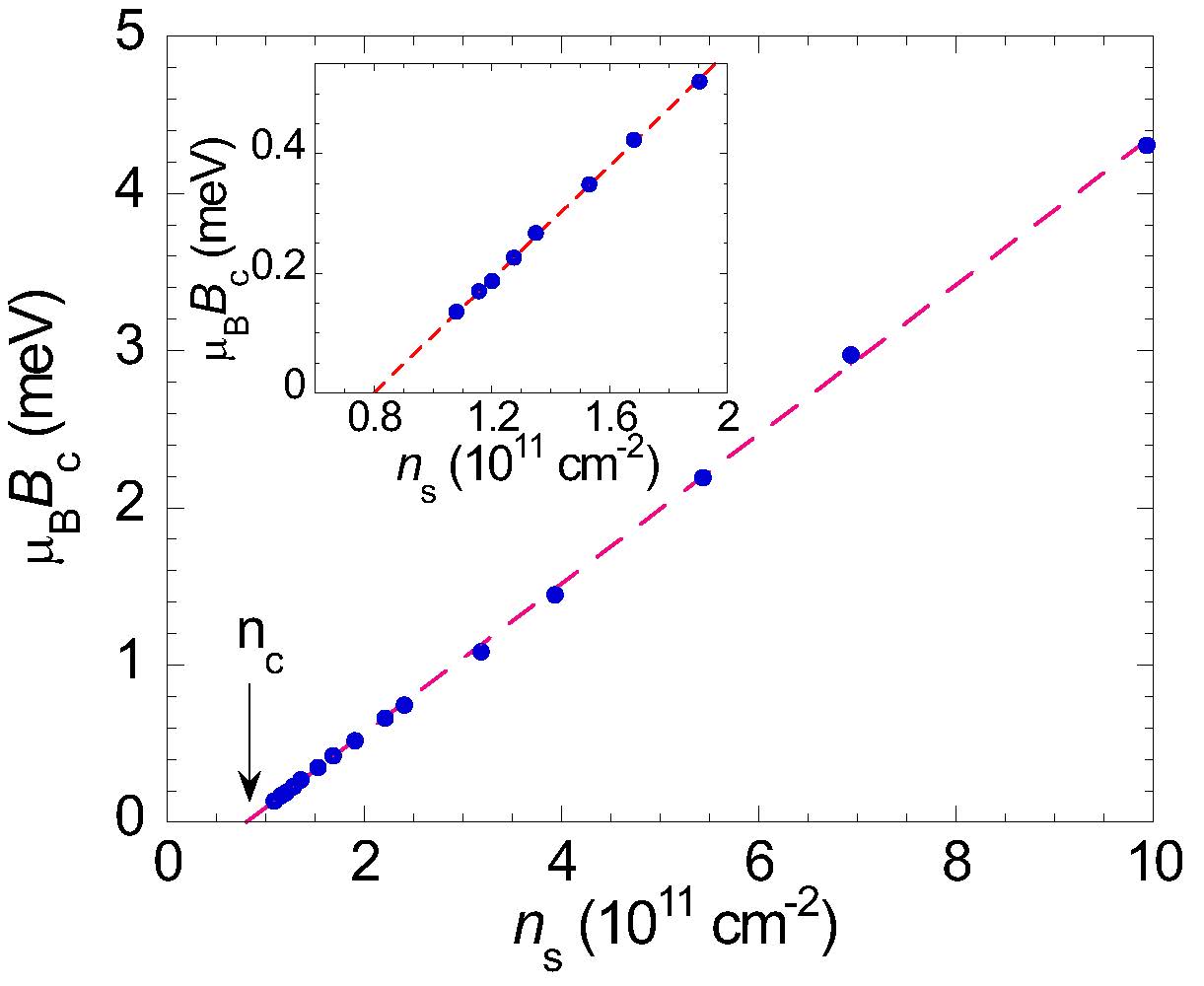}
\caption{Dependence of the field $B_{sat}$ on electron density in the least disordered samples. The dashed line is a linear fit that extrapolates to the critical electron density for the $B=0$ metal-insulator transition. A close-up view of the region near $n_c$ is displayed in the inset. Adapted from Ref.~\cite{shashkin2001indication}.}
\label{Bc}
\end{figure}

\begin{figure}
\centering
\includegraphics[width=8 cm]{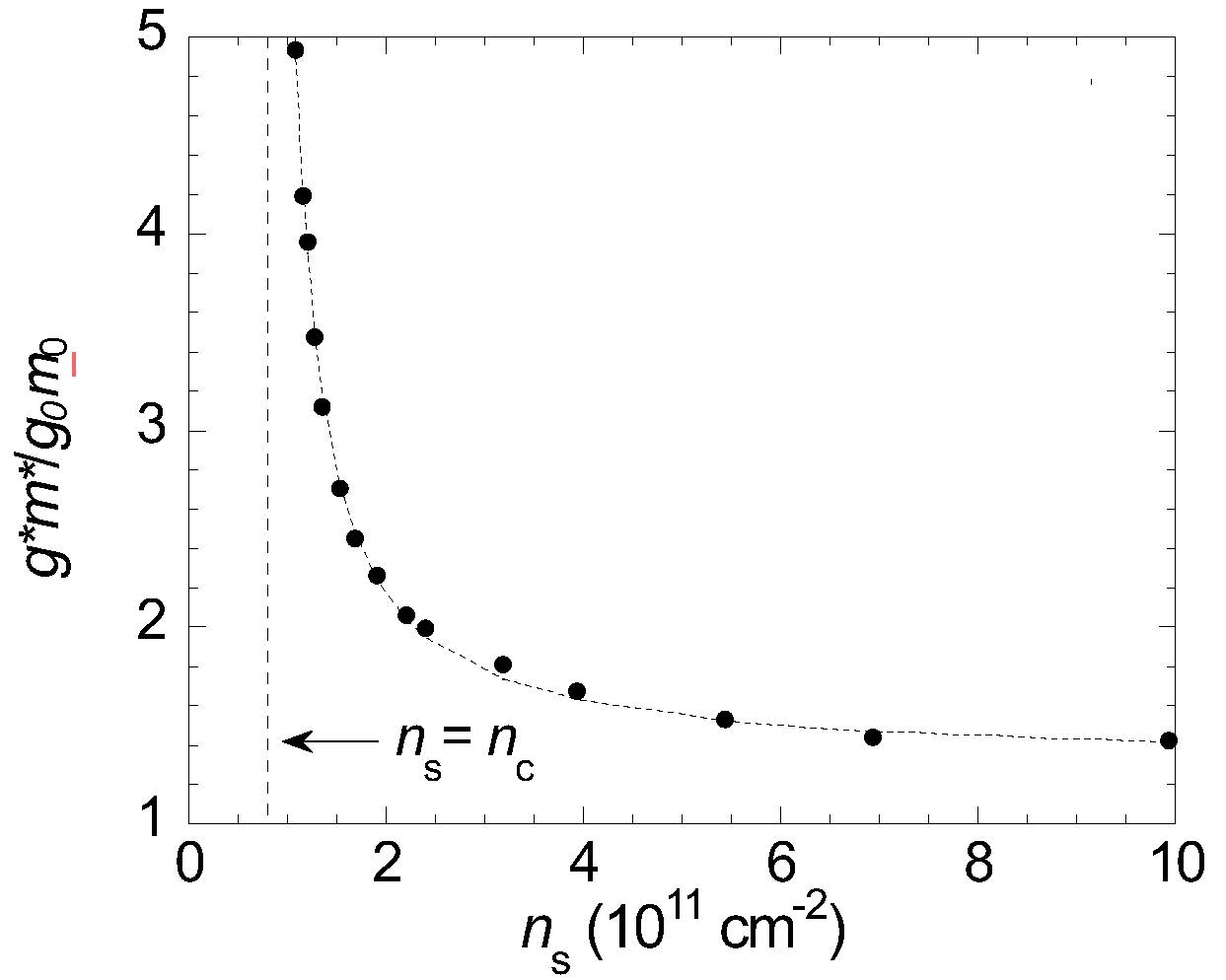}
\caption{The ratio $g^* m^*/g_0m_0$ vs. $n_s$ recalculated from the data shown in Fig.~\ref{Bc}. The dashed line is a guide to the eye. From Ref.~\cite{shashkin2001indication}.}
\label{gm}
\end{figure}

In Fig.~\ref{Bc}, it is shown that with high precision, $B_{sat}$ is proportional to the deviation of the electron density from its critical value, i.e., to $\delta n_s$, over a wide range of electron densities. In other words, the field at which the magnetoresistance saturates and the complete spin polarization is reached tends to vanish at $n_c$. Note that at $n_s$ above $2.4\times 10^{15}$~m$^{-2}$, the saturation of the resistance is not reached in the available magnetic field range. Still, the high precision of the collapse of the high-density experimental curves onto the same scaling curve as the low-density data allows one to draw conclusions about the validity of the obtained law $B_{sat}(n_s)\propto \delta n_s$ over a much wider range of electron densities.

The fact that the saturation (or polarization) field tends to vanish at $n_c$ would be trivial if the density of the delocalized electrons approached zero at the metal-insulator transition. However, the studied least-disordered samples are remarkable for the smallness of the band tail of localized electrons down to electron densities $n_s\approx1\times10^{11}$~cm$^{-2}$. This follows from the coincidence of the full spin polarization field obtained from parallel-field magnetotransport and that obtained from the analysis of the Shubnikov-de Haas oscillations, because the former is influenced by a possible band tail of localized electrons including singly-occupied states \cite{gold2002on} that leads to shifting the $B_{sat}(n_s)$ dependence to higher densities, whereas the latter is not \cite{gold2002on, pudalov2002weak, dolgopolov2002comment, kravchenko2004metal, shashkin2005metal}. In samples with a relatively high critical density $n_c$ for the $B = 0$ MIT, the field $B_{sat}$ decreases, and the band tail of localized electrons increases. In this case, the crucial region of low electron densities falls within the insulating regime where the physics of local moments dominates. Indeed, in those samples, a sublinear magnetic-field dependence of the magnetization characteristic of local moments has been predicted, and the extracted spin susceptibility is expected to have a Curie temperature dependence \cite{gold2002on}. This dependence holds even at high electron densities, where metallic behavior might be expected. Note that the band tail of localized electrons contributes to the measured magnetization owing to the presence of empty states below the Fermi level. 

Based on the data shown in Fig.~\ref{Bc} and the above form of the spin polarization parameter $\xi$, the ratio of the product $g^* m^*$ to $g_0m_0$ can be calculated (here $g_0=2$ and $m_0=0.19m_e$ is the band mass in Si). The dependence $g^* m^*/g_0m_0(n_s)$ is shown in Fig.~\ref{gm}.  Upon approaching the critical density $n_c$, the spin susceptibility increases by a factor of about five compared to its ``noninteracting'' value $2m_0$.

\begin{figure}
\centering
\includegraphics[width=8 cm]{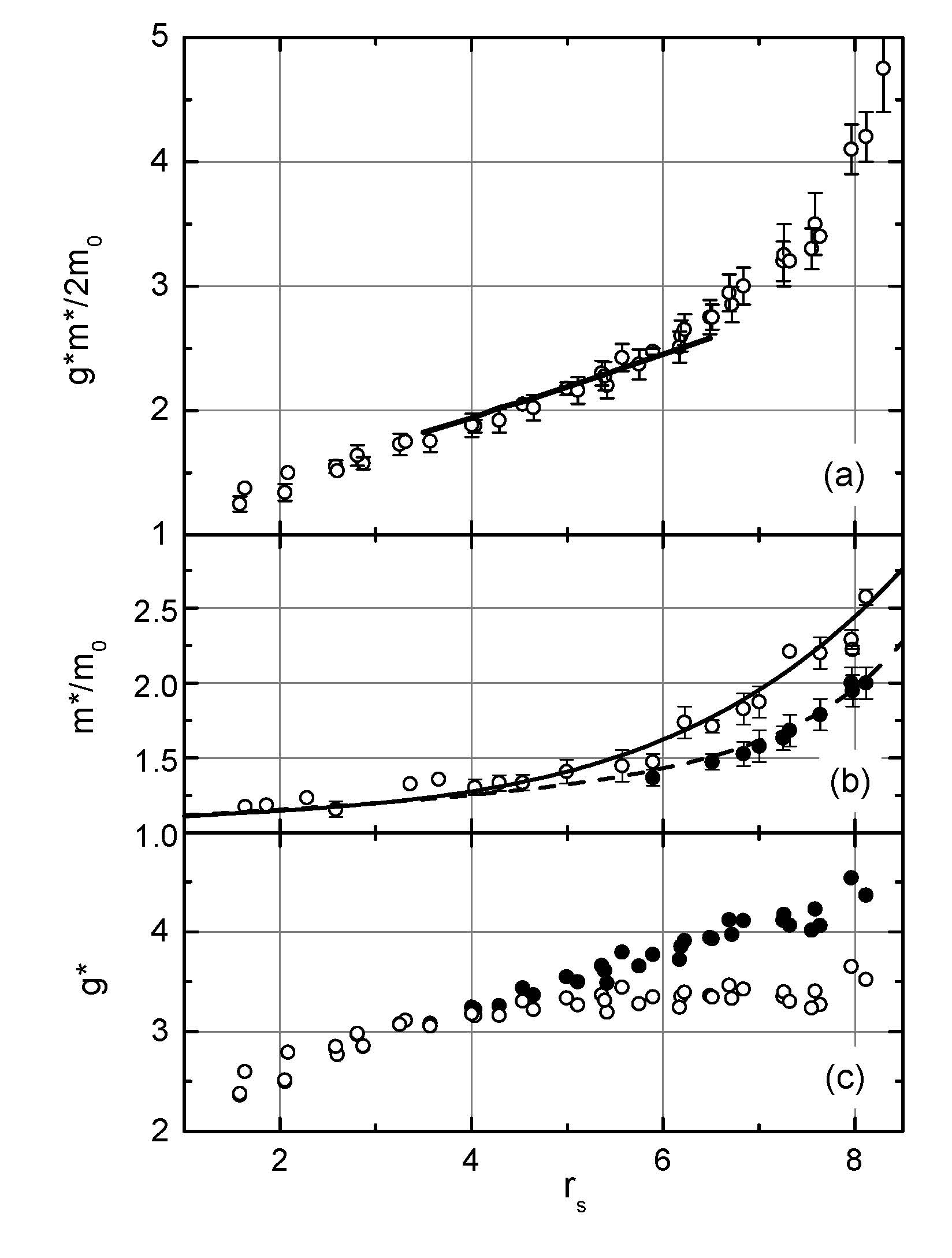}
\caption{$g^*m^*/g_0m_0$, $m^*$, and $g^*$ for different Si MOSFET samples as a function of $r_s$ (dots).  The solid line in (a) shows the data by Okamoto {\em et al.} \cite{okamoto1999spin}. The solid and open circles in (b) and (c) correspond to two different methods of finding $m^*$ described in detail in Ref.~\cite{pudalov2002low}. The solid and dashed lines in (b) are polynomial fits for the two dependences $m^*(r_s)$. The values of $g^*$ shown in (c) were obtained by dividing the $g^*m^*$ data by the smooth approximations of the experimental dependences $m^*(r_s)$ shown in (b). From Ref.~\cite{pudalov2002low}.}
\label{gmrs}
\end{figure}

Another approach to determining the spin susceptibility in Si MOSFET samples was proposed in Ref.~\cite{pudalov2002low}.  The authors measured Shubnikov-de Haas oscillations in superimposed, independently controlled parallel and perpendicular magnetic fields across a wide range of electron densities, including the vicinity of the MIT, and determined the spin susceptibility, the effective mass, and the $g$-factor of mobile electrons.  The results are displayed in Fig.~\ref{gmrs}.  For small $r_s$, $g^* m^*/g_0m_0$ values agree with the earlier data by Okamoto et al. \cite{okamoto1999spin}. For $r_s \geq6$, the ratio $g^*m^*/g_0m_0$ increases with $r_s$ faster than it might be expected from extrapolation of the results of Ref.~\cite{okamoto1999spin} and reaches $\sim5$, consistent with the results of Refs.~\cite{kravchenko2000shubnikov} and \cite{shashkin2001indication}.  Depending on two different methods for determining $m^*$, the renormalized effective mass grows near the MIT by a factor of $\approx2.5$ (open circles in Fig.~\ref{gmrs}) or $\approx2$ (closed circles) compared to its band value, and the renormalized $g$-factor grows by a factor of $\approx1.7$ or $\approx2.2$, respectively.

\begin{figure}
\centering
\includegraphics[width=8 cm]{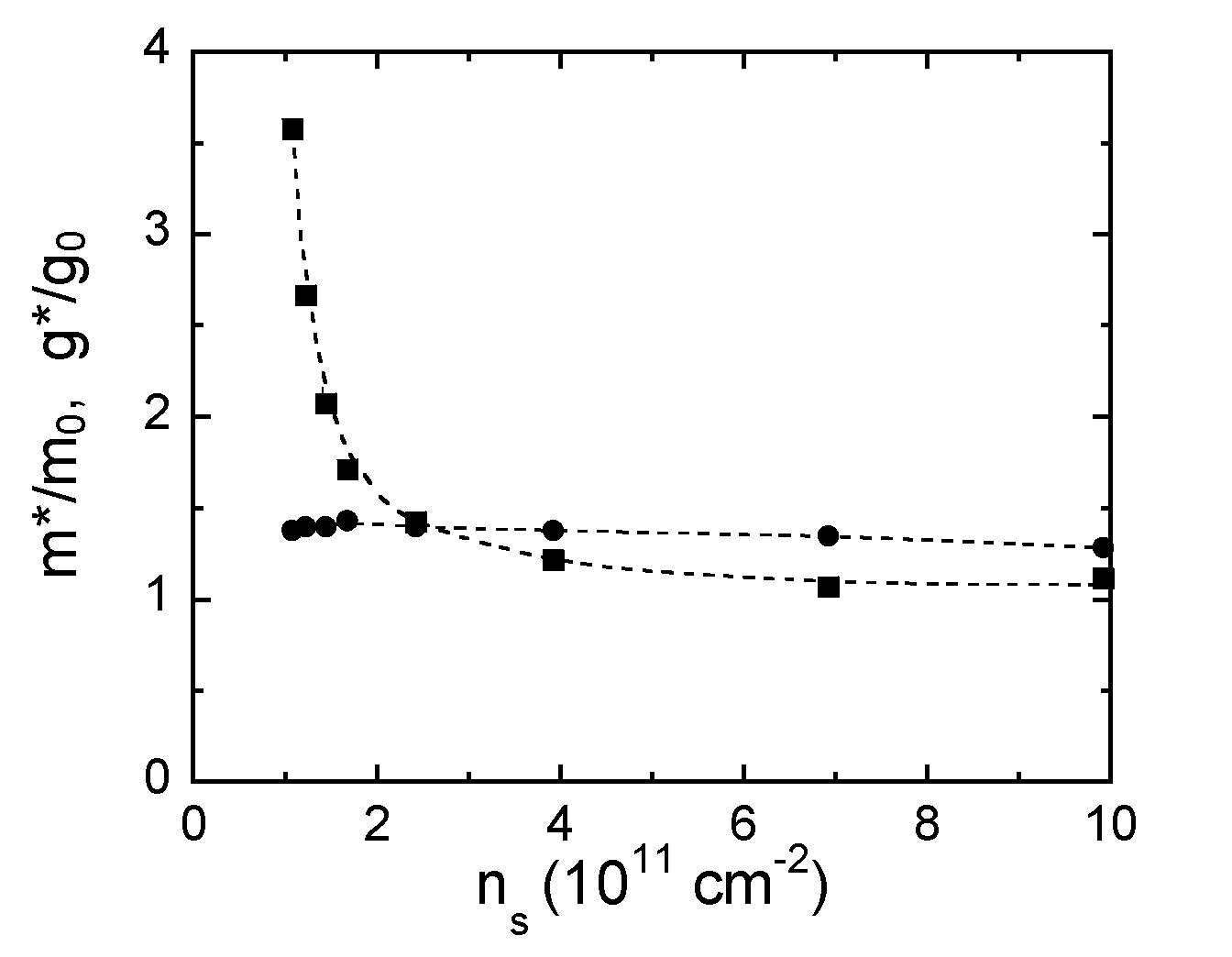}
\caption{Renormalization of the effective mass (squares) and $g$-factor (circles) as a function of the electron density. The dashed lines are guides to the eye.  Adapted from Ref.~\cite{shashkin2002sharp}.}
\label{m and g}
\end{figure}

An independent way to separate $g^*$ and $m^*$ was used in Ref.~\cite{shashkin2002sharp}. Using the theory of interaction-driven corrections to conductivity \cite{zala2001interaction}, the authors extracted from the transport data the Fermi liquid constants, $F_0^a$ and $F_1^s$, responsible for the renormalization of the $g$-factor ($g^*/g_0=1/(1+F_0^a)$) and the effective mass ($m^*/m_0=1+F_1^s$).  It turns out that the effective mass is strongly enhanced as the electron density decreases, while the $g$-factor remains nearly constant. The results are shown in Fig.~\ref{m and g}.  At relatively high electron densities, $n_s\gtrsim4\times10^{11}$~cm$^{-2}$ (lower $r_s$), the enhancement of both $g^*$ and $m^*$ is relatively small, both values slightly increasing with decreasing electron density, in agreement with previous data \cite{ando1982electronic}.  Also, the renormalization of the $g$-factor is dominant compared to that of the effective mass, which is consistent with theoretical results \cite{iwamoto1991static, kwon1994quantum, chen1999exchange}. In contrast, the renormalization in the low $n_s$ (critical) region, where $r_s\gg 1$, is much more striking. As the electron density is decreased, the renormalization of the effective mass overshoots abruptly while that of the
$g$-factor remains relatively small, $g^*\approx g_0$, without tending to increase. Hence, the results of Ref.~\cite{shashkin2002sharp} indicate that it is mainly the effective mass that is responsible for the drastically enhanced $\chi\propto g^*m^*$ near the metal-insulator transition.  At low electron densities, the effective mass determined from analysis of SdH oscillations in tilted magnetic fields is found to be independent of the degree of spin polarization, indicating that the mass enhancement is not related to the electrons' spins
\cite{shashkin2003spin, pudalov2014probing, melnikov2023spin}.

\subsection{Magnetization and magnetocapacitance measurements}

\begin{figure}
\centering
\includegraphics [width=8cm] {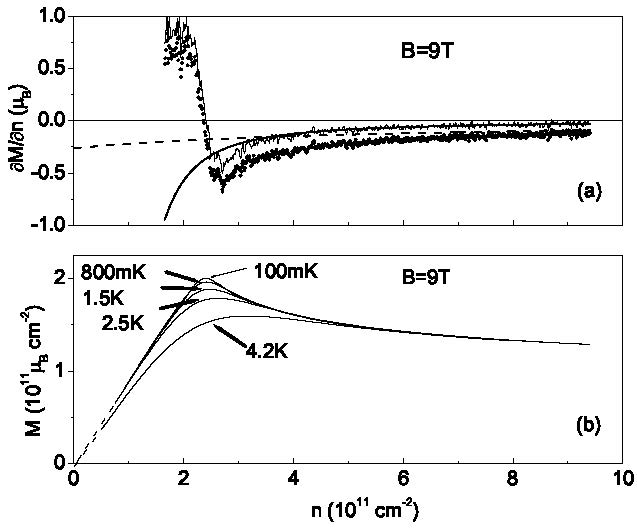}
\caption{(a) Total and spin $\partial M /\partial n$ in a Si MOSFET sample. Dots - measured total $\partial M /\partial n_s$, dashed line - diamagnetic contribution, thick smooth line - spin $\partial M / \partial n_s$ extracted from Sh-dH data~\cite{pudalov2002low}, thin solid line - spin $\partial M / \partial n_s$ extracted by subtraction of the diamagnetic contribution from the total $\partial M /\partial n_s$. (b) Spin magnetization obtained by integration of $\partial M /\partial n_s$ data. From Ref.~\cite{prus2003thermodynamic}.} \label{prus 2}
\end{figure}

A new method for measuring the thermodynamic spin magnetization of a 2D electron system was used in Refs.~\cite{prus2003thermodynamic, shashkin2006pauli, anissimova2006magnetization, reznikov2010thermodynamic}.  The method involved modulating the magnetic field parallel to the 2D plane at a low frequency, thereby modulating the chemical potential of the 2D electron system and measuring the induced current, the imaginary (out-of-phase) component of which is equal to
\begin{equation}
    i=(2\pi fCB_{mod}/e)\,\partial\mu/\partial B,
\end{equation}
where $C$ is the capacitance of the gated structure and $\mu$ is the chemical potential. Applying the Maxwell relation $\partial M/\partial n_s=-\partial\mu/\partial B$, one can obtain the magnetization $M$ from the measured current.  The measured $\partial M /\partial n$ at 9~T magnetic field and $T=100$~mK is depicted by dots in Fig.~\ref{prus 2}(a). The sample is a Si MOSFET with relatively high $n_c= 1.25\times10^{11}$~cm$^{-2}$ for the $B=0$ MIT \cite{prus2003thermodynamic}. The smooth solid line depicts the same quantity as extracted from Shubnikov-de~Haas data in Si MOSFETs \cite{pudalov2002low}. The difference between the two curves is attributed to the diamagnetic shift and to the presence of localized spins. The spin magnetization over the entire density range is obtained by subtracting the diamagnetic contribution (dashed line) from the measured data (dots); the latter is small compared to the spin contribution.  The spin magnetization (Fig.~\ref{prus 2}(b)) at a given field is obtained by numerical integration of the extracted $\partial M /\partial n_s$ values over $n_s$.

\begin{figure}
\centering
\includegraphics [width=8cm] {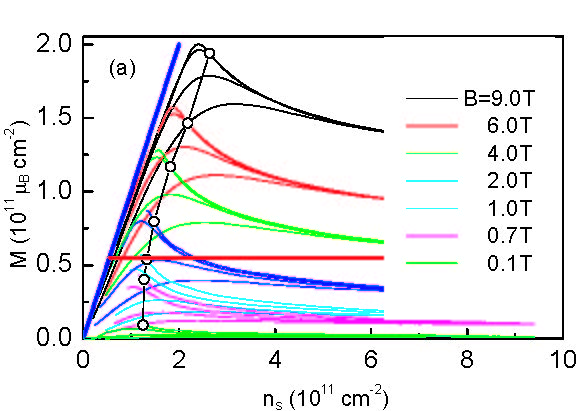}
\caption{(a) Spin magnetization as a function of density at different magnetic fields and temperatures 0.2, 0.8, 2.5, and 4.2~K in a Si MOSFET sample; higher magnetization corresponds to lower temperature. Critical densities, $n_c$, are marked by circles. Thick blue line - full magnetization, thick red line - magnetization of a degenerate ideal electron gas at $B=6$~T. From Ref.~\cite{prus2003thermodynamic}.} \label{prus 3}
\end{figure}

The spin magnetization at various magnetic fields and four temperatures is depicted in Fig.~\ref{prus 3}. For each magnetic field, the curves with higher magnetization values correspond to lower temperatures. The thick blue line corresponds to full polarization of all carriers at a given density, and the thick red line corresponds to the zero-temperature magnetization of a non-interacting degenerate electron gas at $B=6$~T. The empty circles denote, for each magnetic field, the critical density for the MIT, $n_c(B)$. At high magnetic fields, the full spin alignment persists up to densities considerably higher than those predicted for non-interacting electrons.

\begin{figure}[tbp]
\centering
\includegraphics[width=8cm]  {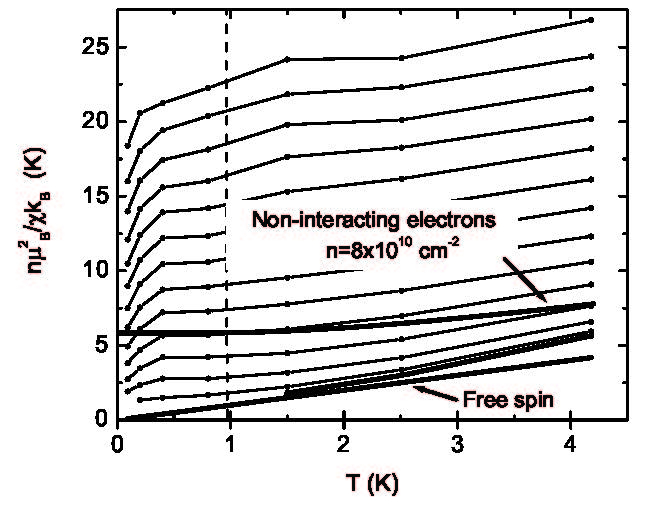}
\caption{Inverse susceptibility in a Si MOSFET sample determined from $M(B)$ at $B=0.7$~T. Experimental points from bottom to top correspond to densities $0.8\div 6\times 10^{11}$cm$^{-2}$ in $4\times10^{10}~{\rm cm^{-2}}$ steps. The thick straight line depicts the Curie law, and the dashed line marks $T=(g\mu_B/k_B)\times 0.7$~T. The experimental points at $n=8\times 10^{10}$ are connected by a thick line for comparison with the expectation for non-interacting electrons of the same density. From Ref.~\cite{prus2003thermodynamic}.} \label{prus 4}
\end{figure}

\begin{figure}[tbp]
\centering
\includegraphics[width=8cm]  {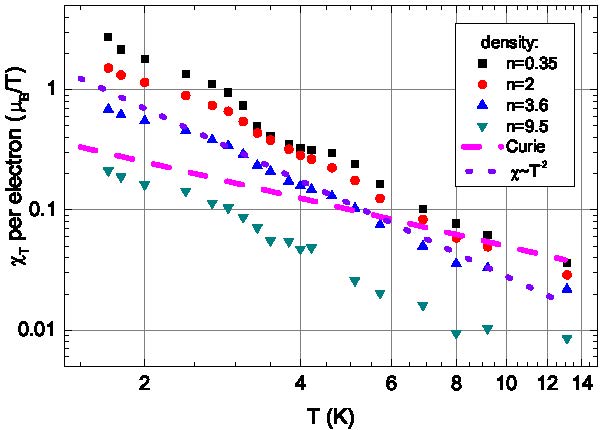}
\caption{Temperature dependence of $\chi_T(T)$ at several electron densities.  $\chi_T\propto T^{-\alpha}$ with $\alpha\approx 2$. $\chi\propto T^{-2}$ and $\chi\propto T^{-1}$ (the Curie law) are shown for comparison. The fact that $\alpha>1$ means that the electrons, even at the lowest density, cannot be considered as noninteracting localized ones. From Ref.~\cite{pudalov2018probing}.} \label{Pudalov droplets}
\end{figure}

The inverse susceptibility, determined from $B=0.7$~T spin magnetization, normalized by the electron density, is depicted as a function of temperature in Fig.~\ref{prus 4}.  The susceptibility measured at densities just below $n_{c}$ approaches the free-spin value (Curie law).  Note that $\chi$ in the metallic phase depends on temperature down to 0.1~K. This dependence indicates the existence of a relevant energy scale considerably smaller than the Zeeman one (dashed vertical line in Fig.~\ref{prus 4}). The authors suggested that such an energy scale may arise from localized spins that interact very weakly with each other.

\begin{figure}[tbp]
\includegraphics[width=7.2cm] {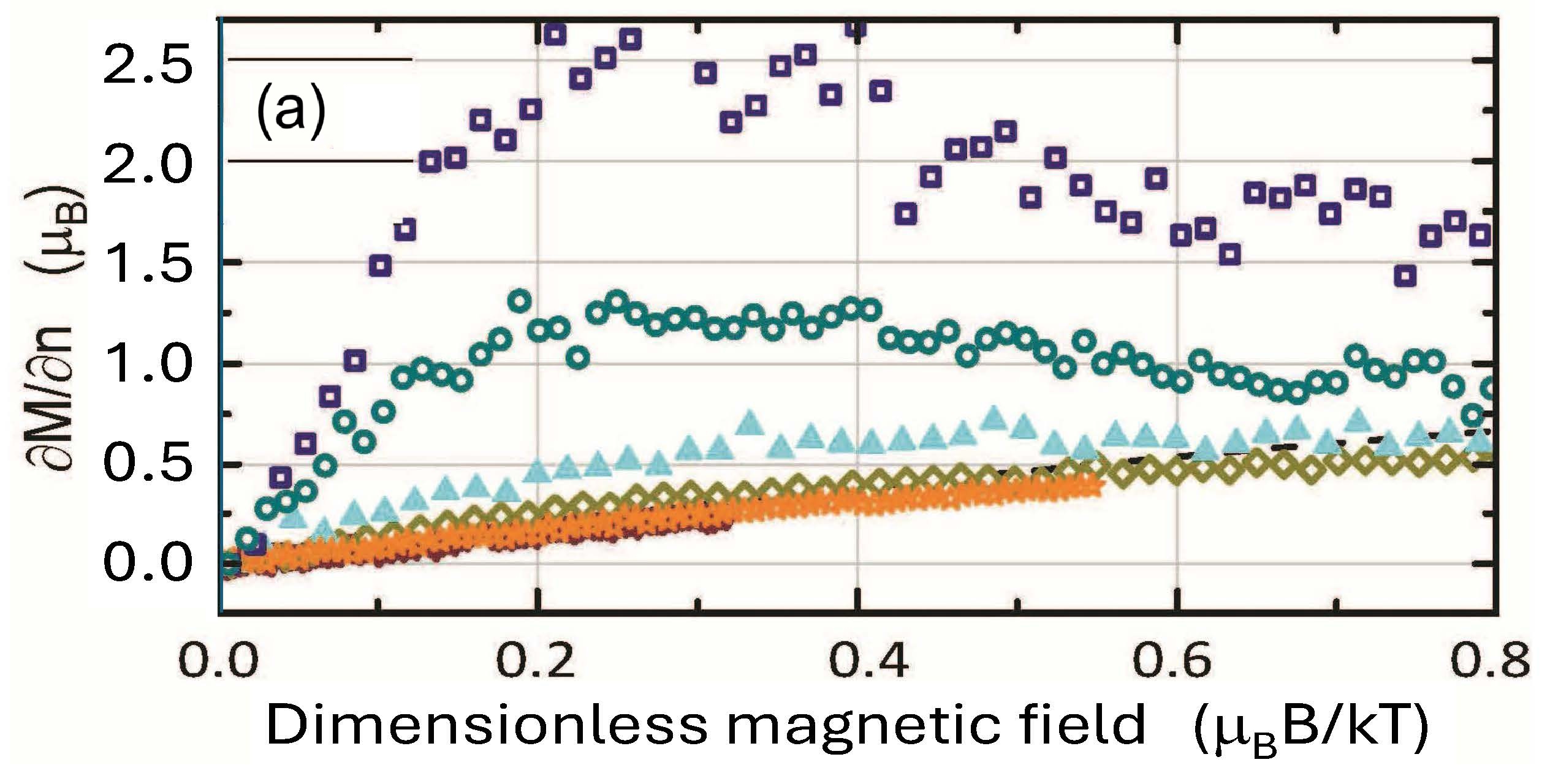}
\includegraphics[width=6cm]  {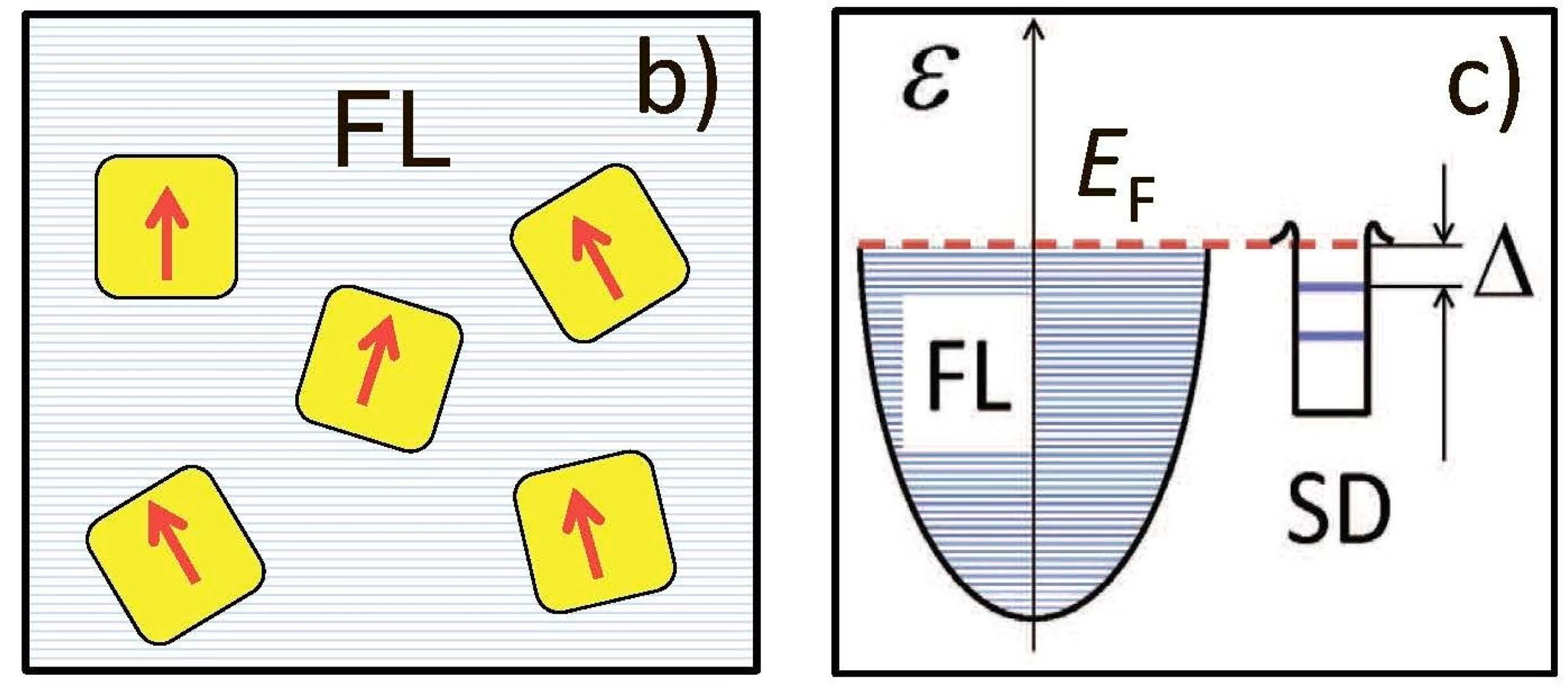}
\includegraphics[width=7.5cm] {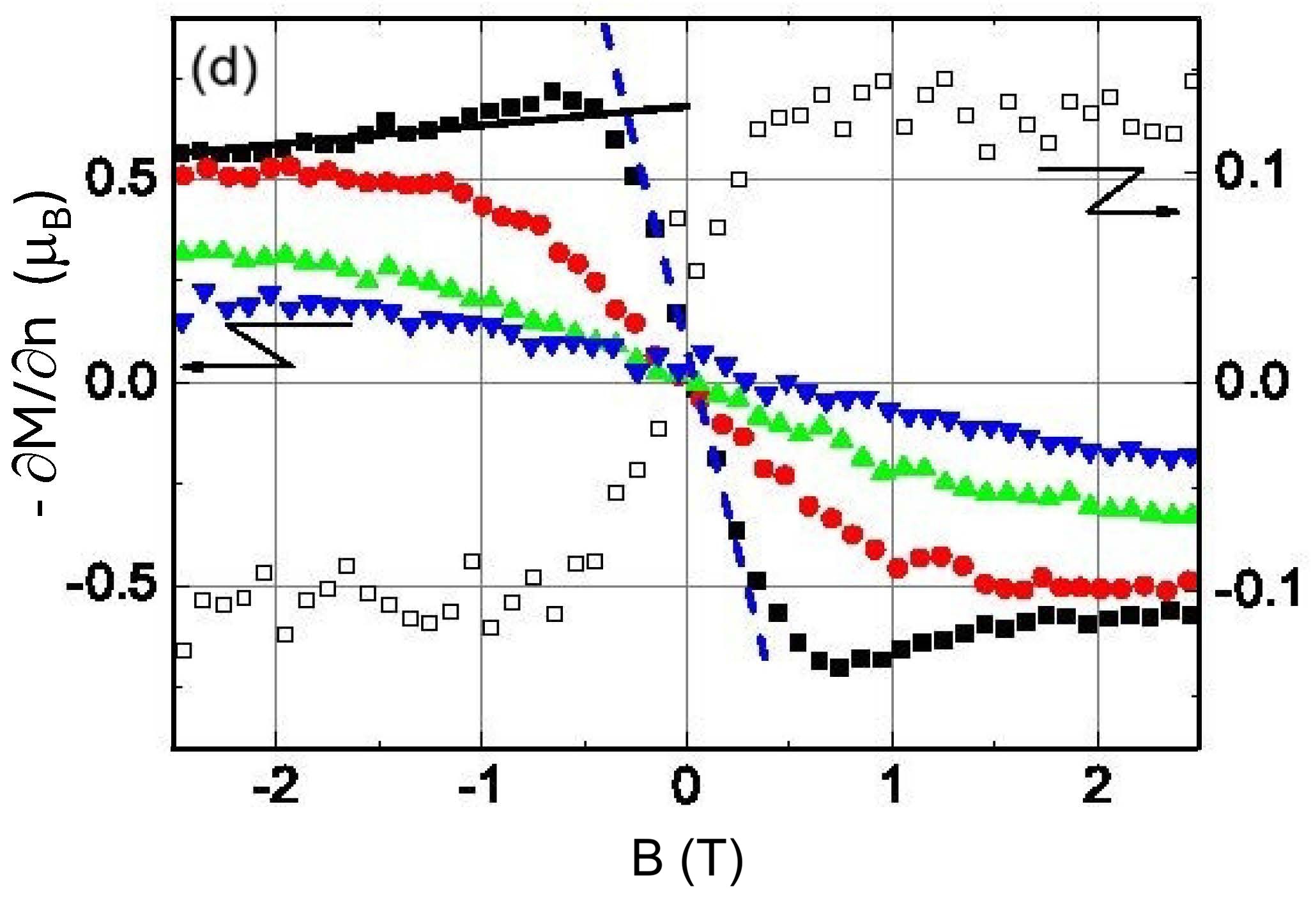}
\caption {(a)~The magnetization per electron, represented as \(\partial M/\partial n_s\), is plotted against the normalized magnetic field for a carrier density of \(5 \times 10^{10} \, \text{cm}^{-2}\) at various temperatures (from top to bottom: \(T = 0.8\), 1.2, 1.8, 4.2, 7, 10, and 24~K). Data are taken from Ref.~\cite{teneh2012spin}. Lower panels: (b)~A schematic representation of the spatial arrangement of the two-phase state and (c)~an assumed energy band diagram of the two-phase system. From Ref.~\cite{pudalov2022phase}. (d)~$\partial\mu/\partial B = -\partial M/\partial n_s$. Left scale -  at $n_s = 5 \times 10^{10}$~cm$^{-2}$, deep in the insulating regime at temperatures (from top to bottom) 1.7, 3.4, 4.6, and 6.8~K.  Right scale  - at $n_s = 2 \times 10^{11}$~cm$^{-2}$, deep in the metallic regime at $T=1.7$~K.  The slope of the dashed line gives the susceptibility per electron, $- \partial \chi_T/\partial n_s$. From Ref.~\cite{reznikov2010thermodynamic}).
 }
\label{Pudalov chi 2022}
\end{figure}

The thermodynamic spin susceptibility $\chi_T$ probed in the low-field regime $B<k_BT/g\mu_B$ increases upon cooling (see Fig.~\ref{prus 4}), which is clearly inconsistent with the spin susceptibility of the Fermi gas (thick black line) or of a Fermi liquid. The increase persists across the overall range of densities and indicates the involvement of localized states both well above and below the critical density $n_c$. The localized states occupy an energy interval at the bottom of the potential well, below the band of extended states (band tail) \cite{mott1974metal, gold2002on}. When these states are treated as free spins, the susceptibility is expected to be proportional to $1/T$.  As shown in Ref.~\cite{bhatt1986magnetic}, interactions between localized spins lead to antiferromagnetic coupling and to a susceptibility that grows slower than $1/T$. These expectations  contradict the experimental results presented in Fig.~\ref{Pudalov droplets}, where  $\chi_T$ varies as $1/T^\alpha$ with $\alpha=1 - 2$ \cite{prus2003thermodynamic, teneh2012spin}. Subsequent measurements \cite{reznikov2010thermodynamic} have shown that the low-field slope of $\partial M/\partial n_s \propto \partial\chi_T/\partial n_s$ changes sign between the regions $n_s > n_c$ and $n_s < n_c$ (dashed line in Fig.~\ref{Pudalov chi 2022}(d)). In higher fields, the linear field dependence of $\partial\mu /\partial B =-\partial M/\partial n_s$ at $n_s=2\times10^{11}$~cm$^{-2}$ saturates at $B^*\approx0.5$~T (see Fig.~\ref{Pudalov chi 2022}(d)).

\begin{figure}[tbp]
\centering
\includegraphics[width=8cm]  {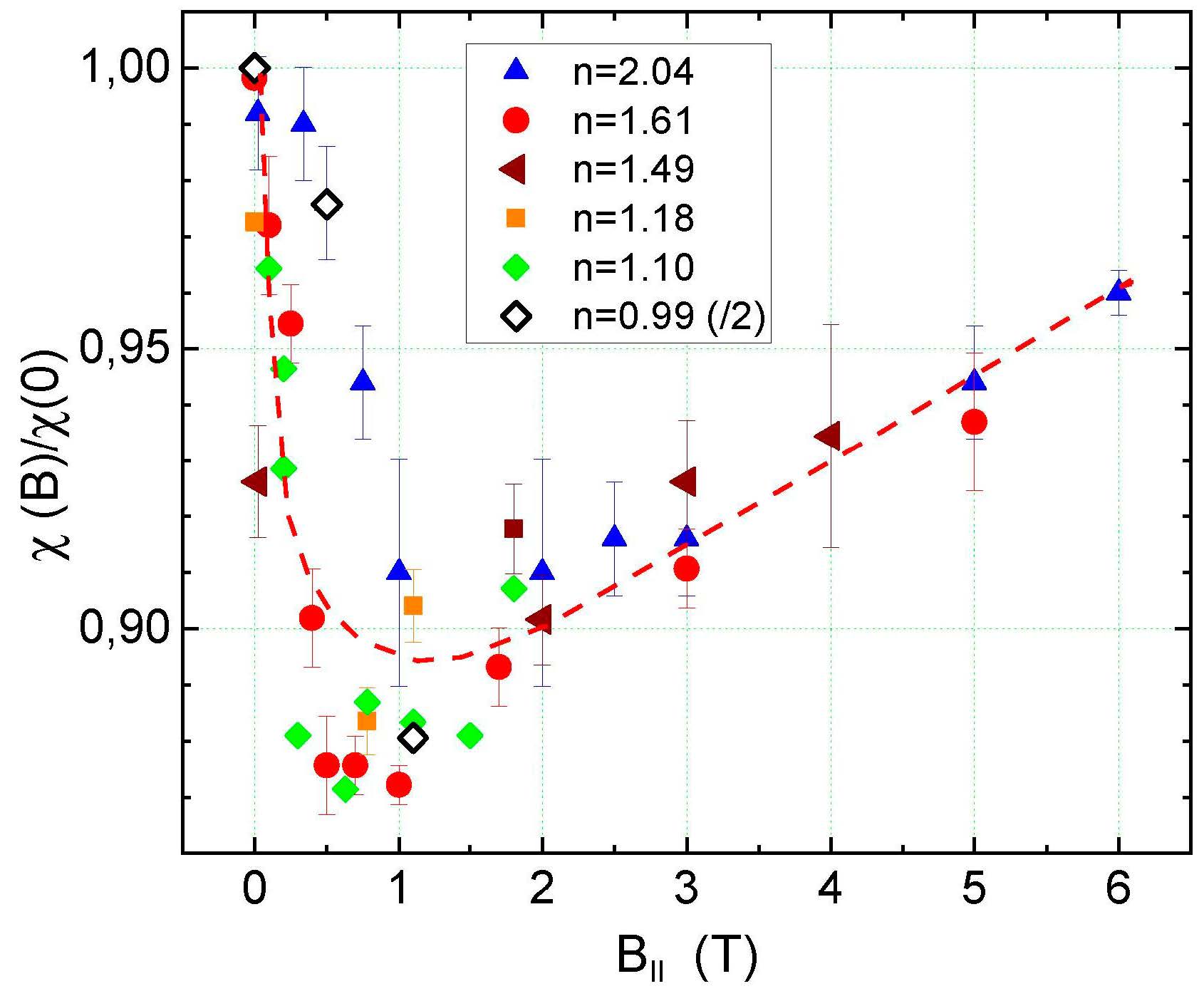}
\caption{$\chi(B_\parallel)/\chi(0)$ data versus $B_\parallel$  for two Si MOSFET samples and for several densities. For the lowest density $n_s=0.99$~cm$^{-2}$, the $\chi(B)/\chi(0)$ variations are scaled down by a factor of 2. The density is indicated in units of $10^{11}$~cm$^{-2}$,  $T=0.1$\,K. From Ref.~\cite{pudalov2021magnetic}.} \label{Pudalov chi}
\end{figure}

It was later confirmed experimentally \cite{teneh2012spin, pudalov2022phase} and supported theoretically \cite{stadnik2013interacting} that the strong temperature dependence of the thermodynamic susceptibility, $\chi_T(T)$, arises from localized spin droplets. To avoid confusion, we note that these are not related to the droplets introduced by Spivak and Kivelson \cite{spivak2004phases}.

\subsubsection{Measurements in low $B_\|$ fields}
More detailed low-field magnetization measurements \cite{teneh2012spin} have shown that the magnetically responsive localized states have an average spin $J\approx2$ (see 
Fig.~\ref{Pudalov chi}) and may thus be viewed as collective spin-droplets comprising four electrons. The ability to recharge these states down to the lowest temperatures (0.4~K in Refs.~\cite{prus2003thermodynamic, teneh2012spin}) without any signs of temperature-activated behavior already indicates that these states lie very close to the Fermi energy, according to Pudalov \cite{pudalov2022phase}. It was established in Ref.~\cite{pudalov2021magnetic} that charge carriers are exchanged between collective localized states and mobile states --- at least down to 0.1~K and in a temperature-independent manner --- via tunneling (for further details, see Figs.~\ref{Pudalov chi 2022} and \ref{Pudalov chi}, and also Section 10.4). This finding provides strong evidence that collective localized states coexist with mobile states at the Fermi level, as illustrated in the schematic diagram in Fig.~\ref{Pudalov chi 2022}(c) (for more detail, see Section 10.4). The authors of Ref.~\cite{teneh2012spin} concluded that the temperature-dependent susceptibility, \(\chi_T\), arises from easily polarizable spin droplets, increases faster than $1/T$ as temperature decreases, and that the number of spin droplets also increases as $\sim 1/T$.

In turn, the spin susceptibility of itinerant electrons, extracted from Shubnikov-de Haas oscillations in low fields ($B < k_B T / g^* \mu_B$), exhibits a weak temperature dependence that is consistent with Fermi liquid theory, according to Pudalov et al. \cite{pudalov2018probing}. Furthermore, from the SdH measurements \cite{pudalov2021magnetic}, $\chi$ was found to vary sharply with the in-plane magnetic field, as Fig.~\ref{Pudalov chi} illustrates. The variations of $\delta\chi(B)/\chi(0)$ range from approximately 25\% at low densities to around 6\% at high densities (specifically, $10 \times 10^{11}$~cm$^{-2}$). The data presented in Ref.~\cite{pudalov2021magnetic} align with the $\chi(B=0)$ values reported in Refs.~\cite{klimov2008interaction, pudalov2021magnetic} when considering the limit as $B_\parallel$ approaches 0.  The minimum of $\chi(B)$ occurs at $B_\parallel \approx 1$~T, for electron densities between $(1.1 - 2) \times 10^{11}$~cm$^{-2}$. This field is significantly weaker than the field of approximately 20 T required to achieve complete spin polarization of the two-dimensional system.

In Figure \ref{Pudalov chi 2022}(a), it can be observed that at low densities, the derivative \(\partial M/\partial n_s\) becomes positive. In all cases, this value is significantly higher than what is expected for the Pauli spin susceptibility. As the magnetic field increases --- while still remaining below the thermal energy condition (\(g\mu_BB < k_BT\)) --- the value of \(\partial M/\partial n_s\) rises sharply. At low temperatures, it even exceeds the Bohr magneton by more than a factor of two. More on this is discussed in Section \ref{sec:chi-nonlinear}.

A potential reason for these variations is the effect of disorder, which is inevitably present in all material systems. Its characteristics (whether long-range or short-range) and magnitude can differ between samples from different manufacturers. Even a robust parameter like the spin susceptibility of itinerant electrons, denoted as \(\chi\), is significantly affected by the nature of the disorder, as established through both experimental \cite{clarke2008impact} and theoretical \cite{depalo2005effects} studies. Weak short-range disorder is known to promote the formation of Wigner crystals by creating a gap in their phonon spectrum \cite{chui1995impurity, eguiluz1983two}. In contrast, long-range disorder tends to increase effective electron-electron interactions, thereby facilitating local Stoner instability \cite{benenti2001spin, andreev1998itinerant, eisenberg1999disorder} and localized Wigner crystallization \cite{reiman2000formation} in confined geometries, such as those found in mesoscopic quantum dots. Additionally, the thickness of the 2D layer and the depletion charge, which can vary across different 2D samples, also influence the renormalization of spin susceptibility \cite{depalo2005effects, zhang2005density}. The same dependence applies to spin magnetization. Interestingly, the nonlinear $\chi(B_\|)$ dependence (Fig.~\ref{Pudalov chi}) holds even at high electron densities, where one might typically expect metallic behavior.

\subsubsection{Measurements in higher $B_\|$ fields}

Such effects have not been observed in much less disordered samples, in which the band tail of localized
electrons is small, and the magnetization in the partially-polarized regime exhibits a proportional dependence on the magnetic field \cite{shashkin2006pauli}, as discussed below.

\begin{figure}
\centering
\scalebox{0.5}{\includegraphics[clip]{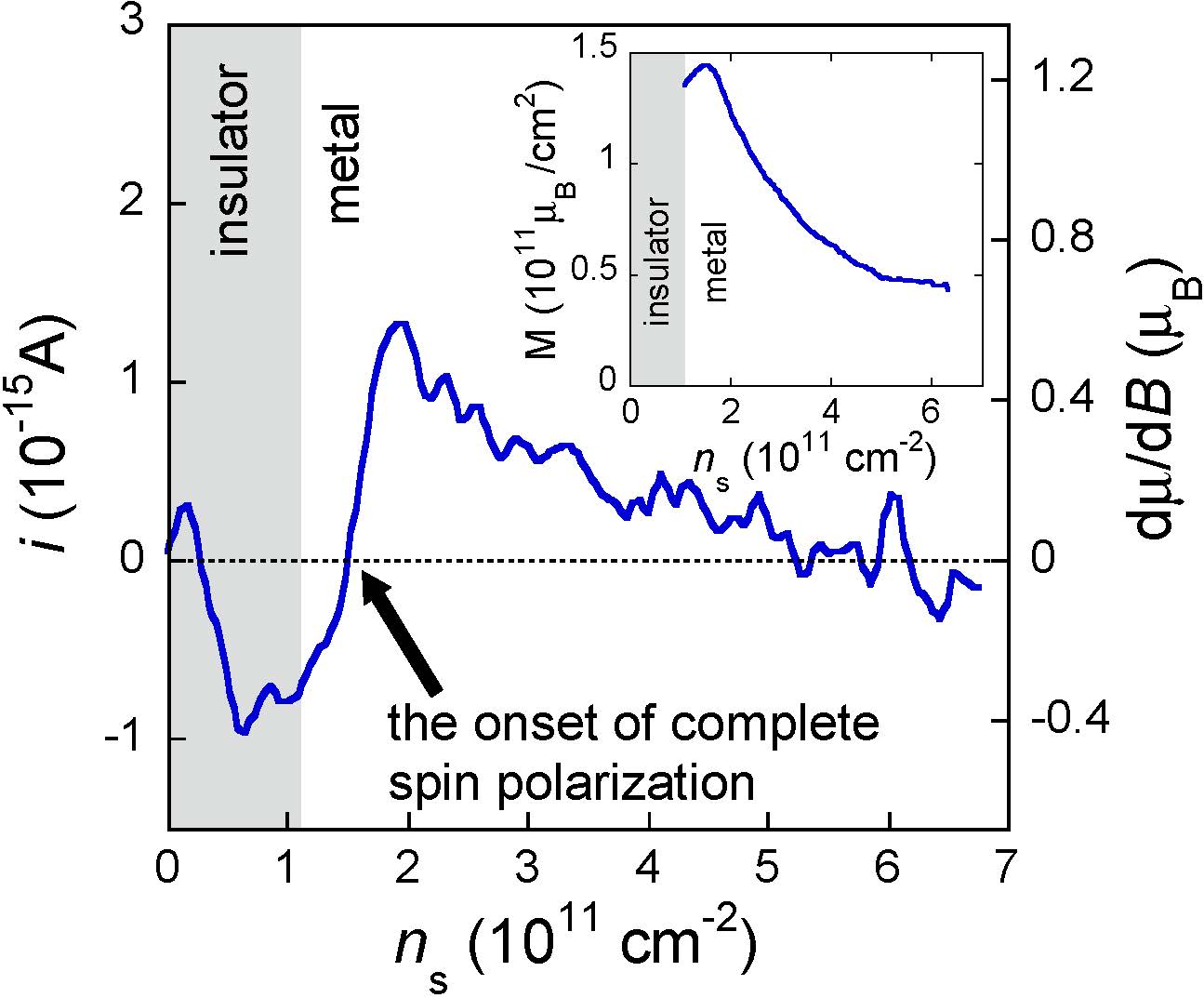}}
\caption{Imaginary current component in the magnetization experiment as a function of the electron density in a magnetic field of 5~T and $T=0.4$~K. The gray area depicts the insulating phase. Magnetization versus $n_s$ is displayed in the inset. From Ref.~\cite{shashkin2006pauli}.}\label{i n}
\end{figure}

A typical experimental trace of $i(n_s)$ in a parallel magnetic field of 5~T, measured in Ref.~\cite{shashkin2006pauli}, is shown in Fig.~\ref{i n}. The inset shows magnetization $M(n_s)$ in the metallic phase obtained by integrating $dM/dn_s=-d\mu/dB$ with the integration constant $M(\infty)=B\chi_0$, where $\chi_0$ is the Pauli spin susceptibility of non-interacting electrons. The onset of complete spin polarization --- the electron density $n_p$ at which the electrons start to fill the upper spin subband --- is given by the condition $d\mu/dB=0$ ($M(n_s)$ reaches a maximum), as indicated by the black arrow in the figure.

\begin{figure}\centering
\scalebox{0.6}{\includegraphics[clip]{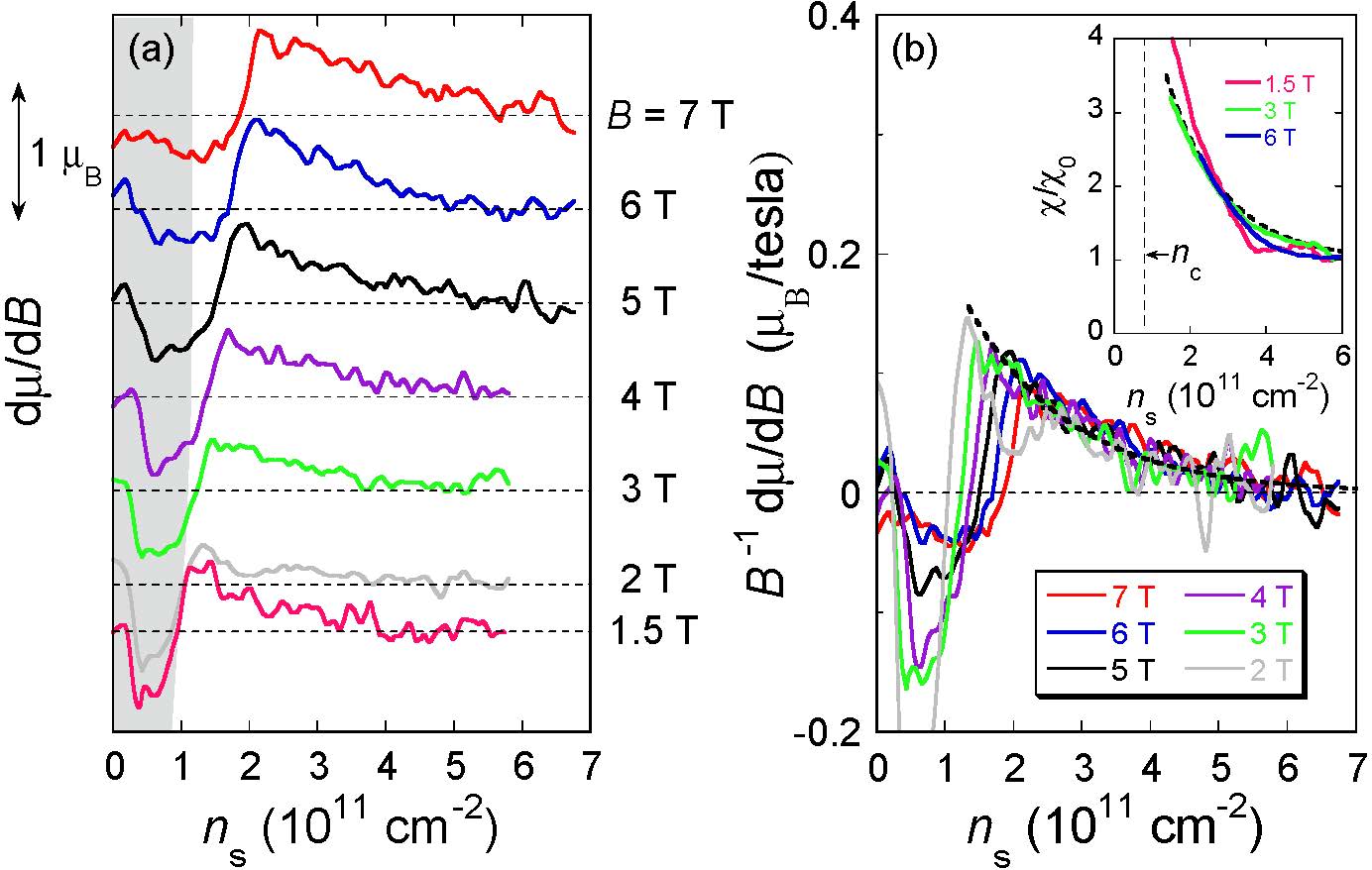}}
\caption{\label{M n} (a)~The experimental $d\mu/dB$ as a function of electron density in different magnetic fields and $T=0.4$~K. The curves are vertically shifted for clarity. The gray area depicts the insulating phase. Note that the onset of full spin polarization in our experiment always takes place in the metallic regime. (b)~Scaling of the $d\mu/dB$ curves, normalized by magnetic field magnitude, at high electron densities. The dashed line represents the ``master curve''. Spin susceptibility obtained by integrating the master curve (dashed line) and raw data at $B=$~1.5, 3, and 6~T is displayed in the inset. From Ref.~\cite{shashkin2006pauli}.}
\end{figure}

\begin{figure}\centering
\scalebox{0.8}{\includegraphics[clip]{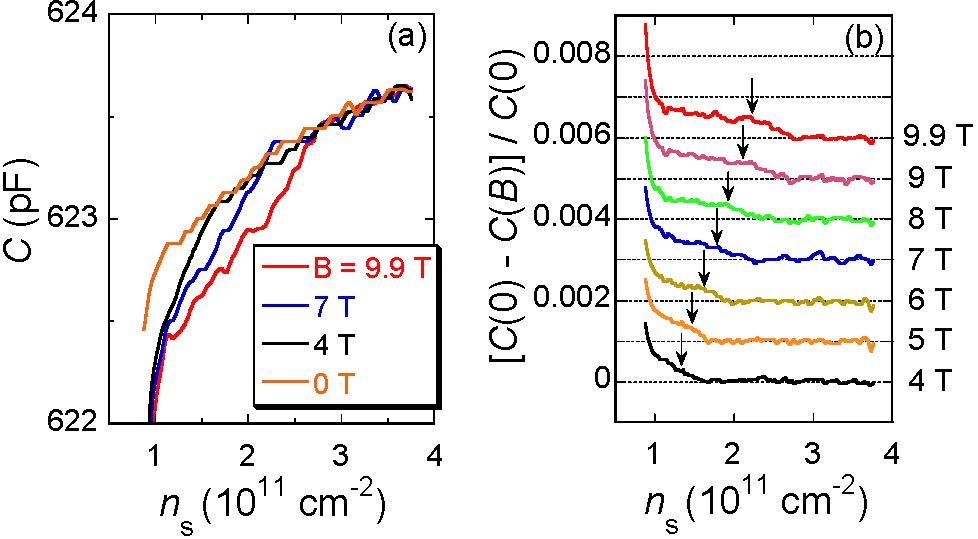}}
\caption{\label{capacitance} (a)~Magnetocapacitance versus electron density for different magnetic fields. (b)~Deviation of the $C(n_s)$ dependences for different magnetic fields from the $B=0$ reference curve. The traces are vertically shifted for clarity. The onset of full spin polarization is indicated by arrows.  From Ref.~\cite{shashkin2006pauli}.}
\end{figure}

In Fig.~\ref{M n}(a), a set of curves for the experimental $d\mu/dB$ versus electron density in different magnetic fields is shown. The onset of full spin polarization shifts to higher electron densities as the magnetic field increases. The gray area represents the insulating phase, which expands somewhat with $B$. The range of magnetic fields is restricted from below by the condition that $d\mu/dB$ crosses zero in the metallic regime. In Fig.~\ref{M n}(b), these curves, normalized by magnetic field, collapse in the partially-polarized regime onto a single ``master curve''. Integration of the master curve over $n_s$ yields spin susceptibility $\chi=M/B$, as shown in the inset to Fig.~\ref{M n}(b). Also shown is the spin susceptibility obtained by integrating the raw curves at $B=$~1.5, 3, and 6 tesla, which, within experimental error, yield the same dependence.

Yet another method for measuring $n_p$ and $\chi$ relies on analyzing the magnetocapacitance, $C$.  Experimental traces $C(n_s)$ are shown in Fig.~\ref{capacitance}(a) for different magnetic fields. As the magnetic field is increased, a step-like feature emerges on the $C(n_s)$ curves and shifts to higher electron densities. This feature corresponds to the thermodynamic density of states abruptly changing when the electrons' spins become completely polarized. To see the step-like feature more clearly, in Fig.~\ref{capacitance}(b), the $C(n_s)$ curves for different magnetic fields are subtracted from the reference $B=0$ curve.

\begin{figure}\centering
\scalebox{0.5}{\includegraphics[clip]{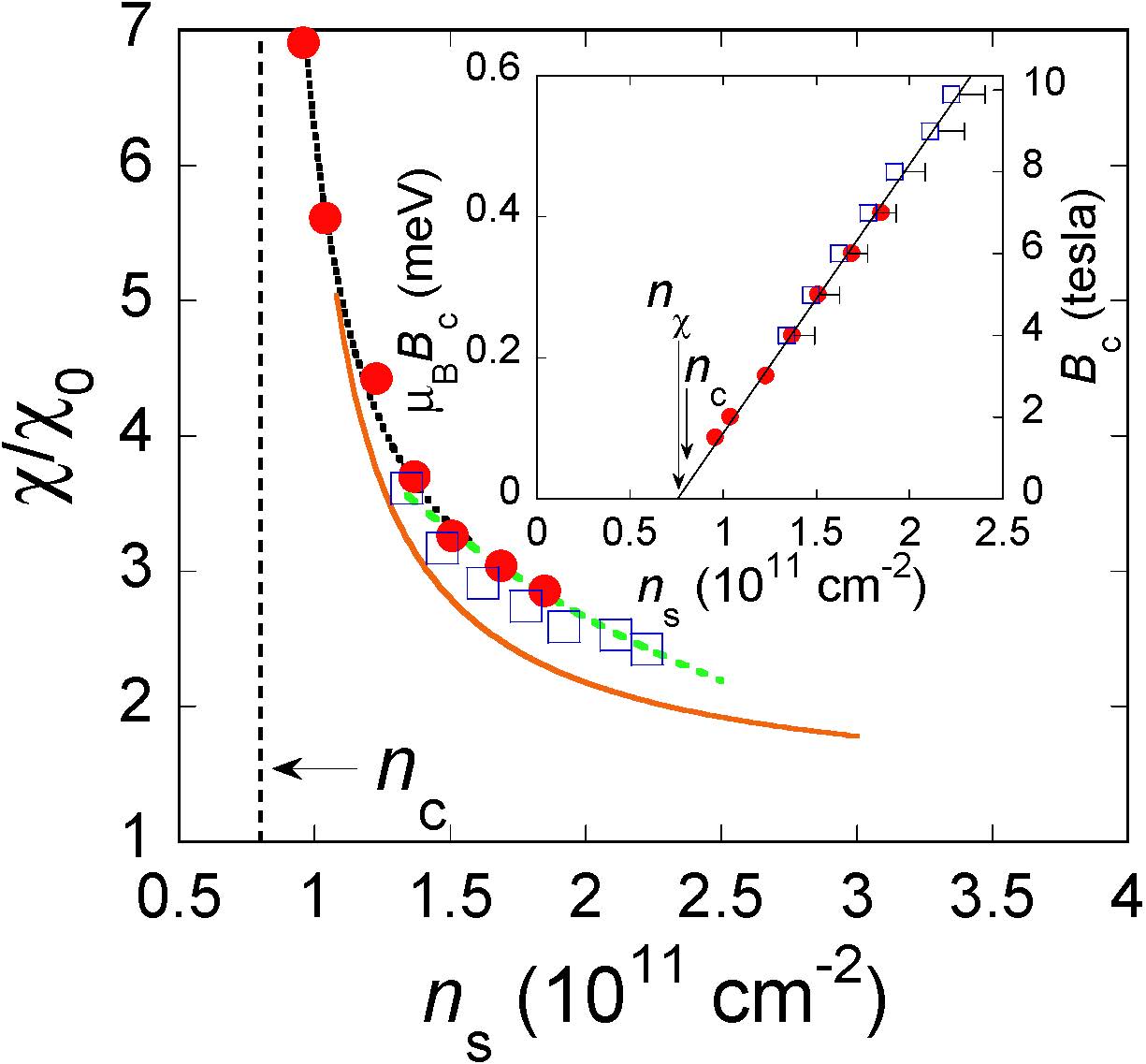}}
\caption{\label{all n} Dependence of the Pauli spin susceptibility on the electron density obtained by integrating the master curve (dashed line), $d\mu/dB=0$ (circles), and magnetocapacitance (squares). The dotted line is a guide to the eye. Also shown by a solid line is the transport data of Ref.~\cite{shashkin2001indication}. Inset: polarization field as a function of the electron density determined from the magnetization (circles) and magnetocapacitance (squares) data. The symbol size in the magnetization data reflects the experimental uncertainty, and the error bars in the magnetocapacitance data extend to the midpoint of the jump in $C$. The data for $B_c$ are described by a linear fit that extrapolates to a density $n_\chi$ close to the critical density $n_c$ for the $B=0$ MIT.  From Ref.~\cite{shashkin2006pauli}.}
\end{figure}

In Fig.~\ref{all n}, we show the summary of the results for the Pauli spin susceptibility as a function of $n_s$, obtained using all three methods described above. There is good agreement between these results and the data obtained from the transport experiments reported in Ref.~\cite{shashkin2001indication}. This adds credibility to the transport data and confirms that full spin polarization occurs at the field $B_c$. However, evidence for the phase transition can only be obtained from thermodynamic measurements. Moreover, the magnetization data extend to lower densities than the transport data, and larger values of $\chi$ are reached, exceeding the ``non-interacting'' value $\chi_0$ by almost an order of magnitude. The Pauli spin susceptibility behaves critically close to the critical density $n_c$ for the $B=0$ metal-insulator transition:
\begin{equation}
\chi\propto n_s/(n_s-n_\chi).
\end{equation}

The same technique was used to study thermodynamic magnetization in perpendicular and tilted magnetic fields \cite{anissimova2006magnetization}. By measuring $\partial\mu/\partial B$ at non-integer filling factors, the authors directly determined the spectral characteristics without fitting or adjustable parameters. The results confirmed that the $g$-factor is weakly enhanced and practically independent of the electron density down to the lowest densities reached ($\approx1.5\times10^{11}$~cm$^{-2}$), whereas the cyclotron mass is strongly enhanced at low $n_s$.

\subsection{Thermopower}

Yet another way to estimate the behavior of the effective mass is to measure the thermoelectric power. Thermopower is defined as the ratio of the thermoelectric voltage to the temperature difference $S=-\Delta V/\Delta T$.  Based on Fermi liquid theory, Dolgopolov and Gold \cite{dolgopolov2011thermoelectric,gold2011temperature} obtained the following expression for the diffusion thermopower of strongly interacting 2D electrons in the low-temperature regime:
\begin{equation}
S=-\beta\frac{2\pi k_B^2m^*T}{3e\hbar^2n_s},\label{eq2}
\end{equation}
where $k_B$ is the Boltzmann constant.  This expression, which resembles the well-known Mott relation for non-interacting electrons, was shown to hold in the strongly interacting case, provided one includes the parameter $\beta$, which depends on both disorder \cite{fletcher1997thermoelectric, faniel2007in, goswami2009highly} and interaction strength \cite{dolgopolov2011thermoelectric, gold2011temperature}. The dependence of $\beta$ on the electron density is rather weak, and the main effect of the electron-electron interactions is to suppress the thermopower $S$.

\begin{figure}
\centering
\includegraphics[width=8 cm]{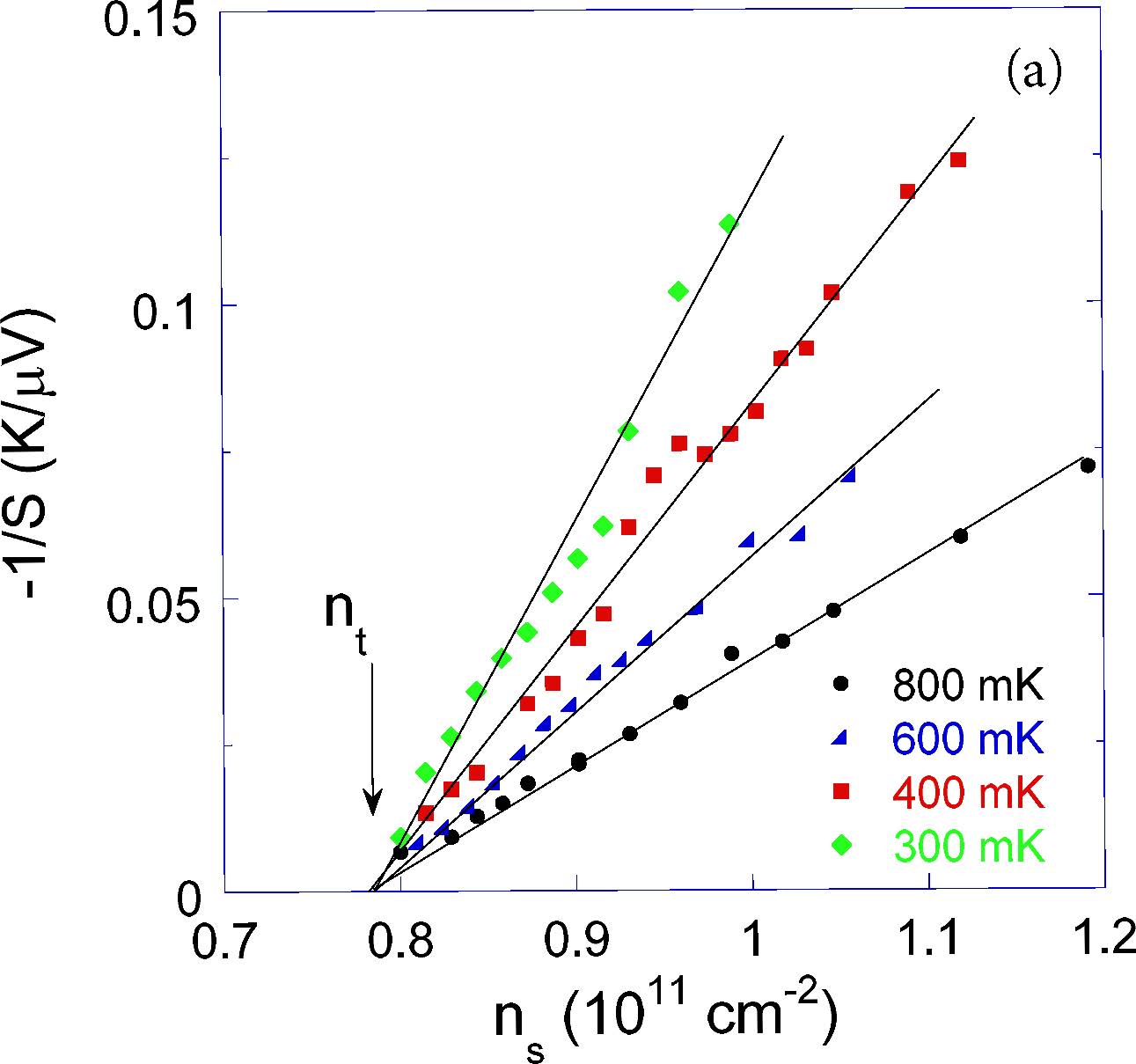}
\includegraphics[width=8 cm]{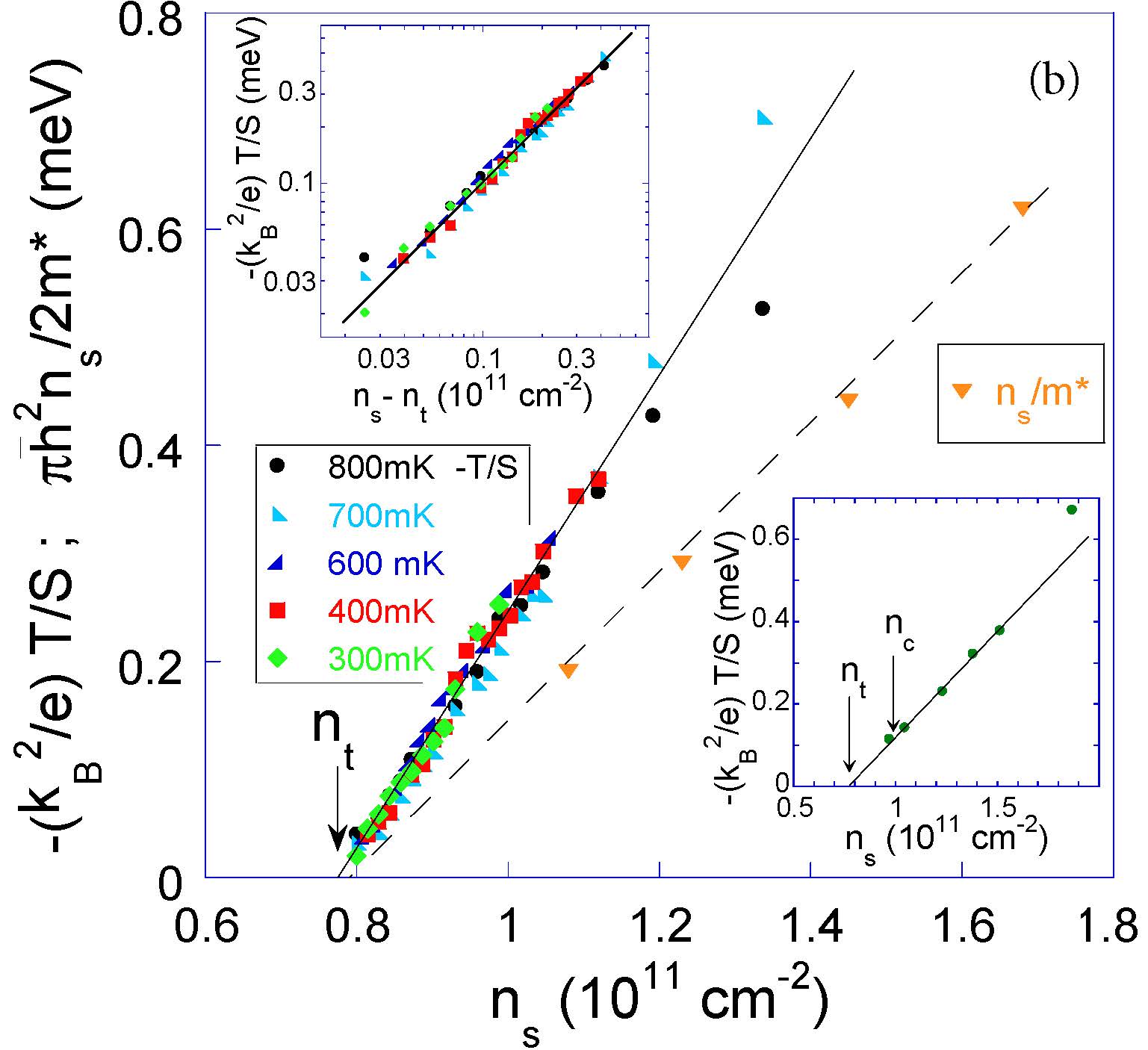}
\caption{(a) The inverse thermopower in the least-disordered sample as a function of electron density at different temperatures. (b): $(-T/S)$ versus electron density for different temperatures. The solid lines in panels (a) and (b) are linear fits that extrapolate to zero at $n_t$. Also shown is the effective mass obtained from transport measurements for the same sample \cite{shashkin2002sharp}. The dashed line is a linear fit.  Upper inset: log-log plot of $(-T/S)$ versus $(n_s-n_t)$, demonstrating a power law approach to the critical density $n_t$. Lower inset: $(-T/S)$ versus density at $T=0.3$~K for a more disordered sample \cite{fletcher2001critical}. The linear fit (solid line) extrapolates to zero at the same density $n_t$. The position of the density $n_c$ for the metal-insulator transition was estimated to be $0.99\pm 0.02\times 10^{11}$~cm$^{-2}$.  From Ref.~\cite{mokashi2012critical}.}
\label{thermo}
\end{figure}

According to Eq.~(\ref{eq2}), $(-S/T)$ is proportional to $m^*/n_s$ and, therefore, the thermopower measurements should yield the effective mass.  The divergent behavior of diffusion thermopower in Si MOSFETs was reported in Ref.~\cite{fletcher2001critical} for a more disordered sample and in Ref.~\cite{mokashi2012critical} for the least-disordered sample, where transport-determined $n_c$ is the lowest across all samples studied. When plotted as the inverse quantity, $(-1/S)$, versus the electron density (see Fig.~\ref{thermo}(a)), in the least-disordered sample it extrapolates to zero at $n_s=n_t = 7.8 \pm 0.1 \times 10^{10}$~cm$^{-2}$, which is close to (or the same as) the density for the metal-insulator transition in this sample ($n_c \approx 8 \times 10^{10}$~cm$^{-2}$), obtained from the resistance measurements.

Figure~\ref{thermo}(b) shows the ratio $(-T/S)$ plotted as a function of $n_s$ in the least-disordered sample. The data collapse onto a single curve, demonstrating that the thermopower $S$ is linear in temperature. In turn, the ratio $(-T/S)$ is a function of electron density $n_s$ of form:
\begin{equation}
(-T/S) \propto (n_s-n_t)^x.\label{eq1}
\end{equation}
Fits to this expression indicate that the thermopower diverges with decreasing electron density with a critical exponent $x=1.0\pm 0.1$. The measured ratio $(-T/S)$, shown in Fig.~\ref{thermo}(b), decreases linearly with decreasing electron density, extrapolating to zero at $n_t$ and indicating an increase of the effective mass by more than an order of magnitude.  The results thus imply divergence of the electron mass at $n_t$: $m^*\propto n_s/(n_s-n_t)$ --- the behavior that is typical in the vicinity of an interaction-induced phase transition.  Importantly, a re-plot of the thermopower data obtained in the more disordered sample \cite{fletcher2001critical}, shown in the lower inset of Fig.~\ref{thermo}(b), demonstrates that $(- T/S)$ extrapolates to the same density $n_t$ as in the least-disordered sample \cite{mokashi2012critical} despite a higher critical density for the MIT obtained from the resistance data.  Thus, while the electron density of the resistive transition, $n_c$, varies with the level of disorder, the divergence of the thermopower and the effective mass occurs at a density $n_t$ that is independent of the disorder and is governed by the strength of the inter-electron interactions.  The thermopower experiments on the least-disordered samples \cite{mokashi2012critical} include data for electron densities much closer to the critical point than those accessible with the techniques described above.  The effective electron mass \( m^* \approx 5\; m_e \),
obtained from thermopower measurements in Ref.~\cite{mokashi2012critical}, significantly exceeds values from
previous transport experiments.

The thermopower, as well as the conductivity, provides a measure of the mass at the Fermi level, while the Zeeman field $B_{sat}$ required to fully polarize the spins measures the mass associated with the bandwidth, which is the Fermi energy measured from the band bottom. For $n_s\geq 10^{11}$~cm$^{-2}$, the mass determined by different methods was found to be essentially the same \cite{kravchenko2004metal, shashkin2005metal, pudalov2014probing}. On the other hand, the
behavior is different at the densities reached in the experiment in the very close vicinity of the critical point $n_t$ ($n_s <10^{11}$~cm$^{-2}$) where the bandwidth-related mass was found to increase by only a factor $\approx4$. This follows from the fact that the Shubnikov-de~Haas oscillations in the dilute 2D electron system in silicon reveal one switch from cyclotron to spin minima (the ratio of the spin and cyclotron splittings reaches $\approx1$) as the electron density is decreased \cite{kravchenko2000shubnikov}, the spin minima surviving down to $n_s\approx n_c$ \cite{diorio1990magnetic, pudalov2018probing, pudalov2020experimental}. In effect, while the bandwidth does not decrease appreciably in the close vicinity of the critical point $n_t$ and the effective mass obtained from such measurements does not exhibit a true divergence, the thermopower measurements yield the effective mass at the Fermi energy, which does indeed diverge
(see also Section 10.4).

The divergence of the effective mass has been predicted by a number of theories: using Gutzwiller's theory \cite{dolgopolov2002on}; using an analogy with He$^3$ near the onset of Wigner crystallization \cite{spivak2003phase,spivak2004phases}; extending the Fermi liquid concept to the strongly-interacting limit \cite{khodel2008topology}; solving an extended Hubbard model using dynamical mean-field theory \cite{pankov2008self}; from a renormalization group analysis for multi-valley 2D systems \cite{punnoose2005metal}; by Monte-Carlo simulations \cite{marchi2009correlation,fleury2010energy}. Some theories predict that the disorder is important for the mass enhancement \cite{punnoose2005metal, marchi2009correlation, fleury2010energy}. In contrast with most theories that assume a parabolic spectrum, the authors of Ref.~\cite{khodel2008topology} stress that there is a clear distinction between the mass at the Fermi level and the bandwidth-related mass. In this respect, our conclusions are consistent with the model of Ref.~\cite{khodel2008topology} in which a flattening at the Fermi energy in the spectrum leads to a diverging effective mass. This Fermi-liquid-based model implies the existence of an intermediate phase preceding Wigner crystallization.

\subsection{Entropy}\label{sec:entropy}

The effective electron mass in 2D electron systems can also be extracted from the measurements of the entropy per electron, $\partial S^*/\partial n_s$.  According to the Maxwell relation
\begin{equation}
(\partial S^*/\partial n)_T=-(\partial \mu/\partial T)_{n_s},
\end{equation}
the sought-for entropy can be found from the temperature derivative of the chemical potential, $\mu$. In order to measure $\partial \mu/\partial T$  of the 2D electron layer, a gated structure, where the 2D electrons and the gate act as two plates of a capacitor, was used \cite{kuntsevich2015strongly}. When the temperature is modulated at a frequency $\omega$, the chemical potential of the 2D layer varies, causing the capacitor to recharge.  The induced current is equal to
\begin{equation}
i(t)= \frac{\partial \mu}{\partial T}\, \Delta T \,\frac{ \omega\, C}{e}\sin(\omega t).
\end{equation}

\begin{figure}
\centering
\includegraphics[width=8 cm]{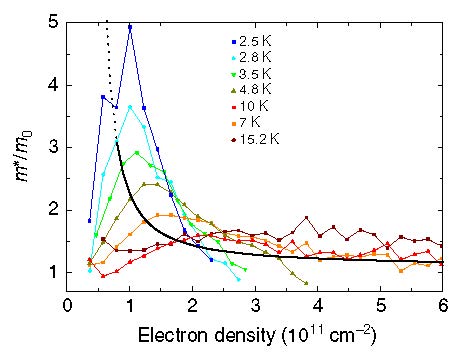}
\caption{The effective mass versus electron density in a Si MOSFET sample. The black curve corresponds to the approximation of $m^*$ from the Shubnikov-de-Haas measurements \cite{pudalov2002low}. Symbols are the $m^*(n_s,T)$ data determined from the measured $\partial S/\partial n_s$ values. Different symbols correspond to different temperatures as indicated. Adapted from Ref.~\cite{kuntsevich2015strongly}.}
\label{entropy}
\end{figure}

The extracted effective mass for different temperatures is shown in Fig.~\ref{entropy}.  In the high-temperature limit $T\gg E_F$, the kinetic energy of electrons is set by temperature, and the 2D electron gas turns out to be weakly interacting, with the density-independent effective mass close to the band mass $m_0$.  For a given temperature, the effective mass exhibits a reentrant behavior: as density decreases, $m^*$  first grows, then passes through a maximum, and falls down approaching a value of the order of
$m_0$
in the low-density non-degenerate ($E_F\lesssim T$) regime. The lower the temperature, the higher the maximum effective mass, reaching approximately $5\,m_0$. The enhanced effective mass is in qualitative agreement with the low-temperature Shubnikov-de~Haas measurements \cite{pudalov2002low} shown by a solid curve in Fig.~\ref{entropy}. The similarity of the data from different measurement methods justifies the validity of the analysis employed. It is still unclear why the $m^*(n_s)$ peak exceeds the SdH data by a factor of up to 2.5. This ``overshooting'' could indicate the beginning of band flattening (see section \ref{sec:flat}), but this interpretation has not yet been explored.

\subsection{Indication of band flattening}\label{sec:flat}

The experimental data obtained in strongly correlated 2D electron systems can be divided into two groups: (i) data describing the electron system as a whole, like thermodynamic density of states, magnetization of the electron system, or the magnetic field required to fully polarize electron spins, and (ii) data related only to the electrons at the Fermi level, like the amplitude of the Shubnikov-de~Haas oscillations.  This yields the effective mass $m^*_F$ and Land\'e $g$-factor $g^*_F$ at the Fermi level.

Usually, the data in the first group are interpreted using the language of quasiparticles, in which the energy-averaged effective mass, $m^*$, and Land\'e $g$-factor, $g^*$, are used. To determine the values, the formulas for non-interacting electrons are used. Although this approach is not ideologically correct, the results for $m^*$ and $g^*$ are often the same as, or close to, those for $m^*_F$ and $g^*_F$. For example, a simultaneous increase of the energy-averaged effective mass and that at the Fermi level in Si MOSFETs was reported in publications described in previous sections \cite{kravchenko2004metal,shashkin2005metal,pudalov2006metal}.

\begin{figure}
\centering
\includegraphics[width=8 cm]{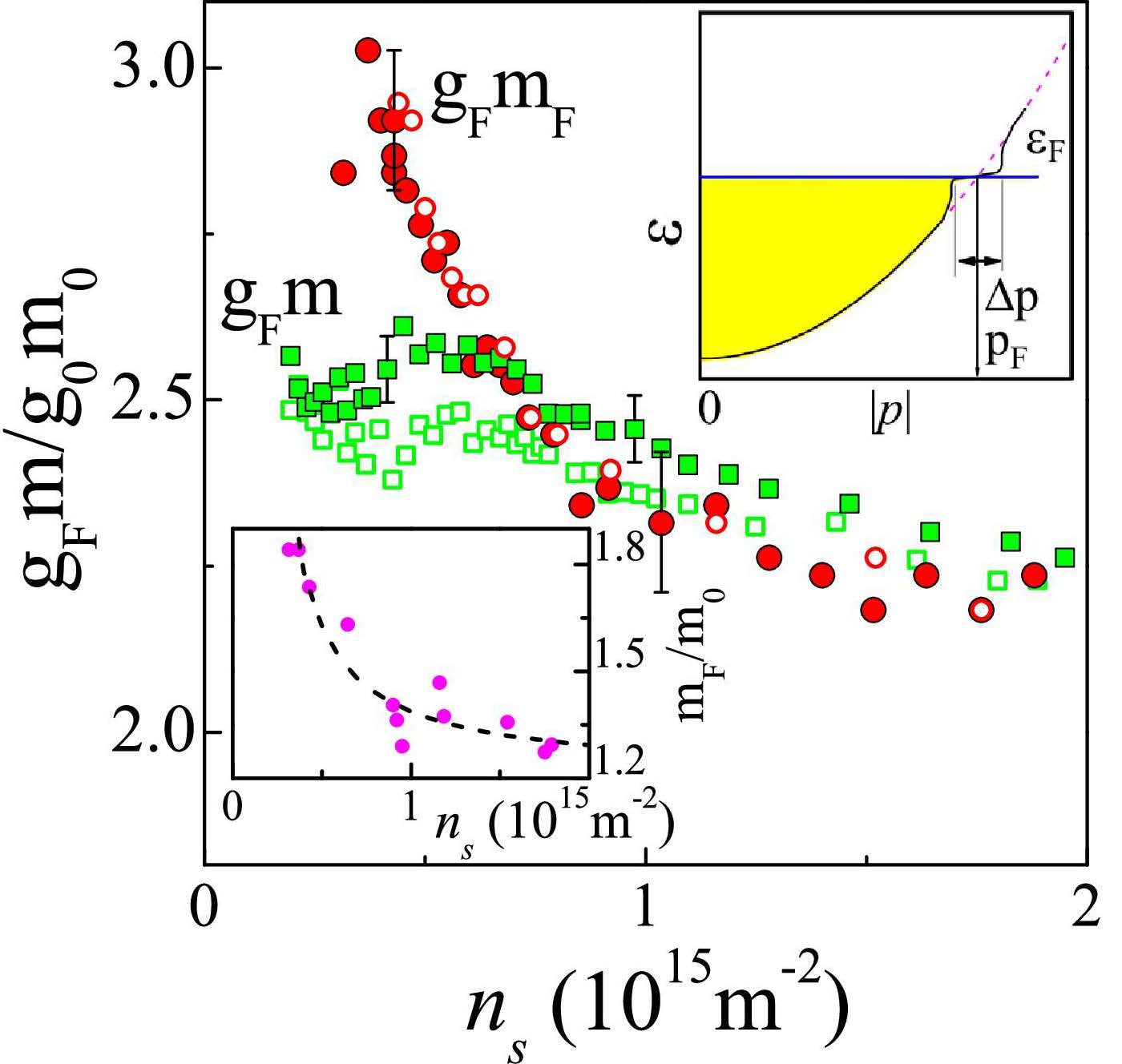}
\caption{Product of the Land\'e $g$-factor and effective mass as a function of electron density in a SiGe/Si/SiGe quantum well determined by measurements of the field of full spin polarization $B^*$ (squares) and Shubnikov-de Haas oscillations (circles) at $T\approx 30$~mK ($g_0=2$ and $m_0=0.19m_e$ are the values for noninteracting electrons). The empty and filled symbols correspond to two samples. The top inset schematically shows the single-particle spectrum of the electron system in a state preceding band flattening at the Fermi level (solid black line). The dashed violet line corresponds to an ordinary parabolic spectrum. The occupied electron states at $T=0$ are indicated by the shaded area. Bottom inset: the effective mass $m^*_F$ versus electron density determined by analysis of the temperature dependence of the amplitude of Shubnikov-de Haas oscillations. The dashed line is a guide to the eye. From Ref.~\cite{melnikov2017indication}.}\label{condensation}
\end{figure}

In contrast, the situation is different in ultra-clean SiGe/Si/SiGe samples \cite{melnikov2017indication}. The behavior of the product $g^*_F m^*$ of the Land\'e factor at the Fermi level (where $g^*_F\approx2$) and the average effective mass differs from that of the product $g^*_Fm^*_F$  of the Land\'e factor and the effective mass at the Fermi level. This difference is illustrated in Fig.~\ref{condensation}. At high electron densities, the dependences of $g^*_F m^*(n_s)$ and $g^*_F m^*_F(n_s)$ are nearly identical. However, they diverge significantly at low densities: while $g^*_F m^*(n_s)$ saturates, $g^*_F m^*_F(n_s)$ increases monotonically as the electron density decreases across the entire range of $n_s$, with a tendency to diverge at low densities.

These experimental results can be interpreted within the concept of the condensation of fermions \cite{khodel1990superfluidity,nozieres1992properties,zverev2012microscopic}.  The fermion condensation occurs at the Fermi level in a range of momenta, unlike the condensation of bosons. As the strength of electron-electron interactions increases, the single-particle spectrum flattens in a region $\Delta p$ near the Fermi momentum $p_{F}$ (top inset to Fig.~\ref{condensation}). At relatively high electron densities ($n_{s}>0.7\times 10^{15}$~m$^{-2}$), this effect is not important because the single-particle spectrum does not change noticeably in the interval $\Delta p$, and the behaviors of the energy-averaged effective mass and that at the Fermi level are similar. However, decreasing the electron density in the range $n_{s}<0.7\times 10^{15}$~m$^{-2}$ stimulates the flattening of the spectrum so that the effective mass at the Fermi level, $m_{F}=p_{F}/V_{F}$, continues to increase (here $V_{F}$ is the Fermi velocity). In contrast, the energy-averaged effective mass does not increase and is not particularly sensitive to this flattening.

\section{The insulating side of the MIT: transport evidence for sliding Wigner solid}\label{sec:WS}

As described in previous sections, experimental investigations into the transport and thermodynamic properties of 2D electrons in semiconductors have indicated that these systems may approach a phase transition to an unknown collective state at low electron densities. This new state could be a quantum Wigner crystal or a precursor \cite{wigner1934on, chaplik1972possible, tanatar1989ground, shashkin2001indication, shashkin2002sharp, attaccalite2002correlation, spivak2004phases, shashkin2006pauli, mokashi2012critical, melnikov2017indication, kagalovsky2020hartree}. Weak disorder is known to facilitate the formation of the pinned Wigner solid \cite{chui1995impurity, thakur1996frozen}; quantum Monte Carlo calculations \cite{chui1995impurity} that account for realistic disorder have predicted a critical $r_s$ significantly lower than $r_s=37$ predicted for the ideal clean 2D system \cite{tanatar1989ground}. The phase transition point in the least-disordered 2D electron systems in semiconductors is near the critical electron density for the metal-insulator transition. Although the insulating regime of the metal-insulator transition has been extensively studied \cite{andrei1988observation, williams1991conduction, goldman1990evidence, jiang1990quantum, jiang1991magnetotransport, diorio1992reentrant, pudalov1993zero, giamarchi2003electronic, qiu2012connecting, knighton2014reentrant, qiu2018new, knighton2018evidence, falson2022competing, hossain2022anisotropic, madathil2023moving}, there has been no definitive conclusion regarding the origin of the low-density state. Single-threshold current-voltage ($I-V$) curves have been observed, which can be interpreted either as a sign of the depinning of an electron solid \cite{goldman1990evidence,williams1991conduction,diorio1992reentrant,pudalov1993zero} or as a breakdown of the insulating phase, consistent with traditional scenarios like strong electric field Efros-Shklovskii variable range hopping \cite{marianer1992effective} or percolation (see, \textit{e.g}., Refs.~\cite{jiang1991magnetotransport, dolgopolov1992metal, shashkin1994insulating, shashkin2005metal} and the references therein). Attempts to observe broadband voltage noise at the threshold of these $I-V$ curves and to probe the low-density state in perpendicular magnetic fields have not yielded sufficient information to clearly differentiate between the depinning of the electron solid and traditional mechanisms \cite{jiang1991magnetotransport, shashkin1994insulating, shashkin2005metal}. It is important to note that much confusion has arisen because many authors have chosen to interpret their data in the context of the Wigner crystal, overlooking conventional interpretations.

A significant breakthrough in understanding the origin of the low-density state in strongly interacting 2D electron systems has been reported in Ref.~\cite{brussarski2018transport}. Two-threshold \( V-I \) characteristics have been observed, marked by a dramatic increase in noise between the two threshold voltages at the breakdown of the insulating low-electron-density state. This noise manifests itself as strong current fluctuations; it rises significantly above the first threshold voltage \( V_{th1} \) and essentially disappears above the second threshold voltage \( V_{th2} \).

\begin{figure}
\centering
\includegraphics[width=8 cm]{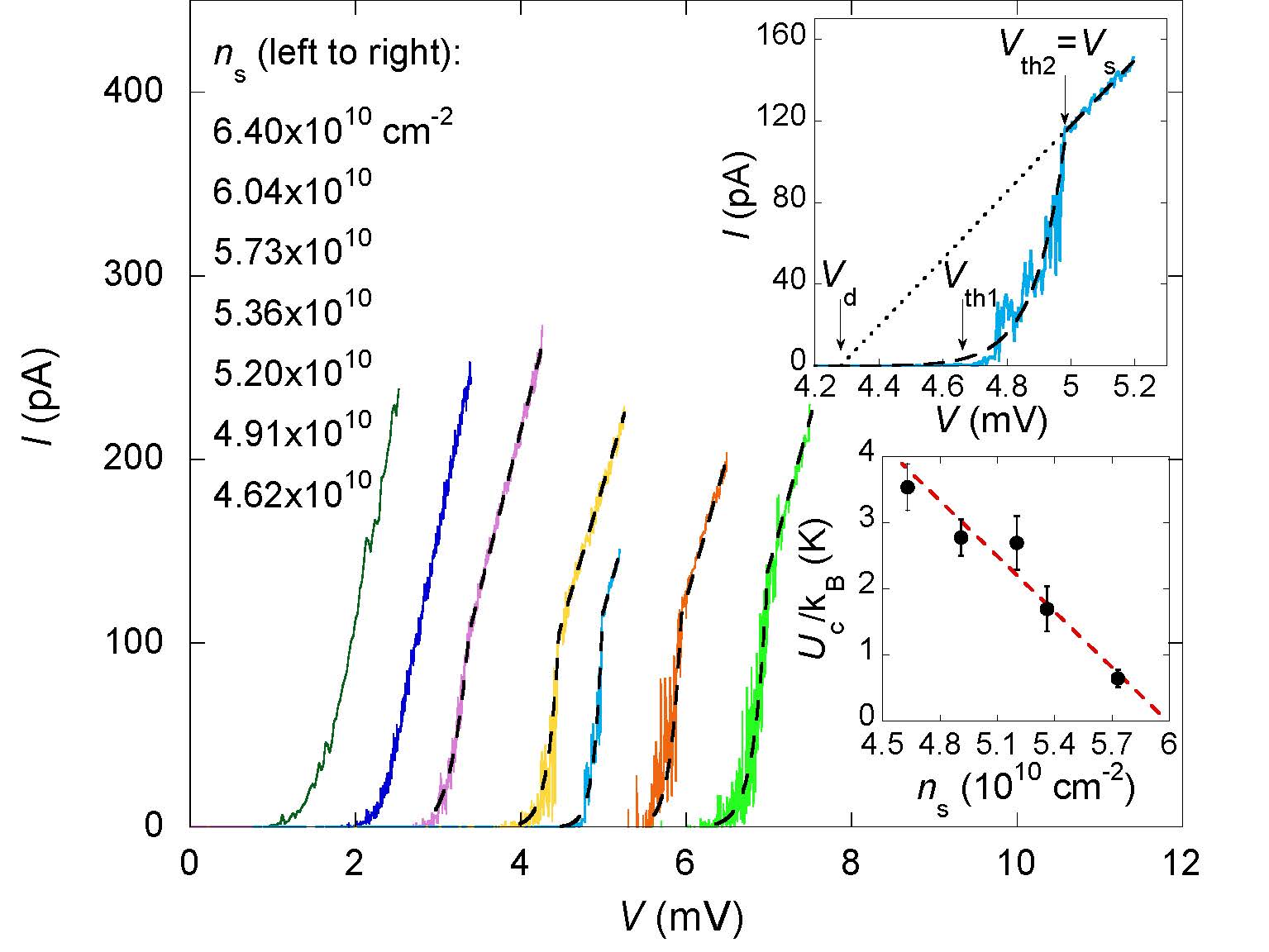}
\caption{$V-I$ characteristics measured at different electron densities in the insulating state at a temperature of 60~mK in a Si MOSFET sample. The dashed lines are fits to the data using Eq.~(\ref{I}). In the top inset, the $V-I$ curve is shown for $n_{s}=5.20\times 10^{10}$~cm$^{-2}$ on an expanded scale; the threshold voltages $V_{{th1}}$ and $V_{{th2}}$, the static threshold $V_{s}=V_{{th2}}$, and the dynamic threshold $V_{d}$ are indicated. In the bottom inset, the activation energy $U_{c}$ is plotted \textit{vs}.\ the electron density. The dashed line is a linear fit.  Adapted from Ref.~\cite{brussarski2018transport}.}\label{VI Si}
\end{figure}

Figure~\ref{VI Si} illustrates a series of low-temperature voltage-current curves measured at different electron densities within the insulating regime in a Si MOSFET sample \cite{brussarski2018transport}. The critical density for the MIT in this electron system is approximately \(n_{c} \approx 8 \times 10^{10} \, \text{cm}^{-2}\) which corresponds to the interaction parameter
\(r_{s} \sim 20\).
For electron densities below approximately \(6 \times 10^{10} \, \text{cm}^{-2}\), two threshold voltages are observed. As the applied voltage increases, the current remains near zero until it reaches a first threshold voltage, \(V_{{th1}}\). Once \(V_{{th1}}\) is surpassed, the current rises sharply until a second threshold voltage, \(V_{{th2}}\), is reached. Beyond \(V_{{th2}}\), the slope of the voltage-current curve decreases significantly, and the behavior becomes linear, although it is not ohmic (see also the top inset of Fig.~\ref{VI Si}). As the electron density is increased, the value of \(V_{{th1}}\) decreases, while the second threshold becomes less pronounced and eventually disappears.

\begin{figure}
\centering
\includegraphics[width=6 cm]{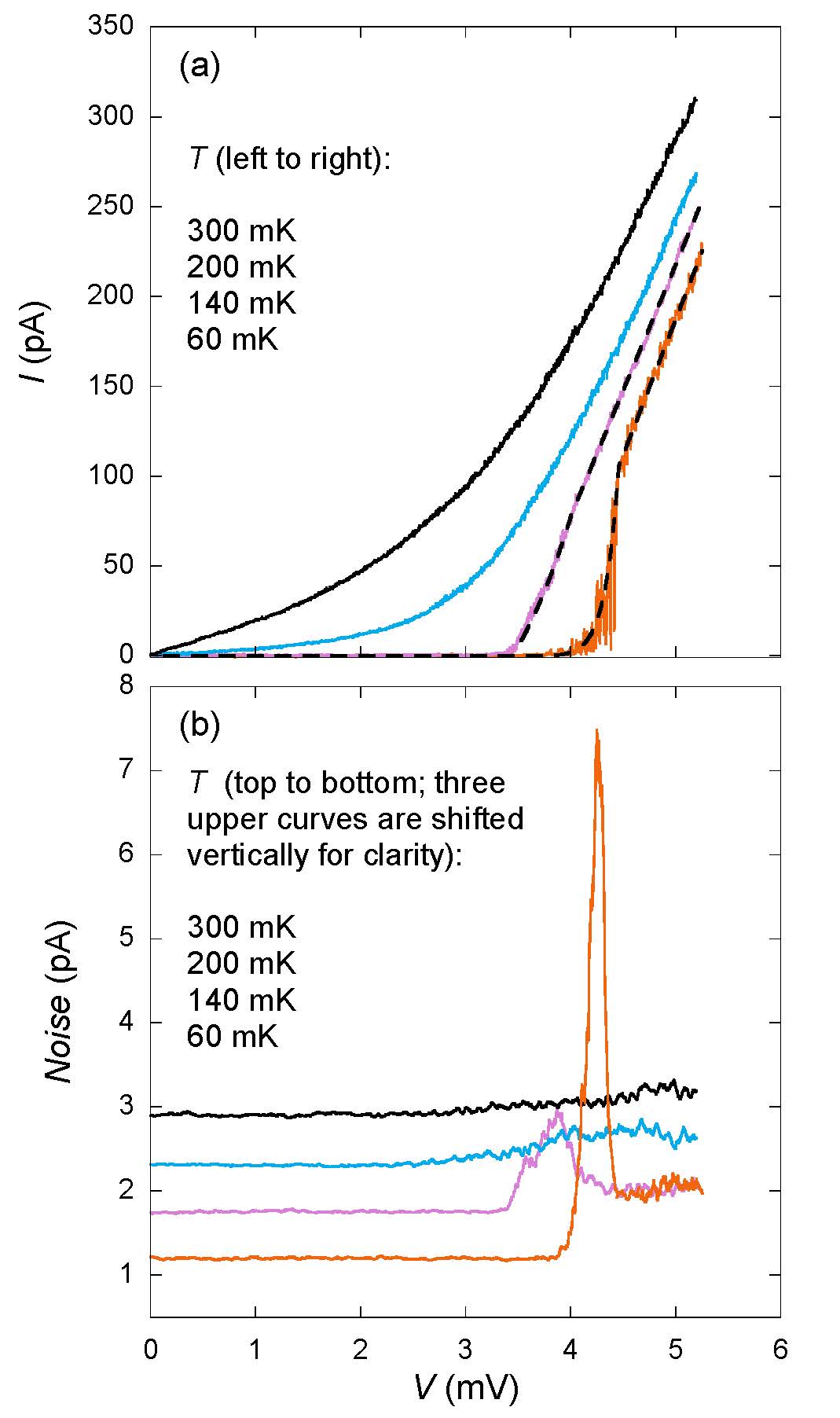}
\caption{(a) $V-I$ characteristics measured at $n_{s}=5.36\times 10^{10}$~cm$^{-2}$ and different temperatures in a Si MOSFET sample. The dashed lines are fits to the data using Eq.~(\ref{I}). (b)~The broadband noise as a function of voltage for the same electron density and temperatures. Adapted from Ref.~\cite{brussarski2018transport}.}\label{noise Si}
\end{figure}

Figure~\ref{noise Si}(a) displays the $V-I$ characteristics for the electron density of $n_{s}=5.36\times 10^{10}$~cm$^{-2}$ at various temperatures. As the temperature $T$ increases, the second threshold voltage, $V_{{th2}}$, becomes less distinct, and the threshold behavior of the $V-I$ curves blurs.

The measured broadband noise is presented as a function of voltage in Fig. \ref{noise Si}(b) for various temperatures, with an electron density of \( n_{s} = 5.36 \times 10^{10} \, \text{cm}^{-2} \). A significant increase in noise is observed between the thresholds \( V_{{th1}} \) and \( V_{{th2}} \) at the lowest temperature. This substantial noise decreases rapidly as the temperature increases, which is consistent with the two-threshold behavior seen in the \( V-I \) curves shown in Fig. \ref{noise Si}(a).

There is a striking similarity between the double-threshold $V-I$ dependences in the insulating state of Si MOSFETs and those (with the voltage and current axes interchanged) known for the depinning of the vortex lattice in type-II superconductors (see, for example, Refs.~\cite{yeh1991flux,blatter1994vortices,bullard2008vortex}). The physics of the vortex lattice in type-II superconductors, in which the existence of two thresholds is well known, can be adapted for the case of an electron solid. The transient region between the dynamic ($V_{d}$) and static ($V_{s}$) thresholds corresponds to the collective pinning of the solid. In this region, the pinning occurs at the centers with different energies, and the current is thermally activated:

\begin{equation}
I\propto\exp\left[-\frac{U(V)}{k_{B}T}\right],
\label{exp}
\end{equation}
where $U(V)$ is the activation energy. The static threshold $V_{s}$ signals the onset of the regime of solid motion with friction. This corresponds to the condition

\begin{equation}
eEL=U_{c},
\label{Uc}
\end{equation}
where $E$ is the electric field and $L$ is the characteristic distance between the pinning centers with maximal activation energy $U_{c}$. From the balance of the electric, pinning, and friction forces in the regime of solid motion with friction, one expects a linear $V-I$ characteristic that is offset by the threshold $V_{d}$ corresponding to the pinning force

\begin{equation}
I=\sigma_0(V-V_{d}),
\label{linear}
\end{equation}
where $\sigma_0$ is a coefficient. Assuming that the activation energy for the Wigner solid is equal to

\begin{equation}
U(V)=U_{c}-eEL=U_{c}(1-V/V_{s}),
\label{U}
\end{equation}
one obtains the expression for the current

\begin{equation}
I=\left\{\begin{array}{c}
\sigma_0(V-V_{d}) {\text{ if }} V>V_{s}\\
\sigma_0(V-V_{d})\exp\left[-\frac{U_{c}(1-V/V_{s})}{k_{B}T}\right] {\text{ if }} V\leq V_{s}.
\end{array}\right.\label{I}
\end{equation}
The fits to the data using Eq.~(\ref{I}) are shown by dashed lines in Figs.~\ref{VI Si} and \ref{noise Si}(a). As the figures show, the experimental two-threshold $V-I$ characteristics are described well by Eq.~(\ref{I}). The value of $U_{c}$ decreases approximately linearly with electron density and tends to zero at $n_{s}\approx 6\times 10^{10}$~cm$^{-2}$ (the bottom inset of Fig.~\ref{VI Si}). This is in contrast to the vanishing activation energy of electron-hole pairs at $n_{c}$ obtained by measurements of the resistance in the limit of zero voltages/currents \cite{shashkin2001metal}. Presumably, the vanishing $U_{c}$ is related to the minimum number of the strong pinning centers for which the collective pinning is still possible. The fact that the coefficient $\sigma_0$ is approximately constant ($\sigma_0\approx 1.6\times 10^{-7}$~Ohm$^{-1}$) indicates that weak pinning centers control the solid motion with friction \cite{blatter1994vortices}. Thus, the physics of the vortex lattice, adapted to the case of an electron solid, is relevant to the insulating state of a 2D electron system in silicon.

\begin{figure*}
\centering
\includegraphics[width=5 cm]{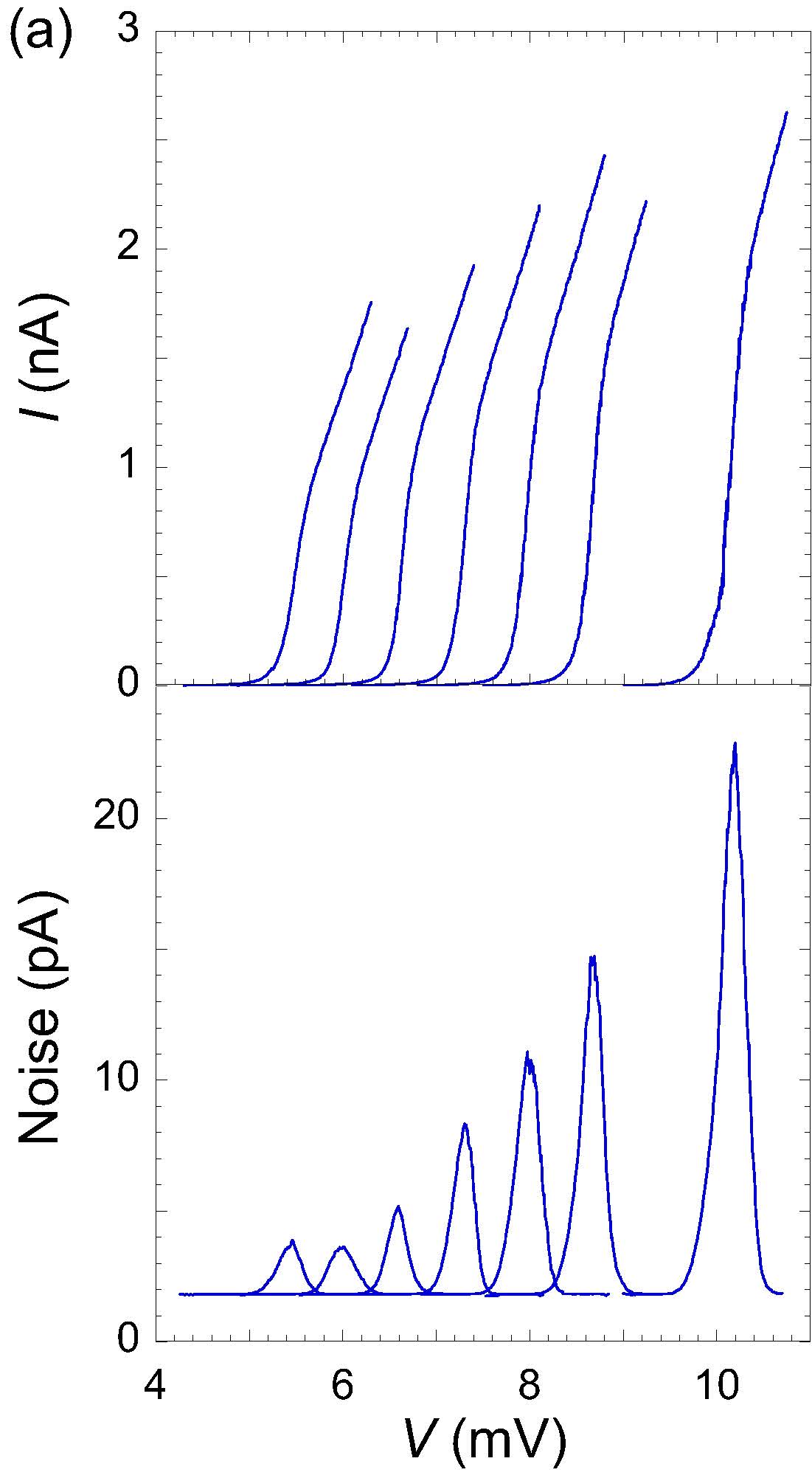}
\includegraphics[width=5 cm]{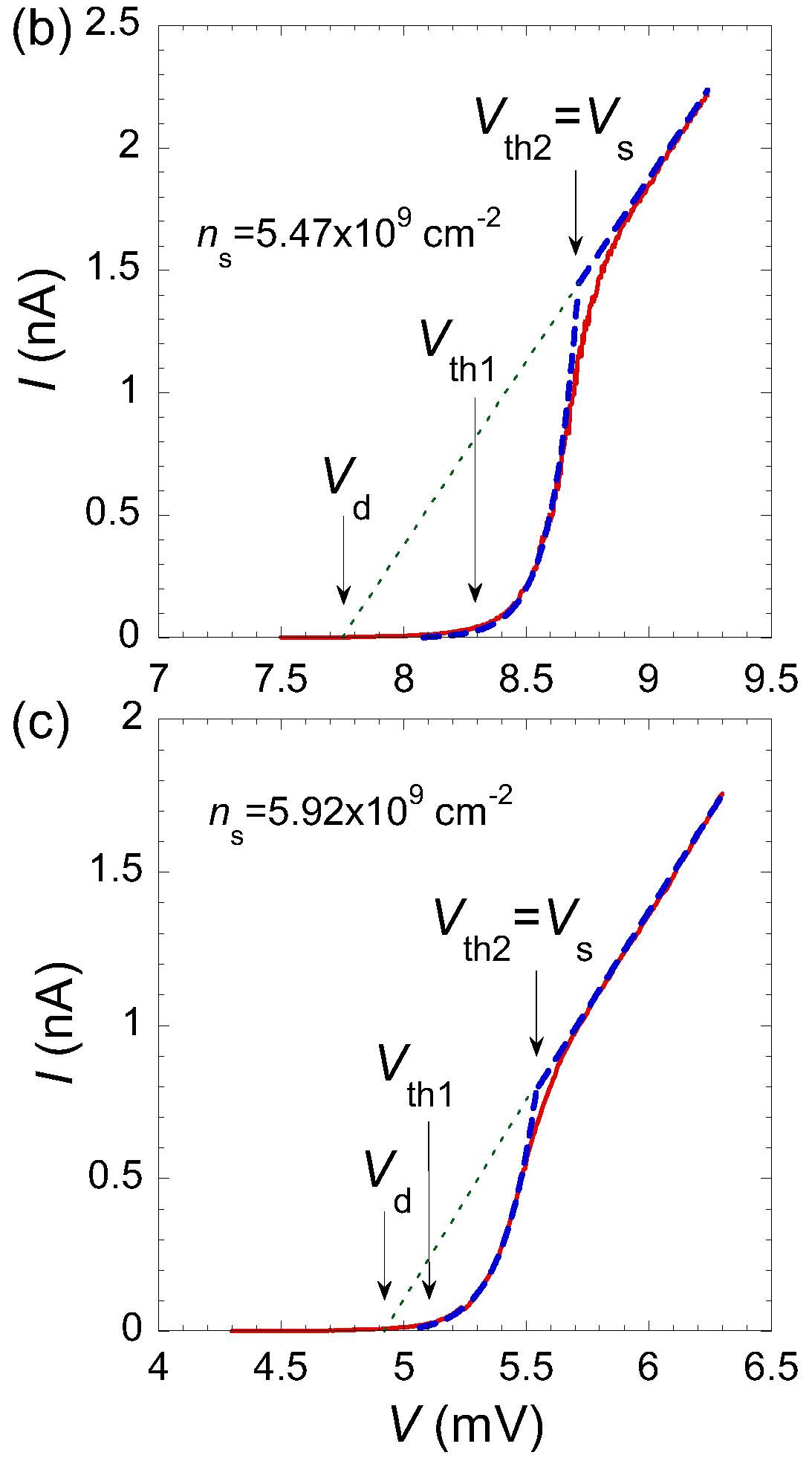}
\caption{(a) The $V$-$I$ characteristics measured in ultra-high mobility SiGe/Si/SiGe heterostructure are shown in the upper panel and the broadband noise in the lower panel at a temperature of approximately 30~mK and various electron densities (from left to right): 5.92, 5.83, 5.74, 5.65, 5.56, 5.47, and $5.29\times10^9$~cm$^{-2}$.  Panels (b) and (c) display the $V$-$I$ characteristics for two specific electron densities on an expanded scale.  Also indicated are the threshold voltages, $V_{{th1}}$ and $V_{{th2}}$, the dynamic threshold $V_{d}$, which is obtained by extrapolating the linear portion of the $V$-$I$ curves to zero current, and the static threshold $V_{s}=V_{{th2}}$.  The dashed lines are fits to the data using Eq.~(\ref{I}). From Ref.~\cite{melnikov2024collective}.}
\label{VI SiGe}
\end{figure*}

\begin{figure}
\centering
\includegraphics[width=5 cm]{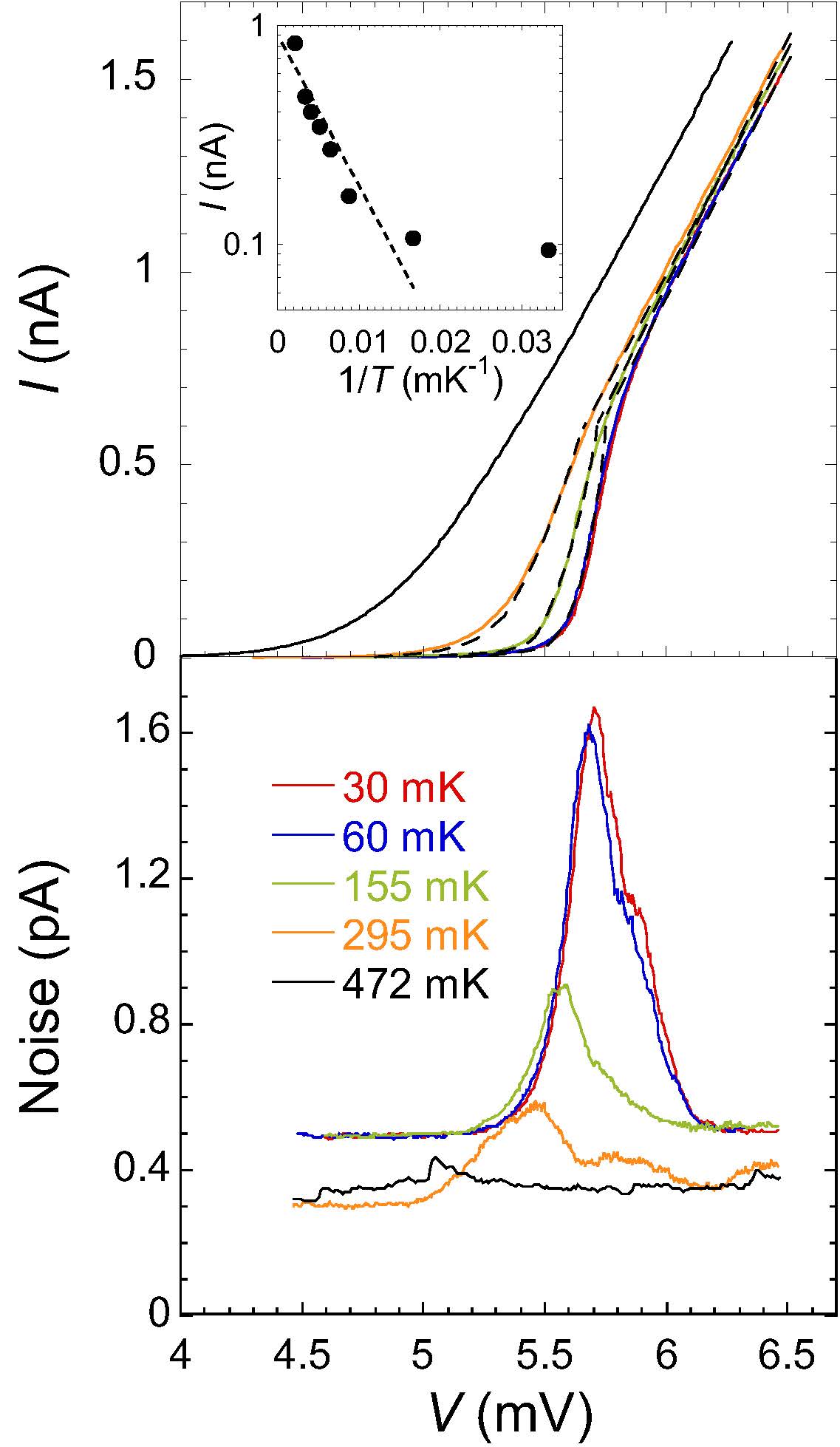}
\caption{The current (upper panel) and broadband noise (lower panel) in another SiGe/Si/SiGe sample are shown as a function of voltage at an electron density of \(n_{s} = 5.83 \times 10^9 \, \text{cm}^{-2}\) for different temperatures. The curves in the upper panel are color-coded to correspond with the same temperatures presented in the lower panel. The dashed lines represent fits to the data using Eq.~(\ref{I}). The inset includes an Arrhenius plot of \(I(T)\) at \(n_{s} = 5.83 \times 10^9 \, \text{cm}^{-2}\) and \(V = 5.6 \, \text{mV}\). The dashed line is a linear fit that excludes the data point at 30 mK. From Ref.~\cite{melnikov2024collective}.}\label{SiGe T}
\end{figure}

Similar effects are observed in much less disordered ultra-high-mobility SiGe/Si/SiGe heterostructures, which belong to a different class of 2D systems as inferred from both the level and character of the disorder potential.  In Fig.~\ref{VI SiGe}(a), typical voltage-current characteristics (upper panel) and the generated noise (lower panel), measured at a temperature of approximately 30 mK, are presented. The measurements were conducted across different electron densities within the insulating regime \cite{melnikov2024collective} (see also Fig.~\ref{VI SiGe}(b, c) for a more detailed view). These densities are below, but not too close to, the critical density for the MIT \( n_{c} \approx 8.8 \times 10^9 \, \text{cm}^{-2} \). At these specific values of \( n_s \), the interaction parameter \( r_{s} \) exceeds 20. Between the two threshold voltages, a peak in broadband current noise is observed, as shown in the lower panel of Fig.~\ref{VI SiGe}(a).  Figure~\ref{SiGe T} shows how the voltage-current characteristics and noise levels change with temperature. As the temperature rises, the voltage-current characteristics become less steep between the two threshold voltages, and the curves shift to lower voltages until the behavior associated with these two thresholds disappears. Similarly, the noise peak diminishes and eventually disappears as temperature increases. The inset of the upper panel in Fig.~\ref{SiGe T}(a) features an Arrhenius plot of \( I(T) \), which illustrates the activation temperature dependence down to approximately 60 mK.

As in silicon MOSFETs, the observed results can be explained by a phenomenological theory of collective depinning of elastic structures, which naturally produces a peak in broadband current noise between the dynamic and static thresholds, transitioning to the sliding of the solid over a pinning barrier once the static threshold is exceeded. These findings provide evidence for the formation of a quantum electron solid within this electron system and demonstrate the generality of the effect across different classes of electron systems.

\section{Modern theoretical approaches to the $B=0$ MIT}

\subsection{Metallic temperature dependence of the resistivity}

In the mid-80s, the quasi-linear ``metallic'' temperature dependence of conductivity observed in the ballistic interaction regime ($k_BT\tau/\hbar>1$)) was attributed \cite{stern1980calculated, gold1986temperature, dassarma1986theory} to the weakening of screening of the scattering potential with increasing temperature.   More recently, a theory was developed that accounted for all interaction contributions to the conductivity, including exchange contributions \cite{zala2001interaction, zala2001interaction_a}. This theory offers a unified approach to both ballistic and diffusive ($k_BT\tau/\hbar< 1$) interaction regimes by considering quantum interference between electron waves scattered on Friedel oscillations of the electron density. The theories mentioned above provide a quantitative description of the linear-in-$T$ corrections to the conductivity in the ballistic regime \cite{shashkin2004comment}. The theory \cite{zala2001interaction, zala2001interaction_a} was extended to the case of a long-range scattering potential \cite{gornyi2004interaction}.  The theories \cite{zala2001interaction, zala2001interaction_a, gornyi2004interaction} naturally incorporate the Altshuler-Aronov results for the interaction corrections to the conductivity in the diffusive regime \cite{altshuler1980interaction}.

\subsection{Two-parameter scaling theory}

 The low-temperature physics of the disordered electron liquid, combined with the large $r_s$ values near the MIT, is determined by the properties of interacting, diffusing electrons \cite{finkelstein1983influence,castellani1984interaction,finkelstein1984weak}.  A modern scaling theory for the 2D MIT in a disordered two-dimensional electron gas was presented in Ref.~\cite{punnoose2005metal}. A quantum critical point has been identified that separates the metallic phase, stabilized by electron-electron interactions, from the insulating phase, where disorder prevails over interactions. According to Ref.~\cite{punnoose2005metal}, the existence of a quantum critical point leads to a divergence in the density of states at the transition, causing the thermodynamic properties to exhibit critical behavior as the transition is approached.

\begin{figure}
\centering \includegraphics[width=.45\textwidth]{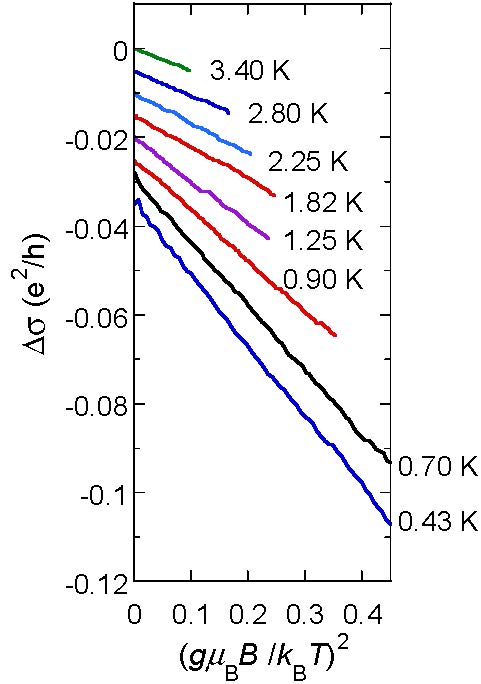}
\caption{Magnetoconductivity in a Si MOSFET sample for electron density $n_s=9.14\times10^{10}$~cm$^{-2}$ measured at different temperatures plotted as a function of $(g\protect\mu_B B/k_BT)^2$. The curves are vertically shifted for clarity. From Ref.~\cite{anissimova2007flow}.} \label{comparison 2}
\end{figure}

\begin{figure}
\centering\includegraphics[width=.6\textwidth]{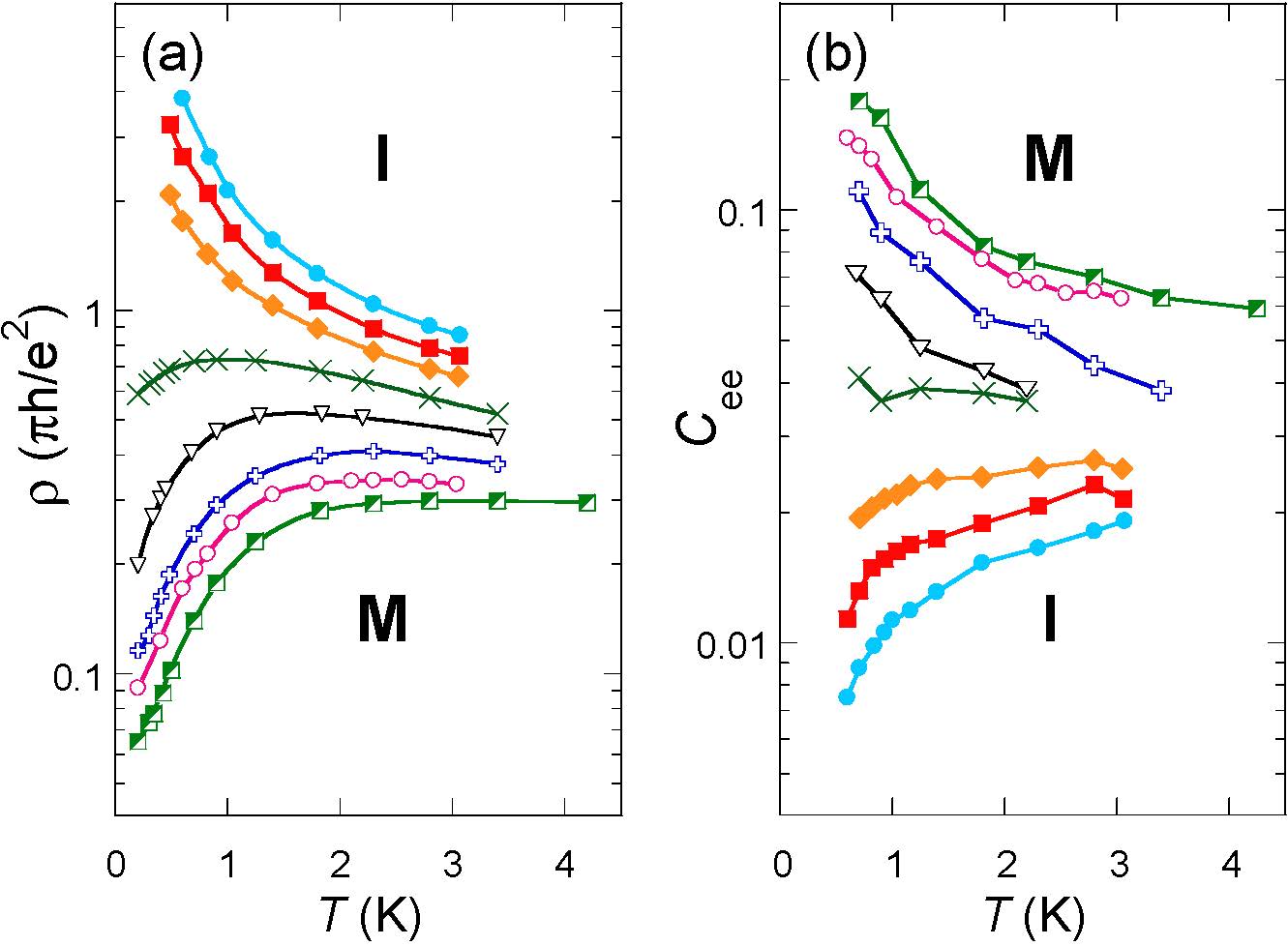}
\caption{Temperature dependences of the resistivity and of the strength of spin-related interactions for electron densities across the MIT in a Si MOSFET sample. (a): $\protect\rho(T)$ traces at $B=0$.  (b): $C_{ee}(T)$, extracted from the slope of the magnetoconductance (see text).   Both $\rho(T)$ and $C_{ee}(T)$ display a fan-like structure across the MIT. The densities (in units of $10^{10}$~cm$^{-2}$) are: 7.24, 7.53, 7.83, 8.26, 8.70, 9.14, 9.58, and 9.87.  In panel (a), the densities increase from top to bottom, while in panel (b), they increase from bottom to top, \textit{i.e.}, the interaction increases in the metallic phase and decreases in the insulating phase as the temperature is lowered. \textbf{M} and \textbf{I} regions indicate metallic and insulating phases, respectively.  From Ref.~\cite{anissimova2007flow}.} \label{comparison 1}
\end{figure}

The experimental test of this theory was done in Refs.~\cite{anissimova2007flow, knyazev2008metal}.  The authors have performed simultaneous measurements of the resistivity, $\rho$, and in-plane magnetoconductance, $\sigma(B)$.  Since the origin of the magnetoconductance lies in the interaction corrections to the conductance involving different spin projections (Hartree-like), the corresponding interaction strength can be determined from the slope of the magnetic field dependence of conductivity $\sigma(B)$. When a magnetic field $B$ is applied parallel to the 2D plane in the diffusive regime, $k_BT<\hbar/\tau<E_F$ \cite{lee1982magnetoresistance,lee1985disordered}, the magnetoconductance $\Delta \sigma (B,T)\equiv \sigma (B,T)-\sigma (0,T)$ is proportional to $b^2$ in the weak field limit $b\equiv g\mu _{B}B/k_{B}T\ll 1$. On the other hand, at higher temperatures $\hbar /\tau <k_{B}T<E_{F}$, referred to as the ballistic regime, $\Delta \sigma (B,T)$ is expected to be proportional to $T b^{2}$ \cite{zala2001interaction_a}. Due to the significant difference in magnetoconductivities between these two regions, measuring $\Delta \sigma(B,T)$ provides a method to identify the diffusive regime, although some accompanying difficulties were reported in Ref.~\cite{morgun2016novel}.

The explicit form of the slope of $\Delta\sigma(b^2)$ for the single valley case was derived in Refs.~\cite{lee1982magnetoresistance,castellani1998metallic} and generalized to include the valley degrees of freedom in Ref.~\cite{anissimova2007flow}. In the case when there are $n_{v}$ degenerate valleys ($n_{v}=2$ for silicon), the authors obtained for $\Delta\sigma(b^2)$, in the limit $b\ll 1$, the expression
\begin{equation}
\Delta \sigma (b^2)=-0.091\frac{e^{2}}{\pi h}n_{v}^{2}\gamma_{2}\left(\gamma_{2}+1\right)b^{2}.
\label{eqn:linear}
\end{equation}
Here, $\gamma_{2}$ is the effective electron-electron interaction amplitude in the spin-density channel.

Eq.~(\ref{eqn:linear}), as was shown in Ref.~\cite{castellani1998metallic}, incorporates renormalization group (RG) corrections to first order in $\rho$ and is therefore strictly valid only deep in the metallic region. However, for general $\rho$, the magnetoconductance in the diffusive regime will still retain the $b^2$-form, $\Delta\sigma=-(e^{2}/\pi h)\,C_{ee}(\gamma_2,\rho)\, n_{v}^{2}\,b^{2}$. The coefficient $C_{ee}$ reflects the strength of spin-related interactions at any value of the resistance, as long as $g\mu_BB<k_BT<h/\tau$ \cite{anissimova2007flow}.

The temperature dependences of the parameter $C_{ee}$, extracted by fitting the $\sigma(b^{2})$ traces shown in Fig.~\ref{comparison 2}, are displayed in Fig.~\ref{comparison 1}(b) for various densities across the MIT.  In Fig.~\ref{comparison 1}(a), $\rho(T)$ dependences at zero magnetic field for the same densities are plotted. Fig.~\ref{comparison 1} reveals that not only $\rho$ but also the interaction strength exhibits a fan-like spread as the MIT is
crossed. While the interaction increases in the metallic phase as the temperature decreases, it is suppressed in the insulating phase. The magnitudes of these changes depend on the distance from the MIT, with both $\rho$ and $\gamma_2$ becoming practically temperature-independent for densities near the MIT.

\begin{figure}
\centering \includegraphics[width=.40\textwidth]{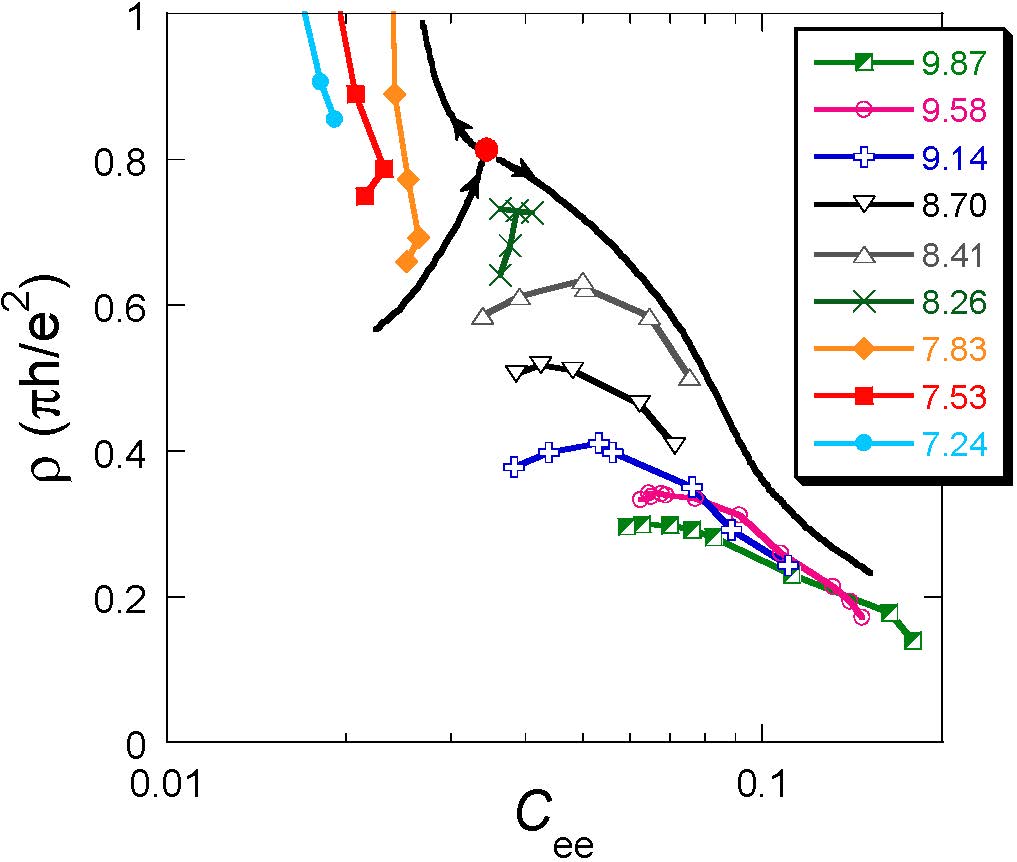} \includegraphics[width=.48\textwidth]{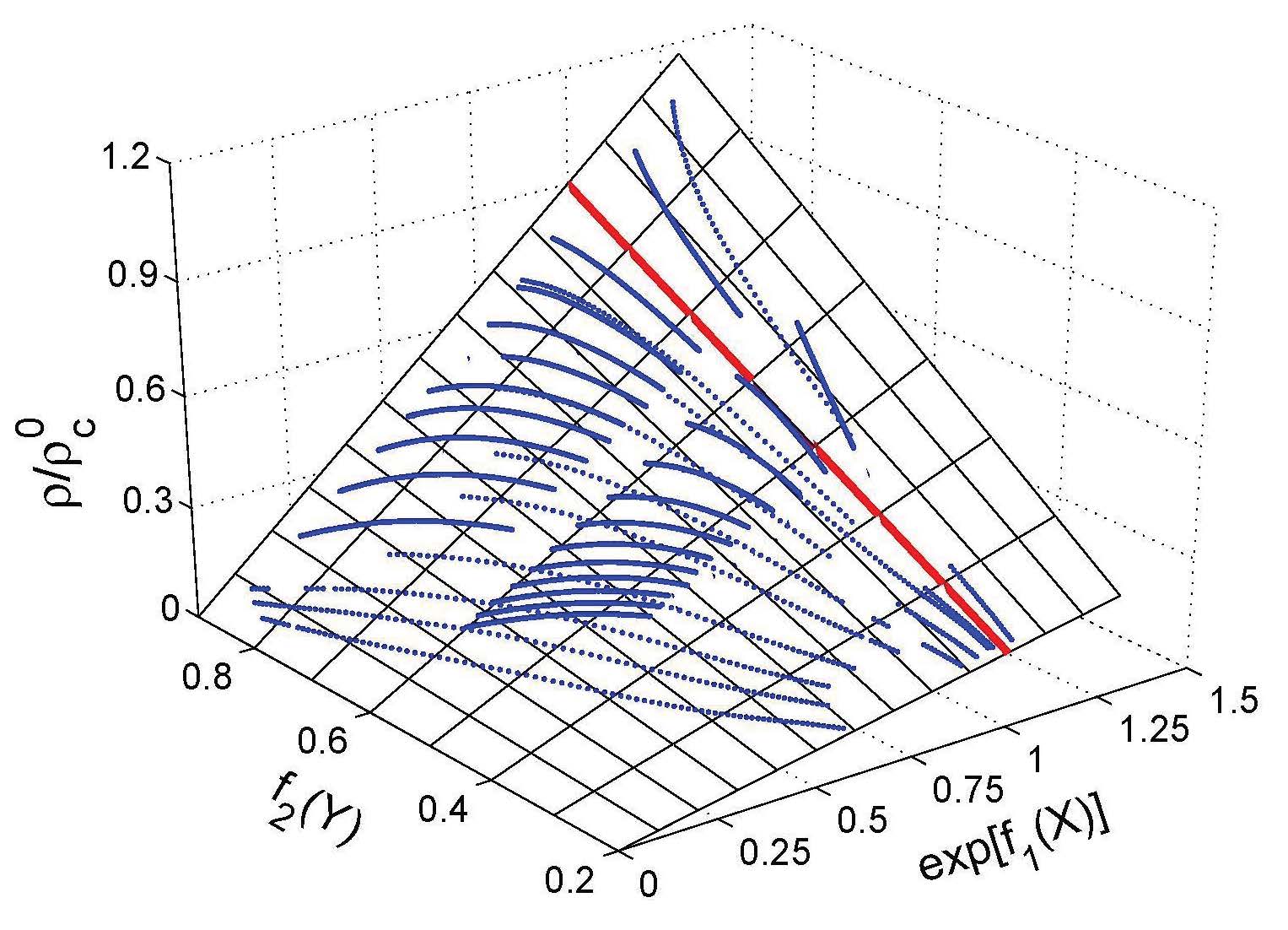}
\caption{Left panel: The disorder-{interaction} ($\rho$~--~$C_{ee}$) flow diagram of the 2D electron liquid in a Si MOSFET. The red dot at $\rho\approx 0.8$ and $C_{ee}\approx 0.035$ indicates the location of the quantum critical point from which the three separatrices (black lines) emanate. Arrows indicate the direction of the flow as the temperature is lowered.  Electron densities are indicated in units of $10^{10}$~cm$^{-2}$. From Ref.~\cite{anissimova2007flow}. Right panel: The empirical two-parameter scaling function $\mathcal{R}(X, Y)$ for five samples (dots) and the model surface $\mathcal{F} = \tilde{X}\tilde{Y}$ (grid). The straight line $\rho/\rho_c^0 = \tilde{Y}$ at $\tilde{X} = 0$ represents the separatrix. From Ref. \cite{knyazev2008metal}} \label{comparison 3}
\end{figure}

To see this in a different way, the authors of Ref.~\cite{anissimova2007flow} combined Figs.~\ref{comparison 1}(a) and (b) and constructed a two-parameter ``flow'' diagram in the disorder--interaction ($C_{ee}-\rho$) space. This flow diagram is presented in Fig.~\ref{comparison 3}. One can see that it clearly indicates the existence of a quantum critical point from which three separatrices, one attractive and two repulsive, can be drawn.

On the metallic side of the flow diagram, a separatrix that forms the envelope of a family of non-monotonic curves is clearly visible in Fig.~\ref{comparison 3} and flows toward the region with $\rho\ll 1$. The theoretical details of the evolution of $\rho$ and $\gamma_2$ in this region with $\rho\ll 1$ were discussed in detail in Ref.~\cite{punnoose2001dilute} in terms of the renormalization group (RG) theory.  We recall the salient features of the theory here. The theory predicts that:\\
(i)~The amplitude $\gamma_2$ increases monotonically as the temperature is reduced.\\
(ii)~The resistance, on the other hand, has a characteristic non-monotonic form changing from insulating behavior ($d\rho/dT<0$) at high temperatures to metallic behavior ($d\rho/dT>0$) at low temperatures, with the maximum value $\rho_{max}$ occurring at a crossover temperature $T=T_{max}$.\\
(iii)~The value of the amplitude $\gamma_2$ at $T=T_{max}$ is universal in the limit $\rho_{max}\ll 1$,
depending only on $n_v$; for $n_{v}=1$, it is $2.08$, while for $n_{v}=2$, it has the much lower value $0.45$.\\
(iv)~Although the values of $\rho_{max}$ and $T_{max}$ are not universal and depend on the system, the behaviors of $\rho(T)/\rho_{max}$ and $\gamma_{2}(T)$ are universal when plotted as functions of $\rho_{max}$ln$(T/T_{max})$.

\begin{figure}
\centering \includegraphics[width=0.8\textwidth]{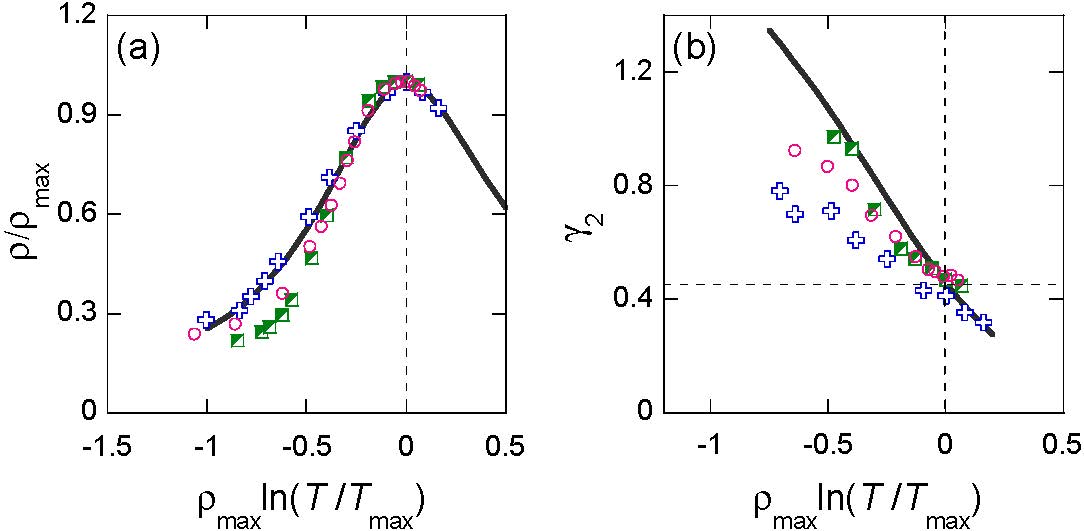}\\
\caption{Comparison between theory (lines) and experiment (symbols).  (a): $\rho/\rho_{max}$ as a function of $\rho_{max}\;\text{ln}(T/T_{max})$.  (b): $\gamma _{2}$ as a function of $\rho_{max}\;\text{ln}(T/T_{max})$. Vertical dashed lines correspond to $T=T_{max}$, the temperature at which $\rho(T)$ reaches maximum.  Note that at this temperature, the interaction amplitude $\gamma_2\approx 0.45$ (indicated by the horizontal dashed line in (b)), is in excellent agreement with theory. Symbols are the same as in Figs.~\ref{comparison 1} and \ref{comparison 3} and correspond to $n_{s}=9.87$, $9.58$ and $9.14\times 10^{10}$~cm$^{-2}$. From Ref.~\cite{anissimova2007flow}.} \label{comparison 4}
\end{figure}

The results of the comparison between experiment and theory are presented in Figs.~\ref{comparison 4}(a) and (b), for the three largest densities in the diffusive regime with $\rho_{max}\lesssim 0.4~(\pi h/e^2)$. At these densities, the resistance is small enough, and one can safely extract the interaction amplitudes $\gamma_2$ from $\Delta\sigma(b^2)$ using Eq.~(\ref{eqn:linear}), \textit{i.e.}, $C_{ee}=0.091 \gamma_2(1+\gamma_2)$. (As the density increases, the maximum in $\rho(T)$ becomes less pronounced, and the non-monotonicity eventually disappears at $n_{s}\gtrsim 1\times 10^{11}$~cm$^{-2}$, as the ballistic region is approached.) The solid lines are the universal theoretical curves for $\rho(T)/\rho_{max}$ and $\gamma_2(T)$ plotted as a function of $\rho_{max}\ln(T/T_{max})$. One sees from Fig.~\ref{comparison 4} that at $T=T_{max}$, the value of the interaction amplitude $\gamma_2\approx 0.45$, which is in remarkable agreement with theory for $n_v=2$. The agreement between theory and experiment is especially striking given that the theory has no adjustable parameters. Systematic deviations from the universal curve occur at lower densities as higher-order corrections in $\rho$ become important.

Parameter $\gamma_2$ may also be determined from the renormalized $g^*$-factor in the diffusive regime, using, e.g., SdH oscillations:
\begin{equation}
\gamma_2(\gamma_2+1) \approx \frac{g^*}{2}(1-\frac{g^*}{2}).
\label{eq:gamma2_vs_g}
\end{equation}
The temperature dependence of the g-factor, and consequently, $\gamma_2$, was determined from weak-field SdH oscillations in the study by Pudalov et al.\ \cite{pudalov2018probing} within a temperature range of 0.1 to 1 K. This analysis was conducted while decreasing the electron density from approximately $7\, n_c$ to $n_c$. The dependence of $\gamma_2$ on temperature was found to be weak, with no significant increase at lower temperatures as predicted by the renormalization group (RG) theory. The authors of the study concluded that a key prediction of RG theory --- a substantial increase in $\gamma_2$ with cooling --- was not supported by the SdH measurements.

It has been predicted \cite{punnoose2005metal} that the effective mass should dramatically increase at disorder-dependent density for the metal-insulator transition while the g-factor remains nearly intact. However, the prediction of disorder-dependent effective mass is not confirmed by the experimental data \cite{kravchenko2006thermodynamic}.

Another method for testing the two-parameter scaling theory is illustrated in the right panel of Fig.~\ref{comparison 3}. In this context, the data should be described by a universal scaling function \(\mathcal{R}(X,Y)\), which represents a two-dimensional surface plotted in the \((X,Y,\rho)\) space \cite{knyazev2008metal}. Here, \(X\) and \(Y\) correspond to the normalized deviations in the density \((n_s-n_c)\) and the normalized temperature, respectively \cite{knyazev2008metal}. The scaling function was reconstructed from experimental data over a wide range of \(X\) and \(Y\) values. The data from five different samples were found to collapse onto a single surface in the \((X,Y,\rho)\) space. This relationship can be expressed as \(\frac{\rho(T,n)}{\rho_c^0} = \exp[f_1(X)] f_2(Y)\). The data are very close to the model surface; maximal deviation of the data from the model surface in the $\rho$ direction does not exceed $0.05\rho_c^0$. Clearly, the surface is curved in both directions, whereas the separatrix $\rho_c(T)$ is represented by the straight line on the model surface.

\subsection{Dynamical mean-field theory}

The two-parameter scaling theory, which has been very successful in describing experimental results in the diffusive regime (roughly $k_{B}T\tau/\hbar<1$), is applicable only to a narrow range of electron densities near the critical density for the metal-insulator transition.  In contrast, the strong changes in the resistivity with temperature are experimentally observed in a wide range of electron densities: up to five times the critical density, including the ballistic regime (roughly, $k_{B}T\tau/\hbar>1$), where the scaling theory is no longer applicable (we stress that the ballistic regime introduced in Ref.~\cite{zala2001interaction} is not related to the well-known ballistic transport, or Knudsen regime, where the mean free path is larger than the sample dimensions).

According to Ref.~\cite{zala2001interaction}, a similar physical mechanism, namely, the elastic but temperature-dependent scattering of electrons by the self-consistent potential created by all other electrons, works in principle in both diffusive and ballistic regimes.  However, small corrections calculated in the ballistic regime \cite{zala2001interaction} cannot convincingly account for the order-of-magnitude changes in the resistivity observed in the experiment.

An alternative interpretation of the temperature dependence of resistivity is the so-called Wigner-Mott scenario, which emphasizes the role of strong electron-electron interactions. The simplest theoretical approach to non-perturbatively tackle the interactions as the main driving force for the metal-insulator transition is based on dynamical mean-field theory (DMFT) methods of Refs.~\cite{camjayi2008coulomb,radonjic2012wigner,dobrosavljevic2017wigner} using the Hubbard model at half filling.  On the metallic side near the metal-insulator transition, the resistivity was predicted to initially increase as the temperature is reduced, reach a maximum at temperature $T_{max}\sim T_{F}$, and then decrease as $T\rightarrow0$.  It was also shown that the resistivity change $\rho(T)-\rho(0)$, normalized by its maximum value, is a universal function of $T/T_{{max}}$.

\begin{figure}
\scalebox{0.45}{\includegraphics[angle=0]{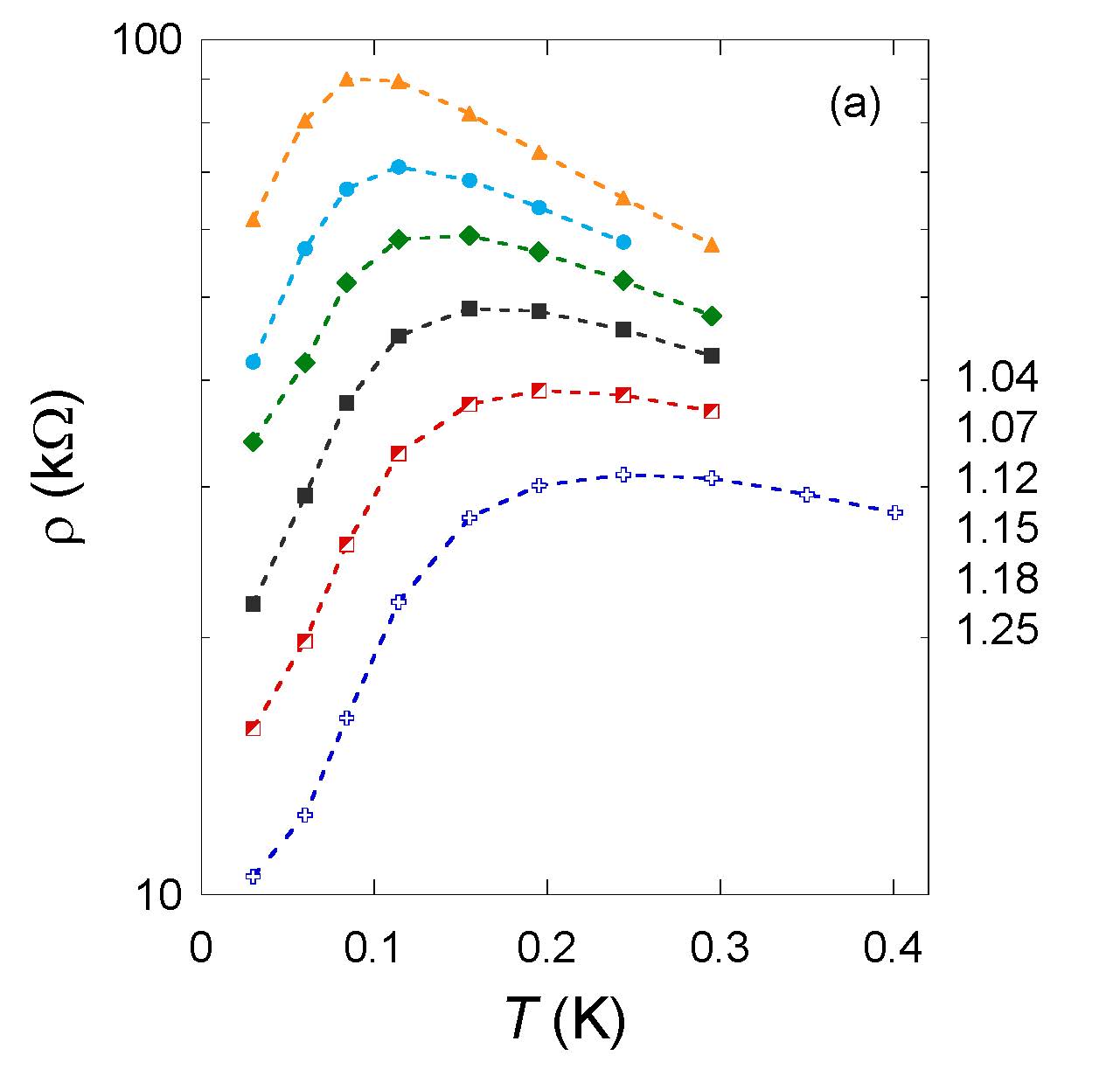}}
\scalebox{0.415}{\includegraphics[angle=0]{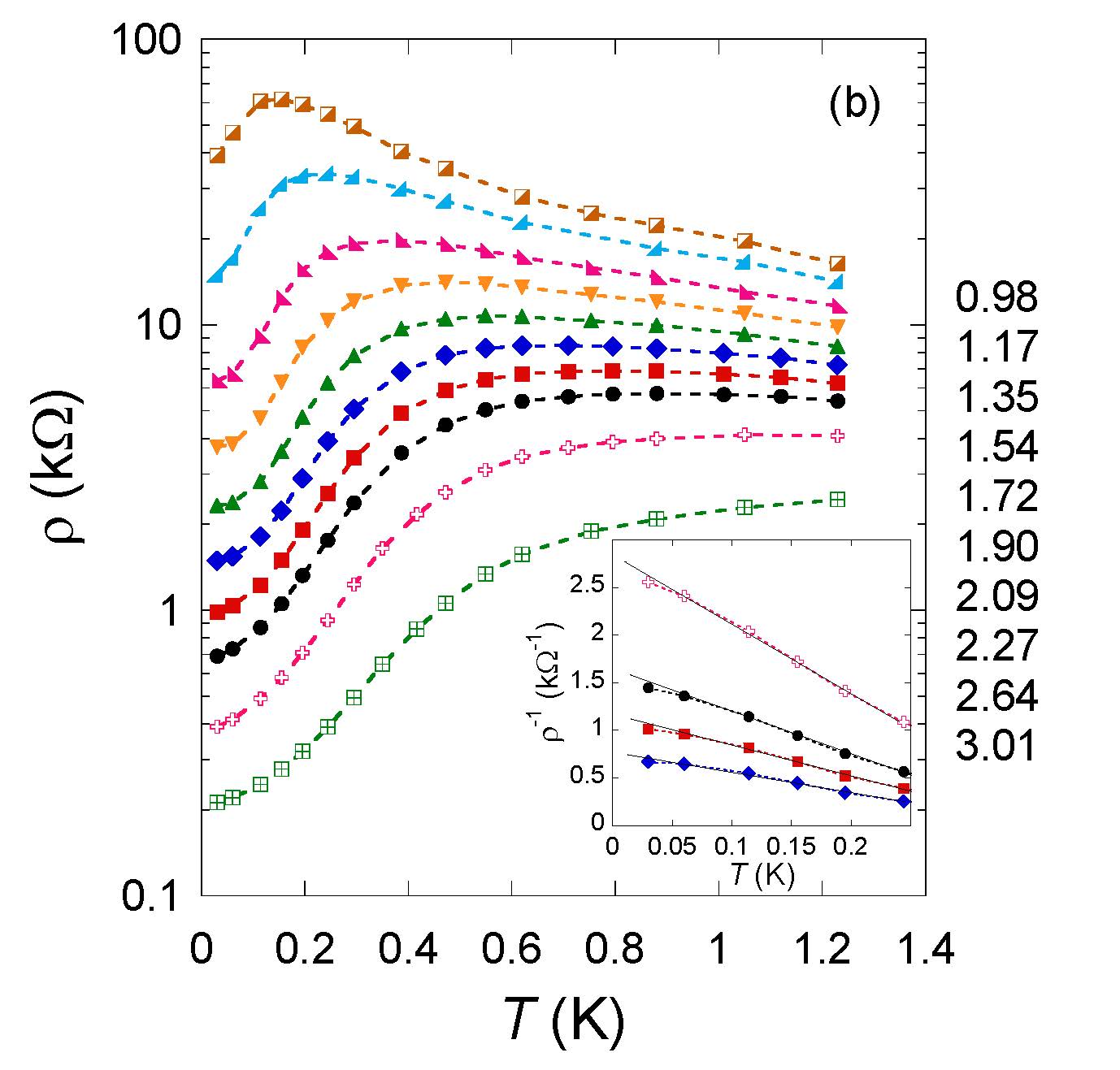}}
\caption{\label{DMFT 1} Non-monotonic temperature dependences of the resistivity of the 2D electron system in SiGe/Si/SiGe quantum wells on the metallic side near the metal-insulator transition for two samples A~(a) and B~(b).  The electron densities are indicated in units of $10^{10}$~cm$^{-2}$.  The inset in (b) shows $\rho^{-1}(T)$ dependences for four electron densities in sample B (the symbols are the same as in the main figure).  The solid lines are linear fits to the data.  From Ref.~\cite{shashkin2020manifestation}.}
\end{figure}

\begin{figure}
\centering
\scalebox{0.45}{\includegraphics[angle=0]{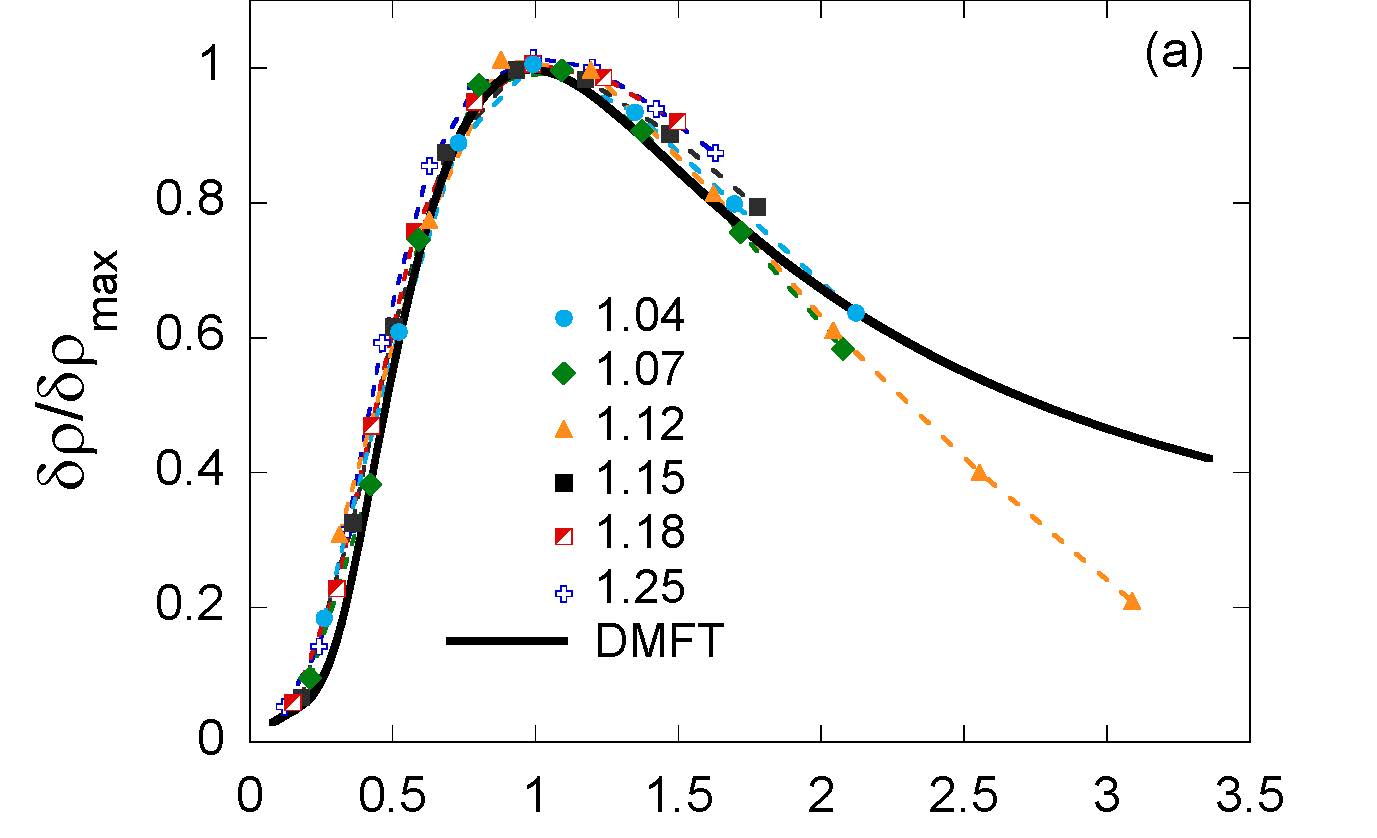}}\\
\scalebox{0.45}{\includegraphics[angle=0]{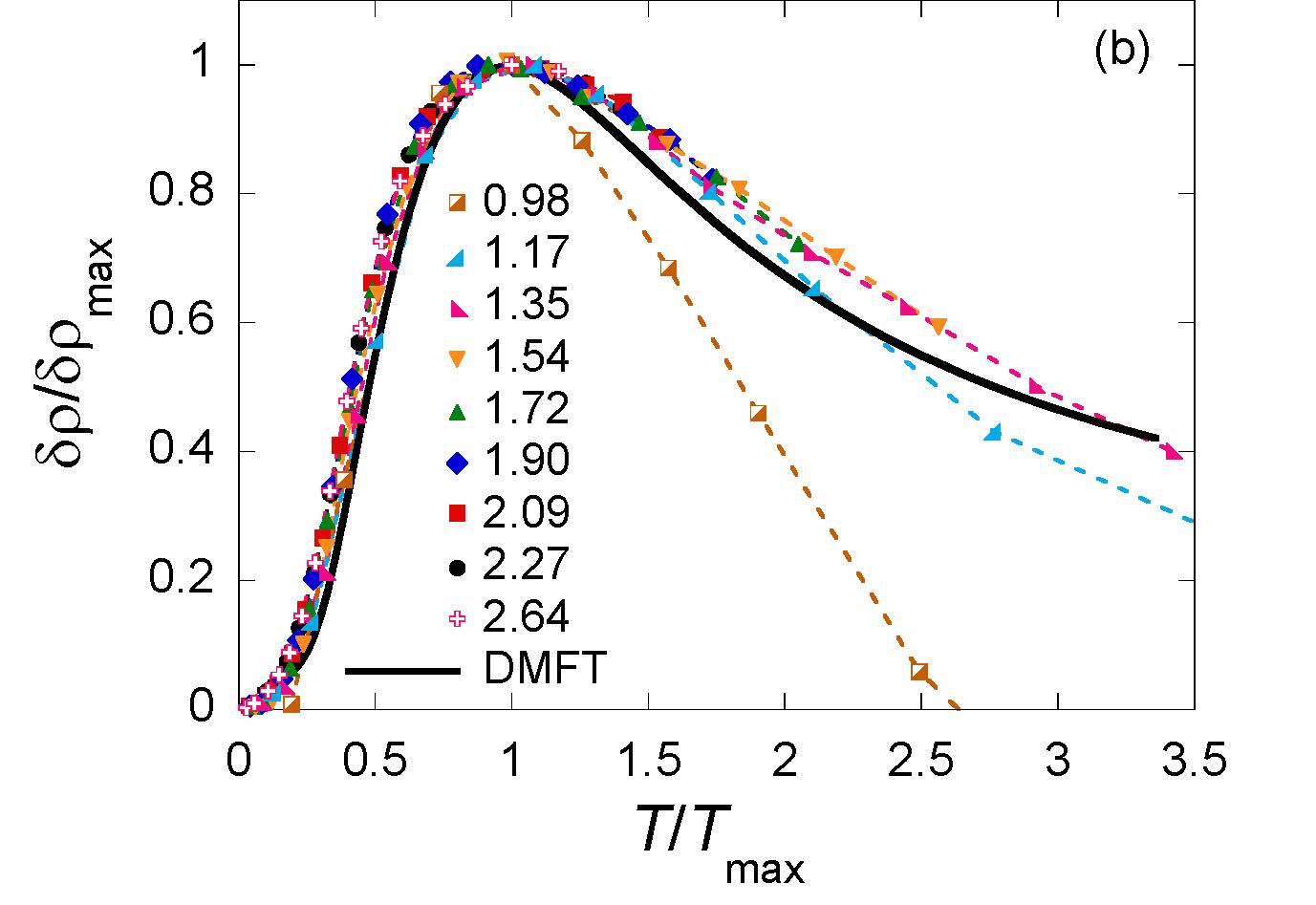}}
\caption{\label{DMFT 2} The ratio $(\rho(T)-\rho(0))/(\rho_{{max}}-\rho(0))$ as a function of $T/T_{{max}}$ for samples A~(a) and B~(b).  Solid lines show the results of the dynamical mean-field theory in the weak-disorder limit \cite{camjayi2008coulomb,radonjic2012wigner,dobrosavljevic2017wigner}. The electron densities are indicated in units of $10^{10}$~cm$^{-2}$.  From Ref.~\cite{shashkin2020manifestation}.}
\end{figure}

\begin{figure}
\centering
\scalebox{0.459}{\includegraphics[angle=0]{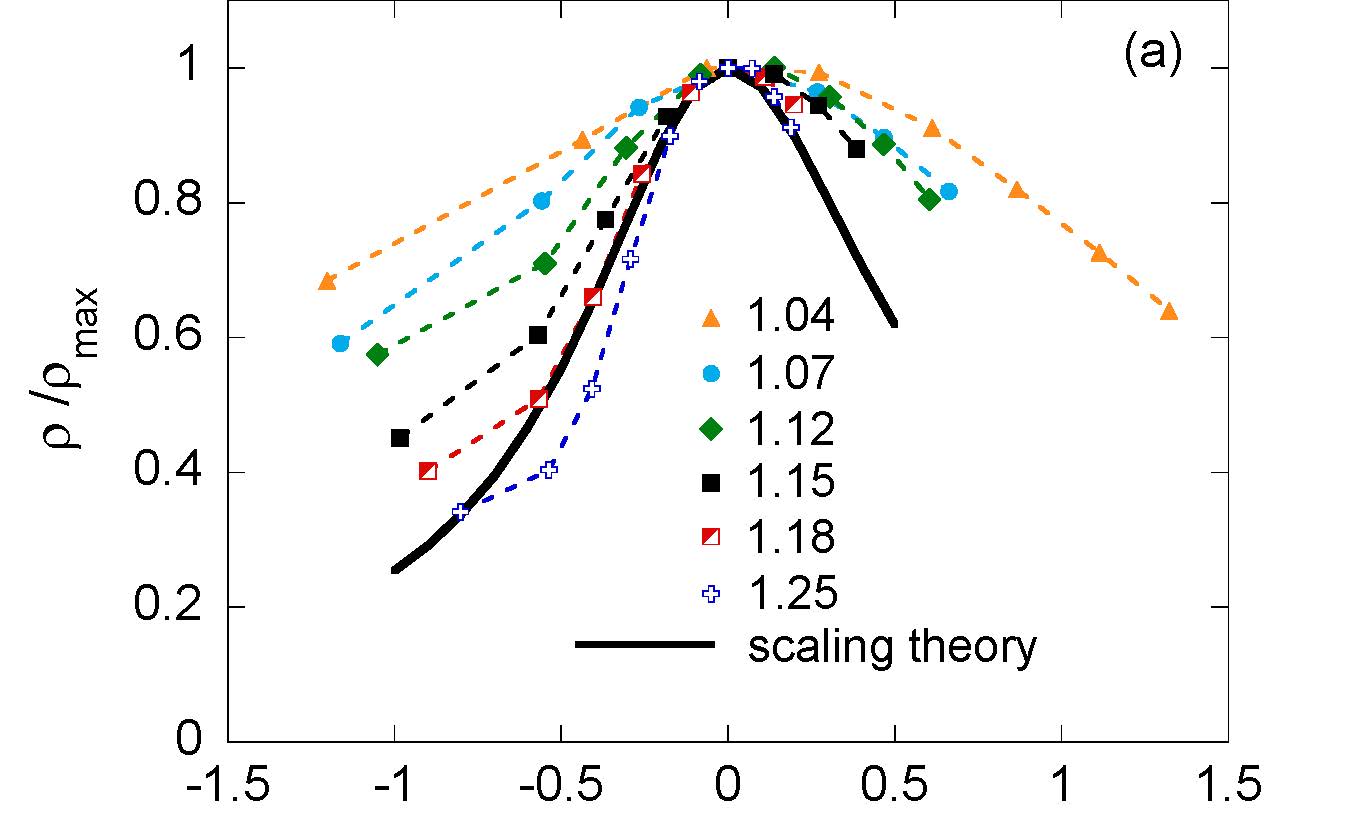}}\\
\scalebox{0.459}{\includegraphics[angle=0]{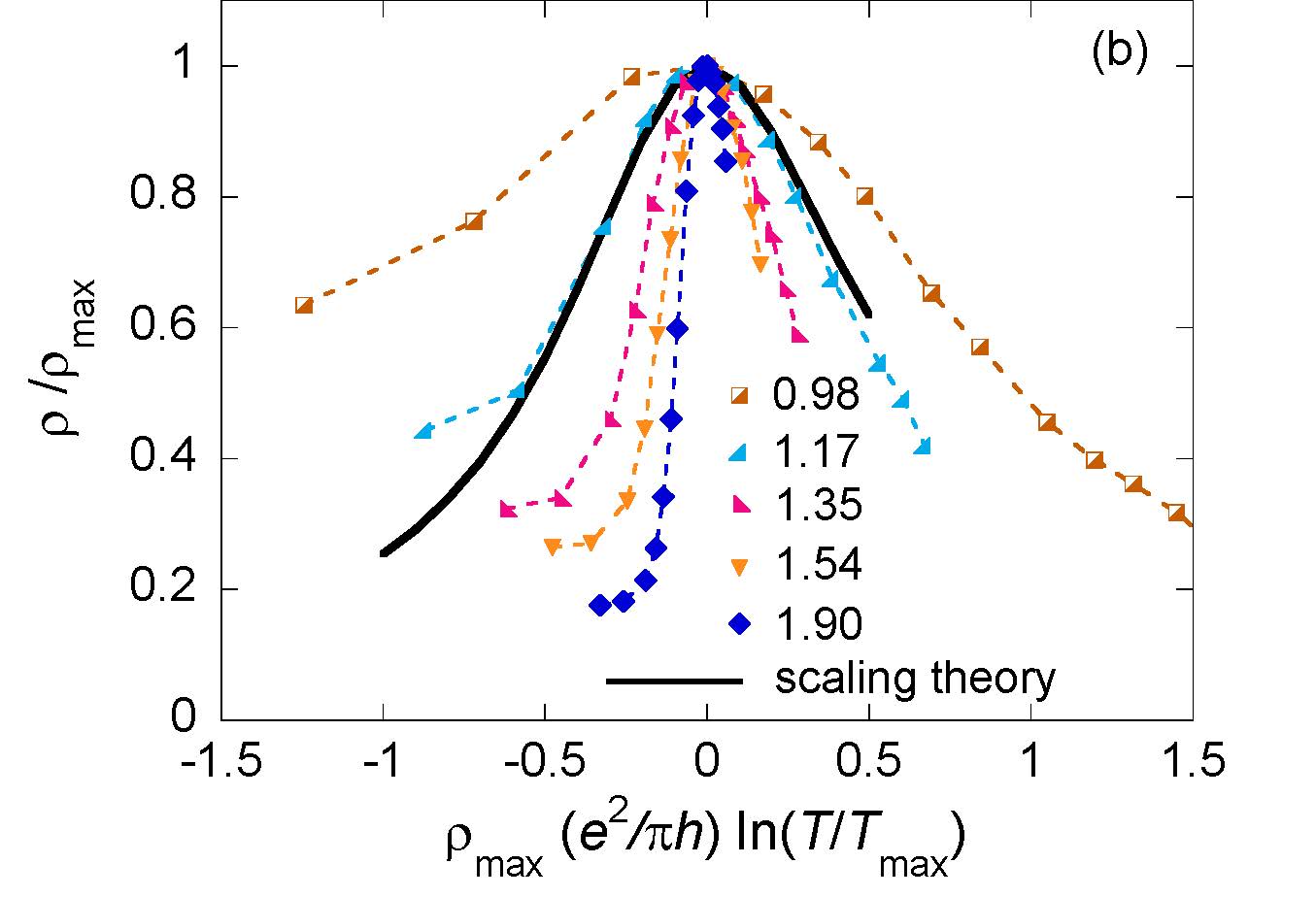}}
\caption{\label{DMFT 3} The ratio $\rho/\rho_{{max}}$ as a function of the product $\rho_{{max}} \ln(T/T_{{max}})$ for samples A~(a) and B~(b).  Solid lines are the result of the renormalization-group scaling theory \cite{punnoose2001dilute,punnoose2005metal}.  The electron densities are indicated in units of $10^{10}$~cm$^{-2}$.  From Ref.~\cite{shashkin2020manifestation}.}
\end{figure}

Temperature dependences of the resistivity for two samples in the metallic regime are shown in Fig.~\ref{DMFT 1} in the range of electron densities where the $\rho(T)$ curves are non-monotonic: at temperatures below a density-dependent temperature $T_{{max}}$, they exhibit metallic temperature behavior ($\mathrm{d}\rho/\mathrm{d}T>0$), while above $T_{{max}}$, their behavior is insulating ($\mathrm{d}\rho/\mathrm{d}T<0$) \cite{shashkin2020manifestation}. Note that the changes in the resistivity with temperature at $T<T_{{max}}$ are strong and may exceed an order of magnitude (more than a factor of 12 for the lowest curve in Fig.~\ref{DMFT 1}(b)). The data recalculated into the conductivity as a function of temperature are displayed in the inset of Fig.~\ref{DMFT 1}(b). Also shown are linear fits to the data. The observed linear temperature dependence is consistent with the ballistic regime not too close to $n_{c}$. As inferred from the temperature dependence of the conductivity, the transient region between ballistic and diffusive regimes corresponds to electron densities around $\approx 1.1\times10^{10}$~cm$^{-2}$.

The results of the scaling of these data for two samples in the spirit of dynamical mean-field theory \cite{camjayi2008coulomb,radonjic2012wigner,dobrosavljevic2017wigner} are shown in Fig.~\ref{DMFT 2}.  The data scale perfectly over a wide range of electron densities and are well described by theory in the weak-disorder limit; it should be emphasized that at some electron densities, the changes in resistivity with temperature exceed an order of magnitude.  Deviations from the theoretical curve arise in the high-temperature limit in the transient region and become pronounced for $T>T_{{max}}$ at electron densities within $\sim10$\% of the critical value, which in these samples is close to $n_{c}\approx 0.88\times10^{10}$~cm$^{-2}$. The fact that in the low-temperature limit the same data display linear-in-$T$ corrections to the conductivity (see the inset in Fig.~\ref{DMFT 1}(b)), which are in agreement with both the theory of interaction corrections \cite{zala2001interaction} and the generalized screening theory \cite{shashkin2004comment}, reveals the consistency of these theories and the DMFT.

For proper perspective and comparison, the authors of Ref.~\cite{shashkin2020manifestation} have also performed a scaling analysis in the spirit of the renormalization-group scaling theory \cite{punnoose2001dilute,punnoose2005metal}, according to which the normalized resistivity $\rho/\rho_{{max}}$ should be a universal function of the product $\rho_{{max}}\ln(T/T_{{max}})$ as described in the previous subsection.  The results are plotted in Fig.~\ref{DMFT 3}.  In both samples, only the data obtained at $n_{s}=1.18\times10^{10}$~cm$^{-2}$ for sample A (Fig.~\ref{DMFT 3}(a)) and at $n_{s}=1.17\times10^{10}$~cm$^{-2}$ for sample B (Fig.~\ref{DMFT 3}(b)) coincide nearly perfectly with the theoretical curve, although some deviations occur at the lowest temperature.  Pronounced deviations from the theory are evident at both higher and lower $n_{s}$.  At lower electron densities, the scaled experimental curves are wider than the theoretical curve, and at higher densities, they are narrower.  A similar shrinkage of the scaled curves with increasing $n_{s}$ was reported earlier in Refs.~\cite{punnoose2001dilute,anissimova2007flow,radonjic2012wigner}.  One should take into account, however, that theory \cite{punnoose2001dilute,punnoose2005metal} has been developed for 2D electron systems that, on the one hand, are in the diffusive regime and, on the other hand, their resistivities are low compared to $\pi h/e^2$: at higher values of $\rho$, higher-order corrections become important and cause deviations from the universal scaling curve.  As a result, the parameter space becomes very narrow.

A natural question arises about the connection between DMFT and scaling theory.  Both theories predict non-monotonic temperature dependences of the resistivity.  Within the renormalization-group scaling theory \cite{punnoose2001dilute,punnoose2005metal}, the maximum in the $\rho(T)$ dependence occurs at a temperature well below $T_{F}$, at which the temperature-dependent interactions become strong enough to stabilize the metallic state and overcome the effect of the quantum localization.  Within the DMFT, in contrast, the maximum corresponds to the quasiparticle coherence temperature $T^*\sim T_{F}$: below this temperature, the elastic electron-electron scattering corresponds to coherent transport, while at higher temperatures the inelastic electron-electron scattering becomes strong and gives rise to a fully incoherent transport.  Even though the theoretical estimates of the positions of the maxima may be crude, the origins of the maxima are clearly different within these two theories in view of the role of the disorder. It should be stressed, on the other hand, that the functional forms of $\rho(T)$ dependences, including the maximum at $T_{{max}}\sim T_{F}$, expected from both the screening theory in its general form and DMFT, are similar. In particular, the linear temperature dependence of the conductivity at $T\ll T_F$ following from the generalized screening theory \cite{shashkin2004comment} and from the theory of the corrections to the conductivity due to the scattering on Friedel oscillations in the ballistic regime \cite{zala2001interaction} is consistent with the prediction of the DMFT.  The similarity of the theoretical predictions adds confidence to both theories and suggests that the underlying microscopic mechanism may be the same, \textit{i.e.}, electron-impurity or impurity-mediated electron-electron scattering in the strongly interacting case, as mentioned above.

Yet another approach to treat the strongly-interacting 2D electron systems, focused on the Pomeranchuk effect expected within a phase coexistence region between the Wigner crystal and a Fermi liquid, was proposed in Refs.~\cite{spivak2003phase,spivak2004phases,spivak2006transport}. The predicted $\rho(T)$ dependence is also non-monotonic: the resistivity increases with decreasing temperature at $T\gtrsim T_{F}$ and decreases at lower temperatures.  However, no quantitative treatment of this problem, capable of quantitative comparison with experiment, currently exists.

In Ref.~\cite{kagalovsky2024quantum}, a Gaussian approximation to the mean-field theory of the second-order phase transition was suggested to describe the 2D MIT. The authors have defined the effective distance to the quantum critical point as the cube of the difference between the critical and actual electron densities. Using this assumption, they analytically found the critical exponent for the divergence of the correlation length, which agrees with the experimental data \cite{kravchenko1995scaling,mason1995experimental} on both sides of the transition. The phase diagram in the $n_s$--$T$ plane was described as typical of a system exhibiting a quantum phase transition.

\subsection{Phase separation in strongly correlated 2D systems}
\label{sec:Phase Separation}

In two-dimensional systems, instability can occur in charge- or spin-exchange channels, leading to the formation of either ``ice floes" in the Wigner phase or ``spin droplets" in the Stoner phase. Theoretically, Spivak highlighted the importance of phase separation by drawing an analogy with liquid helium-3 \cite{spivak2003phase}. His phenomenological model aligns qualitatively with several experimental observations, including a maximum in resistivity versus temperature, negative compressibility, and the coexistence of spatially separated localized and extended phases in equilibrium.

Various scenarios for a two-phase state have been explored, wherein such phases can emerge in restricted geometries due to local Wigner instability \cite{spivak2003phase, spivak2004phases}, local Stoner instability \cite{narozhny2000magnetic, benenti2001spin, sloggett2005electron}, or changes in the topology of the Fermi surface \cite{khodel1990superfluidity, khodel2008topology, dolgopolov2022band}. It has been suggested in previous studies \cite{spivak2003phase, spivak2006transport} and further developed in references \cite{jamei2005universal, ortix2006frustrated, ortix2007screening, ortix2009universality} that the first-order transition from a Fermi liquid to a Wigner crystal is replaced by a series of intermediate ``microemulsion'' phases. In these phases, the two distinct states coexist within spatially structured domains.

Phase separation signatures in Si MOSFETs were observed through thermodynamic spin-magnetization measurements \cite{teneh2012spin}, transport measurements \cite{morgun2016novel, pudalov2022phase}, spin susceptibility via oscillatory magnetotransport \cite{pudalov2020experimental}, and quantum oscillations at the electron solid/liquid boundary \cite{diorio1990magnetic, diorio1992reentrant}.

In 1997, Pudalov suggested \cite{pudalov1997unconventional} that the additive resistivity functional form (Eq.~\ref{eq:rho-exp}) holds over a wide density range. The function $\rho_1(n_s, B)$ decays slowly with respect to $n_s$, while $\rho_0(n_s, T)$ incorporates both Drude resistivity and quantum corrections, arising from single-particle interference and interactions. Although this empirical form of resistivity was proposed in Ref.~\cite{pudalov1997unconventional} based on different principles, it aligns well with the $\rho(T)$ dependence observed in several material systems \cite{pudalov1997unconventional, hanein1998the, papadakis1998apparent, hormann2008pronounced, zhang2014correlation, raghavan2015two}. This empirical additive form of $\rho(T)$ meets the general criteria for transport behavior near a critical point \cite{altshuler2001metal, knyazev2008metal}. On the insulating side of the metal-insulator transition, the resistivity exhibits an exponential temperature dependence. Therefore, this formulation inherently maintains the mirror symmetry of the $\rho(n_s, T)$ scaling in the vicinity of the MIT. Additionally, this form suggests two-channel scattering, consistent with the two-phase state of low-density two-dimensional electronic systems and in accordance with Matthiessen's rule.

\begin{figure}[ht]
\centering
\scalebox{0.5}{\includegraphics{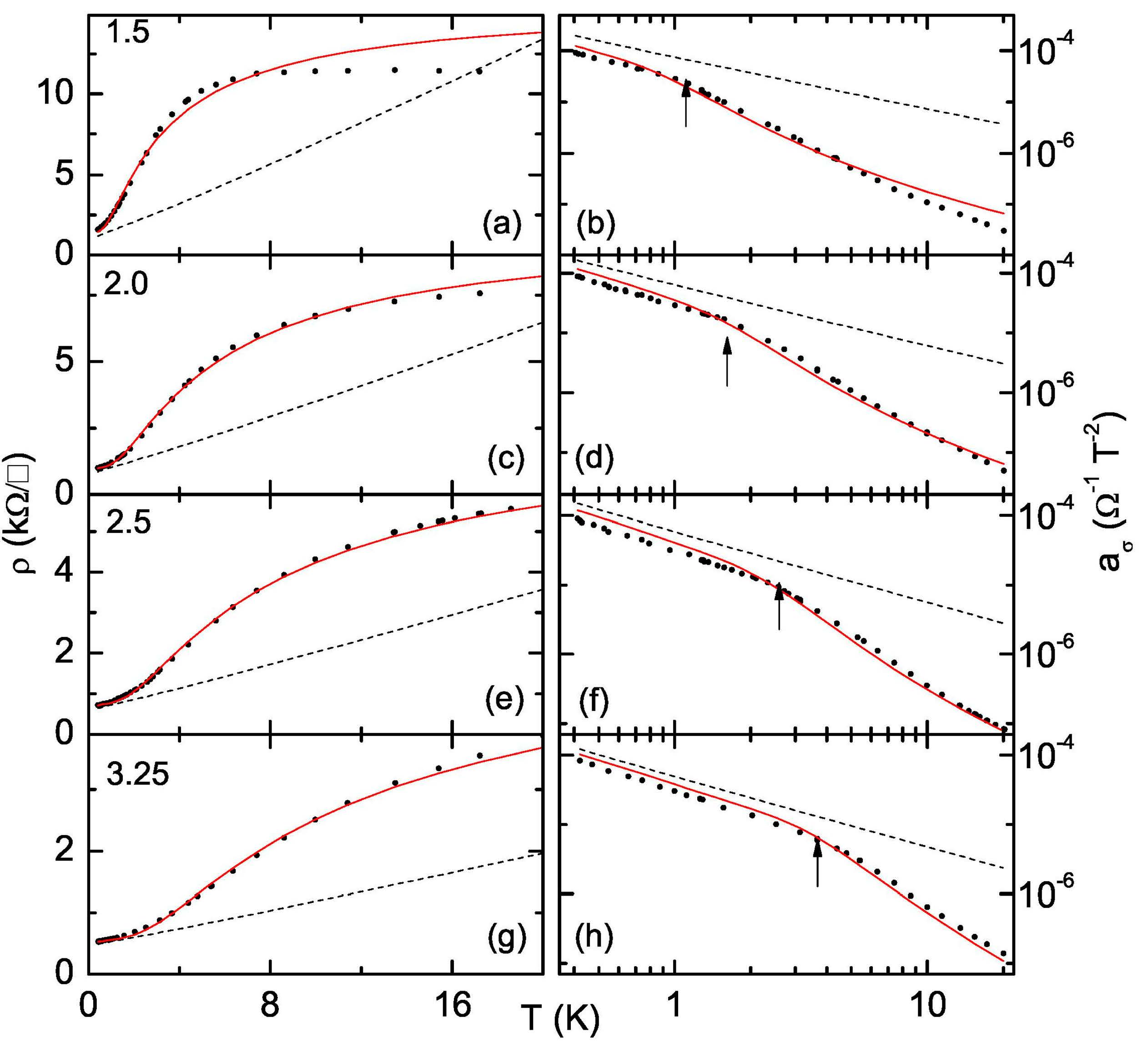}}
	\caption{Fitting dependencies of $\rho(T,B = 0)$ (left) and $a_\sigma (T)$ (right) using the same set of fitting parameters. Carrier densities are shown from top to bottom as $n = 1.5, 2.0, 2.5$, and $3.25\times10^{11}$ cm$^{-2}$.
    Dashed lines represent interaction corrections calculated in ballistic interaction regime.
Vertical arrows indicate the kink positions. From Ref.~\cite{morgun2016novel}.	}
	\label{fig:dualFit}
\end{figure}

The experimental data for both $\rho(T)$ and $\sigma(T)$, as shown in Fig.~\ref{fig:dualFit}, exhibit variations so significant that first-order corrections in \(T\) are insufficient to describe them. The simplest functional form represented by Eq.~(\ref{eq:rho-exp}) effectively captures the inflection point in \(\rho(T)\) and shows the linear relationship between density and the inflection temperature \cite{pudalov1997unconventional}. In this model, the inflection temperature is defined as \(T_{infl} = \Delta/2\). To account for the effects of the magnetic field, all even-order terms in \(B\) and the lowest-order terms in \(B/T\) have been included, as follows:
\begin{equation}
	\Delta(T,B,n)/T =\Delta_0(n)/T - \beta(n) B^2/T  - \xi(n) B^2/T^2,
	\label{eq:nc(B)}
\end{equation}
with $\Delta_0=\alpha[n-n_c(0)]$.

Equations~(\ref{eq:rho-exp}) and (\ref{eq:nc(B)}) link the magnetoconductance with the zero-field $\rho(T)$ temperature  dependence. Combining  equations (\ref{eq:rho-exp}) and (\ref{eq:nc(B)}), the following $\rho(T,B)$ dependence is obtained:
\begin{eqnarray}
	&\rho(B,T) = \left[\sigma_D -
	\delta\sigma \cdot \exp \left(- T/T_B \right)\right]^{-1} \nonumber \\
	&+ \rho_1 \exp \left( - \alpha \frac{n-n_c(0)}{T} - \beta \frac{B^2}{T} - \xi \frac{B^2}{T^2}\right)
	\label{eq:r(B&n)}
\end{eqnarray}
The term in the square brackets includes the Drude conductivity and interaction quantum corrections \cite{zala2001interaction, zala2001interaction_a}, which are smoothly cut off above $T=T_B\approx \Delta/2$. $\delta\sigma(T)$ was calculated in Ref.~\cite{morgun2016novel} using experimentally determined Fermi-liquid coupling constants $F_0^\sigma (n_s)$ \cite{pudalov2002low, klimov2008interaction}, and $\sigma_D$ found by a conventional  procedure \cite{pudalov2002low, klimov2008interaction}.

From Eq.~(\ref{eq:r(B&n)}), the prefactor \( a_\sigma = -\frac{1}{2}\frac{\partial^2\sigma}{\partial B^2} \) is calculated straightforwardly and is compared with experimental data in Fig.~\ref{fig:dualFit}(b, d, f, h). In the fitting of \(\rho(T)\) shown in Figs.~\ref{fig:dualFit}\,(a, c, e, g), there is essentially only one adjustable parameter, \(\rho_1(n_s)\), for each density. Notably, \(n_c(0)\) is determined using conventional scaling analysis at \(B=0\) \cite{knyazev2008metal}.

Both the electrical resistivity, \(\rho(T)\), and \( a_\sigma = -\frac{1}{2}\frac{\partial^2\sigma}{\partial B^2} \), are well captured by the model, which reproduces the key features of the data. It accurately captures the steep rise in \(\rho(T)\), including the inflection point, as well as the kink in the dependence of \(a_\sigma(T)\). This kink marks a transition from a low-temperature regime characterized by magnetoconductance, where interactions yield a linear temperature dependence of \(\sigma(T)\), allowing us to neglect the exponential term. In contrast, the high-temperature regime is characterized by a steep exponential rise in \(\rho(T)\), which has a semiclassical origin.

It should be emphasized that neither regime is related to the diffusive regime of interactions. This conclusion casts doubt on early attempts to use two-parameter scaling to describe the magnetoresistance $\sigma(B)$ and the temperature dependence $\rho(T)$ within the renormalization group approach.

Thus, within the framework of a phenomenological two-phase model with two scattering channels, it is possible to explain all the observed features in transport and magnetotransport in a parallel field. It is important that their characteristic temperatures are close to the crossover temperature $T_{dM/dn_s}(n_s)$, where the spin magnetization per electron changes sign \cite{teneh2012spin}. Physically, this means that at temperatures below $T_{dM/dn_s}(n_s)$, collective droplets with a large spin (minority phase) ``melt'' with increasing density. In other words, the electrons added to the Fermi liquid improve screening, thereby promoting the disappearance of spin droplets. At temperatures above $T_{dM/dn_s}(n_s)$, on the contrary, the number of spin droplets increases with increasing density; in this case, the electrons added to the 2D system prefer to combine and form new spin droplets. In Ref.~\cite{morgun2016novel}, it was concluded that $T^*$ may be related to the averaged energy spectrum of the collective localized states.

\subsubsection{Phase separation effect in spin susceptibility}
\label{sec:chi-nonlinear}
Using a vector magnetic field technique with two independent superconducting coils, Pudalov et al. \cite{pudalov2002stiff, pudalov2021magnetic} precisely measured and analyzed Shubnikov-de~Haas oscillations in various in-plane fields. Earlier \cite{pudalov2002low, pudalov2014probing, pudalov2018probing}, the oscillatory component $\delta \rho_{xx}$ was shown to be well fitted with conventional Lifshits-Kosevich formula \cite{lifshits1955on, isihara1986density, pudalov2014probing,  pudalov2018probing}; this enables accurate extraction of the spin susceptibility $\chi$ and density of mobile carriers $n_{SdH}$ from the beating of oscillations. In particular, $\chi$ values have been determined with an accuracy of $\sim (1 - 2)\%$ as a function of the in-plane field.

\begin{figure}
\centering \includegraphics[width=0.525\textwidth]{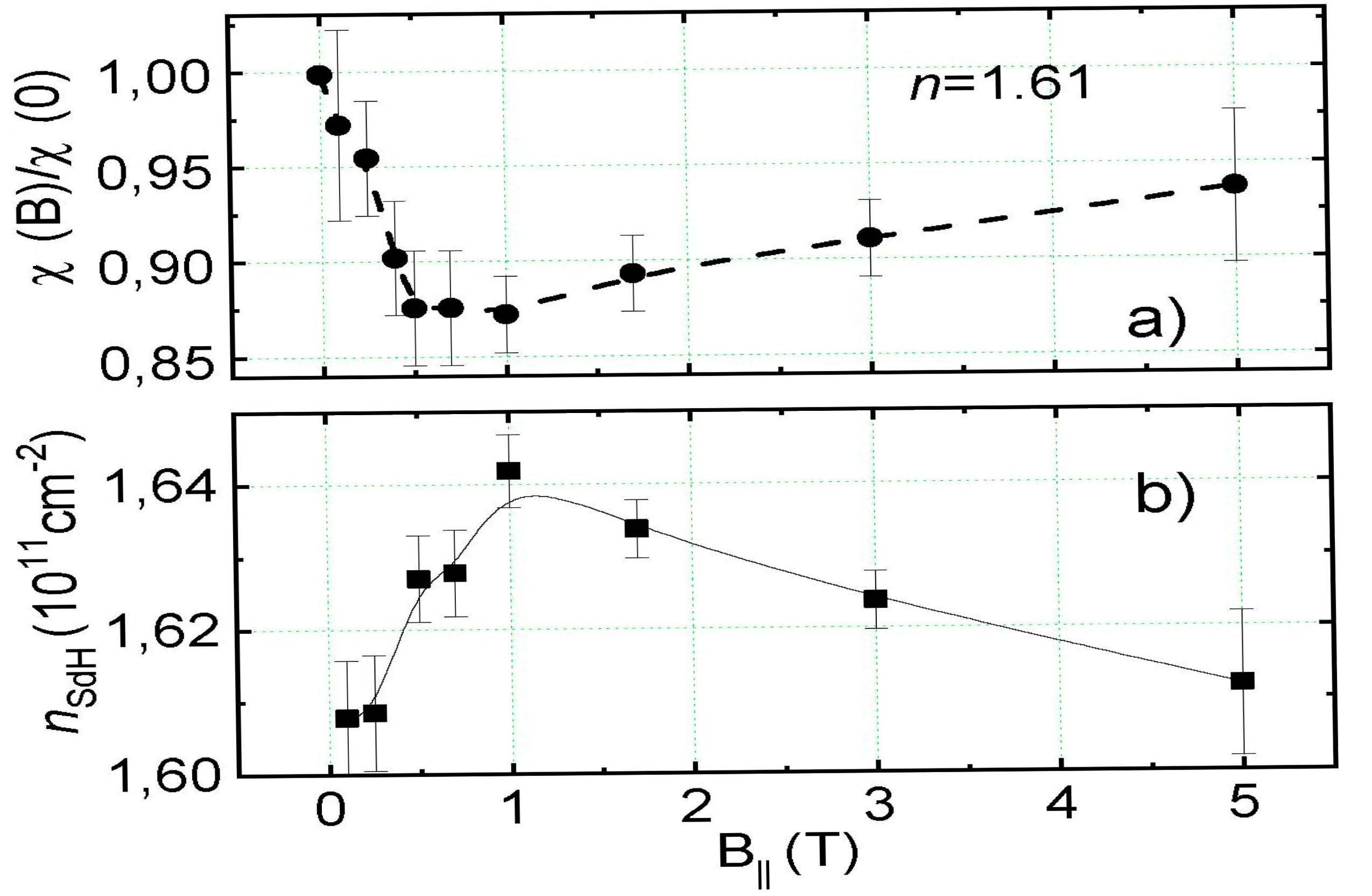}
	\includegraphics[width=0.45\textwidth]{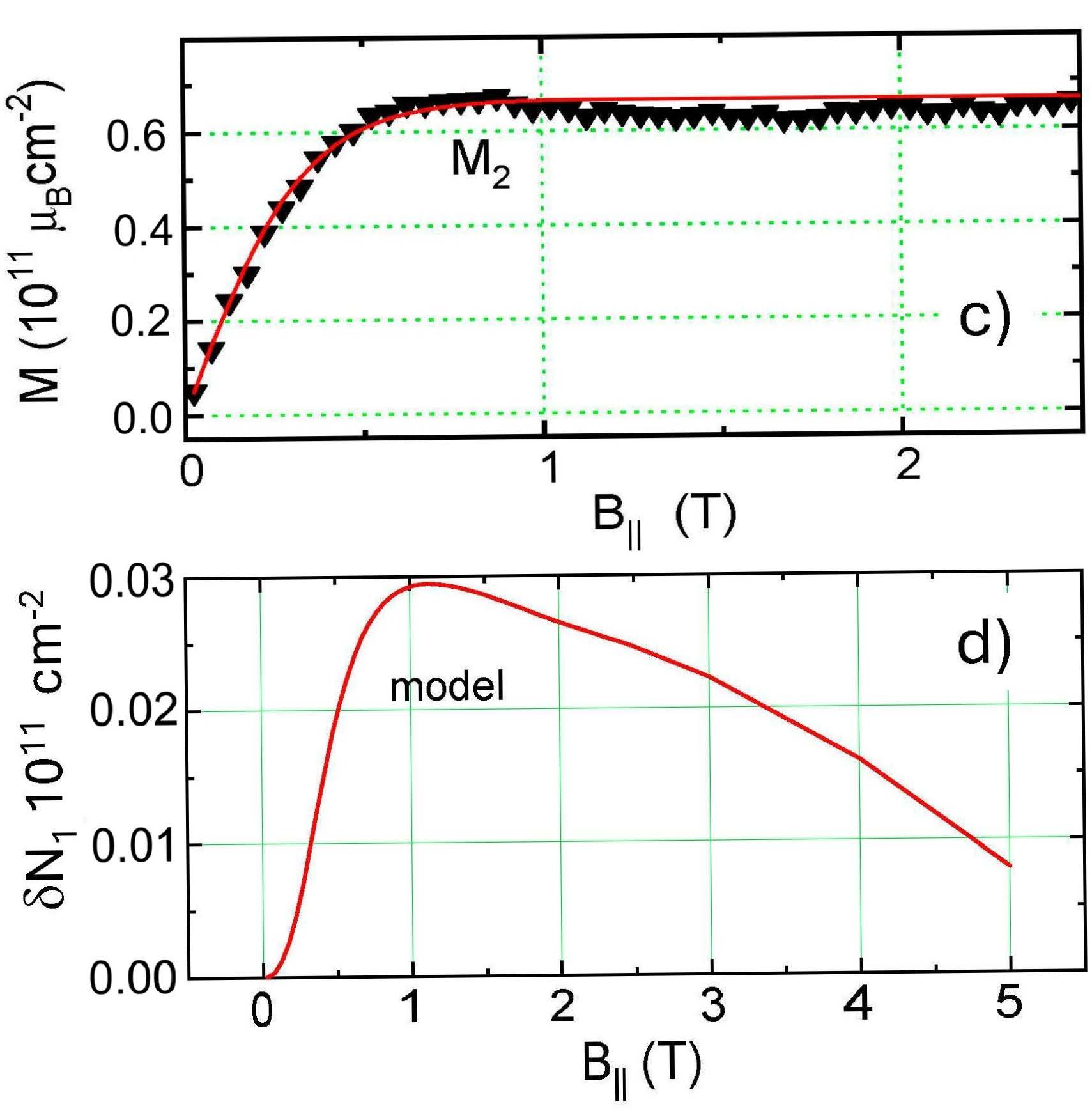}
	\caption{Correlation between  the in-plane field dependence of (a) $\chi(B)/\chi(0)$,
    and (b) density $n_{SdH}(B)$. c) Experimental $M(B\|)$ data  (symbols) and the model curve (red line), showing $\tanh(\mu_B B/k_B T)$ is a fit of the experimental $M(B)$ data. d) Model curve $\delta N_1(B_\parallel)$ calculated from experimental data as described in the text. The zero-field densities are  $n_s=1.61\times 10^{11}$~cm$^{-2}$ for (a) and (b), and $1.4\times 10^{11}$~cm$^{-2}$ for (c). Temperature $T=0.1$\,K. From Ref.~\cite{pudalov2021magnetic}}.\label{Pudalov chi(B) abcd}
    \end{figure}

Figures \ref{Pudalov chi} and \ref{Pudalov chi(B) abcd} show a sharp nonmonotonic dependence of $\chi$ on the in-plane field. The characteristic $\delta\chi(B)/\chi(0)$ variations  are in the range from  $\sim 25\%$ at low densities to  $\sim 6\% $ at high densities $10\times 10^{11}$cm$^{-2}$. The data reported in \cite{pudalov2021magnetic} coincide in the $B_\parallel \rightarrow 0$ limit with the $\chi(B=0)$ values reported in Refs.~\cite{pudalov2002low, klimov2008interaction}. The characteristic field of the  $\chi(B)$-minimum, $B_\parallel \sim 1$\, T for $n_s=(1.1-2)\times 10^{11}$cm$^{-2}$, is much weaker than the field of complete spin polarization of the 2D system ($B\sim 20$~T). Evidently, in a homogeneous single-phase Fermi liquid the only characteristic field is the field of complete spin polarization.

Pudalov et al.~\cite{pudalov2021magnetic} proposed a straightforward thermodynamic model (schematically illustrated in Figs.~\ref{Pudalov chi 2022}(b,c)) featuring two phases coexisting in equilibrium to explain the observed data. This model has effectively accounted for the results both qualitatively and quantitatively, with certain parameters derived from experimental data. Specifically, the dependence of \( n_{FL}(B) \) calculated using this model for a representative density of \( 1.4 \times 10^{11} \)~cm\(^{-2}\) is shown in Fig.~\ref{Pudalov chi(B) abcd}(d). The results closely align with the direct experimental data presented in Fig.~\ref{Pudalov chi(B) abcd}(b), supporting the validity of the two-phase thermodynamic approach. Recently, in the context of Maxwell's construction for addressing the electrostatics of the two-phase state \cite{sandeep2023upper}, an upper bound for charge redistribution was estimated to be \( \delta n_s/n_s \sim 0.1\% \), which is reasonably consistent with the experimental findings illustrated in Fig.~\ref{Pudalov chi(B) abcd}(b). It was also noted in Ref.~\cite{sandeep2026disorder} that the amount of redistributed charge is influenced by the nature of the random potential.

\subsubsection{Evidence of the phase separation in thermodynamic spin magnetization}

In reference \cite{teneh2012spin}, measurements were conducted in weak magnetic fields that were less than the temperature; specifically, \( g\mu_B B \leq k_B T \). As shown in Fig.~\ref{Pudalov chi 2022}(a), at low densities, the derivative of magnetization with respect to the carrier density, \( \partial M/\partial n_s \), becomes positive and is significantly larger than what is expected for the Pauli spin susceptibility. As the magnetic field increases (while still remaining below the temperature, \( g\mu_B B < k_B T \)), \( \partial M/\partial n_s \) exhibits a sharp increase, exceeding the value of the Bohr magneton by more than a factor of two at low temperatures (see Fig.~\ref{Pudalov chi 2022}(a)). The fact that \( \partial M/\partial n_s \) surpasses the Bohr magneton indicates a ferromagnetic ordering of the electron spins.

The magnetization curves \( \partial M/\partial n_s \) (Fig.~\ref{Pudalov chi 2022}(a)) saturate at a field strength \( b = \mu_B B/(k_B T) \sim 0.25 \), suggesting that the particles responding to the field modulation have spins of \( 1/(2b) \approx 2 \), rather than the expected value of 1/2. According to Pudalov \cite{teneh2012spin, pudalov2022phase}, this finding offers compelling evidence for the presence of ferromagnetic domains within the two-dimensional paramagnetic Fermi liquid, which are known as ``spin droplets". These spin droplets consist of approximately 4 or more electrons and have a total spin of approximately 2. This interpretation of the experimental data has received theoretical support \cite{stadnik2013interacting}. Spin droplets have been conceptualized as a type of ``quantum dot'' embedded within a two-dimensional system, coexisting with it in thermodynamic equilibrium (see Fig.~\ref{Pudalov chi 2022}(b,c)).

\subsubsection{Phase separation effects in reentrant integer QHE-insulator transitions: Quantum oscillations of the electron solid/liquid boundary}

\begin{figure}[ht]
\centering
\includegraphics[width=0.54\textwidth]{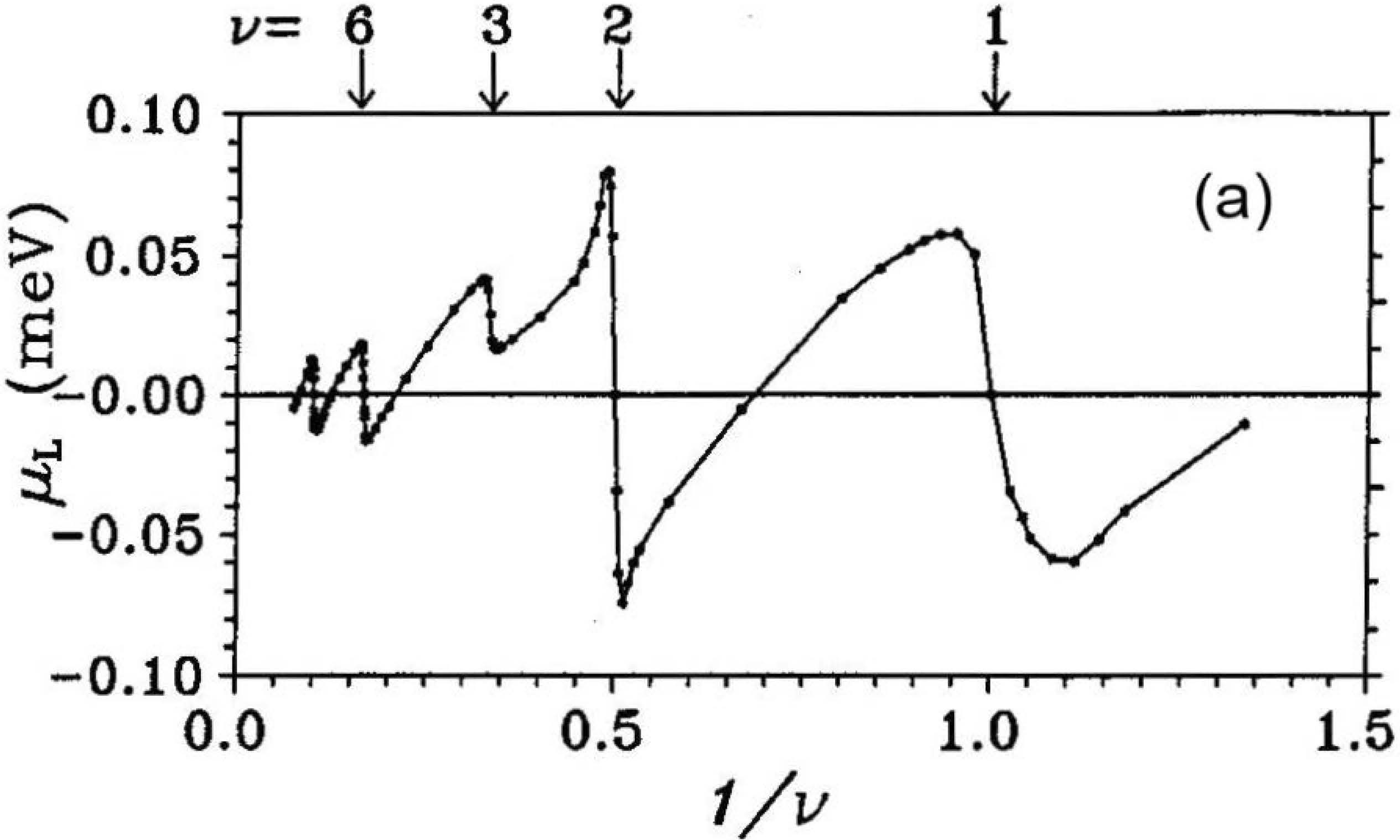}
	\includegraphics[width=0.4\textwidth]{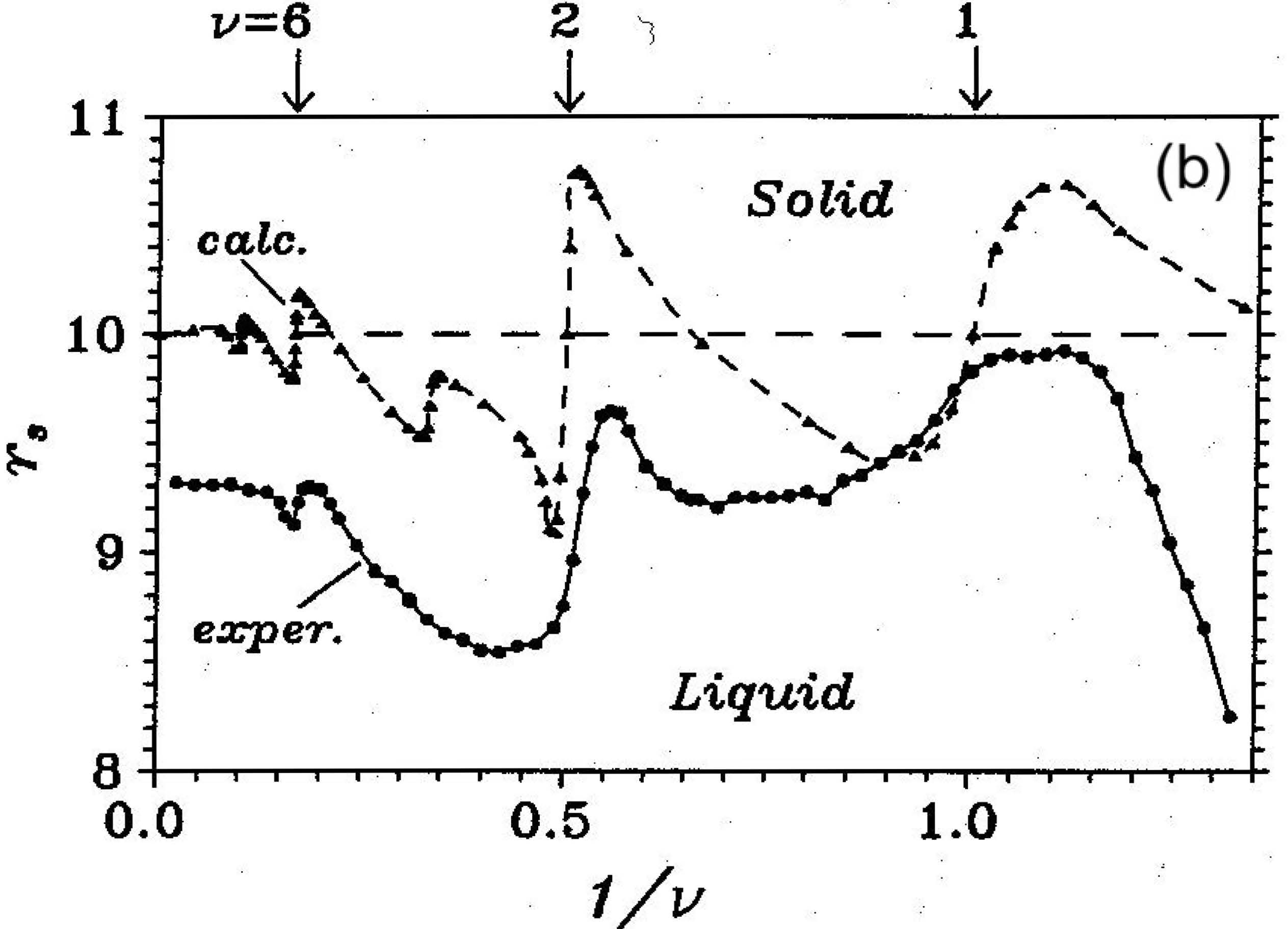}
	\caption{(a) Quantum oscillations of the chemical potential, \(\mu_L\), are displayed as a function of the inverse filling factor, \(1/\nu \propto B\). This plot has been generated by extrapolating results from studies by Pudalov et al.\ \cite{pudalov1985oscillations, pudalov1986quantum} to a lower density of \(n = 8 \times 10^{10} \, \text{cm}^{-2}\). Panel~(b) shows the quantum oscillations of the two-phase boundary, calculated using the data presented in panel (a) as described in the text. From Ref.~\cite{pudalov1994electron}.	
		}
	\label{fig:quantosc-boundary}
\end{figure}

The reentrant transition between the electron liquid and electron solid states, which is marked by resistivity changes spanning several orders of magnitude (see Figs.~\ref{Pudalov Surface Science}, \ref{Gao reentrant}), was discovered in 1990 \cite{diorio1990magnetic}. This effect can also be naturally explained within the framework of the two-phase model.  As the perpendicular magnetic field increases, the chemical potential $\mu_{L}$ in the liquid phase exhibits quantum oscillations \cite{pudalov1985valley, pudalov1985oscillations, pudalov1986quantum} as a function of the magnetic field. To determine the chemical potential in the dilute liquid phase, in Ref.~\cite{pudalov1994metal} the data on the energy spectrum in silicon at high carrier density and high magnetic field \cite{pudalov1985oscillations, pudalov1986quantum} were extrapolated to the low-density range.  This extrapolation is based on established relationships of spin, cyclotron, and valley splittings as functions of the magnetic field, along with considerations of level broadening \cite{pudalov1985valley}. Figure \ref{fig:quantosc-boundary}(a) depicts the resulting model of the chemical potential in the liquid phase, $\mu_L$, as a function of the magnetic field.

At the same time, the chemical potential in the solid (localized) phase remains constant or changes slowly in response to variations in the magnetic field. To ensure that the chemical potential (\(\mu\)) is equal in both the solid and liquid phases, the equilibrium average density --- especially the critical electron density (\(r_c\)) --- will necessarily show quantum oscillations when the magnetic field is varied.

 Pudalov \cite{pudalov1994metal} modeled the relationship between the equilibrium boundary, or critical density, and the perpendicular magnetic field, \( B_\perp \), by relating the chemical potential to the ground state energy per particle, denoted as \( E^{g} = E/n_s \). By substituting \( n_s \) with \( r_s = (a_B)^{-1}(\pi n_s)^{-1/2} \), one obtains the expression
\[
\left(\frac{d\mu_{L}}{dB}\right)\delta B = \frac{1}{2} \left[\frac{\partial (E^{g}_{S} - E^{g}_{L})}{\partial r_s}\right]\delta r_s.
\]

In this context, \( E^{g}_{S} \) and \( E^{g}_{L} \) represent the ground state energies in the solid and liquid phases, respectively, while \( \mu_{S} \) and \( \mu_{L} \) denote their corresponding chemical potentials. Eventually,
\begin{equation}
	\delta r_{s} = 2 \delta \mu_{L} [\partial (E^{g}_{S}-
	E^{g}_{L})/\partial r_{s}]^{-1}.
	\label{eq:drs}
\end{equation}

The last equation allows one to calculate small variations in the critical $r_s$ value from the known dependence of $\mu_{L}(B)$.  The latter (Fig.~\ref{fig:quantosc-boundary}(a)) has been extrapolated from the $\mu_{L}(B)$ oscillations measured in \cite{pudalov1985oscillations, pudalov1986quantum} at higher densities, and scaled according to the numerical calculations \cite{chui1995impurity} of the ground state energy: $\partial (E^{g}_{S}-E^{g}_{L})/\partial r_s \approx -0.174 {\textrm meV}$, at the critical value $r_{s}$=10, for a Si MOSFET sample.

The results of the modeling are plotted in Fig.~\ref{fig:quantosc-boundary}(b) by the short dashed curve. The dots with an interpolating continuous line show direct experimental data taken from \cite{diorio1992reentrant} and recalculated in the units of $r_{s}$. The saw-tooth shape of the modeling boundary between the liquid and solid phases (Fig.~\ref{fig:quantosc-boundary}(b)) explains the observed reentrant QHE-to-insulator transitions. Indeed, when the average carrier density in a 2D system is fixed (by the constant gate voltage) near $n_c$ and the magnetic field varies, the system crosses the critical boundary and enters the liquid and solid phase space. As a result, resistance varies dramatically at low temperatures, from exponentially low minima in the QHE regime to exponentially high values in the insulator \cite{diorio1992reentrant}.

In this framework, the minima of \( R_{xx} \) observed during the reentrant insulating transitions originate from the regular quantum Hall effect. As noted by Pudalov \cite{pudalov1994metal}, these minima indicate changes in the chemical potential rather than the energy gap (see Fig.~\ref{fig:quantosc-boundary}). As a result, these minima are shifted from their positions at integer filling factors towards higher magnetic fields. This observation is consistent with experimental findings \cite{diorio1992reentrant}.

Since calculations \cite{pudalov1994metal} give only the variation in $r_c$, the vertical position of the model boundary is poorly defined; in Fig.~\ref{fig:quantosc-boundary} it was chosen following the Monte-Carlo result of Refs.~\cite{chui1995impurity, reddy2026quantum};  $r_{c} =10$ at $B=0$. In the calculation of the dashed curve in Fig.~\ref{fig:quantosc-boundary}, no fitting parameters were used; nevertheless, the agreement with direct measurements of $r_{c}$ is quite good, both in the magnitude and phase of the quantum oscillations of the $r_{s}$. This is one more argument in favor of the approach based on the two-phase state.  The above empirical approach was supported by recent variational Monte-Carlo calculations of the ground state energy of Wigner crystal \cite{reddy2026quantum}. Their results confirm that quantum Hall states penetrate the Wigner crystal region of the $\nu-B$ phase diagram at integer $\nu$ values.

\section*{Conclusions}

The discovery of the metal-insulator transition in a two-dimensional electron system in high-mobility Si MOSFETs gave a powerful impetus to experimental and theoretical research. The subsequent 30 years have seen great progress in the theory of strongly correlated 2D systems, accompanied by experimental observations of numerous novel effects. It is the interactions between carriers that stabilize the metallic state and lead to a significant increase in the spin susceptibility at low densities, with a tendency to diverge, due to the increase in the effective mass on the metallic side near the MIT. We also review indications of band flattening in ultra-clean 2D systems in SiGe/Si/SiGe heterostructures on the metallic side and the formation of a Wigner solid on the insulating side of the MIT. In the last section, modern concepts and the tendency toward phase separation are discussed. The key question is whether metallic conduction and delocalized states persist as the temperature approaches absolute zero and falls below all relevant relaxation rates, including spin relaxation. We conclude that the promising area of strongly correlated 2D electron systems remains open to future research.

\section*{Disclosure statement}

The authors declare that they have no conflict of interest.

\section*{Funding}

This investigation was supported by the RF state task.


\begin{thebibliography}{100}
\providecommand{\url}[1]{\normalfont{#1}}
\providecommand{\urlprefix}{Available from: }

\bibitem{mott1974metal}
Mott~NF. Metal-insulator transitions. London: Taylor \& Francis; 1974.

\bibitem{anderson1958absence}
Anderson~PW. Absence of diffusion in certain random lattices. Phys Rev. 1958;\hspace{0pt}109:1492--1505.

\bibitem{abrahams1979scaling}
Abrahams~E, Anderson~PW, Licciardello~DC, et~al. Scaling theory of localization: Absence of quantum diffusion in two dimensions. Phys Rev Lett. 1979;\hspace{0pt}42:673--676.

\bibitem{wegner1979disordered}
Wegner~FJ. Disordered system with $n$ orbitals per site: $n=\ensuremath{\infty}$ limit. Phys Rev B. 1979 Jan;\hspace{0pt}19:783--792.

\bibitem{gorkov1979particle}
{Gor'kov}~LP, {Larkin}~AI, {Khmel'nitski{\v{i}}}~DE. {Particle conductivity in a two-dimensional random potential}. JETP Lett. 1979 Aug;\hspace{0pt}30:248.

\bibitem{altshuler1979contribution}
Al'tshuler~BL, Aronov~AG. Contribution to the theory of disordered metals in strongly doped semiconductors. Sov Phys JETP. 1979;\hspace{0pt}50:968.

\bibitem{altshuler1980interaction}
Altshuler~BL, Aronov~AG, Lee~PA. Interaction effects in disordered {F}ermi systems in two dimensions. Phys Rev Lett. 1980;\hspace{0pt}44:1288--1291.

\bibitem{lee1985disordered}
Lee~PA, Ramakrishnan~TV. Disordered electronic systems. Rev Mod Phys. 1985 Apr;\hspace{0pt}57:287--337.

\bibitem{finkelstein1983influence}
Finkel'stein~AM. Influence of {C}oulomb interaction on the properties of disordered metals. Sov Phys JETP. 1983;\hspace{0pt}57:97--108.

\bibitem{finkelstein1984weak}
Finkel'stein~AM. Weak localization and {C}oulomb interaction in disordered systems. Z Phys B. 1984;\hspace{0pt}56:189--196.

\bibitem{castellani1984interaction}
Castellani~C, {Di~Castro}~C, Lee~PA, et~al. Interaction-driven metal-insulator transitions in disordered fermion systems. Phys Rev B. 1984;\hspace{0pt}30:527--543.

\bibitem{dolan1979nonmetallic}
Dolan~GJ, Osheroff~DD. Nonmetallic conduction in thin metal films at low temperatures. Phys Rev Lett. 1979;\hspace{0pt}43:721--724.

\bibitem{uren1980the}
Uren~MJ, Davies~RA, Pepper~M. The observation of interaction and localisation effects in a two-dimensional electron gas at low temperatures. Journal of Physics C: Solid State Physics. 1980;\hspace{0pt}13:L985.

\bibitem{uren1981it}
Uren~MJ, Davies~RA, Kaveh~M, et~al. Logarithmic corrections to two-dimensional transport in silicon inversion layers. Journal of Physics C: Solid State Physics. 1981;\hspace{0pt}14:5737.

\bibitem{efros1975coulomb}
Efros~AL, Shklovskii~BI. Coulomb gap and low temperature conductivity of disordered systems. J Phys C: Solid State Phys. 1975 feb;\hspace{0pt}8(4):L49--L51.

\bibitem{bishop1980nonmetallic}
Bishop~DJ, Tsui~DC, Dynes~RC. Nonmetallic conduction in electron inversion layers at low temperatures. Phys Rev Lett. 1980;\hspace{0pt}44:1153--1156.

\bibitem{burdis1988anomalous}
Burdis~MS, Dean~CC. Anomalous values of interaction constants in the two-dimensional electron gas of a silicon metal-oxide-semiconductor field-effect transistor measured by parallel- and perpendicular-field magnetoconductivity. Phys Rev B. 1988 Aug;\hspace{0pt}38:3269--3275.

\bibitem{cham1980temperature}
Cham~KM, Wheeler~RG. Temperature-dependent resistivities in silicon inversion layers at low temperatures. Phys Rev Lett. 1980 Jun;\hspace{0pt}44:1472--1475.

\bibitem{smith1986temperature}
Smith~R, Stiles~P. Temperature dependence of the conductivity of a silicon inversion layer at low temperatures. Solid State Commun. 1986;\hspace{0pt}58(8):511--514.

\bibitem{vyrodov1988temperature}
Vyrodov~EA, Dolgopolov~VT, Dorozhkin~SI, et~al. Temperature dependence of conductivity in high mobility {MIS} structures on a base of (001) silicon. Sov Phys JETP. 1988 May;\hspace{0pt}67:998--1001.

\bibitem{zavaritskaya1987metal}
Zavaritskaya~TN, Zavaritskaya~{\'E~I}. Metal-insulator transition in inversion channels of silicon {MOS} structures. JETP Lett. 1987;\hspace{0pt}45:609--613.

\bibitem{pudalov1993hall}
Pudalov~VM, D'lorio~M, Campbell~JW. Hall resistance and quantized {H}all effect to insulator transitions in a {2D} electron system. JETP Lett. 1993;\hspace{0pt}57:608.

\bibitem{pudalov1994termination}
Pudalov~VM, D'Iorio~M, Campbell~JW. Termination of the quantized {H}all effect by the electron solid. Surf Sci. 1994;\hspace{0pt}305(1):107--114.

\bibitem{kravchenko1995global}
Kravchenko~SV, Mason~W, Furneaux~JE, et~al. Global phase diagram for the quantum {H}all effect: An experimental picture. Phys Rev Lett. 1995 Jul;\hspace{0pt}75:910--913.

\bibitem{shashkin1993floating}
Shashkin~AA, Kravchenko~GV, Dolgopolov~VT. Floating up of the extended states of {L}andau levels in a two-dimensional electron gas in silicon {MOSFET's}. JETP Lett. 1993;\hspace{0pt}58:220.

\bibitem{shashkin1994insulating}
Shashkin~AA, Dolgopolov~VT, Kravchenko~GV. Insulating phases in a two-dimensional electron system of high-mobility {Si MOSFET's}. Phys Rev B. 1994 May;\hspace{0pt}49:14486--14495.

\bibitem{shashkin1994percolation}
Shashkin~AA, Dolgopolov~VT, Kravchenko~GV, et~al. Percolation metal-insulator transitions in the two-dimensional electron system of {AlGaAs/GaAs} heterostructures. Phys Rev Lett. 1994 Dec;\hspace{0pt}73:3141--3144.

\bibitem{khmelnitskii1984quantum}
Khmelnitskii~DE. Quantum {H}all effect and additional oscillations of conductivity in weak magnetic fields. Phys Lett A. 1984;\hspace{0pt}106(4):182--183.

\bibitem{laughlin1984levitation}
Laughlin~RB. Levitation of extended-state bands in a strong magnetic field. Phys Rev Lett. 1984 Jun;\hspace{0pt}52:2304--2304.

\bibitem{kravchenko1994possible}
Kravchenko~SV, Kravchenko~GV, Furneaux~JE, et~al. Possible metal-insulator transition at {$B=0$} in two dimensions. Phys Rev B. 1994;\hspace{0pt}50:8039--8042.

\bibitem{kravchenko1995scaling}
Kravchenko~SV, Mason~WE, Bowker~GE, et~al. Scaling of an anomalous metal-insulator transition in a two-dimensional system in silicon at {$B = 0$}. Phys Rev B. 1995;\hspace{0pt}51:7038--7045.

\bibitem{pudalov1998metal}
Pudalov~V, Brunthaler~G, Prinz~A, et~al. Metal-insulator transition in two dimensions. Physica E: Low-dimensional Systems and Nanostructures. 1998;\hspace{0pt}3(1):79--88.

\bibitem{gold1986temperature}
Gold~A, Dolgopolov~VT. Temperature dependence of the conductivity for the two-dimensional electron gas: Analytical results for low temperatures. Phys Rev B. 1986 Jan;\hspace{0pt}33:1076--1084.

\bibitem{proskuryakov2002hole}
Proskuryakov~YY, Savchenko~AK, Safonov~SS, et~al. Hole-hole interaction effect in the conductance of the two-dimensional hole gas in the ballistic regime. Phys Rev Lett. 2002 Jul;\hspace{0pt}89:076406.

\bibitem{kuntsevich2007intervalley}
Kuntsevich~AY, Klimov~NN, Tarasenko~SA, et~al. Intervalley scattering and weak localization in {S}i-based two-dimensional structures. Phys Rev B. 2007 May;\hspace{0pt}75:195330.

\bibitem{klimov2008interaction}
Klimov~NN, Knyazev~DA, Omel'yanovskii~OE, et~al. Interaction effects in conductivity of a two-valley electron system in high-mobility {S}i inversion layers. Phys Rev B. 2008 Nov;\hspace{0pt}78:195308.

\bibitem{liu1991scaling}
Liu~Y, McGreer~KA, Nease~B, et~al. Scaling of the insulator-to-superconductor transition in ultrathin amorphous {B}i films. Phys Rev Lett. 1991 Oct;\hspace{0pt}67:2068--2071.

\bibitem{sondhi1997continuous}
Sondhi~SL, Girvin~SM, Carini~JP, et~al. Continuous quantum phase transitions. Rev Mod Phys. 1997 Jan;\hspace{0pt}69:315--333.

\bibitem{kravchenko1996electric}
Kravchenko~SV, Simonian~D, Sarachik~MP, et~al. Electric field scaling at a {$B=0$} metal-insulator transition in two dimensions. Phys Rev Lett. 1996;\hspace{0pt}77:4938--4941.

\bibitem{pudalov1997unconventional}
Pudalov~VM. Unconventional metallic state in a two-dimensional system with broken inversion symmetry. JETP Lett. 1997 Aug;\hspace{0pt}66(3):175--181.

\bibitem{shahar1995universal}
Shahar~D, Tsui~DC, Shayegan~M, et~al. Universal conductivity at the quantum {H}all liquid to insulator transition. Phys Rev Lett. 1995 May;\hspace{0pt}74:4511--4514.

\bibitem{yazdani1995superconducting}
Yazdani~A, Kapitulnik~A. Superconducting-insulating transition in two-dimensional $\mathit{a}$-{MoGe} thin films. Phys Rev Lett. 1995 Apr;\hspace{0pt}74:3037--3040.

\bibitem{wei1994current}
Wei~HP, Engel~LW, Tsui~DC. Current scaling in the integer quantum {H}all effect. Phys Rev B. 1994 Nov;\hspace{0pt}50:14609--14612.

\bibitem{popovic1997metal}
Popovi\'c~D, Fowler~AB, Washburn~S. Metal-insulator transition in two dimensions: Effects of disorder and magnetic field. Phys Rev Lett. 1997;\hspace{0pt}79:1543--1546.

\bibitem{coleridge1997metal}
Coleridge~PT, Williams~RL, Feng~Y, et~al. Metal-insulator transition at {$B=0$} in $p$-type {SiGe}. Phys Rev B. 1997 Nov;\hspace{0pt}56:R12764--R12767.

\bibitem{hanein1998the}
Hanein~Y, Meirav~U, Shahar~D, et~al. The metalliclike conductivity of a two-dimensional hole system. Phys Rev Lett. 1998;\hspace{0pt}80:1288--1291.

\bibitem{yoon1999wigner}
Yoon~J, Li~CC, Shahar~D, et~al. {W}igner crystallization and metal-insulator transition of two-dimensional holes in {GaAs} at $\mathit{B}\phantom{\rule{0ex}{0ex}}=\phantom{\rule{0ex}{0ex}}0$. Phys Rev Lett. 1999 Feb;\hspace{0pt}82:1744--1747.

\bibitem{mills1999nonmonotonic}
Mills~AP, Ramirez~AP, Pfeiffer~LN, et~al. Nonmonotonic temperature-dependent resistance in low density {2D} hole gases. Phys Rev Lett. 1999 Oct;\hspace{0pt}83:2805--2808.

\bibitem{hanein1998observation}
Hanein~Y, Shahar~D, Yoon~J, et~al. Observation of the metal-insulator transition in two-dimensional $n$-type {GaAs}. Phys Rev B. 1998 Nov;\hspace{0pt}58:R13338--R13340.

\bibitem{papadakis1998apparent}
Papadakis~SJ, Shayegan~M. Apparent metallic behavior at {$B=0$} of a two-dimensional electron system in {AlAs}. Phys Rev B. 1998 Jun;\hspace{0pt}57:R15068--R15071.

\bibitem{melnikov2019quantum}
Melnikov~MY, Shashkin~AA, Dolgopolov~VT, et~al. Quantum phase transition in ultrahigh mobility {SiGe/Si/SiGe} two-dimensional electron system. Phys Rev B. 2019;\hspace{0pt}99:081106(R).

\bibitem{falson2022competing}
Falson~J, Sodemann~I, Skinner~B, et~al. Competing correlated states around the zero-field {W}igner crystallization transition of electrons in two dimensions. Nat Mater. 2022 Mar;\hspace{0pt}21(3):311--316.

\bibitem{heemskerk1998nonlinear}
Heemskerk~R, Klapwijk~TM. Nonlinear resistivity at the metal-insulator transition in a two-dimensional electron gas. Phys Rev B. 1998 Jul;\hspace{0pt}58:R1754--R1757.

\bibitem{senz1999metal}
Senz~V, D\"otsch~U, Gennser~U, et~al. Metal-insulator transition in a 2-dimensional system with an easy spin axis. Ann Phys (Leipzig). 1999;\hspace{0pt}8:237--240.

\bibitem{simmons1998metal}
Simmons~MY, Hamilton~AR, Pepper~M, et~al. Metal-insulator transition at $\mathit{B}\phantom{\rule{0ex}{0ex}}=\phantom{\rule{0ex}{0ex}}0$ in a dilute two dimensional {GaAs-AlGaAs} hole gas. Phys Rev Lett. 1998 Feb;\hspace{0pt}80:1292--1295.

\bibitem{kravchenko2000metallic}
Kravchenko~SV, Klapwijk~TM. Metallic low-temperature resistivity in a {2D} electron system over an extended temperature range. Phys Rev Lett. 2000;\hspace{0pt}84:2909--2912.

\bibitem{knyazev2008metal}
Knyazev~DA, Omel'yanovskii~OE, Pudalov~VM, et~al. Metal-insulator transition in two dimensions: Experimental test of the two-parameter scaling. Phys Rev Lett. 2008 Feb;\hspace{0pt}100:046405.

\bibitem{rahimi2003coherent}
Rahimi~M, Anissimova~S, Sakr~MR, et~al. Coherent backscattering near the two-dimensional metal-insulator transition. Phys Rev Lett. 2003 Sep;\hspace{0pt}91:116402.

\bibitem{mills2001suppression}
Mills~AP, Ramirez~AP, Gao~XPA, et~al. Suppression of weak localization effects in low-density metallic 2{D} holes ; 2001. \urlprefix\url{https://arxiv.org/abs/cond-mat/0101020}.

\bibitem{brunthaler2001exclusion}
Brunthaler~G, Prinz~A, Bauer~G, et~al. Exclusion of quantum coherence as the origin of the {2D} metallic state in high-mobility silicon inversion layers. Phys Rev Lett. 2001 Aug;\hspace{0pt}87:096802.

\bibitem{minkov2004magnetoresistance}
Minkov~GM, Germanenko~AV, Gornyi~IV. Magnetoresistance and dephasing in a two-dimensional electron gas at intermediate conductances. Phys Rev B. 2004 Dec;\hspace{0pt}70:245423.

\bibitem{shashkin2001metal}
Shashkin~AA, Kravchenko~SV, Klapwijk~TM. Metal-insulator transition in a {2D} electron gas: Equivalence of two approaches for determining the critical point. Phys Rev Lett. 2001;\hspace{0pt}87:266402.

\bibitem{simonian1997reflection}
Simonian~D, Kravchenko~SV, Sarachik~MP. Reflection symmetry at a {$B=0$} metal-insulator transition in two dimensions. Phys Rev B. 1997;\hspace{0pt}55:R13421--R13423.

\bibitem{sarachik1999novel}
Sarachik~MP, Kravchenko~SV. Novel phenomena in dilute electron systems in two dimensions. Proc Natl Acad Sci USA. 1999;\hspace{0pt}96:5900--5902.

\bibitem{diorio1992reentrant}
D'Iorio~M, Pudalov~VM, Semenchinsky~SG. Reentrant insulating phase in {S}i inversion layers in low magnetic fields. Phys Rev B. 1992;\hspace{0pt}46:15992--16004.

\bibitem{adkins1976the}
Adkins~CJ, Pollitt~S, Pepper~M. The {A}nderson transition in silicon inversion layers. J Phys Colloques. 1976;\hspace{0pt}37:C4--343--C4--347.

\bibitem{pudalov1993zero}
Pudalov~VM, D'Iorio~M, Kravchenko~SV, et~al. Zero-magnetic-field collective insulator phase in a dilute 2{D} electron system. Phys Rev Lett. 1993;\hspace{0pt}70:1866--1869.

\bibitem{pudalov1994character}
Pudalov~VM, Chui~ST. Character of the phase transition at the two-dimensional electron-liquid-to-solid boundary. Phys Rev B. 1994 May;\hspace{0pt}49:14062--14065.

\bibitem{dolgopolov1992properties}
Dolgopolov~VT, Kravchenko~GV, Shashkin~AA, et~al. Properties of electron insulating phase in {Si} inversion layers at low-temperatures. JETP Lett. 1992;\hspace{0pt}55:733--737.

\bibitem{simonian1997magnetic}
Simonian~D, Kravchenko~SV, Sarachik~MP, et~al. Magnetic field suppression of the conducting phase in two dimensions. Phys Rev Lett. 1997;\hspace{0pt}79:2304--2307.

\bibitem{pudalov1997instability}
Pudalov~VM, Brunthaler~G, Prinz~A, et~al. Instability of the two-dimensional metallic phase to a parallel magnetic field. JETP Lett. 1997;\hspace{0pt}65:932--937.

\bibitem{okamoto1999spin}
Okamoto~T, Hosoya~K, Kawaji~S, et~al. Spin degree of freedom in a two-dimensional electron liquid. Phys Rev Lett. 1999;\hspace{0pt}82:3875--3878.

\bibitem{vitkalov2000small}
Vitkalov~SA, Zheng~H, Mertes~KM, et~al. Small-angle {S}hubnikov-de~{H}aas measurements in a {2D} electron system: The effect of a strong in-plane magnetic field. Phys Rev Lett. 2000;\hspace{0pt}85:2164--2167.

\bibitem{yoon2000parallel}
Yoon~J, Li~CC, Shahar~D, et~al. Parallel magnetic field induced transition in transport in the dilute two-dimensional hole system in {GaAs}. Phys Rev Lett. 2000 May;\hspace{0pt}84:4421--4424.

\bibitem{papadakis2000anisotropic}
Papadakis~SJ, De~Poortere~EP, Shayegan~M, et~al. Anisotropic magnetoresistance of two-dimensional holes in {GaAs}. Phys Rev Lett. 2000 Jun;\hspace{0pt}84:5592--5595.

\bibitem{dassarma2000parallel}
Das~Sarma~S, Hwang~EH. Parallel magnetic field induced giant magnetoresistance in low density quasi-two-dimensional layers. Phys Rev Lett. 2000 Jun;\hspace{0pt}84:5596--5599.

\bibitem{abrahams2001metallic}
Abrahams~E, Kravchenko~SV, Sarachik~MP. Metallic behavior and related phenomena in two dimensions. Rev Mod Phys. 2001;\hspace{0pt}73:251--266.

\bibitem{dolgopolov2000magnetoresistance}
Dolgopolov~VT, Gold~A. Magnetoresistance of a two-dimensional electron gas in a parallel magnetic field. JETP Lett. 2000;\hspace{0pt}71:27--30.

\bibitem{dolgopolov2017spin}
Dolgopolov~VT, Shashkin~AA, Kravchenko~SV. Spin polarization and exchange-correlation effects in transport properties of two-dimensional electron systems in silicon. Phys Rev B. 2017;\hspace{0pt}96:075307.

\bibitem{punnoose2001dilute}
Punnoose~A, Finkel'stein~AM. Dilute electron gas near the metal-insulator transition: Role of valleys in silicon inversion layers. Phys Rev Lett. 2001;\hspace{0pt}88:016802.

\bibitem{punnoose2005metal}
Punnoose~A, Finkel'stein~AM. Metal-insulator transition in disordered two-dimensional electron systems. Science. 2005;\hspace{0pt}310:289--291.

\bibitem{fleury2008many}
Fleury~G, Waintal~X. Many-body localization study in low-density electron gases: Do metals exist in two dimensions? Phys Rev Lett. 2008 Nov;\hspace{0pt}101:226803.

\bibitem{melnikov2020metallic}
Melnikov~MY, Shashkin~AA, Dolgopolov~VT, et~al. Metallic state in a strongly interacting spinless two-valley electron system in two dimensions. Phys Rev B. 2020 Jan;\hspace{0pt}101:045302.

\bibitem{diorio1990magnetic}
D'Iorio~M, Pudalov~VM, Semenchinsky~SG. Magnetic field induced transitions between quantized {H}all and insulator states in a dilute 2{D} electron gas. Phys Lett A. 1990;\hspace{0pt}150:422--426.

\bibitem{viehweger1990the}
Viehweger~O, Efetov~KB. The {H}all coefficient of disordered electronic systems in high magnetic fields. J Phys: Condens Matter. 1990 aug;\hspace{0pt}2(33):7049.

\bibitem{kivelson1992global}
Kivelson~S, Lee~DH, Zhang~SC. Global phase diagram in the quantum {H}all effect. Phys Rev B. 1992 Jul;\hspace{0pt}46:2223--2238.

\bibitem{pudalov1999weak}
Pudalov~VM, Brunthaler~G, Prinz~A, et~al. Weak-field {H}all resistance and effective carrier density measurements across the metal-insulator transition in {Si-MOS} structures. JETP Lett. 1999 Jul;\hspace{0pt}70(1):48--53.

\bibitem{qiu2012connecting}
Qiu~RLJ, Gao~XPA, Pfeiffer~LN, et~al. Connecting the reentrant insulating phase and the zero-field metal-insulator transition in a 2{D} hole system. Phys Rev Lett. 2012;\hspace{0pt}108:106404.

\bibitem{dolgopolov1992metal}
Dolgopolov~VT, Kravchenko~GV, Shashkin~AA, et~al. Metal-insulator transition in {Si} inversion layers in the extreme quantum limit. Phys Rev B. 1992;\hspace{0pt}46:13303--13308.

\bibitem{jiang1993observation}
Jiang~HW, Johnson~CE, Wang~KL, et~al. Observation of magnetic-field-induced delocalization: Transition from {A}nderson insulator to quantum {H}all conductor. Phys Rev Lett. 1993 Aug;\hspace{0pt}71:1439--1442.

\bibitem{dolgopolov2018fractional}
Dolgopolov~VT, Melnikov~MY, Shashkin~AA, et~al. Fractional quantum {H}all effect in {SiGe/Si/SiGe} quantum wells in weak quantizing magnetic fields. JETP Lett. 2018;\hspace{0pt}107:794--797.

\bibitem{santos1992observation}
Santos~MB, Suen~YW, Shayegan~M, et~al. Observation of a reentrant insulating phase near the 1/3 fractional quantum {H}all liquid in a two-dimensional hole system. Phys Rev Lett. 1992 Feb;\hspace{0pt}68:1188--1191.

\bibitem{kravchenko2000shubnikov}
Kravchenko~SV, Shashkin~AA, Bloore~DA, et~al. Shubnikov-de~{Haas} oscillations near the metal-insulator transition in a two-dimensional electron system in silicon. Solid State Commun. 2000;\hspace{0pt}116:495--499.

\bibitem{shashkin2001indication}
Shashkin~AA, Kravchenko~SV, Dolgopolov~VT, et~al. Indication of the ferromagnetic instability in a dilute two-dimensional electron system. Phys Rev Lett. 2001;\hspace{0pt}87:086801.

\bibitem{simonian1998h}
Simonian~D, Kravchenko~SV, Sarachik~MP, et~al. {$H/T$} scaling of the magnetoconductance near the conductor-insulator transition in two dimensions. Phys Rev B. 1998 Apr;\hspace{0pt}57:R9420--R9422.

\bibitem{gold2002on}
Gold~A, Dolgopolov~VT. On the role of disorder in transport and magnetic properties of the two-dimensional electron gas. J Phys Condens Matter. 2002 jul;\hspace{0pt}14:7091--7096.

\bibitem{pudalov2002weak}
Pudalov~VM, Brunthaler~G, Prinz~A, et~al. Weak anisotropy and disorder dependence of the in-plane magnetoresistance in high-mobility (100) {Si}-inversion layers. Phys Rev Lett. 2002 Jan;\hspace{0pt}88:076401.

\bibitem{dolgopolov2002comment}
Dolgopolov~VT, Gold~A. Comment on ``{W}eak anisotropy and disorder dependence of the in-plane magnetoresistance in high-mobility (100) {S}i-inversion layers''. Phys Rev Lett. 2002 Aug;\hspace{0pt}89:129701.

\bibitem{kravchenko2004metal}
Kravchenko~SV, Sarachik~MP. Metal-insulator transition in two-dimensional electron systems. Rep Prog Phys. 2004;\hspace{0pt}67:1--44.

\bibitem{shashkin2005metal}
Shashkin~AA. Metal-insulator transitions and the effects of electron-electron interactions in two-dimensional electron systems. Phys Usp. 2005 feb;\hspace{0pt}48:129--149.

\bibitem{prus2003thermodynamic}
Prus~O, Yaish~Y, Reznikov~M, et~al. Thermodynamic spin magnetization of strongly correlated two-dimensional electrons in a silicon inversion layer. Phys Rev B. 2003 May;\hspace{0pt}67:205407.

\bibitem{pudalov2018probing}
Pudalov~VM, Kuntsevich~AY, Gershenson~ME, et~al. Probing spin susceptibility of a correlated two-dimensional electron system by transport and magnetization measurements. Phys Rev B. 2018 Oct;\hspace{0pt}98:155109.

\bibitem{pudalov2021magnetic}
Pudalov~V, Gershenson~M. Magnetic field driven redistribution between extended and localized electronic states in high-mobility {Si MOSFETs} at low temperatures. Physical Review B. 2021 Jul;\hspace{0pt}104.

\bibitem{pudalov2002low}
Pudalov~VM, Gershenson~ME, Kojima~H, et~al. Low-density spin susceptibility and effective mass of mobile electrons in {S}i inversion layers. Phys Rev Lett. 2002 Apr;\hspace{0pt}88:196404.

\bibitem{shashkin2002sharp}
Shashkin~AA, Kravchenko~SV, Dolgopolov~VT, et~al. Sharp increase of the effective mass near the critical density in a metallic two-dimensional electron system. Phys Rev B. 2002;\hspace{0pt}66:073303.

\bibitem{zala2001interaction}
Zala~G, Narozhny~BN, Aleiner~IL. Interaction corrections at intermediate temperatures: {L}ongitudinal conductivity and kinetic equation. Phys Rev B. 2001 Nov;\hspace{0pt}64:214204.

\bibitem{ando1982electronic}
Ando~T, Fowler~AB, Stern~F. Electronic properties of two-dimensional systems. Rev Mod Phys. 1982 Apr;\hspace{0pt}54:437--672.

\bibitem{iwamoto1991static}
Iwamoto~N. Static local-field corrections of two-dimensional electron liquids. Phys Rev B. 1991 Jan;\hspace{0pt}43:2174--2182.

\bibitem{kwon1994quantum}
Kwon~Y, Ceperley~DM, Martin~RM. Quantum {M}onte {C}arlo calculation of the {F}ermi-liquid parameters in the two-dimensional electron gas. Phys Rev B. 1994 Jul;\hspace{0pt}50:1684--1694.

\bibitem{chen1999exchange}
Chen~GH, Raikh~ME. Exchange-induced enhancement of spin-orbit coupling in two-dimensional electronic systems. Phys Rev B. 1999 Aug;\hspace{0pt}60:4826--4833.

\bibitem{shashkin2003spin}
Shashkin~AA, Rahimi~M, Anissimova~S, et~al. Spin-independent origin of the strongly enhanced effective mass in a dilute {2D} electron system. Phys Rev Lett. 2003;\hspace{0pt}91:046403.

\bibitem{pudalov2014probing}
Pudalov~VM, Gershenson~ME, Kojima~H. Probing electron interactions in a two-dimensional system by quantum magneto-oscillations. Phys Rev B. 2014 Aug;\hspace{0pt}90:075147.

\bibitem{melnikov2023spin}
Melnikov~MY, Shakirov~AA, Shashkin~AA, et~al. Spin independence of the strongly enhanced effective mass in ultra-clean {SiGe/Si/SiGe} two-dimensional electron system. Sci Rep. 2023 Oct;\hspace{0pt}13(1):17364.

\bibitem{shashkin2006pauli}
Shashkin~AA, Anissimova~S, Sakr~MR, et~al. Pauli spin susceptibility of a strongly correlated two-dimensional electron liquid. Phys Rev Lett. 2006;\hspace{0pt}96:036403.

\bibitem{anissimova2006magnetization}
Anissimova~S, Venkatesan~A, Shashkin~AA, et~al. Magnetization of a strongly interacting two-dimensional electron system in perpendicular magnetic fields. Phys Rev Lett. 2006;\hspace{0pt}96:046409.

\bibitem{reznikov2010thermodynamic}
Reznikov~M, Kuntsevich~AY, Teneh~N, et~al. Thermodynamic magnetization of two-dimensional electron gas measured over wide range of densities. JETP Letters. 2010 Oct;\hspace{0pt}92(7):470--474.

\bibitem{bhatt1986magnetic}
Bhatt~RN. Magnetic properties of doped semiconductors. Phys Scr. 1986 jan;\hspace{0pt}1986(T14):7.

\bibitem{teneh2012spin}
Teneh~N, Kuntsevich~AY, Pudalov~VM, et~al. Spin-droplet state of an interacting {2D} electron system. Phys Rev Lett. 2012 Nov;\hspace{0pt}109:226403.

\bibitem{pudalov2022phase}
Pudalov~VM. Phase separation in two-dimensional electron systems: Experimental view (brief review). JETP Lett. 2022 Oct;\hspace{0pt}116(7):462--476.

\bibitem{stadnik2013interacting}
Stadnik~YV, Sushkov~OP. Interacting spin droplets and magnetic properties of a low-density two-dimensional electron gas. Phys Rev B. 2013 Sep;\hspace{0pt}88:125402.

\bibitem{spivak2004phases}
Spivak~B, Kivelson~SA. Phases intermediate between a two-dimensional electron liquid and {Wigner} crystal. Phys Rev B. 2004;\hspace{0pt}70:155114.

\bibitem{clarke2008impact}
Clarke~WR, Yasin~CE, Hamilton~AR, et~al. Impact of long- and short-range disorder on the metallic behaviour of two-dimensional systems. Nat Phys. 2008 Jan;\hspace{0pt}4(1):55--59.

\bibitem{depalo2005effects}
De~Palo~S, Botti~M, Moroni~S, et~al. Effects of thickness on the spin susceptibility of the two dimensional electron gas. Phys Rev Lett. 2005 Jun;\hspace{0pt}94:226405.

\bibitem{chui1995impurity}
Chui~ST, Tanatar~B. Impurity effect on the two-dimensional-electron fluid-solid transition in zero field. Phys Rev Lett. 1995;\hspace{0pt}74:458--461.

\bibitem{eguiluz1983two}
Eguiluz~AG, Maradudin~AA, Elliott~RJ. Two-dimensional {W}igner lattice in a magnetic field and in the presence of a random array of pinning centers. Phys Rev B. 1983 Apr;\hspace{0pt}27:4933--4945.

\bibitem{benenti2001spin}
Benenti~G, Caldara~G, Shepelyansky~DL. Spin-polarized ground state for interacting electrons in two dimensions. Phys Rev Lett. 2001 Jun;\hspace{0pt}86:5333--5336.

\bibitem{andreev1998itinerant}
Andreev~AV, Kamenev~A. Itinerant ferromagnetism in disordered metals: A mean-field theory. Phys Rev Lett. 1998 Oct;\hspace{0pt}81:3199--3202.

\bibitem{eisenberg1999disorder}
Eisenberg~E, Berkovits~R. Disorder-induced spin polarization in restricted geometries. Phys Rev B. 1999 Dec;\hspace{0pt}60:15261--15265.

\bibitem{reiman2000formation}
Reimann~SM, Koskinen~M, Manninen~M. Formation of {W}igner molecules in small quantum dots. Phys Rev B. 2000 Sep;\hspace{0pt}62:8108--8113.

\bibitem{zhang2005density}
Zhang~Y, Das~Sarma~S. Density-dependent spin susceptibility and effective mass in interacting quasi-two-dimensional electron systems. Phys Rev B. 2005 Aug;\hspace{0pt}72:075308.

\bibitem{dolgopolov2011thermoelectric}
Dolgopolov~VT, Gold~A. Thermoelectric properties of the interacting two dimensional electron gas in the diffusion regime. JETP Lett. 2011 Nov;\hspace{0pt}94(6):446--450.

\bibitem{gold2011temperature}
Gold~A, Dolgopolov~VT. Temperature dependence of the diffusive thermopower in the two-dimensional interacting electron gas. Europhys Lett. 2011 sep;\hspace{0pt}96(2):27007.

\bibitem{fletcher1997thermoelectric}
Fletcher~R, Pudalov~VM, Feng~Y, et~al. Thermoelectric and hot-electron properties of a silicon inversion layer. Phys Rev B. 1997 Nov;\hspace{0pt}56:12422--12428.

\bibitem{faniel2007in}
Faniel~S, Moldovan~L, Vlad~A, et~al. In-plane magnetic-field-induced metal-insulator transition in {$(311)A$ $\mathrm{GaAs}$} two-dimensional hole systems probed by thermopower. Phys Rev B. 2007 Oct;\hspace{0pt}76:161307.

\bibitem{goswami2009highly}
Goswami~S, Siegert~C, Baenninger~M, et~al. Highly enhanced thermopower in two-dimensional electron systems at millikelvin temperatures. Phys Rev Lett. 2009 Jul;\hspace{0pt}103:026602.

\bibitem{fletcher2001critical}
Fletcher~R, Pudalov~VM, Radcliffe~ADB, et~al. Critical behaviour of thermopower and conductivity at the metal-insulator transition in high-mobility {Si-MOSFETs}. Semicond Sci Technol. 2001 may;\hspace{0pt}16(5):386.

\bibitem{mokashi2012critical}
Mokashi~A, Li~S, Wen~B, et~al. Critical behavior of a strongly interacting {2D} electron system. Phys Rev Lett. 2012;\hspace{0pt}109:096405.

\bibitem{pudalov2020experimental}
Pudalov~VM, Gershenson~ME. Experimental evidence for an inhomogeneous state of the correlated two-dimensional electron system in the vicinity of a metal--insulator transition. JETP Lett. 2020 Feb;\hspace{0pt}111(4):225--229.

\bibitem{dolgopolov2002on}
Dolgopolov~VT. On effective electron mass of silicon field structures at low electron densities. JETP Lett. 2002 Sep;\hspace{0pt}76(6):377--379.

\bibitem{spivak2003phase}
Spivak~B. Phase separation in the two-dimensional electron liquid in {MOSFET}'s. Phys Rev B. 2003 Mar;\hspace{0pt}67:125205.

\bibitem{khodel2008topology}
Khodel~VA, Clark~JW, Zverev~MV. Topology of the fermi surface beyond the quantum critical point. Phys Rev B. 2008 Aug;\hspace{0pt}78:075120.

\bibitem{pankov2008self}
Pankov~S, Dobrosavljevi\ifmmode~\acute{c}\else \'{c}\fi{}~V. Self-doping instability of the {Wigner-Mott} insulator. Phys Rev B. 2008 Feb;\hspace{0pt}77:085104.

\bibitem{marchi2009correlation}
Marchi~M, De~Palo~S, Moroni~S, et~al. Correlation energy and spin susceptibility of a two-valley two-dimensional electron gas. Phys Rev B. 2009 Jul;\hspace{0pt}80:035103.

\bibitem{fleury2010energy}
Fleury~G, Waintal~X. Energy scale behind the metallic behaviors in low-density {Si MOSFETs}. Phys Rev B. 2010 Apr;\hspace{0pt}81:165117.

\bibitem{kuntsevich2015strongly}
Kuntsevich~AY, Tupikov~YV, Pudalov~VM, et~al. Strongly correlated two-dimensional plasma explored from entropy measurements. Nat Commun. 2015 Jun;\hspace{0pt}6:7298.

\bibitem{pudalov2006metal}
Pudalov~VM. Metal-insulator transitions and related phenomena in a strongly correlated two-dimensional electron system. Phys Usp. 2006;\hspace{0pt}49(2):203--208.

\bibitem{melnikov2017indication}
Melnikov~MY, Shashkin~AA, Dolgopolov~VT, et~al. Indication of band flattening at the {F}ermi level in a strongly correlated electron system. Sci Rep. 2017;\hspace{0pt}7:14539.

\bibitem{khodel1990superfluidity}
Khodel~VA, Shaginyan~VR. Superfluidity in system with fermion condensate. JETP Lett. 1990;\hspace{0pt}51:553--555.

\bibitem{nozieres1992properties}
Nozi\`eres~P. Properties of {F}ermi liquids with a finite range interaction. J Phys I (France). 1992;\hspace{0pt}2:443--458.

\bibitem{zverev2012microscopic}
Zverev~MV, Khodel~VA, Pankratov~SS. Microscopic theory of a strongly correlated two-dimensional electron gas. JETP Lett. 2012;\hspace{0pt}96:192--202.

\bibitem{wigner1934on}
Wigner~E. On the interaction of electrons in metals. Phys Rev. 1934;\hspace{0pt}46:1002--1011.

\bibitem{chaplik1972possible}
Chaplik~AV. Possible crystallization of charge carriers in low-density inversion layers. Sov Phys JETP. 1972;\hspace{0pt}35:395--398.

\bibitem{tanatar1989ground}
Tanatar~B, Ceperley~DM. Ground state of the two-dimensional electron gas. Phys Rev B. 1989;\hspace{0pt}39:5005--5016.

\bibitem{attaccalite2002correlation}
Attaccalite~C, Moroni~S, Gori-Giorgi~P, et~al. Correlation energy and spin polarization in the 2{D} electron gas. Phys Rev Lett. 2002;\hspace{0pt}88:256601.

\bibitem{kagalovsky2020hartree}
Kagalovsky~V, Kravchenko~SV, Nemirovsky~D. Hartree-{F}ock description of a {W}igner crystal in two dimensions. Physica E. 2020;\hspace{0pt}119:114016.

\bibitem{thakur1996frozen}
Thakur~JS, Neilson~D. Frozen electron solid in the presence of small concentrations of defects. Phys Rev B. 1996 Sep;\hspace{0pt}54:7674--7677.

\bibitem{andrei1988observation}
Andrei~EY, Deville~G, Glattli~DC, et~al. Observation of a magnetically induced {W}igner solid. Phys Rev Lett. 1988;\hspace{0pt}60:2765--2768.

\bibitem{williams1991conduction}
Williams~FIB, Wright~PA, Clark~RG, et~al. Conduction threshold and pinning frequency of magnetically induced {W}igner solid. Phys Rev Lett. 1991;\hspace{0pt}66:3285--3288.

\bibitem{goldman1990evidence}
Goldman~VJ, Santos~M, Shayegan~M, et~al. Evidence for two-dimensional quantum {W}igner crystal. Phys Rev Lett. 1990;\hspace{0pt}65:2189--2192.

\bibitem{jiang1990quantum}
Jiang~HW, Willett~RL, Stormer~HL, et~al. Quantum liquid versus electron solid around \ensuremath{\nu}=1/5 {L}andau-level filling. Phys Rev Lett. 1990 Jul;\hspace{0pt}65:633--636.

\bibitem{jiang1991magnetotransport}
Jiang~HW, Stormer~HL, Tsui~DC, et~al. Magnetotransport studies of the insulating phase around $\nu = 1/5$ {L}andau-level filling. Phys Rev B. 1991;\hspace{0pt}44:8107--8114.

\bibitem{giamarchi2003electronic}
Giamarchi~T. Electronic glasses. In: Quantum phenomena in mesoscopic systems. IOS Press; 2003. p. 303--339.

\bibitem{knighton2014reentrant}
Knighton~T, Wu~Z, Tarquini~V, et~al. Reentrant insulating phases in the integer quantum {H}all regime. Phys Rev B. 2014 Oct;\hspace{0pt}90:165117.

\bibitem{qiu2018new}
Qiu~R, Liu~CW, Liu~S, et~al. New reentrant insulating phases in strongly interacting {2D} systems with low disorder. Appl Sci. 2018;\hspace{0pt}8:1909.

\bibitem{knighton2018evidence}
Knighton~T, Wu~Z, Huang~J, et~al. Evidence of two-stage melting of {W}igner solids. Phys Rev B. 2018;\hspace{0pt}97:085135.

\bibitem{hossain2022anisotropic}
Hossain~MS, Ma~MK, Villegas-Rosales~KA, et~al. Anisotropic two-dimensional disordered {W}igner solid. Phys Rev Lett. 2022 Jul;\hspace{0pt}129:036601.

\bibitem{madathil2023moving}
Madathil~PT, Rosales~KAV, Chung~YJ, et~al. Moving crystal phases of a quantum {W}igner solid in an ultra-high-quality {2D} electron system. Phys Rev Lett. 2023 Dec;\hspace{0pt}131:236501.

\bibitem{marianer1992effective}
Marianer~S, Shklovskii~BI. Effective temperature of hopping electrons in a strong electric field. Phys Rev B. 1992;\hspace{0pt}46:13100--13103.

\bibitem{brussarski2018transport}
Brussarski~P, Li~S, Kravchenko~SV, et~al. Transport evidence for a sliding two-dimensional quantum electron solid. Nat Commun. 2018;\hspace{0pt}9:3803.

\bibitem{yeh1991flux}
Yeh~WJ, Kao~YH. Flux-flow noise in type-{II} superconductors. Phys Rev B. 1991 Jul;\hspace{0pt}44:360--373.

\bibitem{blatter1994vortices}
Blatter~G, Feigel'man~MY, Geshkenbein~YB, et~al. Vortices in high-temperature superconductors. Rev Mod Phys. 1994;\hspace{0pt}66:1125--1388.

\bibitem{bullard2008vortex}
Bullard~TJ, Das~J, Daquila~GL, et~al. Vortex washboard voltage noise in type-{II} superconductors. Eur Phys J B. 2008 Oct;\hspace{0pt}65(4):469.

\bibitem{melnikov2024collective}
Melnikov~MY, Shashkin~AA, Huang~SH, et~al. Collective depinning and sliding of a quantum {W}igner solid in a two-dimensional electron system. Phys Rev B. 2024 Jan;\hspace{0pt}109:L041114.

\bibitem{stern1980calculated}
Stern~F. Calculated temperature dependence of mobility in silicon inversion layers. Phys Rev Lett. 1980 Jun;\hspace{0pt}44:1469--1472.

\bibitem{dassarma1986theory}
Das~Sarma~S. Theory of finite-temperature screening in a disordered two-dimensional electron gas. Phys Rev B. 1986 Apr;\hspace{0pt}33:5401--5405.

\bibitem{zala2001interaction_a}
Zala~G, Narozhny~BN, Aleiner~IL. Interaction corrections at intermediate temperatures: {M}agnetoresistance in a parallel field. Phys Rev B. 2001 Dec;\hspace{0pt}65:020201.

\bibitem{shashkin2004comment}
Shashkin~AA, Dolgopolov~VT, Kravchenko~SV. Comment on ``{I}nteraction effects in conductivity of {S}i inversion layers at intermediate temperatures''. Phys Rev Lett. 2004 Dec;\hspace{0pt}93:269705.

\bibitem{gornyi2004interaction}
Gornyi~IV, Mirlin~AD. Interaction-induced magnetoresistance in a two-dimensional electron gas. Phys Rev B. 2004 Jan;\hspace{0pt}69:045313.

\bibitem{anissimova2007flow}
Anissimova~S, Kravchenko~SV, Punnoose~A, et~al. Flow diagram of the metal-insulator transition in two dimensions. Nat Phys. 2007;\hspace{0pt}3:707--710.

\bibitem{lee1982magnetoresistance}
Lee~PA, Ramakrishnan~TV. Magnetoresistance of weakly disordered electrons. Phys Rev B. 1982 Oct;\hspace{0pt}26:4009--4012.

\bibitem{morgun2016novel}
Morgun~LA, Kuntsevich~AY, Pudalov~VM. Novel energy scale in the interacting two-dimensional electron system evidenced from transport and thermodynamic measurements. Phys Rev B. 2016 Jun;\hspace{0pt}93:235145.

\bibitem{castellani1998metallic}
Castellani~C, {Di~Castro}~C, Lee~PA. Metallic phase and metal-insulator transition in two-dimensional electronic systems. Phys Rev B. 1998;\hspace{0pt}57:R9381--R9384.

\bibitem{kravchenko2006thermodynamic}
Kravchenko~SV, Shashkin~AA, Anissimova~S, et~al. Thermodynamic magnetization of a strongly correlated two-dimensional electron system. Ann Physics. 2006;\hspace{0pt}321(7):1588--1601. July 2006 Special Issue.

\bibitem{camjayi2008coulomb}
Camjayi~A, Haule~K, Dobrosavljevi\'c~V, et~al. Coulomb correlations and the {W}igner-{M}ott transition. Nat Phys. 2008;\hspace{0pt}4:932--935.

\bibitem{radonjic2012wigner}
Radonji\ifmmode~\acute{c}\else \'{c}\fi{}~MM, Tanaskovi\ifmmode~\acute{c}\else \'{c}\fi{}~D, Dobrosavljevi\ifmmode~\acute{c}\else \'{c}\fi{}~V, et~al. {W}igner-{Mott} scaling of transport near the two-dimensional metal-insulator transition. Phys Rev B. 2012 Feb;\hspace{0pt}85:085133.

\bibitem{dobrosavljevic2017wigner}
Dobrosavljevi\'c~V, Tanaskovi\'c~D. Wigner-{Mott} quantum criticality: From {2D-MIT} to $^3${He} and {Mott} organics. In: Kravchenko~SV, editor. Strongly correlated electrons in two dimensions. Chapter~1. Pan Stanford Publishing; 2017. p. 1--46.

\bibitem{shashkin2020manifestation}
Shashkin~AA, Melnikov~MY, Dolgopolov~VT, et~al. Manifestation of strong correlations in transport in ultraclean {S}i{G}e/{S}i/{S}i{G}e quantum wells. Phys Rev B. 2020 Aug;\hspace{0pt}102:081119(R).

\bibitem{spivak2006transport}
Spivak~B, Kivelson~SA. Transport in two dimensional electronic micro-emulsions. Ann Phys. 2006;\hspace{0pt}321:2071--2115.

\bibitem{kagalovsky2024quantum}
Kagalovsky~V, Kravchenko~SV, Nemirovsky~D. Quantum scaling for the metal--insulator transition in a two-dimensional electron system. Sci Rep. 2024 Jun;\hspace{0pt}14(1):12584.

\bibitem{mason1995experimental}
Mason~W, Kravchenko~SV, Bowker~GE, et~al. Experimental evidence for a {C}oulomb gap in two dimensions. Phys Rev B. 1995;\hspace{0pt}52:7857--7859.

\bibitem{narozhny2000magnetic}
Narozhny~BN, Aleiner~IL, Larkin~AI. Magnetic fluctuations in two-dimensional metals close to the {Stoner} instability. Phys Rev B. 2000 Dec;\hspace{0pt}62:14898--14911.

\bibitem{sloggett2005electron}
Sloggett~C, Sushkov~OP. Electron correlations in two-dimensional small quantum dots. Phys Rev B. 2005 Jun;\hspace{0pt}71:235326.

\bibitem{dolgopolov2022band}
Dolgopolov~VT, Melnikov~MY, Shashkin~AA, et~al. Band flattening and {L}andau level merging in strongly-correlated two-dimensional electron systems. JETP Lett. 2022 Aug;\hspace{0pt}116(3):156--166.

\bibitem{jamei2005universal}
Jamei~R, Kivelson~S, Spivak~B. Universal aspects of {C}oulomb-frustrated phase separation. Phys Rev Lett. 2005 Feb;\hspace{0pt}94:056805.

\bibitem{ortix2006frustrated}
Ortix~C, Lorenzana~J, Di~Castro~C. Frustrated phase separation in two-dimensional charged systems. Phys Rev B. 2006 Jun;\hspace{0pt}73:245117.

\bibitem{ortix2007screening}
Ortix~C, Lorenzana~J, Beccaria~M, et~al. Screening effects in {C}oulomb-frustrated phase separation. Phys Rev B. 2007 May;\hspace{0pt}75:195107.

\bibitem{ortix2009universality}
Ortix~C, Lorenzana~J, Di~Castro~C. Universality classes for {Coulomb} frustrated phase separation. Physica B: Condensed Matter. 2009 03;\hspace{0pt}404:499--502.

\bibitem{hormann2008pronounced}
H\"ormann~T, Brunthaler~G. Pronounced metal-insulator transition according to dipole trap model for two-dimensional {Si-MOS} structures. Physica E: Low-dimensional Systems and Nanostructures. 2008;\hspace{0pt}40(5):1235--1237.

\bibitem{zhang2014correlation}
Zhang~JY, Jackson~CA, Chen~R, et~al. Correlation between metal-insulator transitions and structural distortions in high-electron-density srtio${}_{3}$ quantum wells. Phys Rev B. 2014 Feb;\hspace{0pt}89:075140.

\bibitem{raghavan2015two}
Raghavan~S, Zhang~JY, Stemmer~S. Two-dimensional electron liquid at the (111) {SmTiO3/SrTiO3} interface. Appl Phys Lett. 2015;\hspace{0pt}106:132104.

\bibitem{altshuler2001metal}
Altshuler~BL, Maslov~DL, Pudalov~VM. Metal-insulator transition in 2{D}: resistance in the critical region. Physica E. 2001;\hspace{0pt}9:209--225.

\bibitem{pudalov2002stiff}
{VM Pudalov}. Stiff liquid and solid of electrons in two dimensions. J Phys IV France. 2002;\hspace{0pt}12(9):331--336.

\bibitem{lifshits1955on}
Lifshits~IM, Kosevich~AM. On the theory of magnetic susceptibility of metals at low temperatures. Zh Eksp Teor Fiz. 1955;\hspace{0pt}29:730--742.

\bibitem{isihara1986density}
Isihara~A, Smrcka~L. Density and magnetic field dependences of the conductivity of two-dimensional electron systems. J Phys C: Solid State Phys. 1986 dec;\hspace{0pt}19(34):6777.

\bibitem{sandeep2023upper}
Joy~S, Skinner~B. Upper bound on the window of density occupied by microemulsion phases in two-dimensional electron systems. Phys Rev B. 2023 Dec;\hspace{0pt}108:L241110.

\bibitem{sandeep2026disorder}
Joy~S, Skinner~B. Disorder-induced liquid-solid phase coexistence in two-dimensional electron systems. Phys Rev B. 2026 May;\hspace{0pt}113:L201117.

\bibitem{pudalov1985valley}
Pudalov~VM, Semenchinskii~SG, Edel'man~VS. Valley splitting in the electron spectrum of a {Si} inversion layer. JETP Lett. 1985;\hspace{0pt}41:325--328.

\bibitem{pudalov1985oscillations}
Pudalov~VM, Semenchinskii~SG, Edel'man~VS. Oscillations of the chemical potential and the energy spectrum of electrons in the inversion layer at a silicon surface in a magnetic field. Sov Phys JETP. 1985;\hspace{0pt}62:1079--1086.

\bibitem{pudalov1986quantum}
Pudalov~VM, Semenchinskii~SG. Quantum oscillations of the density and {F}ermi energy of electrons at an inversion layer in a magnetic field. JETP Lett. 1986 12;\hspace{0pt}44:677--681.

\bibitem{pudalov1994metal}
Pudalov~VM. {2D} electron solid in low and zero magnetic field. In: Chui~ST, editor. Physics of quantum solid of electrons. Chapter~4. Intern. Press; 1994. p. 124--178.

\bibitem{pudalov1994electron}
Pudalov~V, D'Iorio~M, Campbell~J. Electron solid at zero magnetic field in a dilute {2D} electron system in {Si}. Physica B: Condens Matter. 1994;\hspace{0pt}194-196:1289--1290.

\bibitem{reddy2026quantum}
Reddy~AP, Fu~L. Quantum melting a {W}igner crystal into {H}all liquids. Phys Rev B. 2026 Apr;\hspace{0pt}113:L161403.

\end{thebibliography}

\end{document}